\documentclass{emulateapj}

\slugcomment{Not to appear in Nonlearned J., 45.}

\shorttitle{Spitzer spectra of Seyfert galaxies}
\shortauthors{Tommasin et al.}

\usepackage{graphicx}
\usepackage{lscape}

\begin{document}

\DeclareGraphicsExtensions{.pdf,.gif,.jpg}


\title{Spitzer-IRS high resolution spectroscopic survey of the 12$\mu$m Seyfert galaxies: \\
  II. Results for the Complete Dataset}


\author{Silvia Tommasin\altaffilmark{1}, Luigi Spinoglio}
\affil{Istituto di Fisica dello Spazio Interplanetario, INAF, Via Fosso del Cavaliere 100, I-00133 Roma, Italy}

\author{Matthew A. Malkan}
\affil{Astronomy Division, University of California, Los Angeles, CA 90095-1547, USA}
\and
\author{Giovanni Fazio}
\affil{Harvard-Smithsonian Center for Astrophysics, 60 Garden Street, Cambridge, MA 02138}


\altaffiltext{1}{also at the Physics Department of Universit\'a di Roma, La Sapienza, Roma, Italy}

\clearpage

\begin{abstract}

We present our {\it Spitzer} IRS  spectroscopic survey from 10$\mu m$ to 37$\mu m$
of the Seyfert galaxies of the 12$\mu m$ Galaxy Sample, collected in high resolution mode (R$\sim$ 600). 
The new spectra of 61 galaxies, together with the data we already published, gives us a
total of 91 12$\mu m$ Seyfert galaxies observed, out of 112.
We discuss the mid-IR emission lines and features of the Seyfert galaxies, using an improved AGN
classification scheme: instead of adopting the usual classes of Seyfert 1's and Seyfert 2's, we use
the spectropolarimetric data from the literature to divide the objects into categories  "AGN 1" and "AGN 2", 
where AGN 1's include all broad-line objects, including the Seyfert 2's showing hidden broad lines in polarized light.
The remaining category, AGN 2's contains only Seyferts with no detectable broad lines in
either direct or polarized spectroscopy.
We present various mid-IR observables, such as ionization-sensitive and density-sensitive line ratios,
the PAH 11.25$\mu$m feature and the H$_2$ S(1) rotational line equivalent widths, 
the (60$\mu m$ - 25$\mu m$) spectral index and the source extendedness at 19$\mu m$, to
characterize similarities and differences in the AGN populations, in terms of AGN dominance versus
star formation dominance.

We find that the mid-IR emission properties characterize all the AGN 1's objects as a single family, with
strongly AGN-dominated spectra. In contrast,
the AGN 2's
can be divided in two groups, the first one with properties similar to
the AGN 1's except without detected broad lines, and the 
second with properties similar to the non-Seyfert galaxies, such as LINERs or starburst
galaxies.

We computed a semianalytical model to estimate the AGN and the starburst contributions 
to the mid-IR galaxy emission at 19$\mu m$. For 59 galaxies with appropriate data, 
we can separate the 19$\mu m$ emission into AGN and 
starburst components using the measured mid-IR spectral features.  We use these to quantify
the brightness thresholds that an AGN must meet to satisfy our classifications:
AGN 1 have an AGN contribution  $\geq$ 73 \% and AGN 2 $\geq$ 45 \% of their total emission at 19$\mu m$.

The detection of [NeV] lines turns out to be an almost perfect signature of energy production
by an AGN.  Only 4 ($\sim$ 7.5\% percent) of 55 AGN 1 and 2 (10\% percent) out of 20 AGN 2
do not have [NeV] 14.3$\mu$m down to a flux limit of $\sim 4 \times 10^{-15} erg s^{-1}cm^{-2}$.
We present mean spectra of the various AGN categories. Passing from AGN-dominated to 
starburst-dominated objects, the continuum steepens, especially at wavelengths shorter than 20$\mu m$, 
while the PAH feature increases in its equivalent width
and the high ionization lines decrease.

We estimate H$_2$ mass and excitation temperature through the measurement of the S(1) rotational line 
of this molecule.  
Finally we derive the first local luminosity functions for the brighest mid-infrared lines and 
the PAH feature at 11.25$\mu m$. No statistical difference is apparent in the space densities for Seyfert 1's and 2's
of a given line luminosity, nor for the new classes of AGN 1's and 2's.
We use the correlation between [Ne V] line and nonstellar infrared continuum luminosity to derive
the global output of accretion-powered galactic nuclei in the local universe.
\end{abstract}


\section{Introduction}\label{intro}

This paper contains the final results of the \textit{Spitzer} high-resolution 
IRS spectroscopic survey of the sample of Seyfert galaxies (hereafter 12MSG)
included in the {\it IRAS} 12$\mu$m galaxy sample  \citep[hereafter RMS]{rms93}.
In \citet{tom08} (hereafter Paper I) we have presented and analyzed the first 30 
high-resolution spectra of 29 Seyfert galaxies of this sample (one IRAS galaxy, Mrk 1034, 
was coincident with a pair, for which we obtained two spectra).
The first spectroscopic observations of active galaxies with the Infrared Spectrometer \citep[IRS]{hou04}
onboard the {\it Spitzer Space Observatory} \citep{wer04} have been collected
on classical AGNs \citep{wee05} and ULIRGs \citep{arm07}. After the work presented in Paper I 
and the referenced works therein, a few more studies have discussed the Spitzer mid-infrared 
spectra of Seyfert galaxies, among these \citet{deo07}, \citet{mel08a} and \citet{wu09}.
As expected, the mid-infrared spectra 
of Seyfert galaxies show forbidden lines originating in the Narrow Line Regions, excited by 
the AGN ionizing flux. This power is thought to be produced by black-hole accretion, i.e., 
ultimately from the conversion of 
gravitational into radiative energy. The fine structure lines of 
[NeV] at 14.32$\mu$m and 24.31$\mu$m originate exclusively in the highly ionized gas (with an
ionization 
potential of 97eV) illuminated by the AGN\footnote{
either currently or some time during the last several hundred years, because,
even if the AGN ionizing continuum could be completely switched off, 
the photoionized NLR could still be detected, due to its
large extension and the long recombination time}.
As discussed in Paper I, the [OIV] line at 25.88$\mu$m 
(ionization potential of 55eV) is most probably excited from the AGN. In fact,  
\citet{mel08b} consider this line as an accurate and truly isotropic indicator of AGN activity, 
even if it could also originate from high excitation starburst emission or in shocks in 
low-metallicty starbursts \citep{lut98}. 

The [NeIII]15.55$\mu$m line (Ne+ and Ne++ have ionization potentials 
of 14eV less than O++ and O+++ respectively) can be excited both from AGN activity \citep{gor07}
and from starbursts \citep{tho00}. 
Superimposed on the AGN spectra, the lines of [NeII]12.81$\mu$m, 
[SIII]18.71$\mu$m and 33.48$\mu$m and [SiII]34.82$\mu$m originate
in gas with moderate ionization and most of their emission is generated 
by young newly formed stars, even if some contribution 
from the AGN is also expected \citep{sm92}. 
The relationship between the [OIV]25.88$\mu$m, 
[NeIII]15.55$\mu$m and [NeII]12.81$\mu$m lines in an eterogeneous sample of Seyfert galaxies 
has been studied by \citet{mel08a}, who found that Seyfert 1's and Seyfert 2's have different
AGN and star formation contributions to the total emission.

The interstellar medium 
produces the pure rotational lines of 
molecular hydrogen, as already shown by the early results of \textit{ISO} spectroscopy 
in \citet{ge98} and \citet{rig02}, respectively. 
\citet{wu09} analyzed the IRS low resolution spectra of 103 Seyfert galaxies from the 12MSG and
measured the policyclic aromatic hydrocarbons (hereafter PAH) emission features and the Silicate absorption strength.
The PAHs have been proposed as
star formation tracers by \citet{pl89}, while the Silicate absorption is sensitive to heavy dust 
obscuration of the nucleus.

According to the simplest
Unified Model for AGN, the Accreting Torus Model (ATM) \citep{mgt}, 
Seyfert 1 and 2 galaxies are the same kind of objects, only viewed from
different angles.  The strongest demonstration of this is detection via optical spectropolarimetry 
of the Broad Line Region emission--the defining characteristic of  Seyfert 1'--in a significant minority of Seyfert 2 galaxies 
\citep{am85,ant93}. 
A different scenario postulates an evolutionary difference: that Seyfert 2 are the early stages of the transition 
of HII/Starburst galaxies into Seyfert 1's. 
Two suggested evolutionary progressions 
are HII $\rightarrow$ Seyfert 2 \citep{kau03,sb01}, 
or a fuller scenario of HII $\rightarrow$ Seyfert 2 $\rightarrow$ Seyfert 1 
\citep{hm99,kro02,lev01}.
Because the radiation due to the star formation processes is roughly isotropic, the ATM
predicts no observational
difference in the star formation tracers between Seyfert 1's and 2's. If, 
on the other hand, star formation is stronger in Seyfert 2's, 
as suggested in \citet{buc06}, then some evolution from Seyfert 2's to Seyfert 1's could be invoked.
If general interstellar extinction towards the center of the galaxy is extremely high,
optical data alone might not always provide the correct (intrinsic) classification.

Our sample is described in \S2, the observations and the data reduction are briefly 
reported in \S3, the direct results of our observations, the estimates of the H$_2$ masses
and temperatures and the measure of the continuum extendedness are presented in \S4, 
the diagnostic diagrams and the semianalytical 
models to interpret them are presented and discussed in \S5.
In \S6 the [NeV] is quantified as an unambiguous AGN activity indicator, 
in \S7 we 
show that the mid-infrared diagnostics differentiate AGN 1's 
from the other populations, and we derive the average spectra for each class of objects 
that can be used as templates also for predictions and comparisons with high redshift populations. 
Finally in \S8 we present the line luminosity functions for our sample, and calculate the total
accretion power generated in the local universe.
The conclusions are summarized in \S9.

\section{The Seyfert galaxies of the 12$\mu$m Galaxy Sample}\label{sample}

From the original Seyfert galaxies list of the RMS, 
we present 91 IRS high-resolution spectra, including one third of them which were published in Paper I.
Our final sample is over 80\% complete,
large enough to give reasonable statistical results, with 41 Seyfert 1's, 
47 Seyfert 2's and 3 galaxies which have been reclassified as optical starburst galaxies, 
according to NED \footnote{NASA Extragalactic Database, IPAC, Caltech Pasadena, http://nedwww.ipac.caltech.edu/}. 

Another improvement of this work is the
classification that we adopt: we reclassify the Seyfert 1's and 2's into more general 
\textit{"AGN 1's"} and \textit{"AGN 2's"}.  We consider AGN 1's to be all those
with broad line regions, including those  Seyfert 2's with \textit{hidden broad 
line regions} (hereafter HBLR), observed in polarized light. The remaining Seyfert 2's 
lacking any broad permitted lines, even in polarized light, are classified as AGN 2's.
Our  classification scheme is an attempt to identify a 'clean' category of intrinsically broad-line AGN.
We follow \citet{tra01b} and \citet{tra03}, 
who made a spectropolarimetric survey of the 12MSG.
Out of the original 47 Seyfert 2's,  they found HBLR in 19, 20 lacking
an HBLR and they reclassify 11 objects 
as LINER, HII or starburst 
galaxies\footnote{Among these 11 objects: three (Mrk897, NGC7496 and NGC7590) were presented in Paper I and
the other eight are: NGC1056, NGC1097, NGC4922, NGC5005, NGC6810, NGC7130, MCG+0-29-23 and CGCG381-051, 
whose IRS spectra are presented in this paper.}. 
We classifiy as non-Sy for all these latter objects and similar ones
(in the diagrams LINER and HII or starburst galaxies will be distinguished).
Exceptions are NGC1097 and MRK897, that we reclassify as being a AGN 1 and an AGN 2 
respectively. We refer to the Appendix \ref{app.A} for the details on the classification of each one of these objects.
\citet{tra03} also adopt a somewhat arbitrary distinction between "bona fide" and "non-bona fide" 
Seyfert 1's, based on the NED classification of Seyfert 1.8, Seyfert 1.9 types and on radio loudness. 
For the remaining 13 objects not considered as "bona fide" Seyfert 1 in \citet{tra03}, we prefer 
instead to carefully classify them on the basis of detection or not of optical broad lines, 
either in direct or polarized spectra. 
We present the details on the classification of these 13 galaxies in Appendix \ref{app.A}, 
while we summarize here 
that we classify as AGN 1 seven Sy1 of the original RMS list (NGC526A, NGC1097, NGC1365, NGC2639, NGC7316, 
ESO545-G13 and ESO362-G18), 
we classify as a HBLR NGC5347 \citep{mor} and
we reclassify as AGN 2 one galaxy (NGC5506), while we consider non-BLR
four objects for which there is no evidence of broad lines, but they lack polarization observations 
(NGC1194, NGC4602, MRK1034 NED02 and MRK897)
and finally we reclassify as non-Sy two galaxies (NGC3511 and MRK1034 NED01).
Although \cite{tra03}  distinguish the radio-loud 3C galaxies, 
we actually classify 3C120, 3C234 and 3C445 as AGN 1, independently of their radio characteristics,
because of the presence of broad line emission.

We note that 7 Seyfert 1's have Balmer lines with relatively small FWHM,
under 2000 km/s but which are nonetheless produced in a BLR \citep{zw06}, 
These are usually classified as "narrow line Seyfert 1's", but our sample contains
too few to define another category.  It will turn out that their infrared
spectra do not appear different from those of normal Seyfert 1's (see Sec. \ref{obs}).

In summary our observed sample of "original" Seyfert galaxies contains 55 
objects showing some evidence of broad lines, either in direct (34 Sy 1) or polarized light 
(21 HBLR) that we classify as AGN 1, 20 AGN 2 (non-HBLR) and 4 non-BLR not included  
in the AGN2 class. We will also consider in the following the 13 non-Sy galaxies, 
however these latter will not be used for any statistical derivation.  

Following the results of \citet{wu09}, who have defined as "20 $\mu$m peakers" the Seyfert 
galaxies having a flux ratio $F_{20\mu m}/F_{30\mu m}\geq 0.95$, we have also identified
these in our sample\footnote{The "20 $\mu$m peakers" in our sample are: MRK335, MRK348, NGC424,
NGC526A, MCG-02-08-039, F03450+0055, ESO033-G002, MRK0006, MRK704, MRK1239, 3C234, MCG-06-30-015, IC4329A,
NGC6860, NGC7213} to search for any difference with our classes of galaxies. We refer to Appendix \ref{20peak}
for the results on these objects.

\section{Observations and data reduction}

Most of the sample galaxies--the 29 galaxies presented in Paper I and the 23 galaxies in this paper--
have been observed within the Spitzer Guaranteed Time Project 30291 (P.I.  Fazio). 
IRS high-resolution observations of another 37 objects, belonging to the 12MSG, 
were extracted from the SSC (Spitzer Science Center) archive. 
For 25 of the latter objects, the observing mode was similar to the one of P30291, 
namely the off-source observations have been collected to allow accurate background subtraction. 
For the remaining 12 sources from the archive (see Tab.\ref{tbl-1}), no off-source observation was taken.
For these objects, we give the line intensities, as for the other galaxies, 
but not  the equivalent widths (EW) of the emission features and lines, nor
do we  present the photometric measurements at 19 and 19.5 $\mu$m for 
computing the extendedness of the sources, because of their large
uncertainty in the level of the continuum. MRK335 was observed by the Program 50253
(P.I. Malkan), in the last cold cycle of Spitzer.  

The whole data reduction was  done using the SMART packages\footnote{SMART is
available on the SSC website and developed by the Infrared Spectrograph (IRS) Instrument Team at Cornell 
University \citep{hig04}}. We refer to Paper I for more details of the data reduction and analysis.

\section{Observational Results}\label{obs}

The journal of the observations of the 61 galaxies presented here is shown 
in Tab.\ref{tbl-1}, giving for each galaxy: 
the equatorial coordinates at the 2000 equinox; the redshift; the original RMS Seyfert class and the new
classification; the IRAS fluxes at 12 and 25$\mu$m; 
the observing date and the number of cycles and integration times per cycle.
For the sources already presented in Paper I, we refer to the table \ref{tbl-11} in the appendix for the adopted
classification.

\begin{figure}
\centerline{\includegraphics[width=8cm]{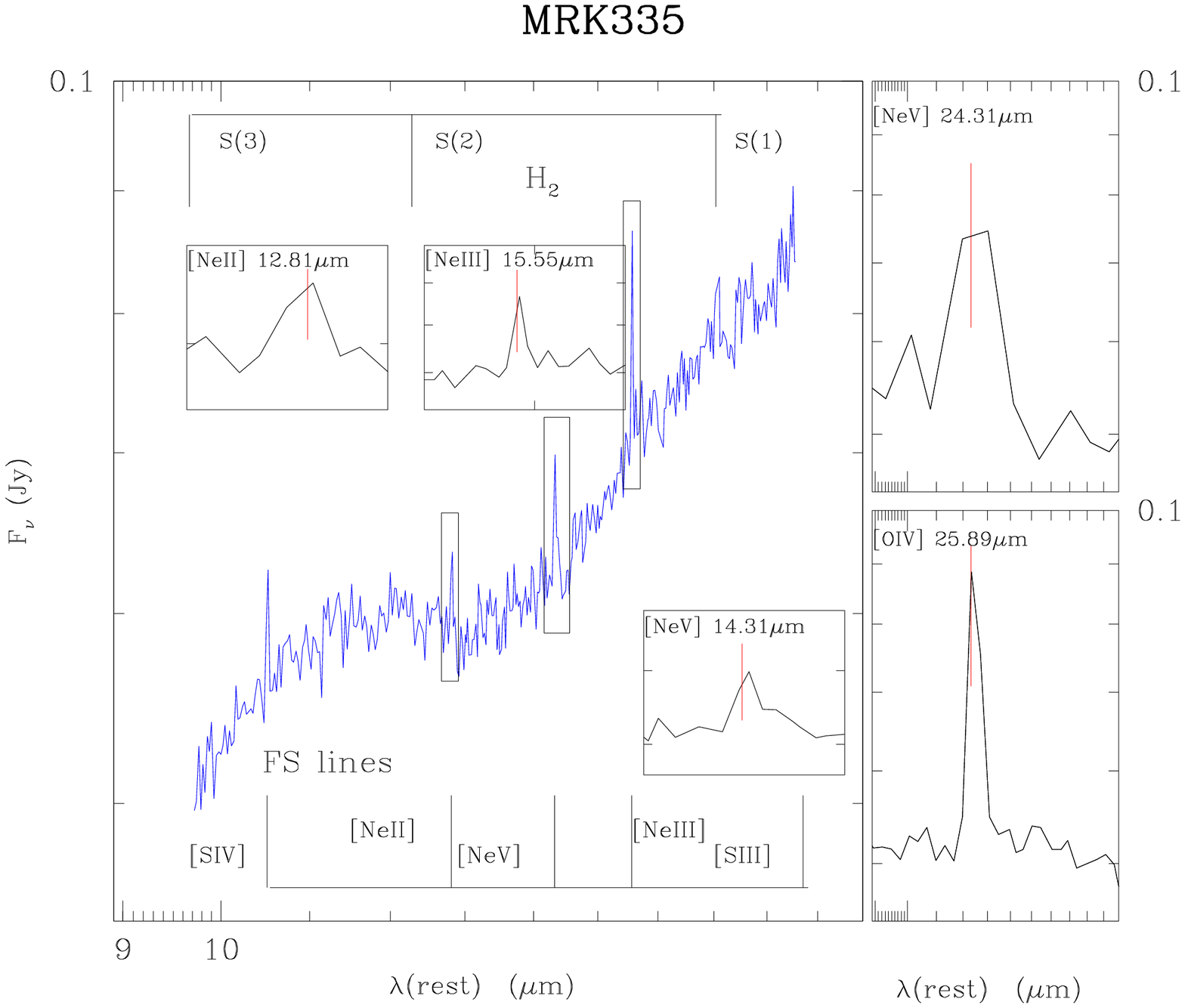}\includegraphics[width=8cm]{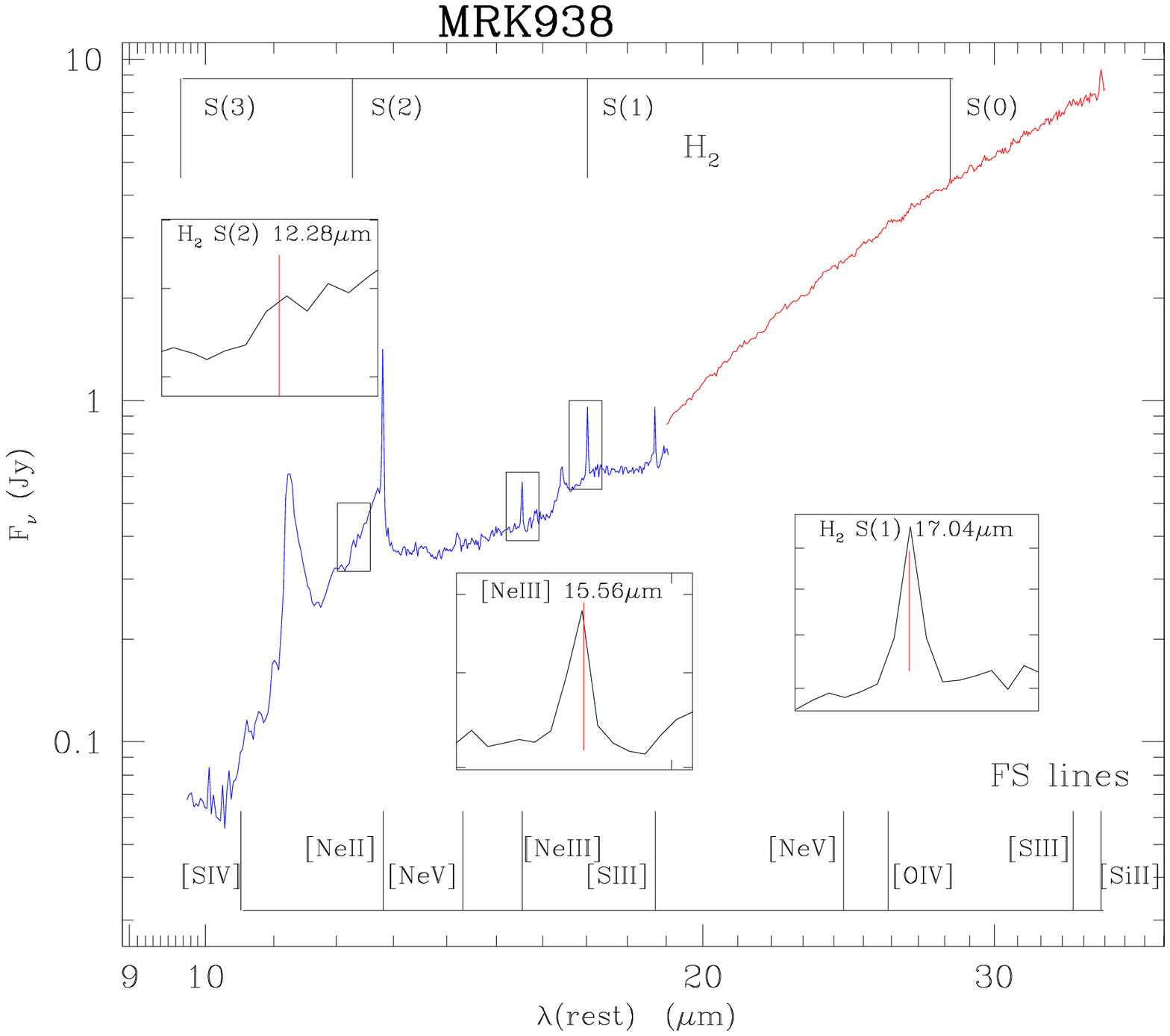}}

\centerline{\includegraphics[width=8cm]{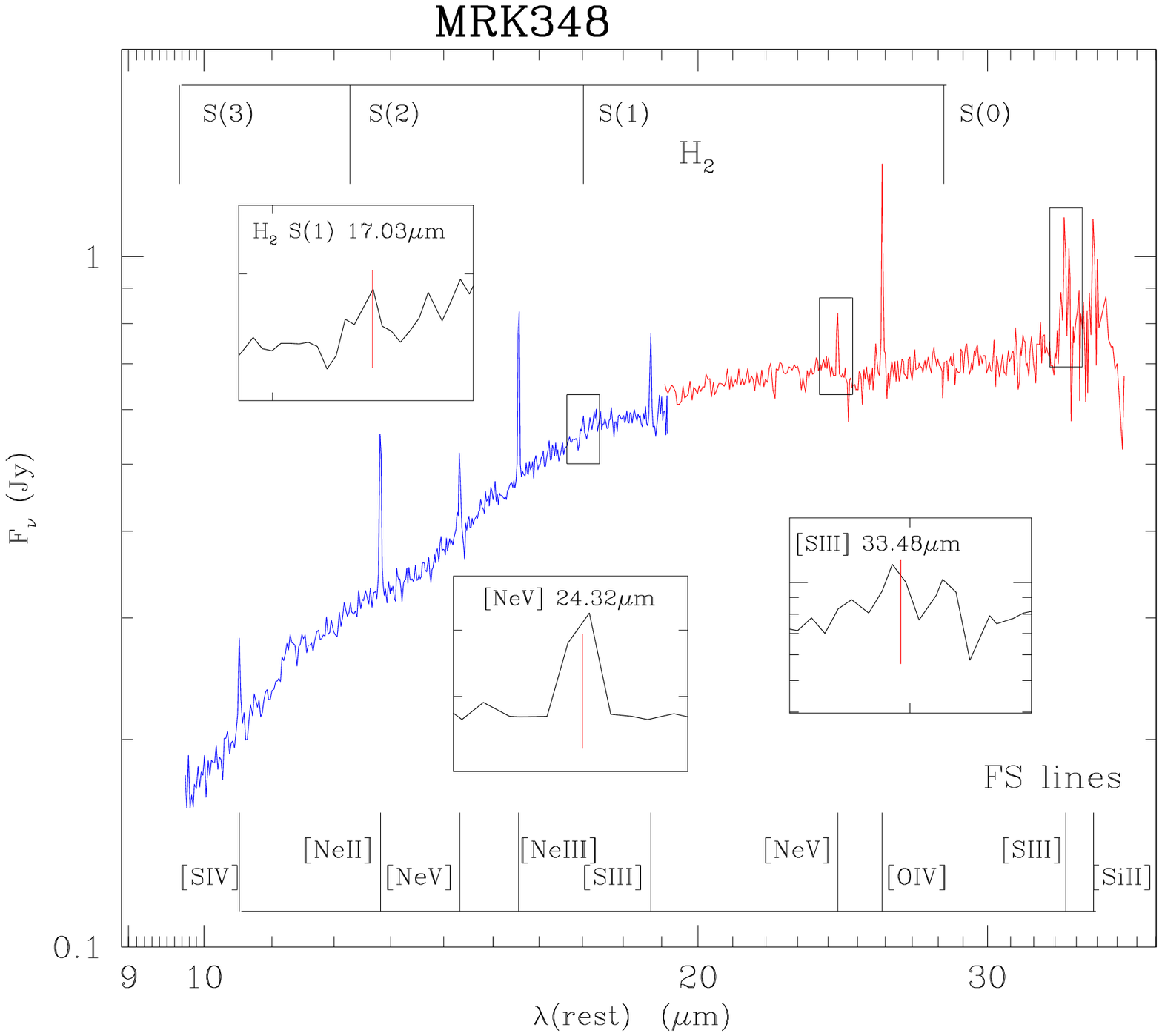}\includegraphics[width=8cm]{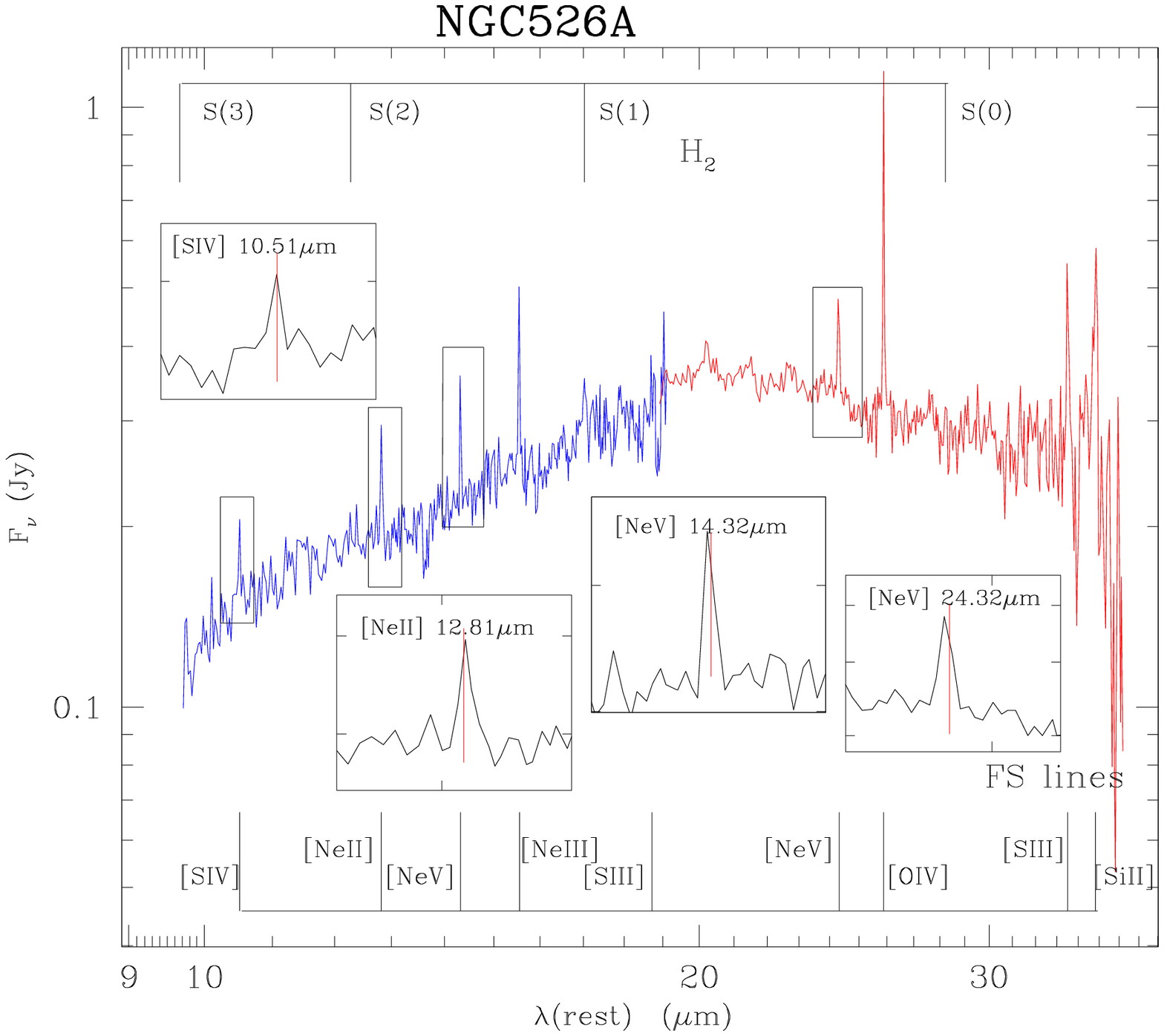}}

\centerline{\includegraphics[width=8cm]{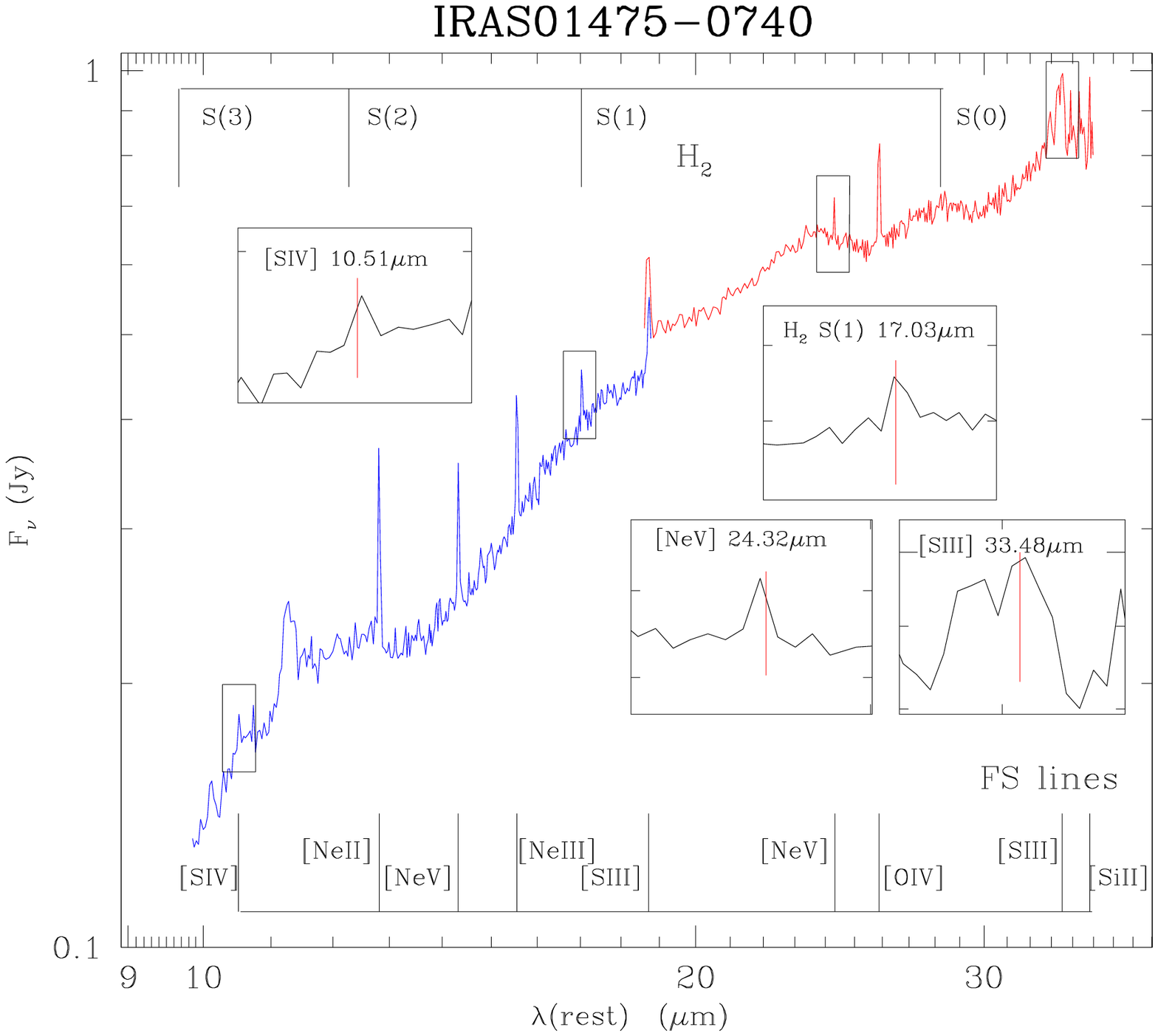}\includegraphics[width=8cm]{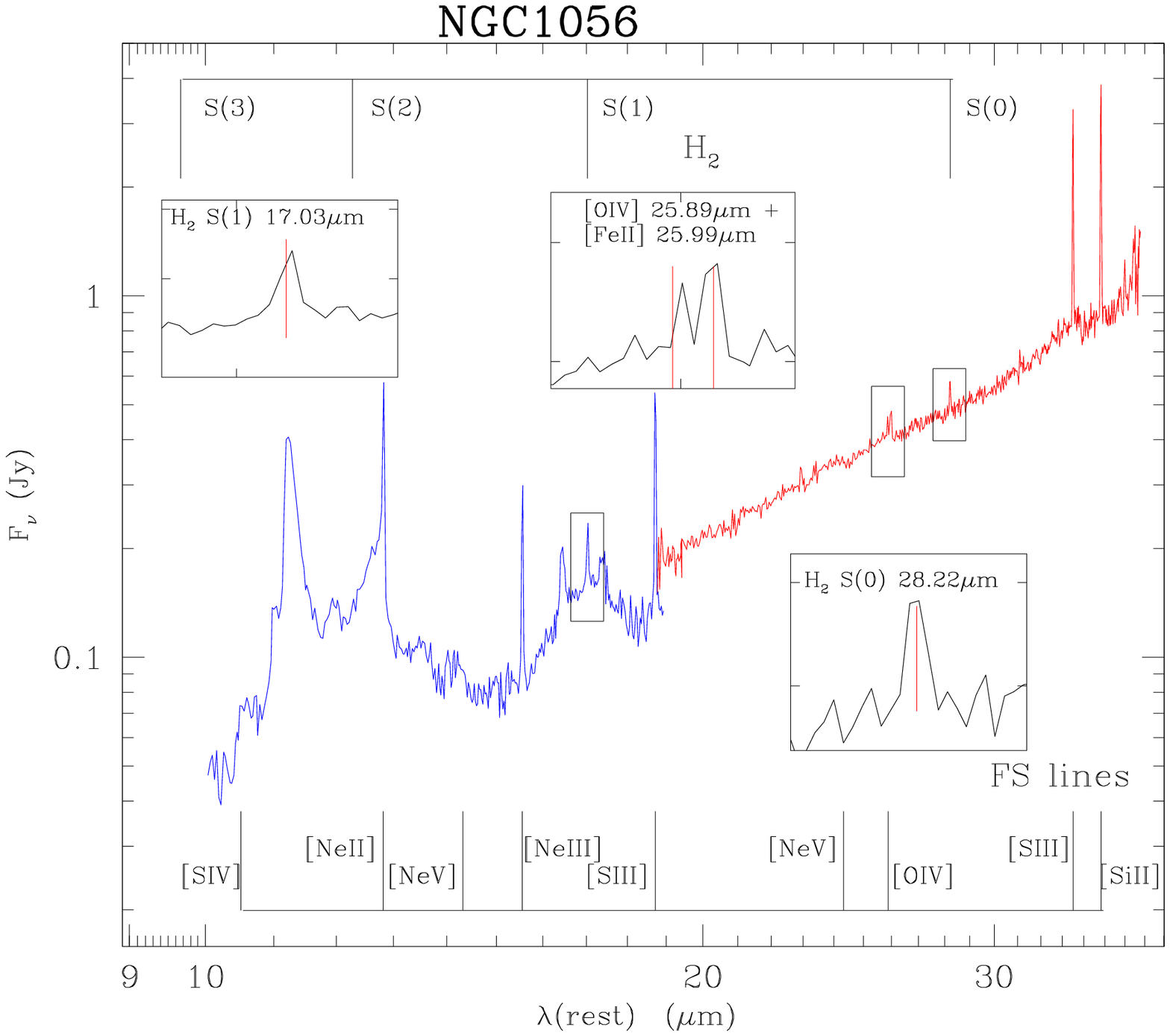}}
\caption{Spitzer IRS SH and LH specra of the observed Seyfert galaxies. Wavelengths have 
been shifted to the galaxies rest frames. For the objects with no off-source observation 
the SH spectrum is shown, because slightly affected from the background emission ($1simeq$ 10 \%), 
together with 
the $>3\sigma$ detected lines in separated boxes. Figures 1.11.61 are available in the online version of the Journal.
\label{fig1}}
\end{figure}

The spectra of the new 61 galaxies are shown in Fig.\ref{fig1}. 
For all galaxies for which the off-source spectra have been subtracted, both the SH and 
the LH spectra are presented. For the galaxies for which 
no off-source observation was available, we show in the figure the on-source SH spectrum only.
This latter is only marginally affected by the lack of background subtraction, because the theoretical 
background, as measured using the background estimator provided by the SSC, is
less than 10\% of the total measured SH emission. For these galaxies, we show only the detected lines 
in the LH range; we do not 
present the whole spectrum, because it is affected by a higher level of background (estimated to be about
20-30\% of the total measured emission).
The spectra of MRK335, F05563-G018, NGC5135, IC4329A, NGC5347 and NGC5506 show only 
the LH detected lines, because after background subtraction, 
the LH orders are not well inter-calibrated and thus the continuum cannot be defined properly.

Tab.\ref{tbl-2} reports the fluxes of the fine structure lines, measured with a Gaussian fit, 
for both the SH and LH spectra. Tab.\ref{tbl-3} gives the fluxes of the molecular H$_2$ rotational lines 
S(0), S(1), S(2) and S(3) and the PAH 11.25$\mu$m integrated flux, measured with a moment fit, 
and its equivalent width. We consider as detections the measurements with a signal to noise 
ratio higher than 3. We also report in Tab. \ref{tbl-11} in the appendix the PAH 11.25$\mu$m 
integrated flux and equivalent width as well as the  H$_2$ masses derived from the H$_2$ 
rotational lines observations for the 30 spectra of the galaxies presented in Paper I. 
We have repeated the measurements of fluxes and equivalent widths of the PAH 11.25$\mu$m to 
be consistent with the following method adopted here, and also because we found that 
some of the equivalent widths given in Paper I were affected by non-traceable errors. 

We measured the PAH fluxes by removing from the spectra the continuum under a baseline traced 
from the continuum 
shortwards of the PAH feature to the continuum longwards of the 
[NeII] line. Such a large interval has been chosen because, in addition 
to the feature at 11.25$\mu$m, two other PAH features are present 
(approximately at 12.0$\mu$m and at 12.5$\mu$m) and they increase the level of the apparent 
continuum under the 11.25$\mu$m feature. The integration range of the PAH emission 
depends on the feature's intensity, for the brightest sources it can be as wide as
0.5 $\mu$m (11.15$\mu m$ - 11.65$\mu m$).
Chosing this large baseline allowed us to avoid the 
other PAH contributions and remove the correct continuum from the galaxy. 
By dividing the resulting PAH integrated flux by the continuum flux density at the 
midpoint wavelength of the baseline, we can obtain the EW of the feature. 

\subsection{H$_2$ excitation diagrams: H$_2$ temperatures and masses}

Using the H$_2$ rotational lines intensities, we can estimate the temperature 
and the mass of the H$_2$ line emitting regions 
(see Paper I for the details of the method used). 
The derivation of the masses presented in Paper I was affected by some errors, therefore we recomputed 
their values, and list them in Table 9. 

In Fig.\ref{fig2} the excitation diagrams of each of the sources are shown. 
Tab.\ref{tbl-4} presents the derived temperatures T(3-2), T(2-1), T(1-0) 
of the gas in those regions where the rotational lines S(3), S(2) and S(1) are emitted, respectively. 
The 0-0 S(1) is the most intense H$_2$ line in every source, and so we estimated the mass of the molecular 
regions from its intensity. We calculated the masses from
the Boltzmann equation using the two transitions of S(0) and S(1),
listing the results  in Tab.\ref{tbl-4}. 
Therefore we obtain the masses only for the 50 galaxies where both these 
rotational lines were detected,
while we give an upper limit to the mass when only the S(1) line was detected. 
The temperature 
values that we derive here are comparable with the ones we report in Paper I.

The estimated H$_2$ masses are of the same order of magnitude as  
those calculated by \citet{rig02} in Seyfert and starburst 
galaxies and by \citet{hig06} in ULIRGs, ranging from 10$^7$ to 10$^9$ M$_{\sun}$.
The average mass (in units of 10$^8$ M$_{\sun}$) for the AGN 1 class is 1.8$\pm$1.3, for AGN 2 is 0.65$\pm$0.47,
excluding from the average the outlier NGC1142 because it has a mass an order of magnitude larger than the others.
For non-Sy galaxies is not possible compile a sensible average, because their masses spread over a wide range of values;
MRK1034 NED1 has a mass of 5.36, CGCG381-051 3.8, NCG3511 0.15 and NCG7590 0.12.
For the two Seyfert galaxies in common with \citet{rig02}, NGC1365 and NGC7582, our results are
in complete agreement, even though
they used ISO spectra, for which the H$_2$ S(1) data were obtained throught an aperture of 14''$\times$27'',  
that is 7 times larger than the aperture of SH. This implies that the H$_2$ emission is concentrated in the
inner $\sim$ 50"$^2$ of these two large (several arcminutes in diameter) galaxies.

In figure \ref{fig3}a,b we present the H${_2}$ flux vs PAH flux and the H${_2}$ 
luminosity vs PAH luminosity. We confirm the results of Paper I with our new classification: 
there are no differing trends which could discriminate between AGN 1
and AGN 2.

\subsection{Source extendedness}\label{ext}

The SH and LH spectra overlap in the range 17-19$\mu$m, allowing us to form an estimate of the extendedness 
of the sources called R, the ratio of the flux measured in LH to the flux measured in SH in 
an adjacent spectral interval.
This parameter can be defined only for sources after an appropriate background subtraction. 
We refer to Paper I for the details. 
The flux densities measured at 19$\mu$m in SH and those at 19.5$\mu$m in LH are reported 
in Tab. \ref{tbl-5}, with the R ratio and the corresponding extendedness class (see Paper I). 
We report the average extendedness factors in the classification table (Tab. \ref{tbl-6}), for
each class of galaxies and also for the group of "20$\mu m$ peakers". 
Both SH and LH spectra were already corrected for light lost outside the slits, assuming a point source:  
this correction takes into account the 
increasing size of the PSF as wavelength increases from the blue to the red end of each spectrum.
In order to better understand the validity of the spectroscopic extendedness ratio R, we analyzed
Spitzer direct images of four compact (R$\sim$1) sources taken with IRAC at 8$\mu m$ and MIPS at 24$\mu m$.
We measured in these images the fluxes through the same apertures as observed by IRS in SH and in LH,
flux-corrected with the prescriptions of the SSC for point-sources. The mean extendedness of these four sources measured through
IRS is $<R>_{IRS}=1.00\pm0.01$, while for IRAC at 8$\mu m$ and MIPS at 24$\mu m$ are respectively $<R>_{IRAC}=1.00\pm0.03$ and 
$<R>_{MIPS}=1.02\pm0.06$. Therefore the IRS estimate of the extendedness is reliable for compact sources.
We have also attempted an analogous test on the extended sources, although this estimate has serious limitations
because the aperture corrections to photometry of extended sources are not well defined for IRAC\footnote{For IRAC, see:  
http://ssc.spitzer.caltech.edu/irac/calib/extcal/index.html} and MIPS.  
We find that, using both IRAC at 8$\mu m$ and MIPS at 24$\mu m$, the extendedness classes we have defined are reproduced
for our sources, although there are some differences in individual values.
This implies a broad agreement among the IRS, IRAC at 8$\mu m$ and MIPS at 24$\mu m$ measurements of the extendedness.  
A paper with the analysis of the IRAC four channels and MIPS at 24$\mu m$ images of the Seyfert galaxies of our sample is in preparation.

\section{AGN diagnostic diagrams: data and models}

One of our aims is to develop a method to disentangle the contributions 
of the AGN and the Starburst to the total infrared emission of the Seyfert galaxies 
of our sample using mid-infrared spectral features.
In this section we use the diagnostic diagrams, together with semi-analytic models, 
to estimate for the sample galaxies 
the AGN contribution, e.g. through the [NeV] to [NeII] line ratio, and the star formation contribution, 
e.g. with the equivalent widths of the 11.25$\mu$m PAH feature and 
of the [NeII] emission line, and to search for differences between AGN 1's compared with AGN 2's.

To estimate the percentage of the contribution of the AGN and the Starburst to the observed emission 
of a galaxy at 19$\mu$m, we computed a mathematically simple analytic model 
to describe both the emission due to the AGN and that due the star formation
in each of the following observed quantities: the extendedness of
the source, the equivalent widths of the PAH at 11.25$\mu$m and of the [NeII] line at 12.81$\mu$m, 
the line ratios [NeV]14.32$\mu$m/[NeII]12.81$\mu$m and [OIV]25.89$\mu$m/[NeII]12.81$\mu$m and 
the spectral index $\alpha$ at (60-25)$\mu$m. We constructed analytic models for these quantities, 
because they provide the best separation of emission from AGN and non-AGN,
and therefore the best estimates of the AGN percentage contribution
to the total mid-infrared emission at 19$\mu m$. The simple equations for each of the models 
are given in the Appendix \ref{mod}. 
In the following plots of those quantities, 
we compare the observations with the semi-analytic models, which are plotted with solid lines. 

\subsection{Line ratios versus PAH equivalent widths}\label{5.1}

In Fig.\ref{fig4}a the line ratio [NeV]14.32$\mu$m/[NeII]12.82$\mu$m is shown as a 
function of the equivalent width of the PAH feature at 11.25$\mu$m. 
This ratio is the best AGN tracer in the IRS wavelength range, because [NeV] 
can be excited only by the AGN ionizing continuum. Its ratio to 
[NeII]12.82$\mu$m is not directly affected 
by abundances. 
76\% of the AGN 1's have an absolute value 
of the EW of the PAH at 11.25$\mu$m of $\vert EW PAH\vert$ $<$ 0.14$\mu$m
and a neon ratio [NeV]/[NeII] $>$ 0.15. On the other hand, 87\% of 
non-Sy and non-bona fide Sy1 have $\vert EW PAH\vert$ $>$ 0.14$\mu$m and [NeV]/[NeII] $<$ 0.5. 
AGN 2's (i.e. non-HBLR) show a wide range of both the EW of the PAH and of the neon line ratio. 
Fig.\ref{fig4}b shows a similar diagram with the PAH EW vs [OIV]/[NeII] line ratio. 
This diagram presents the same characteristics of the former, but with fewer upper 
limits, confirming its results. 
There is a Sy1, NGC7213, that has not been detected in the [OIV] line 
and has been reported to be at an intermediate stage between LINER and Seyfert 
\citep{sta05}.

Both the diagrams of Fig.\ref{fig4} show semi-analytic models which reproduce the
empirical data and estimate the percentage of the AGN and the starburst contributions
to the total infrared emission at 19$\mu m$. 
If we define as an {\it Seyfert-dominated} galaxy any of
the sources with an AGN contribution equal or more than
50\% (see the dashed lines in the diagrams), this corresponds to a the line ratios
[NeV]/[NeII] $>$ 0.054 and [OIV]/[NeII] $>$ 0.28.
Using this simple argument, we 
confirm that all the AGN 2's (non-HBLR's) are Seyfert-dominated, even if they span a wide range 
of AGN percentage in the
diagnostic diagrams. 
We report in the Table \ref{tbl-6} the average PAH EW and [NeV]/[NeII] ratio for each class of galaxies. 

Also the diagrams PAH EW vs [NeV]14.32$\mu$m/[SiII]34.8$\mu$m and PAH EW vs 
[NeIII]15.55$\mu$m/[NeII]12.82$\mu$m, presented in Fig.\ref{fig5}a,b, show 
similar characteristics as the previous two: the line ratio increases when the PAH EW decreases.
All these diagrams (Figs.\ref{fig4}, \ref{fig5}) show the general inverse relation between the AGN dominance, 
as measured from the ionization sensitive line ratios, and the star formation dominance, 
estimated from the PAH EW.

\subsection{Line ratios as density indicators}

In the IRS high-resolution spectral range, there are two fine structure line doublets,
from which the electron density can be derived. These are 
[NeV] at 14.32$\mu$m and 24.31$\mu$m and [SIII] at 18.71$\mu$m and 33.48$\mu$m. 
Following \citet{dud07}, as already reported in paper I, 
we present the diagram of the neon doublet ratio versus the sulfur doublet ratio in Fig.\ref{fig6}.
In the diagram are shown the low-density limits for both line ratios and the estimated electron density
as calculated by \citet{dud07} assuming a temperature T = 10$^4$K. 
The sulfur doublet ratio has been computed correcting the flux of the [SIII]18.71$\mu$m
taken throught the smaller aperture SH, multiplying it by its extendedness factor (see Table \ref{tbl-5})
to be compared with the larger aperture LH.
If the ratio of two lines of the same species is below the low density limit,  either there is
extinction that preferentially reduces the flux of the shorter 
wavelength transition, and/or the assumed temperature is wrong, and 
these doublet ratios could not be used to measure the electron density.

We find that 34$\%$ of the AGN1's 
lie under the neon ratio low density limit and 15$\%$ under the sulfur low density limit,
and that 30$\%$ of the AGN2's lie under the sulfur limit and 30$\%$ under the neon's.
Our current sample is much larger than that of Paper 1, making these results statistically
significant.

Taken at face value, these diagrams indicate low electron densities (10$^{(3-4)}$ cm$^{-3}$) in the [NeV]
emitting region.  For a typical ionizing luminosity of 10$^{43}$ erg s$^{-1}$ and an ionization parameter of 10$^{-2}$, 
this implies a characteristic radius of about 100 parsecs.
Since the recombination time is long, for densities of 10$^{(3-4)}$ cm$^{-3}$ it ranges from 10-100 
years\footnote{A typical recombination time is $\tau_r\sim 10^5/n_e$ years \citep{of06}}, 
'fossil' NLR emission will continue to be detectable
in AGN 2 which could have 'shut down' production of ionizing photons for the last 300 
years.  We suspect this may be the explanation of many of the 'pure Seyfert 2' members of
our AGN 2 class.
 
Thus a significant percentage of each of the Seyfert types falls 
under the low-density limit for either one or the other doublet, therefore, at least for the 
galaxies whose line ratio(s) are below the limit, we cannot estimate correctly 
the electron densities. However all the objects are below the low density limit
only in one line ratio and not in the other line ratio. 
If extinction were responsible, 
both ratios would show this effect.
We do not have any explanation for this unexpected behaviour and the possibility
that some currently adopted atomic parameters for these lines (transition probabilities and/or 
collision strengths) might be inaccurate must be considered. 

In Fig.\ref{fig7}a we show the [SIII] doublet ratio as a function of the [SIV]10.51$\mu$m/[SiII]34.8$\mu$m ratio, 
with the [SIII]18.71$\mu$m flux corrected for the extendedness factor (cf Sec. \ref{ext}). 
These lines are produced either in the narrow-line regions of the active nuclei or in the HII regions. 
In Fig.\ref{fig7}b the [NeV] 
doublet ratio as function of the [NeV]14.32$\mu$m/[NeII]12.82$\mu$m line ratio is given. 
Figures \ref{fig7}a and b show no differences in 
the density-sensitive line ratios between the different Seyfert populations.

\subsection{Ionization Diagrams}\label{ion}

[NeV]14.32$\mu m$ and [OIV]25.89$\mu m$ are the best AGN tracers in the high-resolution IRS spectra. We 
normalized them both to the [NeII]12.81$\mu m$, that is mainly produced in the HII/star-forming regions, to 
produce two ionization-sensitive ratios (in Fig.\ref{fig8}a). All the sources lie 
along the same sequence. 
As expected, the AGN 1's occupy the highest ionization region of the diagram as in the other diagrams.
Most AGN 2's lie in the same region of the AGN 1's.   
This diagram can be used to estimate the ionizing power in the NLR, 
both in AGN 1 and AGN 2 objects.
As mentioned above, we computed a semi-analytic model 
to reproduce the data in this diagram.
We choose as the threshold to consider a source as an 
Seyfert-dominated AGN again a 50\% AGN contribution to the 19$\mu m$ continuum, 
and this includes all the AGN 1's and all the detected AGN 2's. 
A clear implication is that using optical spectroscopy to identify Seyfert nuclei,
as we have done in the 12MGS, will systematically miss out a significant population
of non-HBLR Seyfert 2's (weak AGN 2's) which have less than 50\% AGN contributions to their
infrared emission at 19$\mu m$.

\citet{vei09} present the same diagram for their sample of ULIRGs belonging to the QUEST survey, and compute a model
to estimate the AGN and starburst contributions to the total emission. Their values of the line ratios
are in the same range as ours. However, the percentage of AGN contribution they estimate is 
on average lower than
the percentage we compute. This difference is likely due to the different populations of the samples. In fact ULIRGs 
have a stronger starburst component than Seyfert's and/or more heavily obscured AGN emission, even in the mid-infrared. 

\subsection{Line Equivalent Width Diagnostics}\label{5.2}

The line equivalent widths can also be used to estimate 
the contributions of the AGN and the starburst to the total galaxy emission. 
Since the original IRAS studies of Seyfert galaxies \citep{sm89,rms93,spi95,sam02}, 
we have known that the mid-IR continuum can be dominated by the AGN
continuum (e.g. from 12$\mu$m to 25$\mu$m), while at longer wavelengths the continuum
due to the galactic disk and reprocessed thermal emission from dust around the young stellar populations
increases, and it dominates the total emission at wavelenghts of 60-100$\mu$m.
Therefore the equivalent widths of emission lines in the mid-IR are differently affected by
the underlying continuum, as a function of the wavelength: 
around 12$\mu$m the EW will be more depressed by the strong AGN
continuum, than they would be around 25$\mu$m.
The lines originating in the galactic disk and its star-forming regions, such as
[NeII]12.8$\mu$m  and the H$_2$ S(1) at 17.02$\mu$m, will have a smaller equivalent width 
in objects with a stronger AGN continuum, while the lines tracing the AGN, like 
[NeV]14.32$\mu$m and the [OIV]25.89$\mu$m will not suffer this depression
because they are enhanced in proportion to the strength itself of the AGN,
keeping their equivalent widths more constant.

In our analysis of the equivalent widths, we consider only objects for which
an accurate measurement of the continuumm was possible through background subtraction and 
with detected lines, neglecting the ones with upper limits. The equivalent width 
of the [NeII]12.81$\mu m$, the [NeV]14.32$\mu m$, the [NeIII]15.55$\mu m$, the [OIV]25.89$\mu m$ and 
H$_2$ 17.02$\mu m$ are reported in the Tab.\ref{tbl-7}. 

We present in Fig.\ref{fig8}b the PAH EW as a function of the [NeII]12.82$\mu$m equivalent width.
This diagram is able to disentangle  
Seyfert-dominated AGN from starbursts.
Comparing it with our semi-analytic model, we find that almost all the sources lie in the
region of AGN contribution greater than 50\%. 
We further find all the AGN 1's and 50\% of the AGN 2's
lie in the region with the modelled AGN contribution $\geq$75\%, 
i.e. for $\vert$EW [NeII]$\vert$ $<$ 0.08. 
The likely explanation is that both the HII regions/star-formation tracers--
[NeII] and PAH--happen to have wavelengths in a continuum range that can be strongly enhanced 
by the presence of an AGN contribution. The brighter the nuclear AGN continuum,
the smaller the equivalent width of the spectral feature, as can be seen in the
average spectra of the different populations of Fig.\ref{fig14} in \S\ref{spec}. 
We report in table \ref{tbl-6} the [NeII]EW for each class of galaxies.

%\textbf{
We computed a mean PAH EW at 11.25$\mu m$ for the AGN 1 of
$<\vert$PAH EW$\vert>$ = 0.17$\pm$0.20 (Seyfert 1: 0.17$\pm$0.21 and HBLR: 0.16$\pm$0.19),
and for the AGN 2 $<\vert$PAH EW$\vert>$ = 0.34$\pm$0.28. 
\citet{wu09} find, for the 55 objects that are in common,
a similar mean PAH EW at 11.25$\mu m$.
From their data, we calculated for the AGN 1 class $<\vert$PAH EW$\vert>$ = 0.19$\pm$0.20
(Seyfert 1: 0.20$\pm$0.20 and HBLR: 0.17$\pm$0.21), 
and for the AGN 2 $<\vert$PAH EW$\vert>$ = 0.43$\pm$0.27. 
For the detailed comparison between ours and their data, we refer to Appendix \ref{Wu}.
%}

Using a Kolmogorov-Smirnov test, we derive that the probability that the two kind of AGN 1's--
the Seyfert 1's and the HBLR-Seyfert 2's--  
belong to the same family is 78\%. 
Applying the same test to the Sy 1 and the AGN 2 or to the HBLR-Sy2's and the AGN 2,
results in both cases in a probability of less then 1\% that the two groups are drawn from the same population.
This reinforces our hypothesis that the
HBLR belong with the Sy1, not the other Sy2s.  We will further discuss this topic 
in Par.\ref{stat}. 

\subsubsection{Line equivalent width, extendedness factor and far-IR slope}\label{5.3}

We present in Figures \ref{fig9} - \ref{fig10} the extendedness versus the equivalent
widths of 11.25$\mu$m PAH, [NeII]12.8$\mu$m, and [NeIII]15.5$\mu$m, respectively.
The galaxies with the largest extendedness, which we define as class III, namely NGC1056, NGC1142, 
NGC1667 and NGC5005, are the ones with the largest equivalent widths 
of the PAH at 11.25$\mu$m and of the [NeII]12.8$\mu$m. 
These emission features are generated principally in the interstellar medium 
and in the HII regions of the galaxies, but not much from their active nuclei. Among these four galaxies,
NGC1056 and NGC5005 have no detected [NeV] lines, and are reclassified as non-Sy 
(see tables 1 and 2);
NGC1142, NGC1667 are AGN 2. Including the galaxies of Paper I, among the galaxies of class III 
there is also a non-BLR, NGC4602, and a Sy1, NGC4748. 
We find the same trend as in Paper I, that with increasing size of the 19$\mu$m emitting region, 
both the PAH and [NeII] equivalent widths increase. 
That is, in these diagrams AGN 1's 
occupy the region where the point sources lie. In contrast the non-Sy's 
with a measured extendedness, 
occupy the region of significantly extended sources with large PAH or [NeII] 
equivalent widths. This separation works properly in the diagrams of 
extendedness vs PAH EW and extendedness vs [NeII] EW, shown in Figg.\ref{fig9}a and b.
In fact just 1 of 23 Sy 1's 
and 2 out of 17 HBLR's do not lie in the "AGN 1 region", with extendedness $\leq$1.3. 
But 2/3 of the non-Sy have extendedness $\geq$1.3. 
AGN 2 split almost equally in the two regions: 8 of 18 lie in the AGN 1 region and 10 in the other.

Comparing the diagrams of the extendedness versus EW PAH and versus EW [NeII] with our semi-analytic model,
we find that these diagrams do separate the  
Seyfert-dominated AGN 
from Starburst-dominated ones:
a modelled AGN contribution greater than 75\% corresponds
to $\vert$EW PAH$\vert \leq$0.4$\mu m$ and $\vert$EW [NeII]$\vert \leq$0.08$\mu m$. 

The diagram of [NeIII] equivalent widths versus the 19$\mu$m source extendedness
(Fig.\ref{fig9}) shows a similar trend of the figure with  [NeII], but with a much higher scatter. 
On the contrary, the diagrams of the [NeV] and [OIV] equivalent widths versus 
the 19$\mu$m source extendedness (that we do not show)
do not show any clear trend.  This confirms that
the high-ionization lines are almost totally produced in the active nuclei with
no significant contribution from in extended regions.
[NeIII] is an intermediate case, with contributions from both the Seyfert
nucleus and also the extended star-forming regions.
 
The IRAS 12$\mu$m luminosity and the [NeV]14.32$\mu$m/[NeII]12.82$\mu$m line ratio 
are not correlated for any Seyfert type (see Fig.\ref{fig11}), 
confirming the finding of Paper I. We notice that high-luminosity objects, e.g. those
with $L_{12\mu m} > 10^{44} erg s^{-1}$, have the ratio [NeV]14.32$\mu$m/[NeII]12.82$\mu$m $>$ 0.2,
i.e. are AGN-dominated.
Fig.\ref{fig12}a,b shows the diagrams of [NeV]14.32$\mu$m/[NeII]12.82$\mu$m 
and [NeIII]15.55$\mu$m/[NeII]12.82$\mu$m vs (60-25)$\mu$m spectral index. 
In both diagrams AGN 1's 
lie in the upper right region, with [NeV]/[NeII] $>$ 0.05 and [NeIII]/[NeII] $>$ 0.1, 
and the spectral index  $\alpha_{60-25} > -2.2$. 
The non-Sy's cluster in a region [NeV]/[NeII] $<$ 0.2 
and [NeIII]/[NeII] $<$ 0.5, with a steeper spectral index, $\alpha_{60-25} < -1.2$. 
AGN2 lie across both the regions.
The average spectral indices are given in table \ref{tbl-6}.

\subsection{The AGN and starburst contribution to the 19$\mu m$ flux}

By inverting the semi-analytic models, i.e. solving the analytical expressions with the true values 
of the observed quantities, we obtained a value of $\Re$, defined as starburst to AGN continuum
ratio at 19$\mu$m (see Appendix \ref{mod}) for each of the five observed quantities 
of every source. We then computed the mean value of $\Re$ to estimate the relative percentage 
of AGN and Starburst emissions. From the value of $\Re$ we computed in Tab.\ref{tbl-8} the 
percentages of the Starburst and of the AGN emission.
When, for a particular source, the scatter is greater than half of the mean and 
is due to a single $\Re$ value of one of the models, which does not describe that source 
because of a noisy detection, we removed that discrepant value of $\Re$ and recomputed the mean. 
The sample of sources to which we apply this analysis is reduced from the 
observed sample of 89 because the models depend on 
the extendedness and the PAH equivalent width, which are not always detected (for the PAH) or measurable 
(for the extendedness).
We found that the model can disentangle the AGN and the Starburst emission for 31 AGN1 (17  
bona fide Sy1's and 14 HBLR), 3 non-BLR, 15 AGN 2's and 9 objects re-classified 
as non-Seyferts.  As can be seen from the histogram in Fig.\ref{fig13} Sy1's 
have a mean AGN contribution at 19$\mu m$ of $92\%\pm6\%$; 
HBLR's $92\%\pm8\%$; AGN's $79\%\pm16\%$, 
non-Sy's $69\%\pm16\%$, and non-BLR's $62\%\pm7\%$.
These average percentages are also given in table \ref{tbl-6}.
It is perhaps an uncomfortable surprise that up to {\it half} 
of the 19$\mu$m continuum in the IRS slit can come from a normal AGN, but still not be
strong enough to make it unambiguously classified as an AGN from spectroscopy.
As discussed below, this has lead to substantial confusion in the literature
when Seyfert ''AGN" are discussed and compared, using different selection
observations.

The differences that we find between the AGN 1 and the AGN 2, (even if the latter cannot be considered from our 
data as an homogeneous population), i.e. the lower ionization ratios, the increased
PAH and [NeII] EQW, the extendedness of the 19$\mu$m emission, can all be related to a weaker strength 
of the AGN component with respect to the starburst component.
As a matter of fact, all our AGN 2's have evidence of  an AGN both in the optical (as seen in their
optical line ratios in the BPT diagrams \citep{bpt81}, where we have used the optical spectral observations of \citet{rod09})
and also in the hard X-rays, at 2-10keV \citep{shu06,mal07,bia05}, 2-8keV \citep{car07},
15-136keV \citep{del04} or 40-100keV \citep{bir08}. 

\subsection{Line equivalent widths for the different AGN types}\label{5.6}

The average equivalent width of 
the fine structure lines in the various AGN classes will change depending
on whether the line emission, and also the underlying continuum,
is dominated by the nonstellar nucleus or the hot stars.
Thus the equivalent width of [NeV]14.32$\mu m$ is the same for 
AGN 1's: 0.020$\pm$0.021
and AGN 2's: 0.025$\pm$0.010$\mu m$,  because both the line and the underlying
continuum emission are proportional to the luminosity of the AGN component.
The EW drops in non-Seyfert galaxies because of the complete cutoff of [NeV]
emission--it is detected in two non-Seyfert galaxies which are classified as 
LINER (see Appendix \ref{app.A}), but measured only in NGC7130,
for which we have an off-source measurement.

The [OIV] EW decreases somewhat because the underlying continuum has a larger
starburst contribution passing from AGN1 to non-Sy, in fact the [OIV] EW of
the AGN1 is
 0.116$\pm$0.134$\mu m$, of the AGN2 is 0.093$\pm$0.054$\mu m$ and of the
non-Seyfert galaxies: 0.033$\pm$0.035$\mu m$. In contrast, the [NeII] and the
H$_2$ equivalent widths increase for the same sequence of objects, that is
[NeII] EW for the
AGN1 is  0.028$\pm$0.026$\mu m$, for the  AGN2 it is
0.081$\pm$0.057$\mu m$, for the non-Seyfert galaxies it is
0.116$\pm$0.029$\mu m$
and the average EQW H$_2$ is a factor ten greater in the non-HBLR's and the
normal
galaxies than in Sy1's and HBLR's (AGN1: 0.009$\pm$0.008$\mu m$;
AGN2: 0.046$\pm$0.069$\mu m$; normal galaxies: 0.058$\pm$0.062$\mu m$).
These trends are summarized in the 'classification table' given in Table \ref{tbl-6}.

\section{[NeV] as indicator of AGN activity}\label{ne5}

Because of the very high ionization potential of NeV, 
we consider it as the best emission feature to distinguish active galaxies from starburst 
galaxies. Strong starburst activity could possibly excite some high-ionization lines such as 
[OIV]25.89$\mu$m \citep{lut98}, but the ionizing spectrum of O and B stars is not hard enough
to produce much Ne$^{4+}$.  
Shocks would need to have exceptionally high velocities.
Among the 91 sources we analyzed here and in Paper I classified as Seyfert galaxies, 
16 objects have no [NeV] emission. 
As already mentioned in Sec.\ref{sample}, 10 (NGC1056, MCG+00-23-029, 
NGC5005, NGC6810, CGCG381-051, MCG-03-34-063, NCG7496, NCG7590,NGC3511 and MRK1034 NED01) 
of these 16 galaxies  have already been 
reclassified as LINER, HII or starburst galaxies by \citet{tra01b} and \citet{tra03}.
4 are Sy1's: NGC1097 and NGC2639 lie in the HII region area 
of the BPT diagram based on [NII]6584\AA/H$\alpha$ vs [OIII]5007\AA/H$\beta$ 
\citep{bpt81}; NGC7213, which also lacks the [OIV] 
line detection and has been classified as intermediate between LINER and Seyfert \citep{sta05}.
The other undetected sources are the AGN 2 NGC4501 and MRK938, which also lack a detectable [OIV] line. 
 
In contrast, two of the non-Sy's show [NeV]14.32$\mu$m emission: 
NGC4922 and NGC7130 are LINERs, whose active nuclei can produce highly photo-ionized gas emission  \citep{dsm09}.
There are also 4 sources with detections of [NeV] at 14.32$\mu$m but 
not 24.31$\mu$m. Nevertheless we consider them as [NeV] emitters, even if not 
detected at 24.31$\mu$m, because the noise of LH is greater than the noise of SH. 

In conclusion, we can consider the [NeV] emission lines as a strong indicator for a galaxy to 
be classified as an AGN. In fact it is detected in 88\% of the AGN 1's,
90\% of the AGN 2's, only the 17\% of the non-Sy's. 
This means that deep spectroscopic searches, e.g.,  for [NeV]14$\mu m$, can discover 
relatively weaker AGN with lower luminosities, which only produce less than 45\% of the
total 19$\mu m$ emission observed.

Thus our infrared classification of galactic nuclei turns out to be in very close
agreement to the classifications of them originally made from optical spectroscopy.
The 12MGS AGN sample, which is originally defined based on optical
spectra, does not contain very heavily obscured ("buried") Seyfert nuclei, 
more or less by definition.  Deep searches for [NeV]14$\mu m$ emission in the remaining
''non-Seyfert'' members of 12MGS would be required to find out how
common buried Seyferts are. A first step has recently been taken by \citep{ga09},
who uncovered a significant fraction of [NeV]-emitting
luminous infrared galaxies whose Seyfert nuclei had been missed by previous
optical spectroscopy, due to lack of sensitivity and heavy dust reddening.

\section{AGN 1 statistics}\label{stat}
We give in Tab.\ref{tbl-9}
the probabilities P that the two AGN1 sub-populations, the Sy1's
and the HBLR Sy2's, do not differ significantly from one another,
for each of the observed quantities.
To derive those probabilieties we used a Kolmogorov-Smirnov test,
which calculates the probability of two sets of data values arising from the
same intrinsic distribution. The higher  the probability, the closer are the two sets of data.
We have grouped similar diagrams together to derive an average probability
$<P>$, with its standard deviation. For the first group of the relations combining
fine structure line ratios with PAH equivalent width
(EW PAH vs [NeV]/[NeII], vs [NeIII]/[NeII], vs [OIV]/[NeII],
vs [NeV]/[SiII]), we obtain  $<P>$ = 73\%$\pm$31\% that the AGN1 populations--Sy 1's and HBLR Sy2's--belong in 
the same class.
For the second group, relations of
the extendedness vs the equivalent width of the [NeII], [NeIII], 
[OIV],
PAH  we find $<P>$ = 71\%$\pm$12\%.
For the the relations between the H$_2$ S(1) line
and the PAH flux and luminosity we obtain  $<P>$ = 77\%$\pm$2\%.
Grouping together the relations between the neon ratios ([NeV]/[NeII] and
[NeIII]/[NeII])
vs the spectral index at (60-25)$\mu$m, gives $<P>$ = 99\%$\pm$2\%.
A probability of P = 78\%  
is obtained using the diagrams EW [NeII] vs EW PAH 
and ([NeV]/[NeII].
The high probabilities obtained strengthen our classification which joins both Sy 1's
and HBLR's as members of a single AGN 1 category.
We computed the same statistical tests to calculate the probability that the Sy1 and
AGN2 or 
HBLR and AGN2 objects belong to the same family.  Both of these probabilities
are on average
lower. 
The indistinguishability between Sy 1's and HBLR has been also verified by
plotting the average of the spectra of all
AGN1's with an AGN percentage  greater than 95\%, and the average spectrum of
those AGN1's
with an AGN percentage below that value. The resulting plot  in Figure 14
looks almost identical
to the plot with the average spectra of Sy 1's and HBLR's.

\subsection{Average Spectra for each class}\label{spec}

To study representative spectral features of all classes of galaxies in our sample, 
and provide templates useful for comparisons and predictions of high-redshift
galaxy populations, we computed the average 
spectrum of each of the classes normalized to the flux at
27$\mu m$. All the normalized average spectra are presented together in Fig.\ref{fig14},
and are available electronically. %\textbf{
They are compared to the average spectrum of a sample of starburst galaxies 
from the IRS high resolution data by \citet{bs09}.%}

The slopes of the continua of the average spectra are steepest in starburst and non Seyfert's and 
in AGN 2, with spectral indices\footnote{We define as spectral index 
$\alpha =\frac{dlog(F_{\nu})}{dlog(\nu)}$.} (10-35)$\mu m$ of 
-2.88, -2.25 and -1.65, respectively.
We cannot fit the continuum of the bona fide Sy1's and the HBLR with a 
single power-law.
Instead we must distinguish the (10-18)$\mu m$ and the (18-35)$\mu m$ continuum slopes. 
The shorter wavelength range in the bona fide Sy1's has 
a spectral index of the averaged spectra of -1.21 
and for the HBLR of -1.95, 
while the longer wavelength range has continuum spectral indexes of -0.65
and -0.90, respectively.

\citet{wu09} find an average spectral index of the Seyfert 1's to be -0.85$\pm$0.61 and of the Seyfert 2's to be -1.53$\pm$0.84,
quite similar to the original findings based on IRAS data of \citet{em86}. 

Fig.\ref{fig14} shows a clear sequence. The highly ionized lines, such as [SIV], [NeV] and [OIV] are intense 
in the mean spectra of the Seyfert 1's, HBLR Sy 2's,
and AGN 2s. The PAH feature is stronger in starburst 
galaxies, non Seyfert's and AGN 2, and is weaker in AGN 1.
Thus, going  from Seyfert-dominated to starburst-dominated 
objects, the continua steepen from AGN 1 to AGN 2 to non-Seyfert's and starbursts, 
the higher ionization lines decrease not only in their flux, but also in their equivalent widths (see \S\ref{5.6}), while the PAH
feature remains almost constant in flux, while its equivalent width increases. It seems therefore that the PAH grains
do survive in the highly ionized medium of AGNs, however they appear weaker in the most powerful ones,  
because they are masked by the strong  underlying AGN continuum. The H$_2$ S(1) line has a behavior similar to the PAH,
and in fact its equivalent width increases with decreasing AGN activity (see \S\ref{5.6}) .
The presence of a BLR is closely correlated with strong thermal continuum from hot dust grains
which emit around 10$\mu m$.

\subsection{Line Luminosity Functions}

From our statistically complete sample of galaxies, we are able to compute the first infrared
line luminosity functions for Seyfert 1's and Seyfert 2's. 
The 12MGS includes 53 Seyfert 1's and 63 Seyfert 2's, but only 42 Seyfert 1's and 50 Seyfert 2's have been observed by IRS, 
therefore we have to correct the luminosity function for completeness. 
Tab.\ref{tbl-10} gives the luminosity functions for Seyfert 1's and Seyfert 2's, for several of the
bright emission lines.
In Figg.\ref{fig15}-\ref{fig20} the luminosity functions 
of the [SIV], [NeII], [NeV]14.32$\mu m$, [NeIII], [SIII]18.71$\mu$m,  [NeV]24.31$\mu m$, [OIV], [SIII]33.5$\mu$m, [SiII], 
H$_2$ 17.04$\mu m$ and PAH 11.25$\mu m$ are shown. 
All the fine structure lines have a luminosity in the range 10$^{39}$-10$^{43}$ erg sec$^{-1}$ and the 
corresponding space densities range from 2$\times$10$^{-4}$ to 10$^{-9}$ Mpc$^{-3}$, 
while the H$_2$ rotational line and
the PAH 11.25$\mu m$ feature have luminosities from 10$^{38}$ to 10$^{42}$ and from 10$^{37}$ to 10$^{41}$ respectively.

We do not find any significant difference in the line LFs of Seyfert 1's and Seyfert 2's, 
which are indistinguishable at the 2$\sigma$ level.
We also computed the line luminosity functions separately for the AGN 1's and
the AGN 2's, using our new classification, and find no differences, 
probably because the shift of the 20 HBLR from the Seyfert 2's class into the new AGN 1
class has not a noticeable (statistical) effect on the new AGN 1 luminosity functions,
while it increases the uncertainties in the AGN 2 luminosity function.
We do not present the luminosity functions with the new classification,
because the AGN 1 sample size is 53 but the AGN 2 sample is only 21, making the statistics of these latter quite poor.

%\textbf{
Our sample includes all the galaxies 
selected at 12$\mu m$ that show evidence of Seyfert activity through optical spectroscopy. 
We are aware that some AGN activity can be detected in the mid-IR in optically unidentified AGN.
\citet{ga09}, using a volume limited sample to D$<$ 15Mpc, find that $\sim$ 50\% of the AGN they detect
are not identified using optical spectroscopy. However, these objects are typically
starburst dominated systems hosting modest AGN activity and have low luminosities.
The luminosity of the [NeV]14.32$\mu m$ emission line measured in these galaxies is about 10$^{37-39}$erg/sec,
which is too low to affect our luminosity function (see Fig. 16a). %}
 
\section{The accretion power in the Local Universe}

Because the [NeV]14.3$\mu$m line is uniquely
generated by the accretion process through the ionizing continuum of the central engine,
its luminosity function
can be used to measure the accretion power in the Local Universe.
The first step is to determine the correlation between the luminosity in this line and the 19$\mu$m continuum
luminosity that we measured in 59 of our Seyfert galaxies (see Table \ref{tbl-8}).
This relation is:

$$ Log L^{AGN}_{19\mu m} = 0.9667 \times Log  L([NeV]14.32\mu m) + 4.3263 $$

and is shown in Fig. \ref{fig21}a, for each of the galaxies detected in the [NeV]14.32$\mu$m line
and for which we have a measure of the AGN component.
We can use this relation to convert the [NeV]14.32$\mu$m line luminosity function, of the whole
set of galaxies measured in this line, into a 19$\mu$m AGN luminosity function.
We present this latter in Fig. \ref{fig21}b.
The integration over all luminosities and over the volume defined by the average redshift 
of the 12MSG of z=0.030 (RMS) gives a measure of the accretion power in the local universe at 19$\mu$m
of 2.1 $\times$ 10$^{46}$ erg s$^{-1}$. 

If we want to convert this 
monochromatic power at 19$\mu$m into the bolometric power due to accretion, we simply use the
relations between the bolometric luminosity and the infrared luminosities published in \citet{spi95}.
Taking the typical spectral index for Seyfert galaxies $\alpha(12-19\mu m)$ = -1 and the 12$\mu$m vs bolometric
luminosity correlation, we find a bolometric power of accretion of 8.0 $\times$ 10$^{46}$ erg s$^{-1}$,
over the local volume out to z=0.03.
We are well aware that the relation between 12$\mu$m and bolometric luminosity has been derived from
large-beam (few arcminutes square) IRAS data, however both quantities are affected by possible extended emission
in an analogous way. Moreover all the Seyferts are dominated by AGN emission, which is almost point-like.
In any case, a better estimate of the bolometric correction will be done using the nuclear fluxes of the 
12MSG that we are planning to measure through the Spitzer IRAC and MIPS images (Spinoglio et al. 2009, in prep.).

A more serious concern is that not all of the AGN 2's now emitting [NeV] are generating
accretion power right now, i.e., in this decade, even though they evidently did in some previous
centuries.  This is offset to some degree by the missing population 
of ``weak" AGN 2's which are not optically classified as Seyferts because their AGN component produces
less than 50\% of their infrared emission currently (which we call ''non-Seyfert-dominated").  
The existence of these AGN 2's continues to complicate many discussions about 
''AGN", since they can leave and enter this category depending on details
of the detection observations.
But since the 14.32$\mu$m luminosities of all these AGN 2's 
are well below the characteristic ''knee" in the luminosity function, their inclusion or exclusion
should not alter the luminosity integral substantially.

We compare the derived accretion power with the power which originates in the starburst component
of our Seyfert galaxies. We follow a similar procedure by using the [NeII]12.81$\mu$m vs starburst luminosity
relation, as taken from the starburst percentage of the sample galaxies from Table \ref{tbl-8}: 

$$ Log L^{SB}_{19\mu m} = 0.9897 \times Log  L([NeII]12.81\mu m) + 2.0198 $$

We derive a monochromatic power at 19$\mu$m due to star formation and stellar evolution in Seyfert galaxies of 
2.3 $\times$ 10$^{45}$ erg s$^{-1}$, which corresponds to about 1/10 of the accretion power.
This is quite reasonable, since the host galaxies of unambiguous Seyfert nuclei are only about one ninth of the
12 Micron Galaxy survey. In summary, because of their rarity, the set of galaxies we classify as AGN, namely those
producing NeV line emission,  generate
only a few percent of the total fusion power in the local universe. 

\section{Discussion and conclusions}

Our large sample of Seyfert galaxies, and the accompanying
spectropolarimetric classifications
have improved and extended the analysis of Paper I.
In the IR diagnostic diagrams presented, we find that AGN 1--defined as having broad line emission of
some kind--have high values of ionization-sensitive line ratios,
relative to the strength of star-formation tracers.
In contrast, those galaxies we re-classify as non-Seyfert's
have low ionization-sensitive line ratios and high PAH equivalent widths,
and lie in the opposite regions of the diagnostic diagrams from AGN.
The class of
AGN 2's--those without broad lines--spread across both regions of the diagrams.

The simplest and strongest version of the AGN Unified Model requires that any differences 
between Seyfert 1's and Seyfert 2's should be 
due to the inclination angle of an obscuring structure (e.g. of a torus) with the line of sight
to the observer, the Seyfert 2's being covered by the thick dust in the equator. 
However, in order to salvage this unification model from the results presented here, one has to explain
why AGN 2's show a wider range of mid-IR properties, with
similarities both to AGN 1 and the non-Seyfert's.
Perhaps this could 
derive from the particular geometry of the "torus-BLR-scattering region":
the torus in fact can absorb the radiation not only from the BLR but also from the scattering 
region where the polarized emission could be produced, 
depending, for example,  from the precise inclination angle,
or the intrinsic clumpiness of the torus.
That is the basic suggestion of \citet{hei97}.
However, to suppress high-ionization emission lines we observe in the IR--
which are assumed to be isotropic--
the absorbing structure would also need to block out a significant
fraction of the ionizing photons that would otherwise
illuminate the NLR, even though this does not seem to happen
in Seyferts with any detectable BLR.  But once additional intrinsic
differences other than just the viewing angle are admitted, the
attractive simplicity of ATM unification is lost.

On the contrary, however, there may be ''genuine'' type 2 Seyfert nuclei.  These would be the half
of our AGN2 category which have smaller proportions of AGN emission, and do not have analogs among
the Seyfert 1s.  As we discussed, the remaining NLR emission we detect in the optical and
infrared spectra of these true Sy2's may be the ''fossil'' remnants of a 10--100 parsec extent
of gas that had been ionized by an AGN which was active in previous millenia, but has been effectively
''turned off'' for the last several hundred years.  Given their strong variability over all
timescales, AGN which recently ''turned off'' must exist. Depending
on the duty cycles of power generation through black hole accretion, they could account for up to half
of our AGN2.  This would then imply the existence of a comparable population of ''recently turned on''
AGN.  These could stand out as having unusually strong broad line and near-IR continuum emission, 
relatively to their NLR strength.  The most likely candidates for these ''young'' or more accurately,
''rejuvenated'' AGN1's are those Seyfert 1's having relatively strong FeII emission and low values of
NLR/BLR ratios such as [OIII]5007/H$\beta$ \citep{bg92}. 
Thus it could well turn out that long-term variability is at least as important as viewing angle
in unifying the various observed classes of AGN.

\begin{itemize}
\item[-]The main results of this paper are as follows:
We present  the {\it Spitzer} IRS high-resolution spectra of
almost 80\% of the Seyfert galaxies of the 12$\mu$m galaxy sample, a total of 91 galaxies;
\item[-] We adopted a spectropolarimetric classification, with the "AGN 1" class  broadly defined to include both the Seyfert 1's 
and the "hidden broad line Region" Seyfert 2's, as detected through optical spectropolarimetry.
All of our infrared diagnostics are consistent with Sy 1's and HBLR Sy 2's belonging to the same single class.
The AGN 2 class contains the
remaining Seyfert 2 galaxies without polarized broad lines are likely a mixture of weaker
AGN 1, in which the BLR exists but has not yet been detected, and "true" AGN 2, galaxies that may not
have been producing much hard ionizing radiation for the last several hundred years.
Our AGN 1/2 distinction, based solely on reddening-independent IR data, is
supported by more of the data than the usual traditional classification scheme that divides the Seyfert 1's and 2's
based strictly on the detectability of BLR emission in direct optical spectroscopy.
It appears that the mid-IR observed properties characterize the AGN 1 as an homogeneous class of objects.
The mid-IR behaviour of AGN 2's, instead, shows objects similar to AGN 1's and others 
more similar to non-Seyfert galaxies.

\item[-] Semi-analytic models based on the observed mid-IR spectra are effective in separating the AGN
and starburst components in Seyfert galaxies. We find that for 31 AGN 1 the average AGN percentage
contribution to
the 19$\mu$m luminosity is 92.2$\pm$6.6\%, while for 16 AGN 2 this percentage decreases to 79.3$\pm$15.7\% and
for 9 non-Seyfert's is 67.6$\pm$17.2\%. 
Although with large scatter, there is a clear trend of decreasing AGN 
strength from AGN1 to AGN2 to non-Seyfert's. 
Ionization-sensitive line ratios can discriminate AGN 1's from AGN 2's and non-Seyferts. 
The diagnostic diagram of [NeV]/[NeII] versus [OIV]/[NeII] provides a measure of the AGN strength in both 
AGN 1's and AGN 2's.
The AGN 2 do not appear to be a homogeneous population with respect to 
starburst tracers, such as the EW PAH and the EW of [NeII].
Moreover, some AGN 2 nuclei could either contain more ongoing star formation and/or be covered by more 
dust than any of the AGN 1's.

\item[-] The 0-0 S(1) H$_2$ rotational transition can be used to estimate the mass of the
emitting regions. The average mass value is of the order of 10$^8$ M$_{\sun}$, which is in agreement with
other estimates for Seyfert and starburst galaxies.

\item[-] The density-sensitive line ratios (the [NeV] and [SIII] doublet ratios), 
in about 40\% of the objects imply electron densities below the low density limit. Therefore they
can not be used reliably to  estimate the electron density of these regions. 
The simple interpretation  that the line emission from the NLR and the HII regions 
can be heavily absorbed by dust even in these mid-IR lines \citep{dud07} is not confirmed from our data
because the galaxies of our sample {\it do not} show both line ratios affected at the same time in the same objects.
In addition, the abnormally low ratios occur with the same frequency in all different types of emission line
galaxies.  The other possibility is that some assumed atomic physics parameters need revision for both of these
line ratios.
\item[-] We derived for the first time the line luminosity functions for either classical Seyfert 1's 
and 2's and also for AGN 1 and AGN 2. We do not find significant differences between the various 
populations, in either the shape or the normalization of their line LFs. 

\item[-] The mid-infrared [NeV] lines are unambiguous tracers of the AGN NLR.
We therefore use the [NeV] line luminosity function of all Seyfert galaxies to estimate the accretion power in
the local universe within a volume out to z=0.03. We find that the power originating from accretion at 19$\mu$m is 
$\sim$ 2 $\times$ 10$^{46}$ erg s$^{-1}$ , about 4 times less than the bolometric power. For comparison,
the power related to star formation and stellar evolution in the Seyfert galaxies population
at 19$\mu$m is one tenth of that one from accretion. 

\end{itemize}

\acknowledgments

This work is based on observations made with the Spitzer Space Telescope 
which is operated by the Jet Propulsion Laboratory and Caltech under a 
contract with NASA. 
This research has been funded in Italy by ASI under contract I/05/07/0.
We benefited from helpful discussions with R. Antonucci.
We especially thank Kevin Hainline for help with early stages of the data reduction,
and Howard Smith and members of the IRAC team for contributing 
Guaranteed Time to obtaining an essential portion of these data. We thank
Jeronimo Bernard-Salas for having sent us the data of the average Starburst high-resolution
Spitzer spectrum. We thank the anonymous referee for the useful comments and suggestions. 
We thank Mrs. Erina Pizzi of IFSI-INAF 
for the preparation of the postscript figures of the article.  This research has been funded in Italy
by ASI under contract I/05/07/0 and in the U.S. by NASA contract 59586.


\begin{appendix}
%\appendix

\section{A. Classification: notes on individual objects}\label{app.A}

\begin{itemize}

\item[]{\bf NGC526A}: has no detected broad lines, according to new data from \citet{ben}, 
however broad H$\alpha$ wings were observed more than 20 years ago (Lawrence 2009, priv. com.).
Because of the possible changes in BLR characteristics over timescaales of the order of years, 
we conservatively classify this object as a AGN1.
\item[]{\bf MRK1034 NED1 = Akn80}: this has not a Seyfert like optical spectrum 
\citep{op}. 
We classify it as a non-Sy (IRS spectrum in paper I).
\item[]{\bf MRK1034 NED2 = Akn81}: only narrow lines, slightly wider than the instrumental resolution \citep{op}. 
We classify it as a non-BLR as no polarimetric observations are available (IRS spectrum in paper I).
\item[]{\bf ESO545-G13 = MCG-03-07-011 = MBG 02223-1922}: detected broad H$\alpha$ and H$\beta$ 
$\sim$ 2000 km/s \citep{coz}.
We classify this object as an AGN1 (IRS spectrum in paper I).
\item[]{\bf NGC1056}: this object has been classified as an HII region galaxy 
through optical spectroscopy by \citet{vei95}. 
\item[]{\bf NGC1097}: detected broad H$\alpha$ $\sim$ 10000 km/s FWHM \citep{sb93}.
On the basis of this BLR evidence, we classify this object into the AGN1 
class.
\item[]{\bf NGC1194}: only narrow lines were detected: H$\alpha$ $\sim$230 km/s FWHM and 
[OIII]5007\AA$\sim$400 km/s FWHM \citep{ks90}.
Because no polarization data are available, we classify it as a non-BLR object, but we
are unable to include it in either AGN1 or AGN2 classes.
\item[]{\bf NGC1365}: detected broad H$\beta$ of 1896 km/s FWHM \citep{sh99}.
On the basis of the BLR evidence,  we classify this object into the AGN1 
class. Note: [NeV] is detected.
\item[]{\bf ESO362-G018 = MCG-05-13-017}: H$\beta$ has a witdh of 5240 $\pm$ 500 km/s FWHM, therefore 
is a genuine AGN1 \citep{ben}. 
\item[]{\bf NGC2639}: detected broad H$\alpha$ of 3879 km/s FWZI \citep{ke83}.
On the basis of the BLR evidence, we classify this object into the AGN1 
class.
\item[]{\bf NGC3511}: only narrow lines were detected: H$\alpha$ $\sim$ 135 km/s FWHM, this galaxy is 
classified as HII region galaxy \citep{ks90}.
We classify this object as a non-Sy. 
\item[]{\bf MCG+00-29-23}: this object has been classified as an HII region galaxy 
through optical spectroscopy by \citet{vei95}.
\item[]{\bf NGC4602}: detected only marginal evidence for an H$\alpha$  broad component with 
FWZI$\sim$10000 km/s \citep{kf}. 
We classify it as a non-BLR, as no polarimetric observations are available (IRS spectrum in paper I).
\item[]{\bf NGC4922}: this object has been classified as intermediate between a LINER and an HII region galaxy 
through optical spectroscopy by \citet{vei95}.
\item[]{\bf NGC5005}: this object is classified as a LINER galaxy \citep{gk86,ter00}.
\item[]{\bf NGC5506}: observed in spectropolarization: no broad H$\alpha$ component was 
detected at a level of 1x10$^-15$ cgs \citep{lu04}.
We classify this object as an non-HBLR, and thus AGN2. 
\item[]{\bf NGC6810}: this object is classified as a transition object between an HII region galaxy and an AGN, having an 
[OIII]$\lambda$5007$\AA$ width of 304 km/s FWHM and an H$\alpha$ width of 263 km/s FWHM and the following line ratios:
[NII]/H$\alpha$ = 0.62, [OIII]/H$\beta$ =0.6, [SII]/H$\alpha$ = 0.3 \citep{ks90}. We conservatively classify it an a non-Sy.
\item[]{\bf MRK897}: \citet{rod09} confirm the classification of this galaxy as a Seyfet 2 of \citet{dur}.
Because of the lack of polarization data, we classify it as a non-BLR (IRS spectrum in paper I).
\item[]{\bf NGC7130 = IC5135}: this object is classified as a LINER\citep{vei95}.
\item[]{\bf NGC7314}: detected in spectropolarization at H$\alpha$ : F6563(broad)=5.6x10$^-15$ cgs \citep{lu04}.
We classify this object as an HBLR, and thus AGN1.
\item[]{\bf NGC7496}: the relative strength of the emission lines in the nucleus of this object  is typical
of normal photoionization found in HII regions \citep{pen84} (IRS spectrum in paper I).
\item[]{\bf NGC7590}: the relative strength of the emission lines in the nucleus of this object  is typical
of normal photoionization found in HII regions \citep{sto95} (IRS spectrum in paper I).
\item[]{\bf CGCG381-051}: this object has been classified as an HII region galaxy 
through optical spectroscopy by \citet{deg}.
\end{itemize}

\section{B. Peculiarities of the "20$\mu m$ peakers"}\label{20peak}

We have considered the "20$\mu m$ peakers" identified by Wu et al 2009 in section 2 and
computed the average values of the discussed quantities in the classification Tab. \ref{tbl-6}.
We find in our sample fifteen "20$\mu m$ peakers": 11 Sy1's, 3 HBLR's and 1 AGN2.
The major peculiarities of this group of objects are that: they are almost all compact
at 19$\mu m$, they have very faint PAH 11.25$\mu m$ emission, a flat average spectral index (60-25)$\mu m$,
and low equivalent widths of the lines (see Tab. \ref{tbl-6}).

In particular, among them 11
are compact sources (belonging to the extendedness class I), 1 belongs to class II, 3 do
not have background subtracted spectra and therefore no extendedness measurement.
7 of them do not show any PAH feature at 11.25 $\mu m$ and
the mean PAH EW 11.25 $\mu m$ among the other 8 objects is -0.04$\pm$0.04$\mu m$.
The mean EW of the lines and the feature in Tab. \ref{tbl-6} are consistent with the lowest values for the AGN1.
On the other hand, the AGN contribution is among the highest values in the AGN1's, as it
can be seen from the mean [NeV]14.32$\mu m$/[NeII]12.81$\mu m$ ratio, the mean spectral index and
the average modelled AGN contribution.    

\section{C. Mathematical treatment of the models}\label{mod}

We present here the details of the semi-analytic models.
To model the observed IR quantities, we have found an analytical expression for each of them as a 
function of $\Re$, which is defined as the ratio of the Starburst dust continuum at 19$\mu$m to the AGN flux 
at the same wavelength. $\Re$ varies from zero - emission totally from the AGN - to infinity - 
emission totally from the Starburst.

\begin{equation}
\Re = \frac{F^{gal}_{19\mu m}}{F^{AGN}_{19\mu m}}
\end{equation}

The extendedness is defined as the ratio between the flux measured at 19.5$\mu$m 
from the LH slit to that measured at 19$\mu$m by the SH slit. 
%\textbf{
If we define $F^{gal}_{19\mu m}$ and $F^{AGN}_{19\mu m}$ as the fluxes of galaxy and AGN, respectively, through the SH slit, considering
that at maximum the LH slit can intercept 4 times the $F^{gal}_{19\mu m}$ and that $F^{AGN}_{19\mu m}$ does not depend
on the slit aperture, as can be approximated to a point-like source, then: 
Flux(LH) = $F^{AGN}_{19\mu m} + 4\cdot F^{gal}_{19\mu m}$; Flux(SH) = $F^{AGN}_{19\mu m} + F^{gal}_{19\mu m}$. %}

\begin{equation}
ext=\frac{F^{AGN}_{19\mu m} + 4\cdot F^{gal}_{19\mu m}}{F^{AGN}_{19\mu m} + F^{gal}_{19\mu m}}
=\frac{1 + 4\cdot \frac{F^{gal}_{19\mu m}}{F^{AGN}_{19\mu m}}}{1 + \frac{F^{gal}_{19\mu m}}{F^{AGN}_{19\mu m}}}=\frac{1 + 4\cdot \Re}{1 + \Re}
\end{equation}

We know that the PAH emission feature at 11.25$\mu$m originates in the host galaxy only, 
while its underlying continuum derives from both galaxy and AGN. 
Therefore we can predict its equivalent width as:

\begin{equation}
EWPAH=\frac{A_1 \cdot F^{gal}_{11.25\mu m}}{F^{gal}_{11.25\mu m} + F^{AGN}_{11.25\mu m}}\label{eq1}
\end{equation}

where A$_1$ is a costant depending on the measured features flux and we adopt A$_1$=1.8 for the PAH at 11.25$\mu$m.
From the assumption that we can model the continua with a power law:

\begin{equation}
\nu ^{-\alpha} F_{\nu}=k\label{eq2}
\end{equation}

where $\alpha$ is the spectral index, $k$ is a constant. We adopt $\alpha$=-1.4 for the AGN component 
and $\alpha$=-2 for the  galaxy, these values are in agreement with the values commonly used 
as spectral index for AGN and starburst and availble in literature (cf \citet{em86}).
We derive the F$^{gal}_{11.25\mu m}$ and the F$^{AGN}_{11.25\mu m}$ as functions 
of F$^{gal}_{19\mu m}$ and F$^{AGN}_{19\mu m}$, respectively,
and consequentely as functions of the free parameter $\Re$. From equations \ref{eq1} and \ref{eq2},
the PAH EW becomes:

\begin{equation}
EWPAH=\frac{1.8 \cdot (\frac{k/19\mu m}{k/11.25\mu m})^{(-2 + 1.4)}\cdot \Re}{(\frac{k/19\mu m}{k/11.25\mu m})^{(-2 + 1.4)}\cdot \Re + 1}
\end{equation}

The [NeII] EW has contributions both from the galaxy and the AGN:

\begin{equation}
EW[NeII]=\frac{A_2 \cdot F^{gal}_{12.81\mu m} + B_2 \cdot F^{AGN}_{12.81\mu m}}{F^{gal}_{12.81\mu m} + F^{AGN}_{12.81\mu m}}
\end{equation}
 
Because the galaxy component is stronger than the AGN's in the [NeII] emission, 
we chose the constant A$_2$ two orders of magnitude greater than B$_2$, 
adopting A$_2$=0.25 and B$_2$=0.001. 

The procedure for modeling the line ratios is similar. We modelled the emission of each 
single line and then we simply took their ratio:

\begin{equation}
[NeV]/[NeII]=\frac{A_3 \cdot F^{gal}_{14.32\mu m} + B_3 \cdot F^{AGN}_{14.32\mu m}}{A_2 \cdot F^{gal}_{12.81\mu m} + B_2 \cdot F^{AGN}_{12.81\mu m}},
\end{equation}

\begin{equation}
[OIV]/[NeII]=\frac{A_4 \cdot F^{gal}_{25.89\mu m} + B_4 \cdot F^{AGN}_{25.89\mu m}}{A_2 \cdot F^{gal}_{12.81\mu m} + B_2 \cdot F^{AGN}_{12.81\mu m}},
\end{equation}

with A$_3$ = 0 and B$_3$ = 0.01 and A$_4$ = 0 and B$_4$ = 0.01, because we consider
the [NeV]14.32$\mu m$ and [OIV]25.89$\mu m$ lines to be produced exclusively by the AGN.

To model the $\alpha_{(60-25)\mu m}$ spectral index,  
we use its definition as: 

\begin{equation}
\alpha_{(60-25)\mu m}=\frac{log(\frac{F_\nu (60\mu m)}{F_\nu (25\mu m)})}{log(\frac{\nu_{60\mu m}}{\nu_{25\mu m}})},
\end{equation}

writing F$_\nu (60\mu m)$ and F$_\nu (25\mu m)$ in terms of F$^{gal}$ and F$^{AGN}$,
considering that F$_\nu (60\mu m)$ is due mainly to the galaxy emission, it becomes:

\begin{equation}
\alpha_{(60-25)\mu m}=\frac{log(\frac{F^{gal}_{60\mu m} + 0.001F^{AGN}_{60\mu m}}{F^{gal}_{25\mu m} + F^{AGN}_{25\mu m}})}{log(\frac{\nu_{60\mu m}}{\nu_{25\mu m}})}.
\end{equation}

The treatment of the previous equations is the same of the EW PAH model's. %\textbf{
All the factors A$_i$ and B$_i$ have been found
by fitting the data with the models and they depend on the instrument (spectrograph properties, slits areas, 
spectral extraction methods). These same models can be suitable for other observables and instruments, by scaling those factors.%}

\section{D. PAH 11.25$\mu m$ EW - Low/High resolution comparison}\label{Wu}

%\textbf{
%To compare the PAH 11.25$\mu m$ EW that we measured with those measured by \citet{wu09}, 
We plot in Figs. \ref{fig22}a and b the PAH 11.25$\mu m$ EW measured by \citet{wu09} versus that one measured by us,
to compare the results on the objects in common. In Fig. \ref{fig22}a we present the ratio of the values found 
by us to the values of \citet{wu09} to better show the scatter of the two sets of data. 
In Fig. \ref{fig22}b we plot the PAH 11.25$\mu m$ EW by \citet{wu09} versus ours and we computed their least square fits, 
calculated the mean slopes and their confidence 
intervals by using the bootstrap method (by 1000 resampling). The two sets of data are consistent with each other, 
as it is shown by the slope of the lines reproducing the least square fits of the data: 
considering all the galaxy types, the slope of the fitting line is 1.14$\pm$0.17; if we
consider only the AGN1 and AGN2 the slope is 0.97$\pm$0.18. These two results agree within the errors.
However, the fitting line slope becomes steeper when we add the non-Sy's data, because for these 
we used a different method to measure the PAH emission feature at 11.25$\mu m$. 
\citet{wu09} remove the continuum underlying the range 10.80-11.80$\mu m$ 
and integrate the PAH feature flux inside this window, in each of the sources. 
On the other hand, we measure the flux in the range 11.15-11.65$\mu m$ for all the sources, 
but we subtract the continuum underlying the whole range 11.2-13.0$\mu m$ for
those sources having in this range a broad multicomponent feature. 
This results in a lower continuum baseline measure and 
therefore in a higher PAH feature equivalent width. %}

\section{E. Additional Observational Measurements}\label{aom}

\subsection{Weak ionic fine structure lines}
Besides the ionic lines we presented in \S\ref{obs}, the spectra of seven sources
showed other fine structure lines, e.g. [ArV]13.10$\mu m$, [FeII]25.99$\mu m$ and 24.81$\mu m$,
[FeIII]22.95$\mu m$ and 33.04$\mu m$. The [FeII]25.99$\mu m$ line is often blended with 
the [OIV]25.89$\mu m$ line, because the IRS spectral resolution is not high 
enough to resolve these two lines in all the objects.
These seven galaxies and their respective lines fluxes are reported in Tab.\ref{tbl-12}.  

There are unfortunately not enough good detections of these lines in this sample to base
strong statistical tests on them.
We have nevertheless compared our [FeII]25.99$\mu m$ detections with those of \citet{dal09} 
in the line ratio diagram of [SiII]34.8$\mu$m/[SIII]33.5$\mu$m vs [FeII]25.99$\mu m$/[NeII]12.8$\mu$m.
Although the small number of points, our measurements are in agreement with theirs, as the few AGN 2's
are located in the region of the Seyfert's of \citet{dal09} and the non-Sy are located among their 
normal spiral galaxies.

\subsection{Full Width Half Maximum of the spectral lines}

The lines FWHM are due to the velocity field of the gaseous regions where they are produced. 
The velocity dispersion $\sigma$ of the [OIII]  at 5007\AA, originated in the 
Narrow Line Regions, is related to the black hole mass \citep{nel00,gh05}. 
It is interesting to extend the same relation to the mid-infrared lines of [NeV] at 14.32$\mu$m, 
[NeV] at 24.31$\mu$m and the [OIV] at 25.89$\mu$m, because they suffer less obscuration than 
the optical lines and are mostly originated by the AGN activity, though the [OIV] 
can be contaminated by strong starburst emission.

\citet{das08} claim to have measured resolved spectral widths of the IRS lines.
The IRS resolving power for the high resolution modules is on average (500 +/- 50) 
km/sec along the whole spectrum. We find the FWHM of all the lines to be, within the errors, 
of the same value of the resolving element and to have a scatter of about the 15$\%$ of 
the mean value (the mean value and the standard deviation of the FWHM of all the fine 
structure lines are reported in Tab.\ref{tbl-13}). We cannot conclude that the lines are 
resolved, because we consider a line to be resolved if its FWHM is significantly 
larger than the resolution element. While \citet{das08} consider the 
instrumental velocity dispersion (the one corresponding to the resolving power 
of the instrument) as the ratio between resolution element and 2.35, we compare 
the measured FWHM with the resolution element itself, to be more conservative.    

Moreover, both the typical AGN lines and those originated by starburst emission have nearly
the same mean FWHM (see Tab.\ref{tbl-13}), which does not change with the Seyfert or galaxy type. 
The AGN lines, originated from the NLR, should suffer from more broadening than 
the lines emitted by the galactic 
regions. We conclude that the broadening of the lines is not intrinsic to the gas motion, 
but most probably is due to instrumental effects, otherwise we would measure different widths 
from lines originated from the different regions.

In conclusion, we consider that the FWHM of the lines measured in the IRS spectra do not 
measure the dispersion velocity field in the NLR, and higher spectral resolution 
data would be necessary. 
\end{appendix}




\clearpage
%table 1
\LongTables
\begin{landscape}
%\begin{turnpage}
\begin{deluxetable}{lccclccclccc}
\tabletypesize{\scriptsize} %\rotate
\tablewidth{0pt} 
\tablecaption{Journal of Spitzer IRS observations\label{tbl-1} } 
\tablehead{ & \colhead{R.A.} & \colhead{Dec.}  &  & & &  &  &  & {SH } & {LH}  \\
\colhead{NAME}& (J2000.0) & (J2000.0)    & \colhead{type} & \colhead{new class.}  & \colhead{z}  & \colhead{F$_{12\mu m}$}  &  \colhead{F$_{25\mu m}$} & \colhead{Obs. Date}  &  Int.time  & Int.time \\
& h~m~s    &  $\deg$ ~'~" &  & &  &   (Jy)    &   (Jy)      &    &  (sec.)  & (sec.)}
\startdata
MRK335          &  00:06:19.5   & +20:12:10    &  Sy1& & 0.025785  &  0.27  &  0.45  & 2009/01/12  &   4$\times$30 & 2$\times$60&3\\ 
MRK938          &  00:11:06.5   & -12:06:26    &  Sy2&non-HBLR  & 0.019617  &  0.40  &	2.37  & 2006/12/20  &   2$\times$30 & 4$\times$14&\\
MRK348          &  00:48:47.1   & +31:57:25    &  Sy2&HBLR  & 0.015034  &  0.49  &	1.02  & 2007/09/01  &   2$\times$30 & 4$\times$14&\\
NGC0526A        &  01:23:54.4   & -35:03:56    &  Sy1 &non-BLR  & 0.019097  &  0.23  &   0.48  & 2007/08/01   &   6$\times$7  & 4$\times$14&2,4 \\
IRAS01475-0740  &  01:50:02.7   & -07:25:48    &  Sy2&HBLR  & 0.017666  &  0.31  &	0.96 & 2007/08/04  &   4$\times$30 & 2$\times$60&\\
NGC1056         &  02:42:48.3   & +28:34:27    &  Sy2&non-Sy  & 0.005154  &  0.33  &	4.83  & 2007/09/02  &   4$\times$30 & 2$\times$60&\\
NGC1097  	&  02:46:19.0   & -30:16:30    &  Sy1& & 0.004240  &  2.88  &	7.70  & 2004/01/08  &   4$\times$30 & 2$\times$60&1,2,4\\
NGC1142         &  02:55:12.2   & -00:11:01    &  Sy2&non-HBLR  & 0.028847  &  0.26  &   0.62  & 2006/02/02  &   4$\times$120& 6$\times$60&2\\
\enddata
\tablenotetext{*}{NOTES ON PROGRAMS: 1: Data without off-source measurements; 2: Data from Spitzer Archive; 3: Data from P50253 to be released; 4: see Appendix \ref{app.A} for the source classification.
}
%\tablenotetext{}{}
%\tablenotetext{1}{}
%\tablenotetext{2}{}
\end{deluxetable}
\clearpage
\end{landscape}
%\end{turnpage}

%table2

\clearpage
\LongTables
%\begin{turnpage}
\begin{landscape}
\begin{deluxetable}{lccccccccccc}
\tabletypesize{\scriptsize} %\rotate
\tablecaption{Fine Structure Lines in our Sample\label{tbl-2}}\tablewidth{0pt} 
\tablehead{& \multicolumn{5}{c}{Line fluxes (10$^{-14}~erg~s^{-1} cm^{-2}$) in SH} && \multicolumn{5}{c}{Line fluxes (10$^{-14}~erg~s^{-1} cm^{-2}$) in LH}\\
 \cline{2-6}\cline{8-12}\\
\colhead{NAME} &
\colhead{[SIV]} & \colhead{[NeII]} & \colhead{[NeV] } & \colhead{[NeIII]}&
\colhead{[SIII]} && \colhead{[SIII]} & \colhead{[NeV]} & \colhead{[OIV]} & \colhead{[SIII]} &
\colhead{[SiII]}\\
 &(10.51$\mu m$) &(12.81$\mu m$) &(14.32$\mu m$)&(15.56$\mu m$)&(18.71$\mu m$)&&(18.71$\mu m$)&(24.32$\mu m$)&(25.89$\mu m$)& (33.48$\mu m$)&(34.82$\mu m$)}
\startdata
MRK335     & 0.43$\pm$0.03 &  0.25$\pm$0.05   &   0.38$\pm$0.04 & 0.61$\pm$0.04 & 	\nodata  &&  $<$0.62 & 1.97$\pm$0.18  &  7.24$\pm$0.19 & $<$1.21 & $<$1.45 \\   
MRK938          &     $<$1.46     &  52.1$\pm$1.45   &   $<$2.19       &  6.37$\pm$0.55 & 7.56$\pm$0.64 &&   \nodata       &    $<$0.37    &  $<$0.66   	&  $<$10.7       & 40.5$\pm$4.21 \\ 
MRK348		&  7.16$\pm$0.43   &  16.4$\pm$0.35   &   5.82$\pm$0.35	 &  20.4$\pm$0.38 & 7.00$\pm$0.55 &&  \nodata 	     &  4.95$\pm$0.36 &	 17.6$\pm$0.42	&  12.2$\pm$1.12  & 9.81$\pm$1.21\\ 
NGC526A         & 5.32$\pm$0.58    &  5.77$\pm$0.52   &  6.35$\pm$0.45   &  10.4$\pm$0.71 & $<$2.50 &&  \nodata        &  5.92$\pm$0.29 &  19.3$\pm$0.44  &  5.91$\pm$1.05  &  8.58$\pm$1.19\\
IRAS01475-0740  &  2.14$\pm$0.32   &  13.7$\pm$0.30   &   6.38$\pm$0.32  &  9.95$\pm$0.34 & $<$4.55 &&  \nodata 	     &  1.87$\pm$0.29 &  6.49$\pm$0.31  &  3.12$\pm$0.83  & $<$6.13 \\   
NGC1056         & $<$1.31         &  33.6$\pm$1.04   &   $<$1.80       &  10.4$\pm$0.34 & 18.3$\pm$0.35 &&  \nodata        &  $<$1.23      &  1.40$\pm$0.30  &  36.6$\pm$0.69  &49.1$\pm$0.97 \\ 
NGC1097         & $<$0.85          &  59.1$\pm$0.22   &   $<$0.92        &  7.08$\pm$0.19 & 22.5$\pm$0.39 &&   \nodata       &  $<$2.31       &  6.20$\pm$0.75      &  131$\pm$2.47 &254$\pm$1.25\\
NGC1142		& 1.43$\pm$0.11    &  17.2$\pm$0.11   &   0.92$\pm$0.09  &  5.40$\pm$0.10 & 8.32$\pm$0.12 &&  24.7$\pm$0.25  &  1.82$\pm$0.13 &  5.31$\pm$0.18  &  32.2$\pm$0.67  & 67.0$\pm$0.96\\
\enddata
\end{deluxetable}
%\end{turnpage}
\clearpage
\end{landscape}
								
%table3								
%\clearpage

\begin{deluxetable}{lccccccccc}
\tabletypesize{\scriptsize} %\rotate
\tablecaption{Molecular Hydrogen lines and PAH emission feature at 11.25$\mu$m\label{tbl-3}}\tablewidth{0pt} 
\tablehead{& \multicolumn{5}{c}{Line fluxes (10$^{-14}~erg~s^{-1} cm^{-2}$)} & \\
 \cline{2-6}\\
%\tablehead{ 
\colhead{NAME} &
\colhead{H$_2$ S(3)} & \colhead{ H$_2$ S(2)} & \colhead{H$_2$ S(1)} & \colhead{H$_2$ S(0)}&
\colhead{PAH} & \colhead{EQ.W.} \\
 &(9.67$\mu m$) &(12.28$\mu m$) &(17.04$\mu m$)&(28.22$\mu m$)&(11.25$\mu
 m$)&($\mu m$)\\
  & (1)       & (2)                & (3)             & (4)               & (5)                &(6)  }
\startdata
MRK335         &   \nodata 	 & $<$0.16  &  0.39$\pm$0.05 &   $<$0.72	      &	\nodata   & \nodata \\		
MRK938                  &   \nodata	 & 2.72$\pm$0.38   &  7.06$\pm$0.65 &   $<$1.02      &     292&  -0.687 \\		
MRK348                  &   \nodata	 &    $<$1.05	   &  1.77$\pm$0.48 &	$<$1.14	     & \nodata& \nodata\\		
NGC526A                 &   \nodata      &   $<$1.65       &  2.40$\pm$0.77 &   $<$1.41	     &\nodata & \nodata\\		
IRAS01475-0740          &   \nodata	 &    $<$1.20      &  2.75$\pm$0.30 &   $<$2.18      &	 30.3 & -0.071   \\		
NGC1056                 &   \nodata	 &    $<$2.46      &  3.58$\pm$0.48 & 2.25$\pm$0.27  &    214 &-1.196   \\		
NGC1097                 &   $<$1.69 	 &  6.95$\pm$0.55  &  14.4$\pm$0.24 & 4.95$\pm$0.94  & 285    &  -0.911 \\		
NGC1142                 & 1.88$\pm$0.10	 & 1.09$\pm$0.10   &  4.60$\pm$0.08 &	2.76$\pm$0.21&   73.5 & -0.707  \\
\enddata
\tablenotetext{*}{Columns (1) to (4) give the line fluxes of the H$_2$ rotational lines in  units of 10$^{-14}~erg~s^{-1} cm^{-2}$.
Column (5) and (6) give the flux and the equivalent width of the PAH 11.25 $\mu$m emission feature, respectively.\\ 
}
\end{deluxetable}

\clearpage
%table 4

\begin{deluxetable}{lcccrrr}
\tabletypesize{\scriptsize} %\rotate 
\tablecaption{H$_2$ Estimated Temperatures and Masses \label{tbl-4}} \tablewidth{0pt}
\tablehead{\colhead{NAME} &  \colhead{T(2-3)}  &  \colhead{T(1-2)} & \colhead{T(0-1)} & \colhead{H$_2$ Mass} \\
                          & (K)                &       (K)     &  (K)                  &  (10$^{8}$ M$_\sun$) }
\startdata
MRK938		&	\nodata	    &       304 &    $>$404  &  $<$1.47    \\
MKN348		&	\nodata	    &    $<$379 &    $>$184  &	$<$0.51    \\
NGC526A		& 	\nodata	    &    $<$414 &   190      & 	   1.05    \\
IRAS01475-0740	&	\nodata	    &    $<$322 &    $>$171  &	$<$1.32    \\
NGC1056		&	\nodata	    &    $<$413 &   186      &	   0.12    \\
NGC1097		&            $<$195 & 	    339 &   206      & 	   0.27    \\
NGC1142		&	363 	    &	    249 &   189      &	   4.65    \\
MCG-02-08-039	&	\nodata	    &	    432 &    $>$198  &	$<$1.23  \\
\enddata
\end{deluxetable}

% table5
\clearpage

\begin{deluxetable}{lcccccc}
\tabletypesize{\scriptsize} %\rotate 
\tablecaption{Photometric fluxes and extendedness factors \label{tbl-5}} \tablewidth{0pt}
\tablehead{\colhead{NAME} &  \colhead{F$_{19\mu m}$(SH)}  &  \colhead{F$_{19.5\mu m}$(LH)} & \colhead{R} & \colhead{Class} \\
                                          & (Jy)                       &       (Jy)                &      &      }
\startdata
MRK938			   &	    0.742$\pm$0.006&	  0.885$\pm$0.009  &    1.19    & II  \\  
MRK348        	           &	    0.589$\pm$0.007&	  0.647$\pm$0.003  &    1.10    &  I \\		
NGC526A       	           &	    0.363$\pm$0.017&	  0.361$\pm$0.003  &    0.993   &  I \\	
IRAS01475-0740		   &	    0.462$\pm$0.003&	  0.507$\pm$0.004  &    1.1     &  I   \\  
NGC1056       		   &	    0.130$\pm$0.006&	  0.209$\pm$0.001  &    1.61    & III   \\  
NGC1142			   &	    0.096$\pm$0.001&	  0.184$\pm$0.015  &    1.92    & III   \\   
MCG-02-08-039  		   &	    0.356$\pm$0.005&	  0.379$\pm$0.003  &    1.06    &  I  \\  
NGC1194			   &	    0.336$\pm$0.007&	  0.427$\pm$0.004  &    1.27    & II   \\
\enddata
\end{deluxetable}

\clearpage

\begin{deluxetable}{llcccccc}
\tabletypesize{\scriptsize} 
\tablecaption{CLASSIFICATION TABLE\label{tbl-6}} \tablewidth{0pt}
\tablehead{&\colhead{Sy1} &  \colhead{HBLR}  &  \colhead{AGN1} & \colhead{AGN2}& \colhead{non-Sy}& \colhead{"20$\mu$m peakers"}}
\startdata
Extendedness &1.17$\pm$0.29 & 1.17$\pm$0.22& 1.17$\pm$0.26&  1.39$\pm$0.45& 1.54$\pm$0.58 & 1.08$\pm$0.08\\
EW PAH 11.25$\mu m$&  -0.15$\pm$0.20 & -0.16$\pm$0.19 & -0.15$\pm$0.19 & -0.35$\pm$0.27 & -0.81$\pm$0.42& -0.04$\pm$0.04\\
$\frac{[NeV]}{[NeII]}$& 0.60$\pm$0.51 & 0.73$\pm$0.53 & 0.65$\pm$0.52 & 0.66$\pm$0.84& 0.10$\pm$0.03& 1.03$\pm$0.87\\
EW[NeII]12.81$\mu m$& -0.028$\pm$0.027 & -0.028$\pm$0.026 & -0.028$\pm$0.026 &-0.081$\pm$0.057 &-0.116$\pm$0.029& -0.014$\pm$0.015\\
EW[OIV]25.89$\mu m$& -0.122$\pm$0.127 & -0.082$\pm$0.069 & -0.116$\pm$0.134 &-0.093$\pm$0.054 &-0.033$\pm$0.035 & -0.075$\pm$0.047\\
EW[H$_2$]17.07$\mu m$& -0.009$\pm$0.008 & -0.008$\pm$0.006 & -0.009$\pm$0.008 &-0.046$\pm$0.069 &-0.058$\pm$0.062& -0.008$\pm$0.003\\
\%AGN at 19$\mu m$ & $92\%\pm6\%$ & $92\%\pm8\%$ &$92\%\pm7\%$ &$79\%\pm16\%$&$69\%\pm16\%$& $96.8\%\pm1.5\%$\\
$\alpha_{(60-25)\mu m}$ &  -0.84$\pm$0.62 & -0.86$\pm$0.81 & -0.86$\pm$0.74& -1.65$\pm$0.93& -2.19$\pm$0.65& -0.44$\pm$0.65 \\
\enddata
\tablenotetext{*}{Note: the mean values have been computed without considering the upper limits for all quantities
and the zero values for the PAH EW. }
\end{deluxetable}

\clearpage
%table 6

\begin{deluxetable}{lccccc}
\tabletypesize{\scriptsize} %\rotate 
\tablecaption{Equivalent widths of key lines\label{tbl-7}} 
\tablewidth{0pt}
\tablehead{\colhead{NAME} & \colhead{[NeII]12.81$\mu$m} & \colhead{[NeV]14.32$\mu m$} & \colhead{[NeIII]15.56$\mu m$ } & \colhead{H$_2$ 17.02$\mu$m}& \colhead{[OIV]25.89$\mu m$}  \\
             & ($\mu$m)   & ($\mu$m)   & ($\mu$m)   & ($\mu$m)  &  ($\mu$m)   }
\startdata
MRK335            &     -0.002     &       -0.005    &      -0.010    &     -0.005    &      ...   \\
MRK938			  &     -0.106     &	  \nodata    &	   -0.025    &     -0.025	   &	  ...	    \\  
MRK348        	  &     -0.030     &	   -0.008    &     -0.037    &     -0.004	   &	  -0.063   \\  
NGC526A       	  &     -0.017     &	   -0.022    &     -0.040    &     -0.013	   &	  -0.165   \\  
IRAS01475-0740	  &     -0.028     &	   -0.006    &     -0.023    &     -0.008	   &	  -0.023   \\  
NGC1056       	  &     -0.103     &	  \nodata    &	   -0.093    &     -0.025	   &	  -0.007   \\  
NGC1142			  &     -0.141     &	   -0.016    &     -0.063    &     -0.056	   &	  -0.032   \\  
MCG-02-08-039  	  &     -0.011     &	   -0.022    &      -0.030   &     -0.006	   &	   -0.103  \\
\enddata
\tablenotetext{*}{MRK335, F05563-3820, NGC5135, IC4329A, NGC5347 and NGC5506 do not have calculated [OIV] equivalent width, because 
even if they have background subtracted spectra, the continuum in the (19-35)$\mu m$ range results not well calibrated.}

\end{deluxetable}

\clearpage
%table 8

\begin{deluxetable}{lccc}
\tabletypesize{\scriptsize} %\rotate 
\tablecaption{AGN and Starburst Contribution at 19$\mu$m\label{tbl-8}} \tablewidth{0pt}
\tablehead{\colhead{NAME} & \colhead{AGN \%} & \colhead{Starburst \%} & \colhead{error}\\
&&&\\
\hline
\multicolumn{4}{c}{Seyfert 1}}\\
\hline
& & & \\
\startdata
ESO012-G021	 & 85.4     &   14.6     &   4.5	\\
ESO545-G013	 & 78.4     &   21.6     &   13.5	\\
NGC0931 	 & 97.7     &   2.3	   &   0.7	\\
MKR618   	 & 95.6     &    4.4     &  3.4    \\
ESO362-G018	 & 85.9     &   14.1     &   4.3	\\
MRK0006 	 & 98.2     &   1.8	   &   0.9	\\
MRK0009 	 & 97.6     &   2.4	   &   0.7	\\
MKR79		 & 98       &   2	   &   1.1	\\
\enddata        							 
\end{deluxetable}							 

\clearpage
%table 7

\begin{deluxetable}{lccc}
\tabletypesize{\scriptsize} %\rotate 
\tablecaption{S-K Statistic for AGN families\label{tbl-9}} \tablewidth{0pt}
\tablehead{\colhead{Diagram} &  \colhead{Sy1 + HBLR}  &  \colhead{Sy1 + non-HBLR} & \colhead{HBLR + non-HBLR}}
\startdata
F(H$_2$) vs F(PAH) & 75\% & 18\% & 77\%\\
L(H$_2$) vs L(PAH) & 79\% & 88\% & 20\%\\
EW PAH vs $\frac{[NeIII]}{[NeII]}$ & 100\% & 15\% & 39\%\\
EW PAH vs $\frac{[NeV]}{[NeII]}$& 54\% & 23\% & 54\%\\
EW PAH vs $\frac{[OIV]}{[NeII]}$& 40\% & 8\% & 41\%\\
EW PAH vs $\frac{[NeV]}{[SiII]}$ & 98\% & 9\% & 20\%\\
EW PAH vs EW[NeII] & 78\% & 0.1\% & 0.4\%\\
ext vs EW[NeII] & 60\% & 19\% & 10\%\\
ext vs EW[NeIII] & 69\% & 17\% & 19\%\\
ext vs EW PAH & 84\% & 14\% & 33\%\\
$\frac{[NeIII]}{[NeII]}$ vs $\alpha$ & 97\% & 12\% & 12\%\\
$\frac{[NeV]}{[NeII]}$ vs $\alpha$ & 100\% & 6\% & 5\%\\
\enddata
\end{deluxetable}%
%\end{turnpage}

%figure
\clearpage

\clearpage

\begin{figure}
\centerline{\includegraphics[width=8cm]{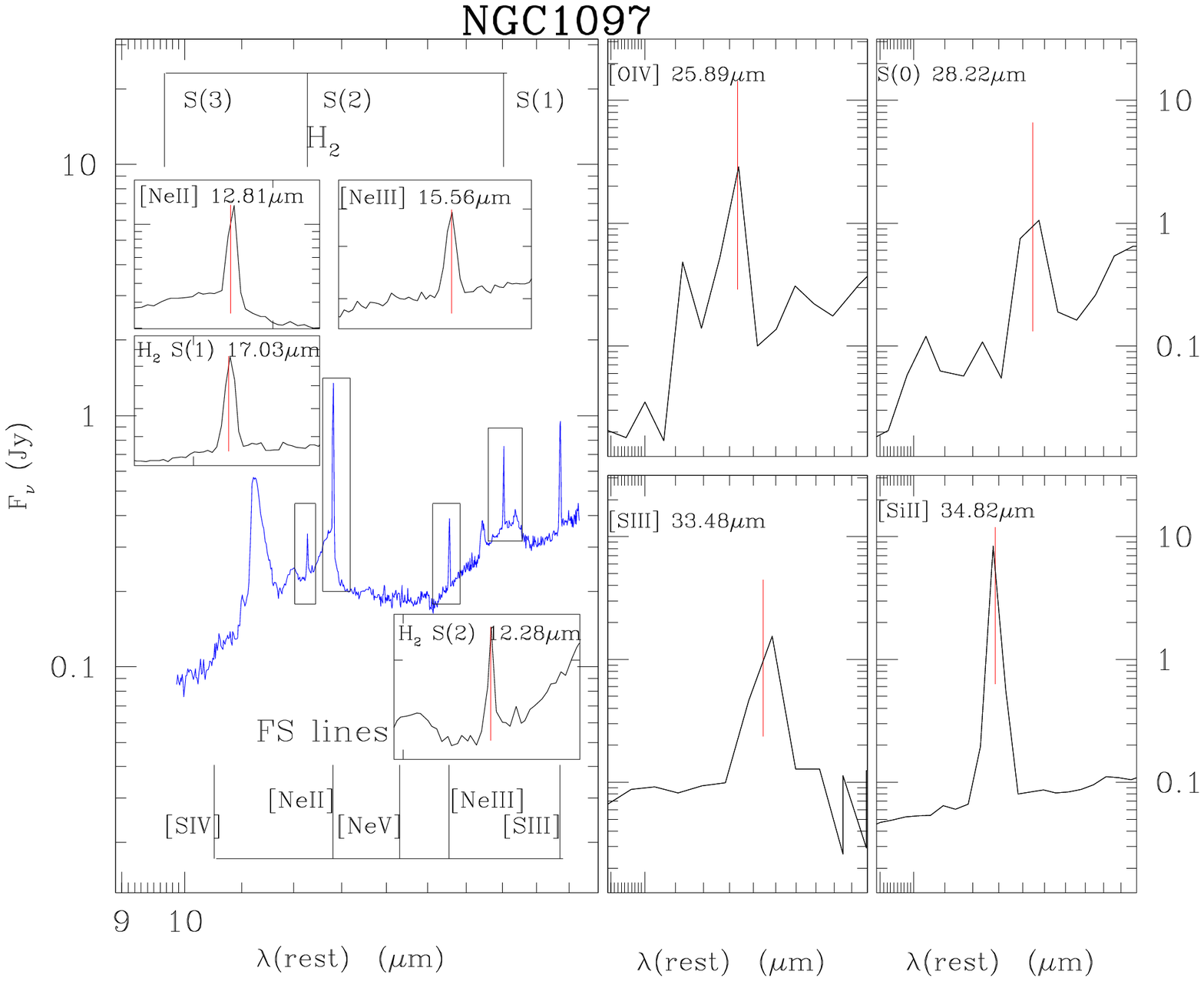}\includegraphics[width=8cm]{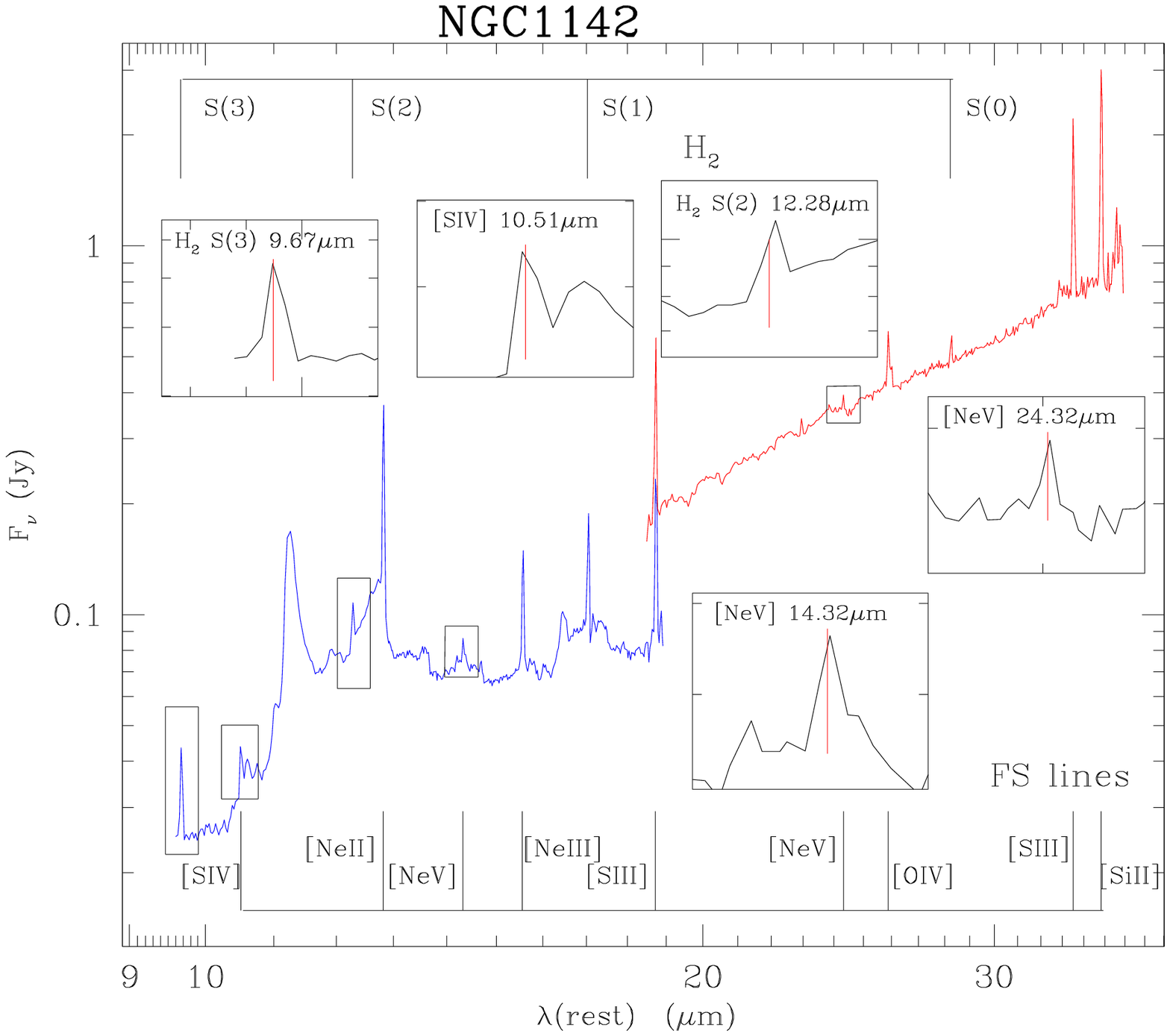}}

\centerline{\includegraphics[width=8cm]{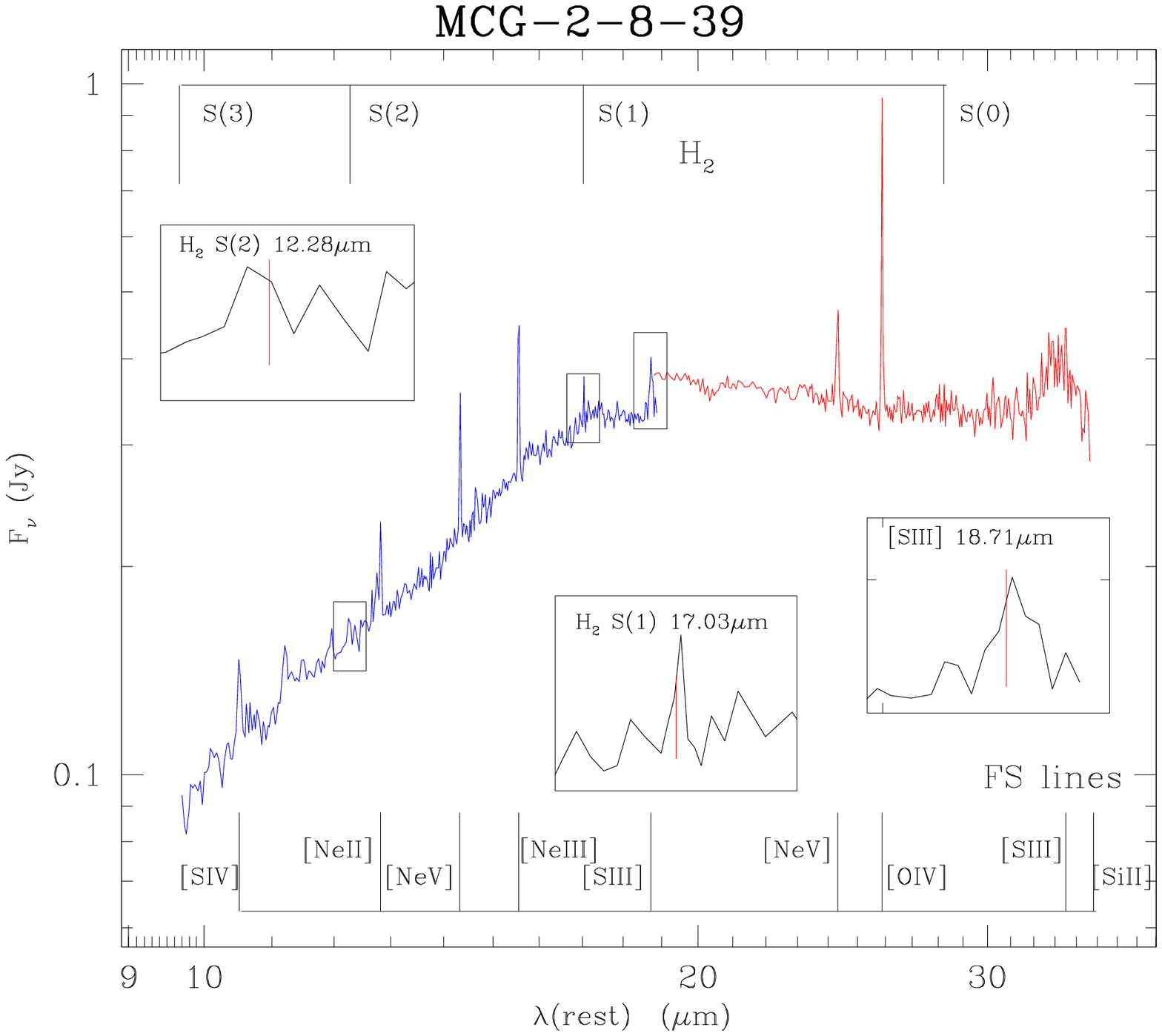}\includegraphics[width=8cm]{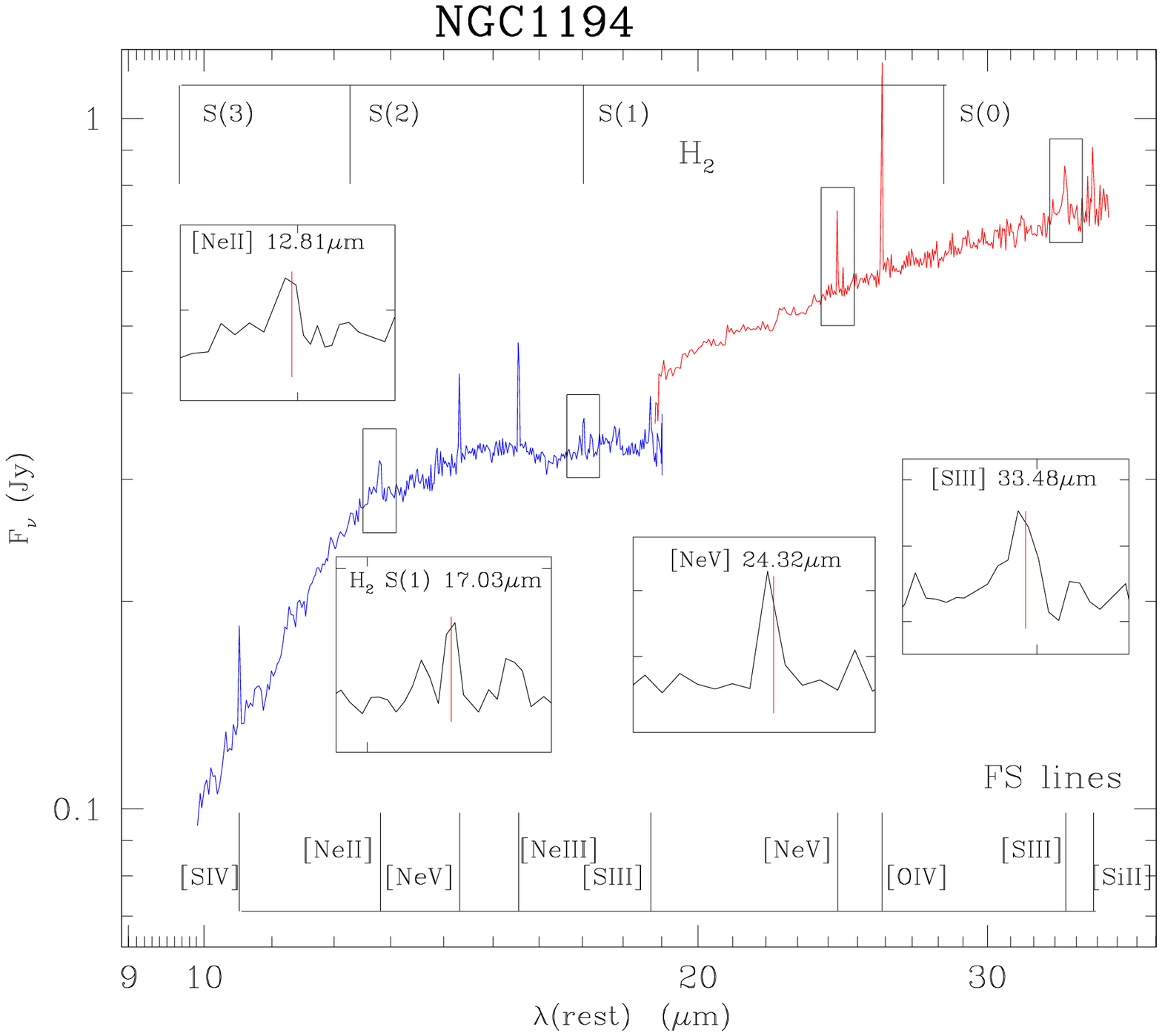}}

\centerline{\includegraphics[width=8cm]{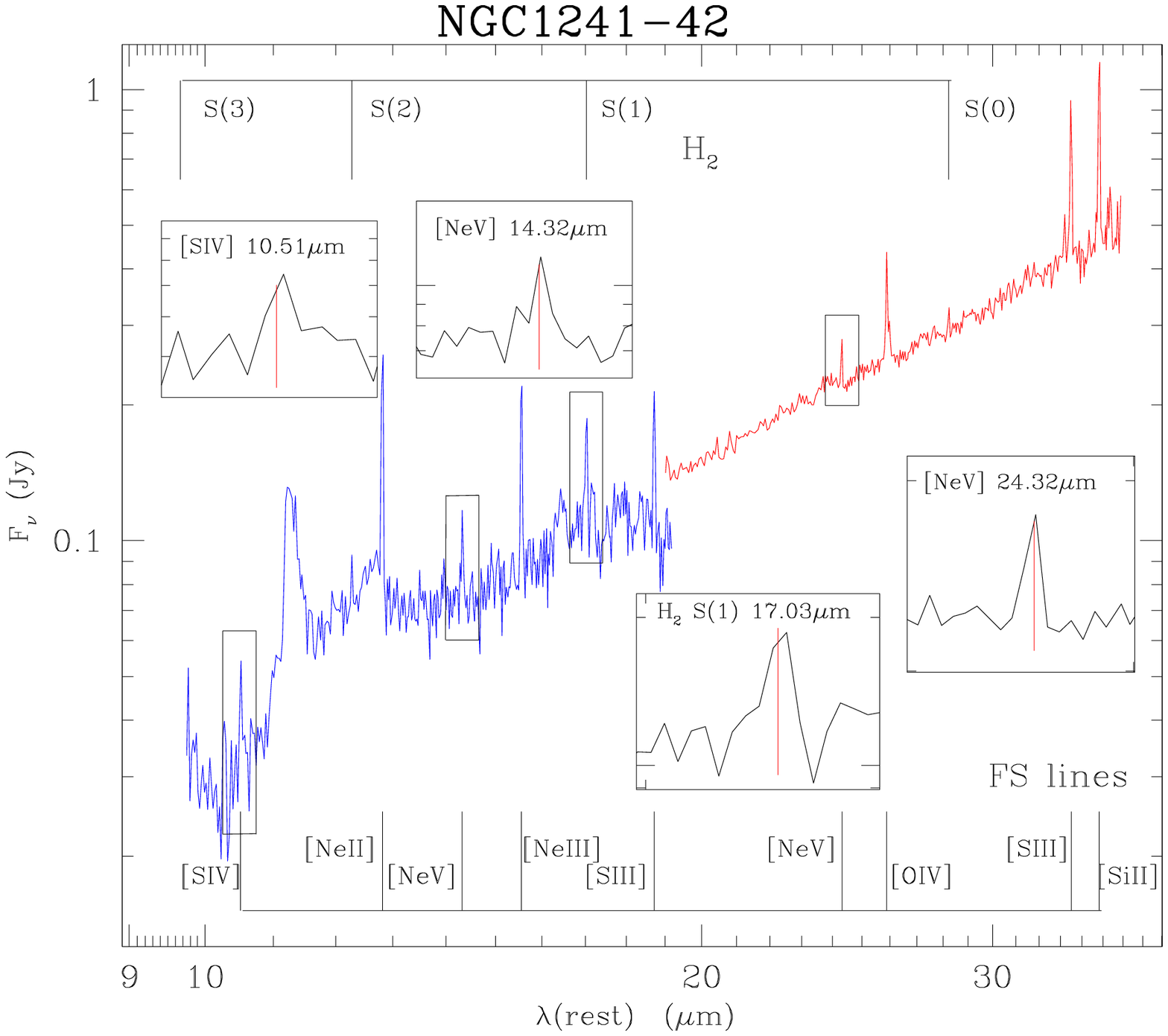}\includegraphics[width=8cm]{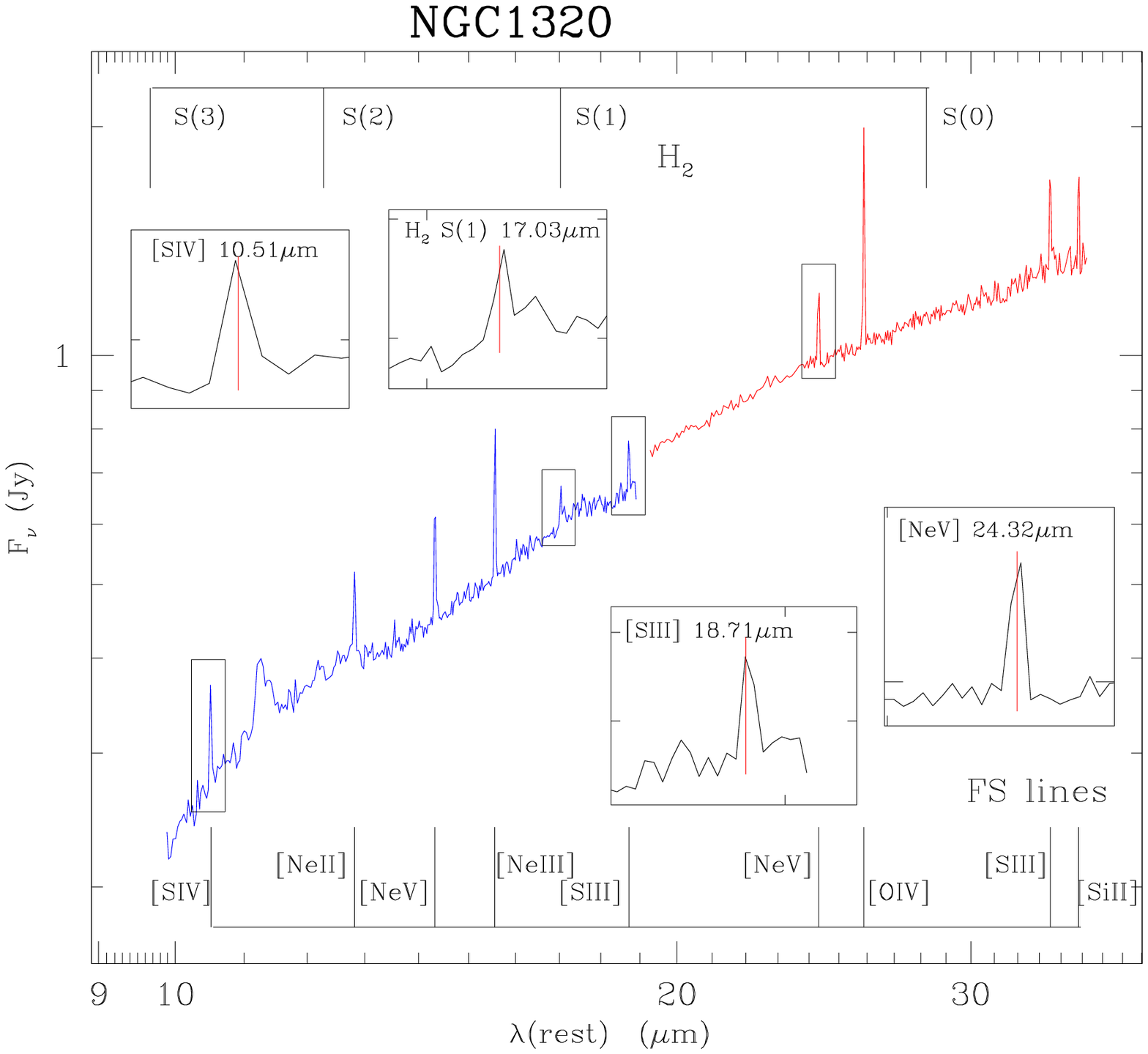}}

\end{figure}
\clearpage

\begin{figure}
\centerline{\includegraphics[width=8cm]{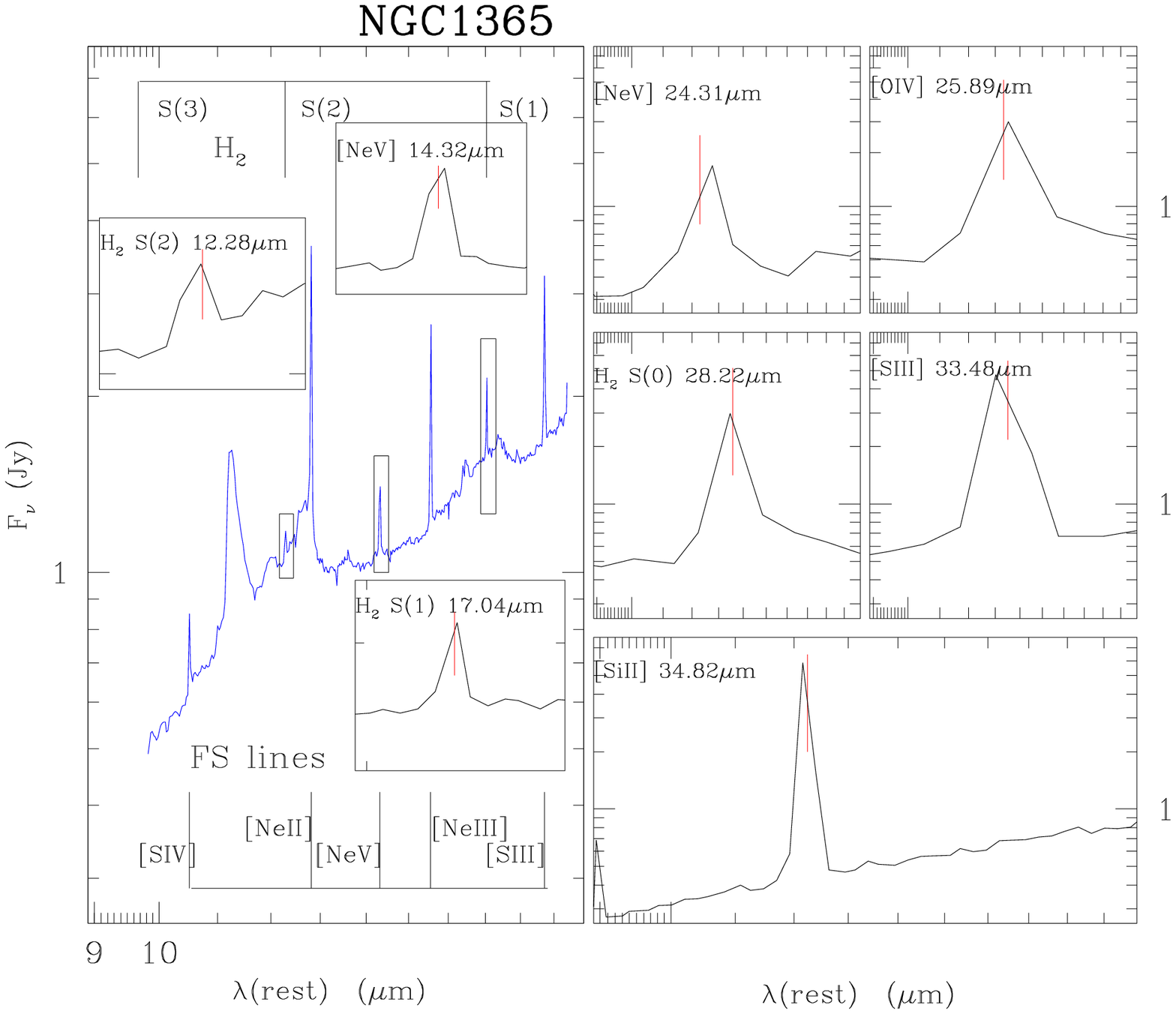}\includegraphics[width=8cm]{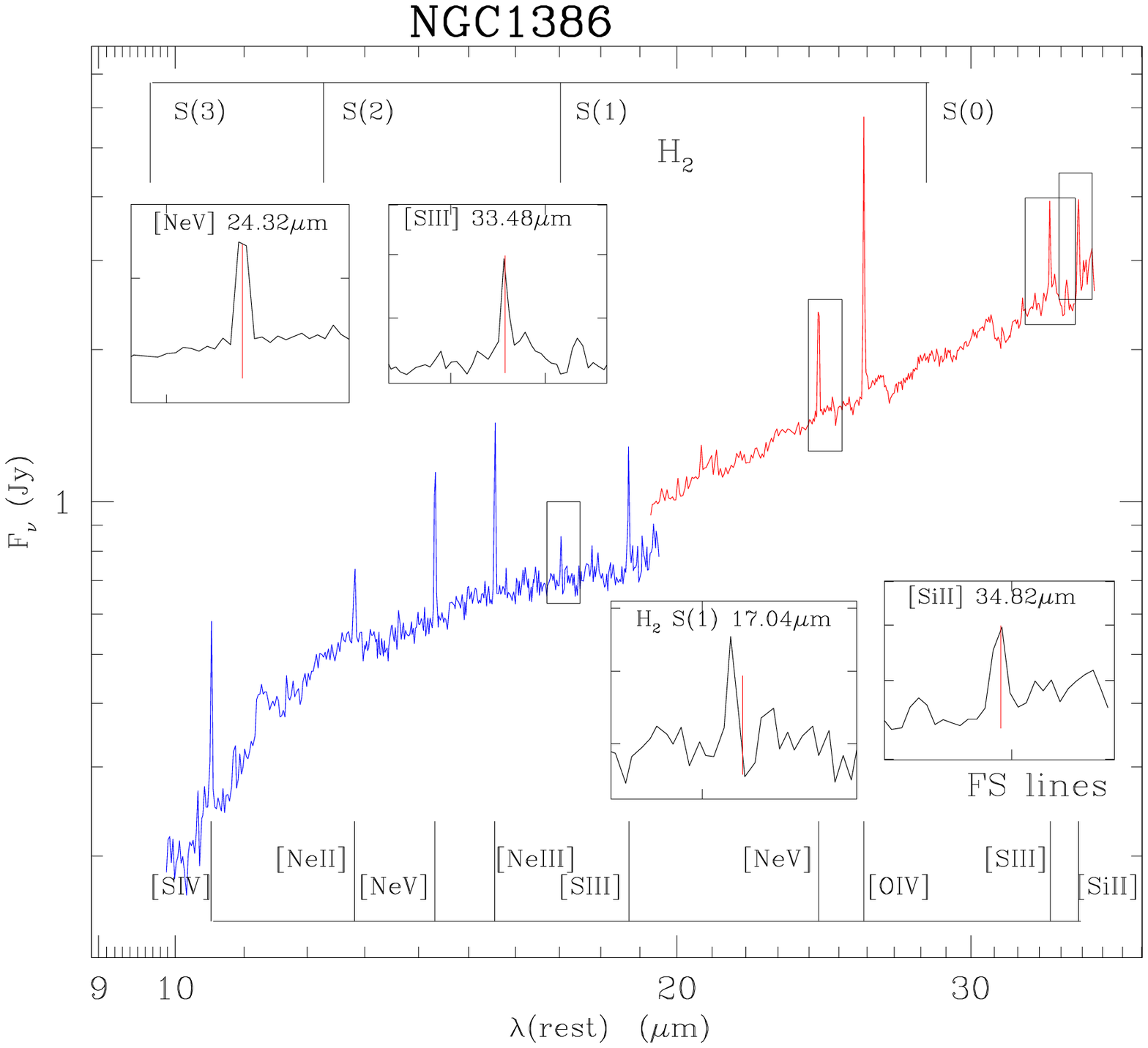}}

\centerline{\includegraphics[width=8cm]{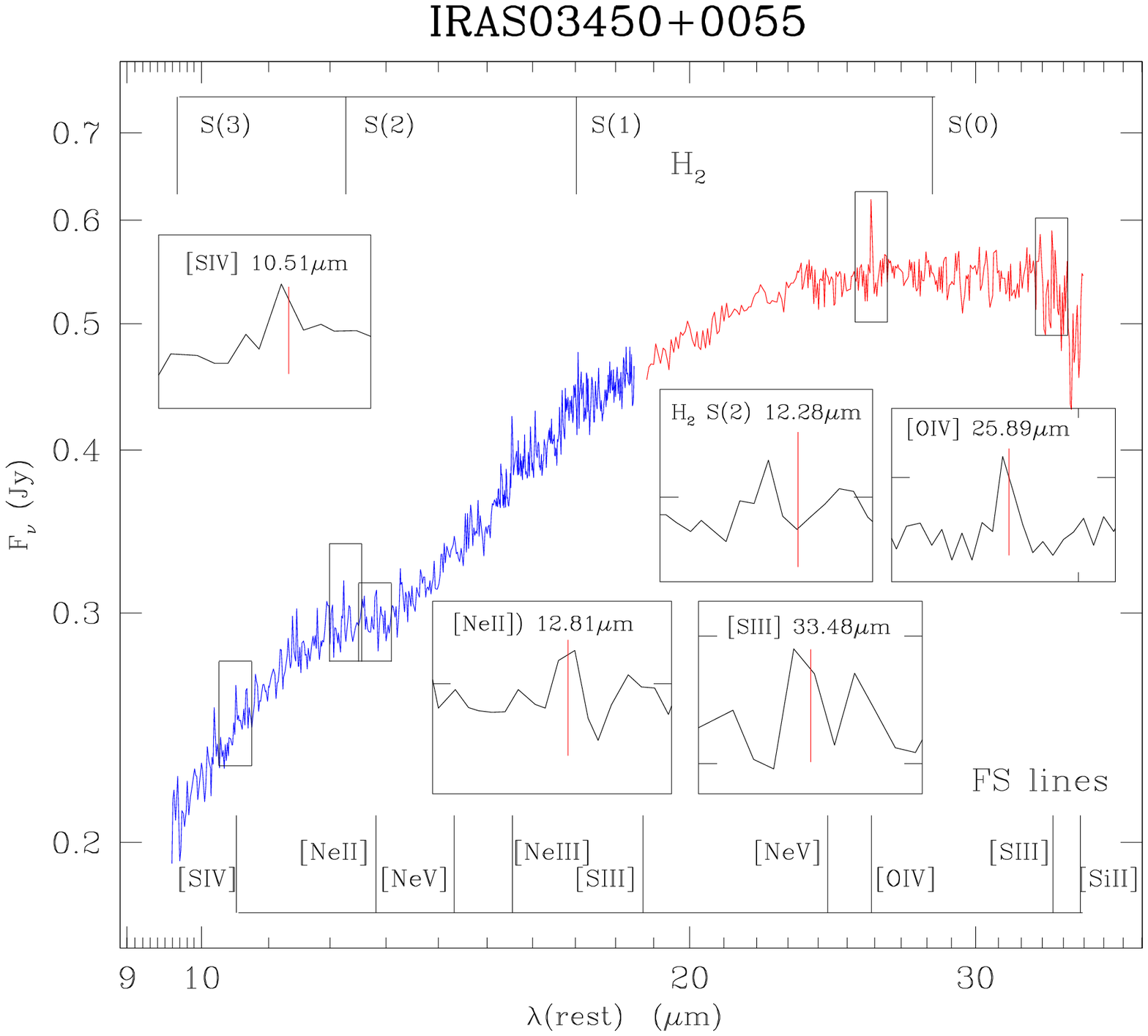}\includegraphics[width=8cm]{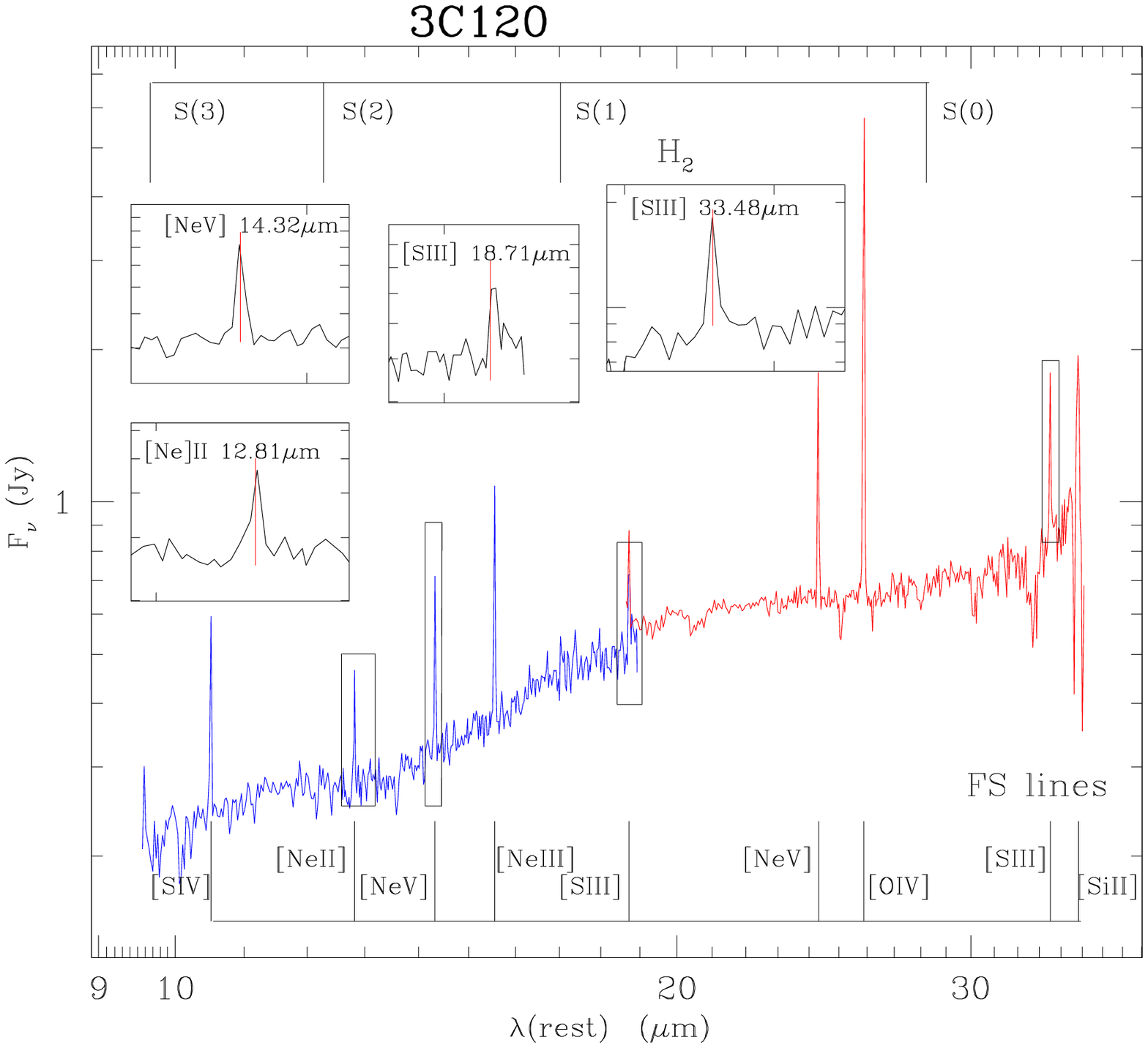}}

\centerline{\includegraphics[width=8cm]{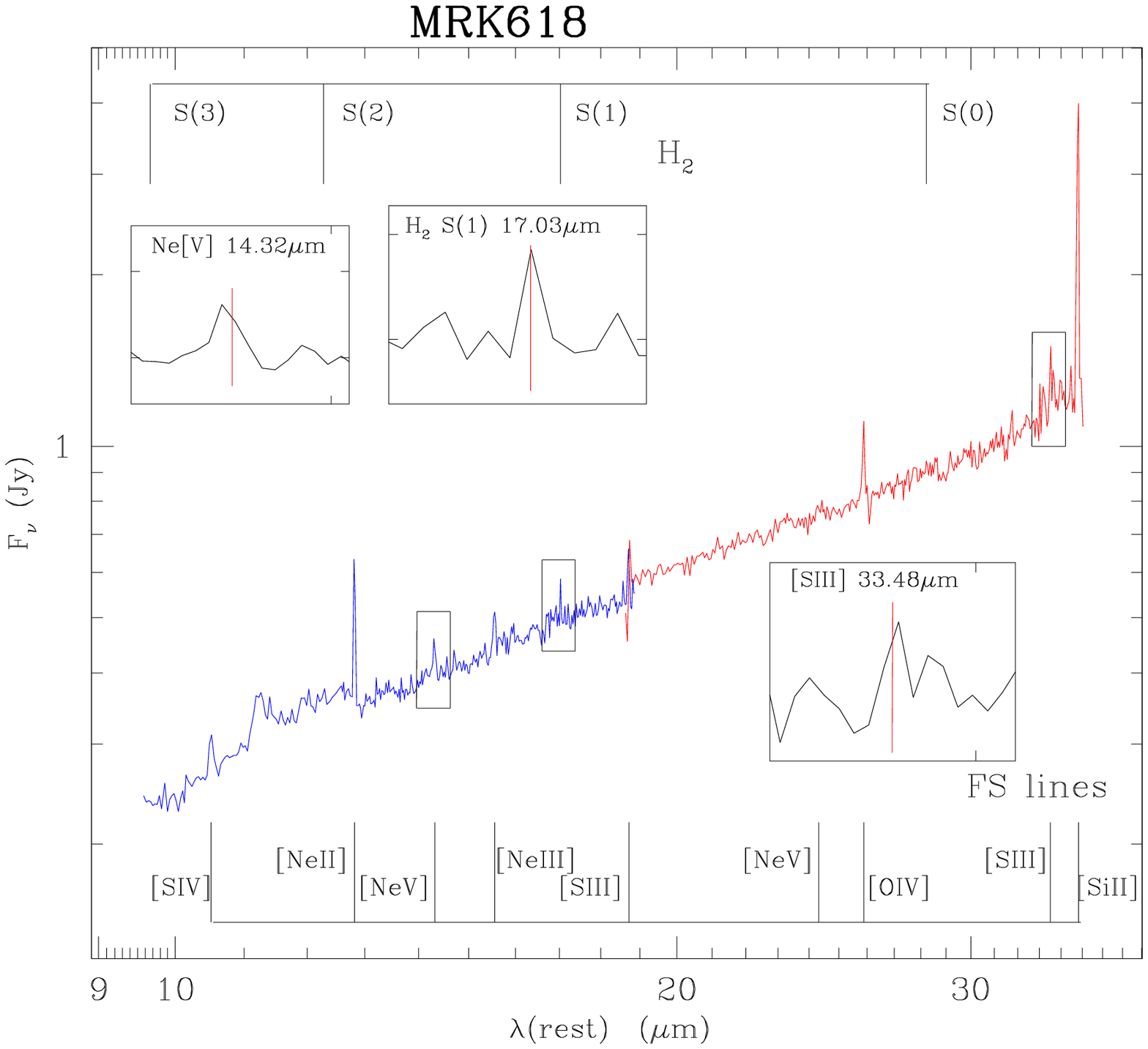}\includegraphics[width=8cm]{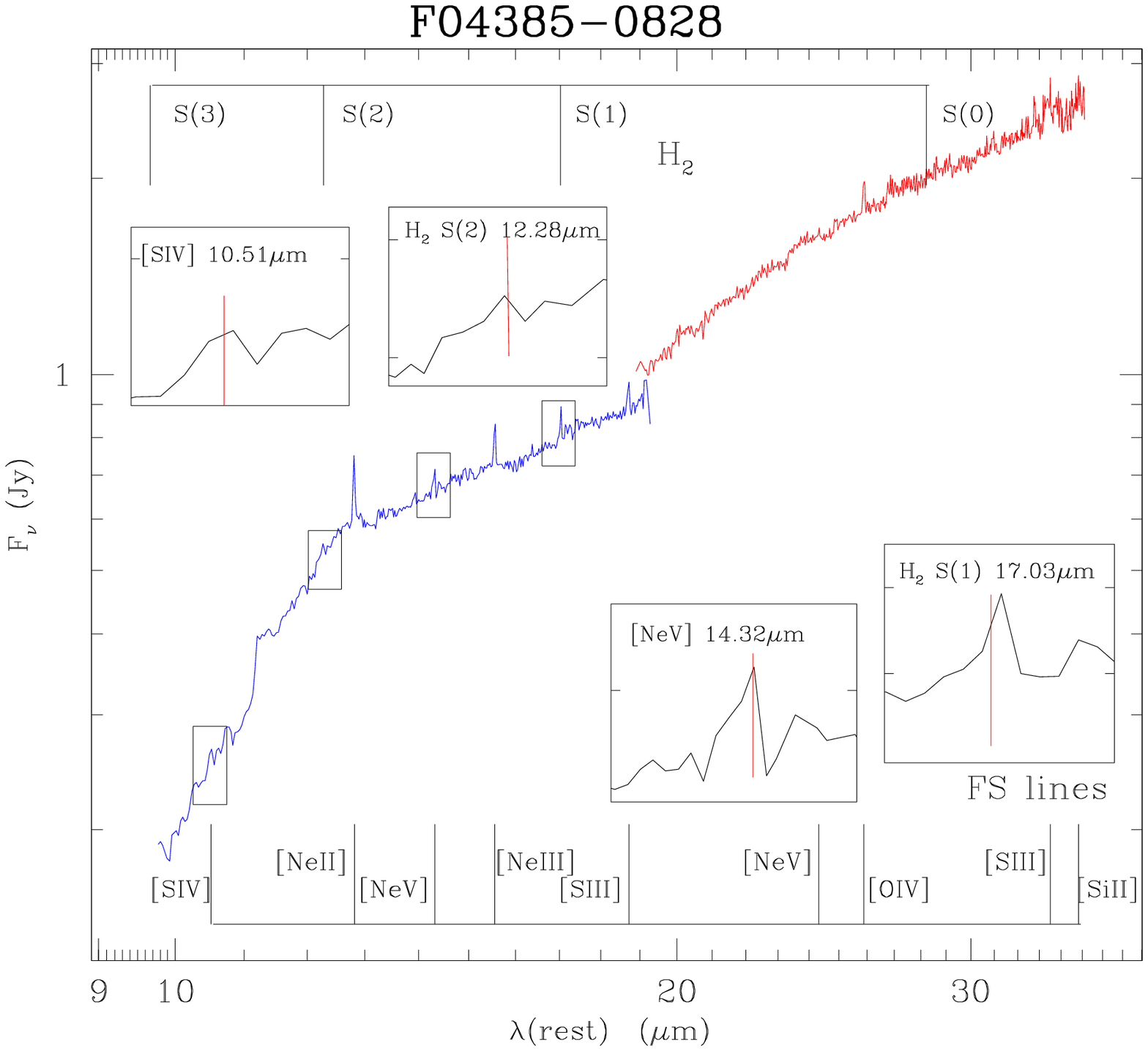}}
\end{figure}
\clearpage

\begin{figure}

\centerline{\includegraphics[width=8cm]{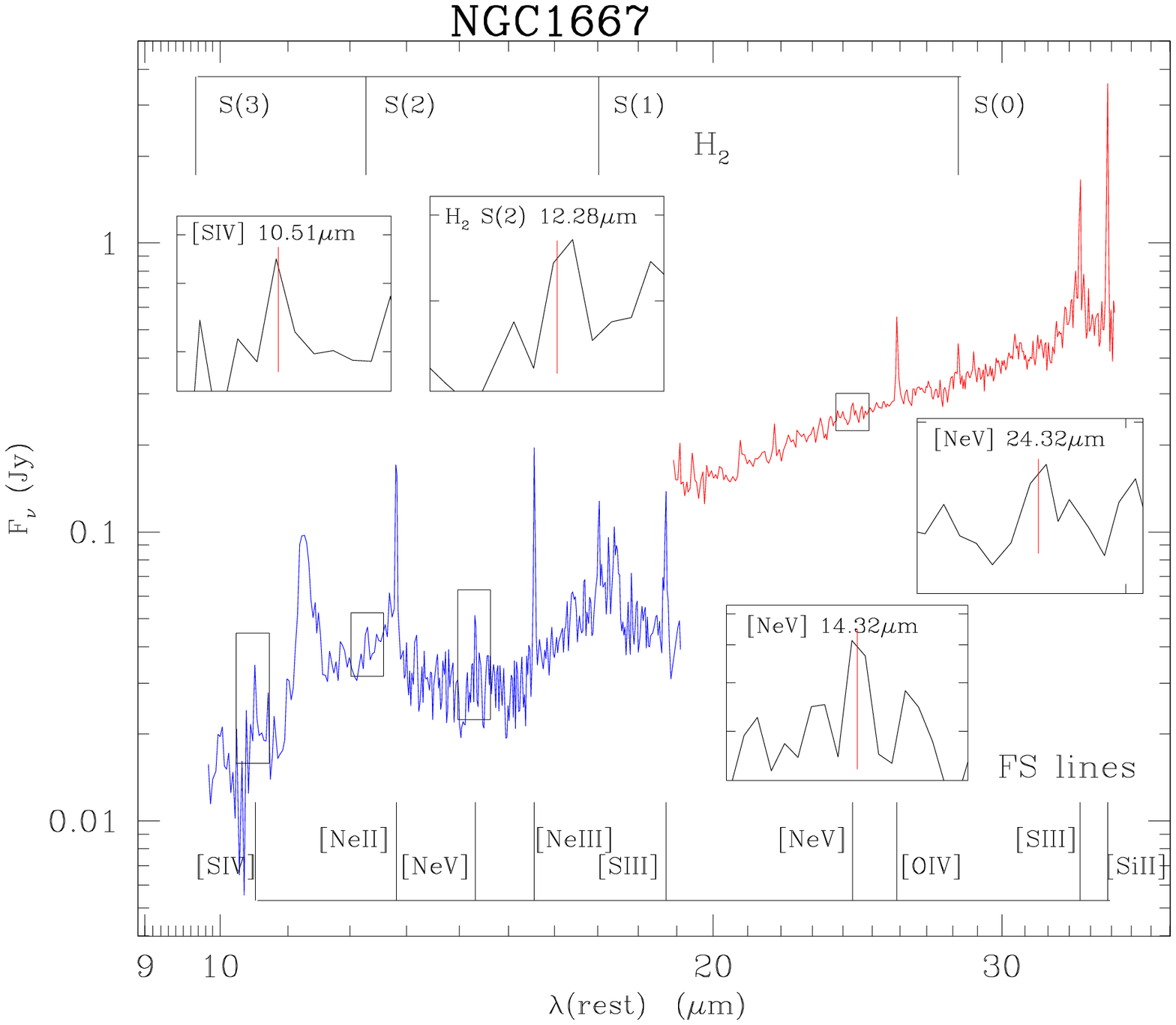}\includegraphics[width=8cm]{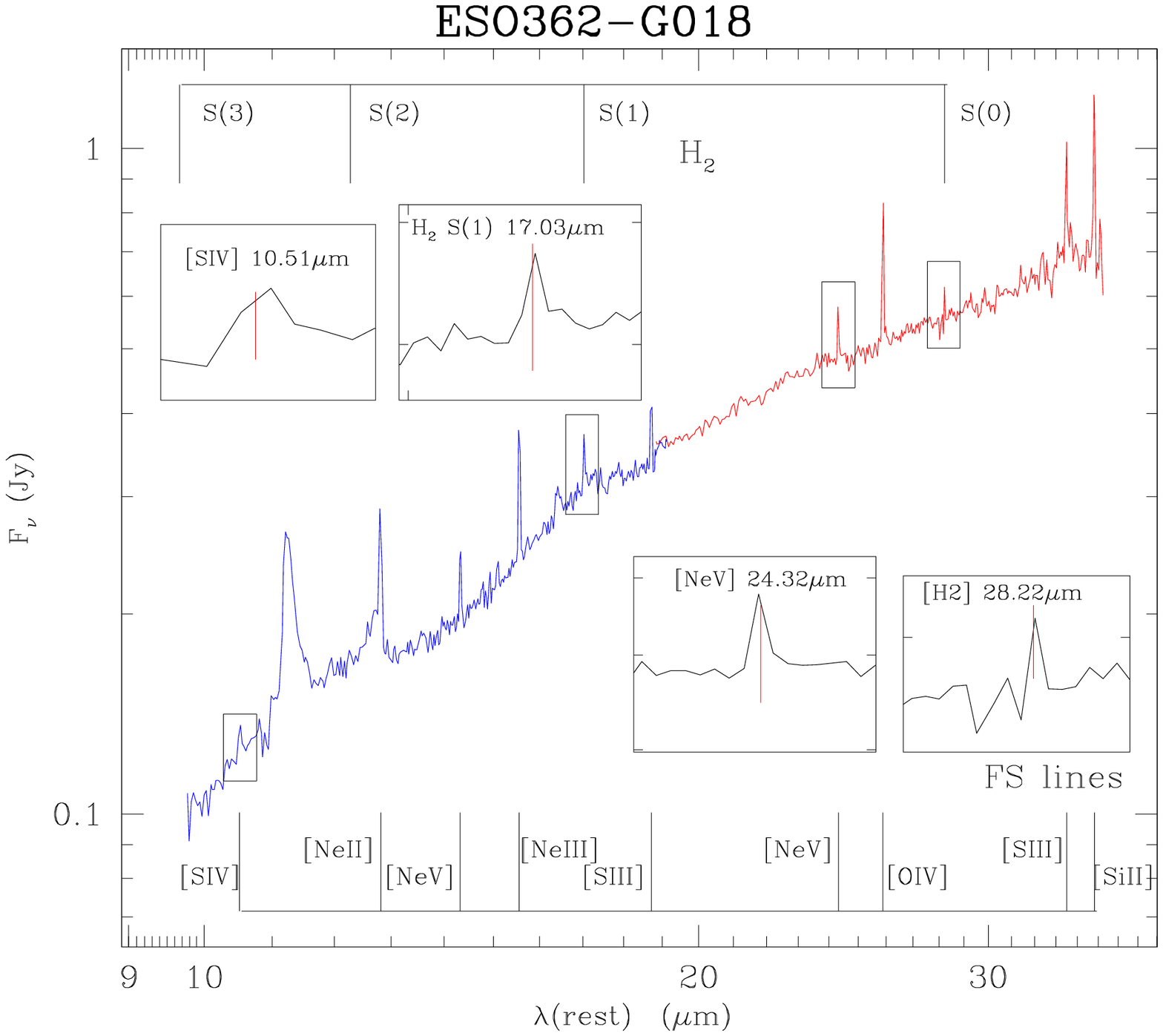}}

\centerline{\includegraphics[width=8cm]{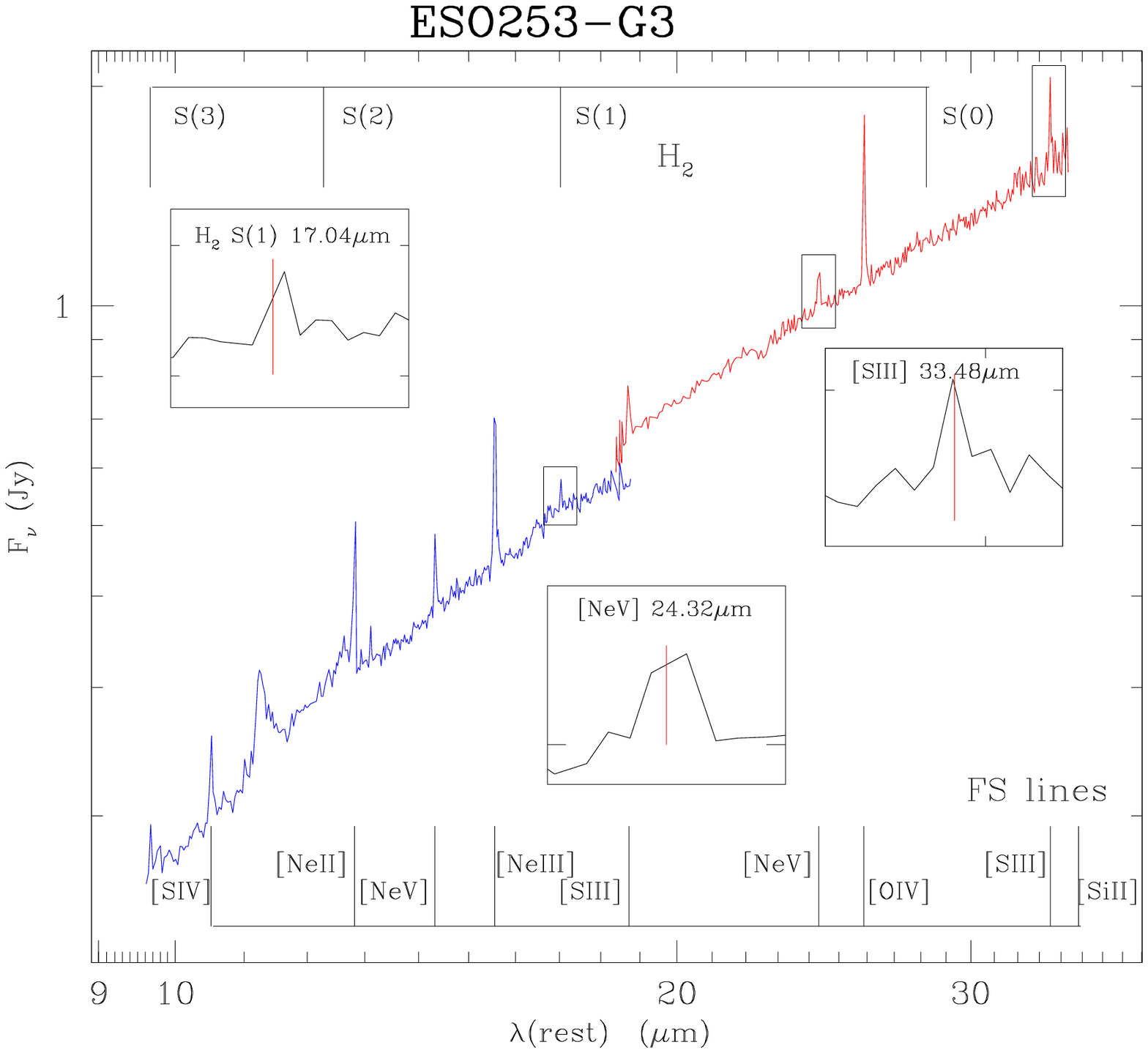}\includegraphics[width=8cm]{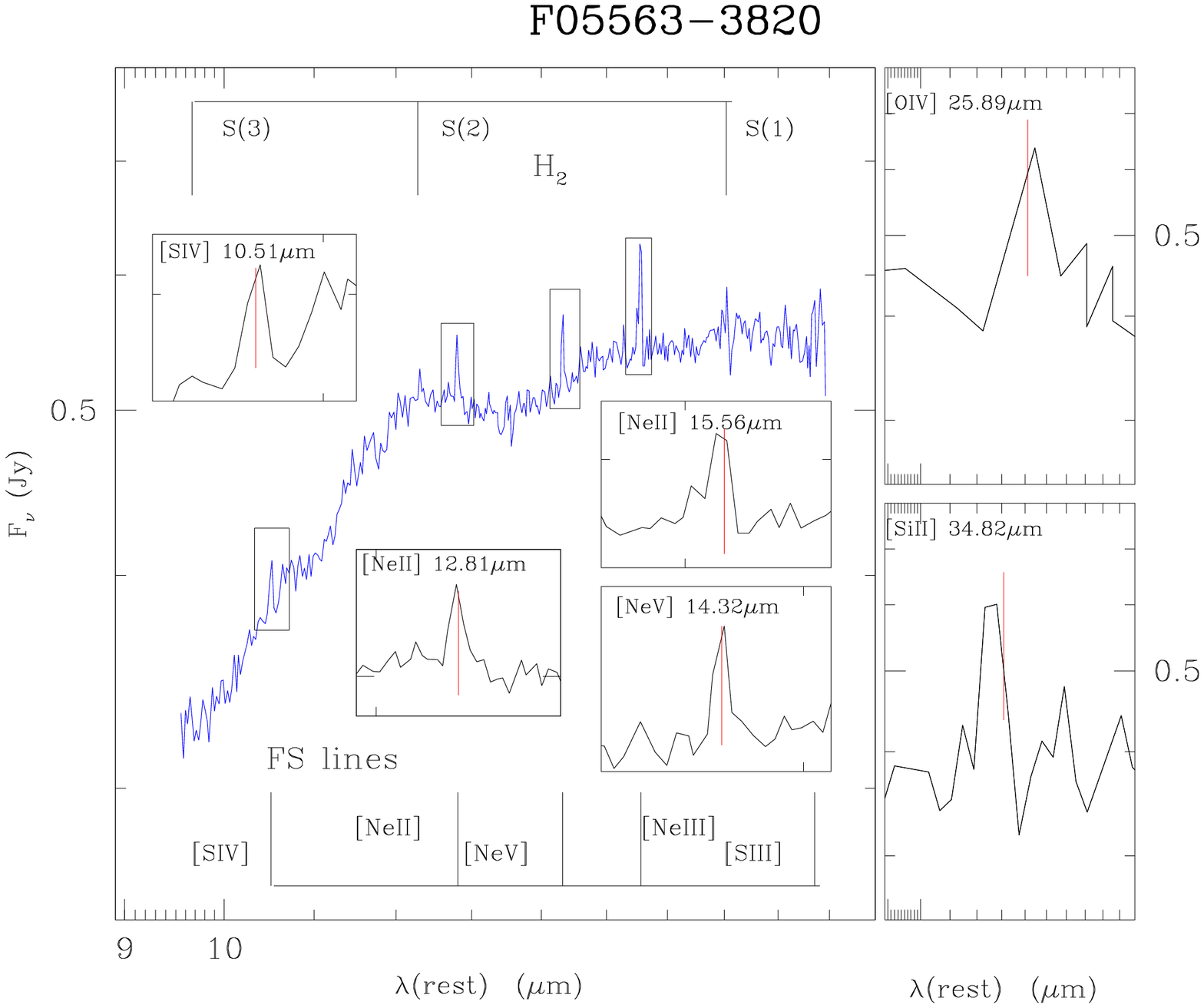}}

\centerline{\includegraphics[width=8cm]{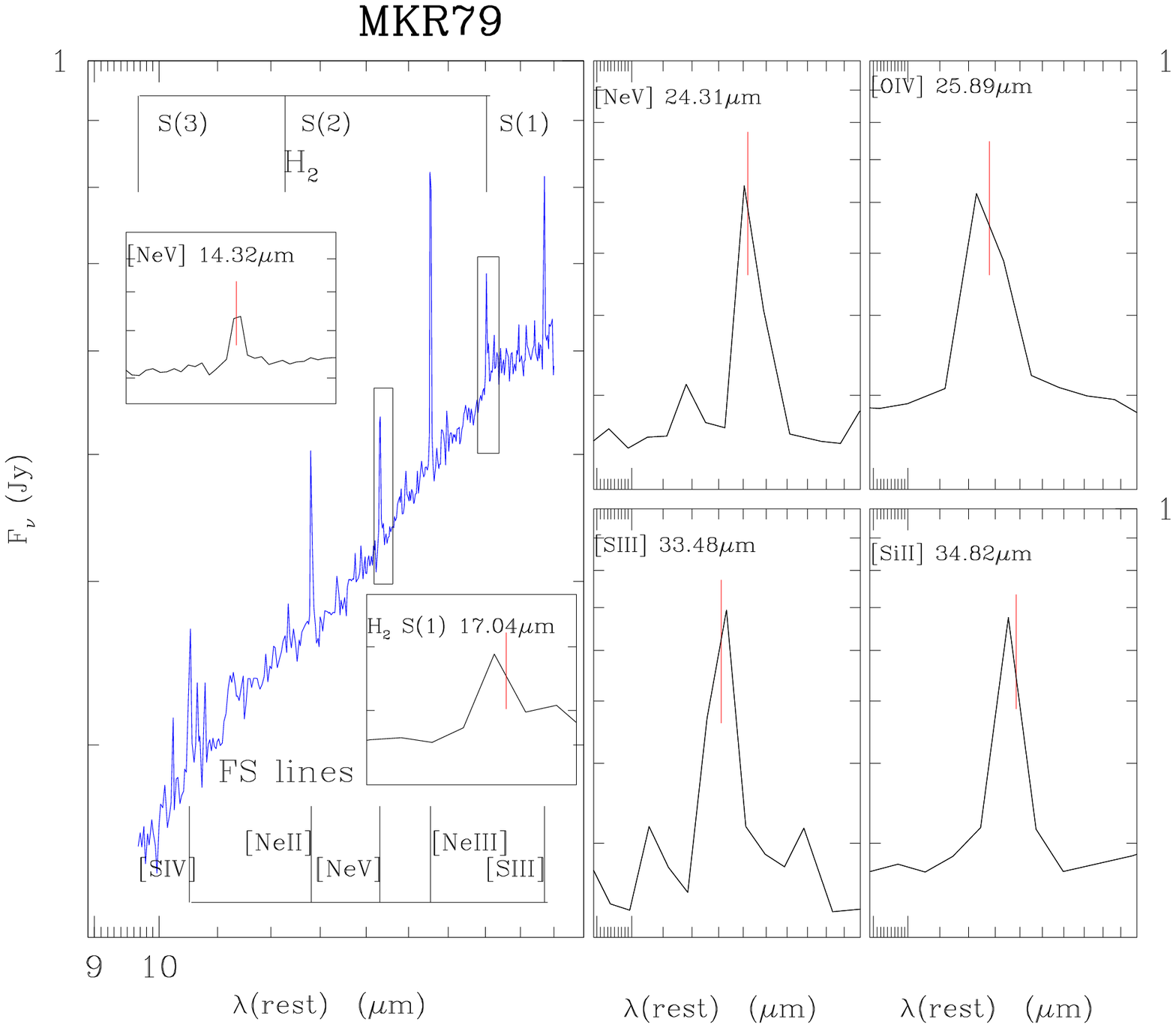}\includegraphics[width=8cm]{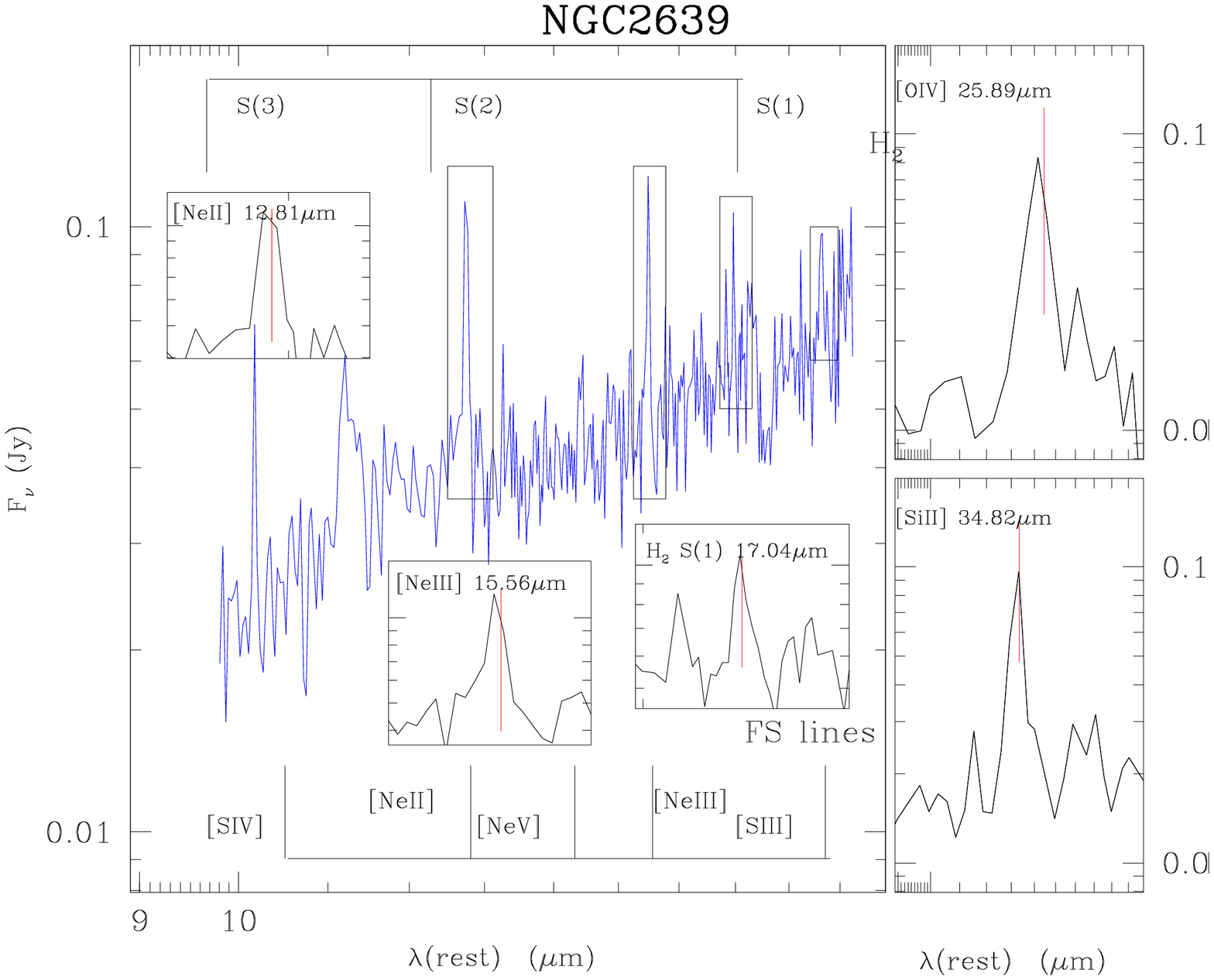}}
\end{figure}
\clearpage

\begin{figure}
\centerline{\includegraphics[width=8cm]{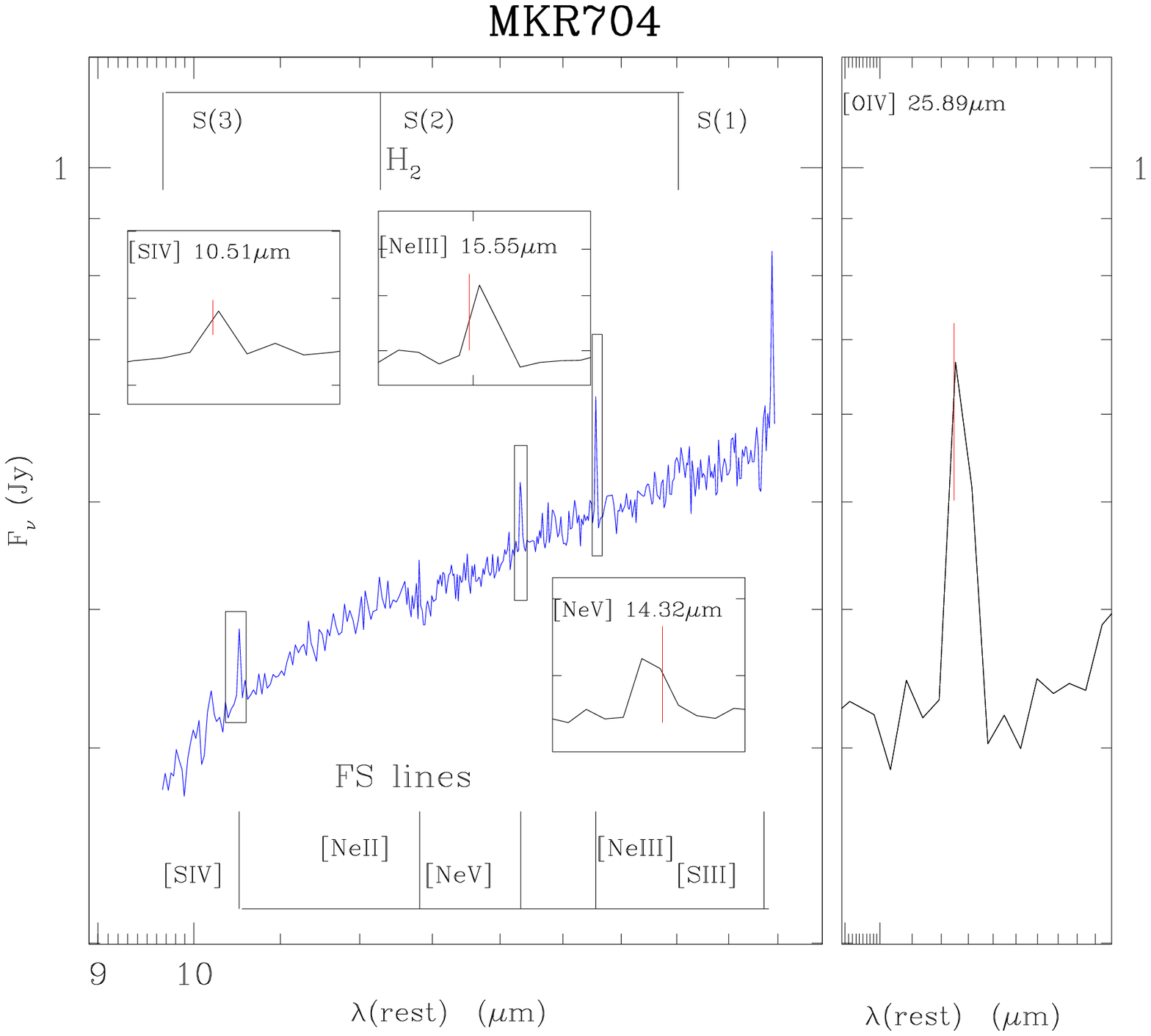}\includegraphics[width=8cm]{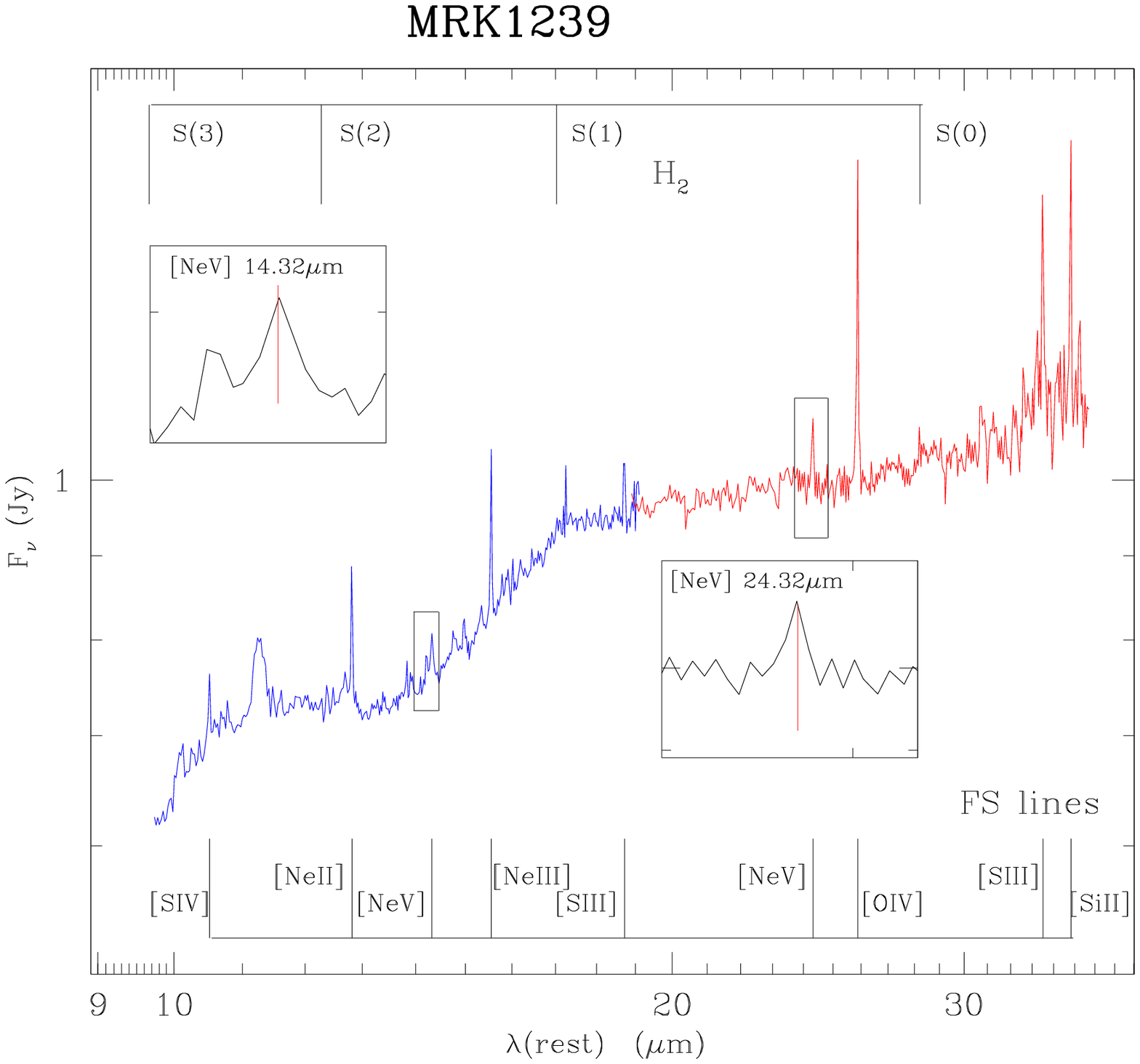}}

\centerline{\includegraphics[width=8cm]{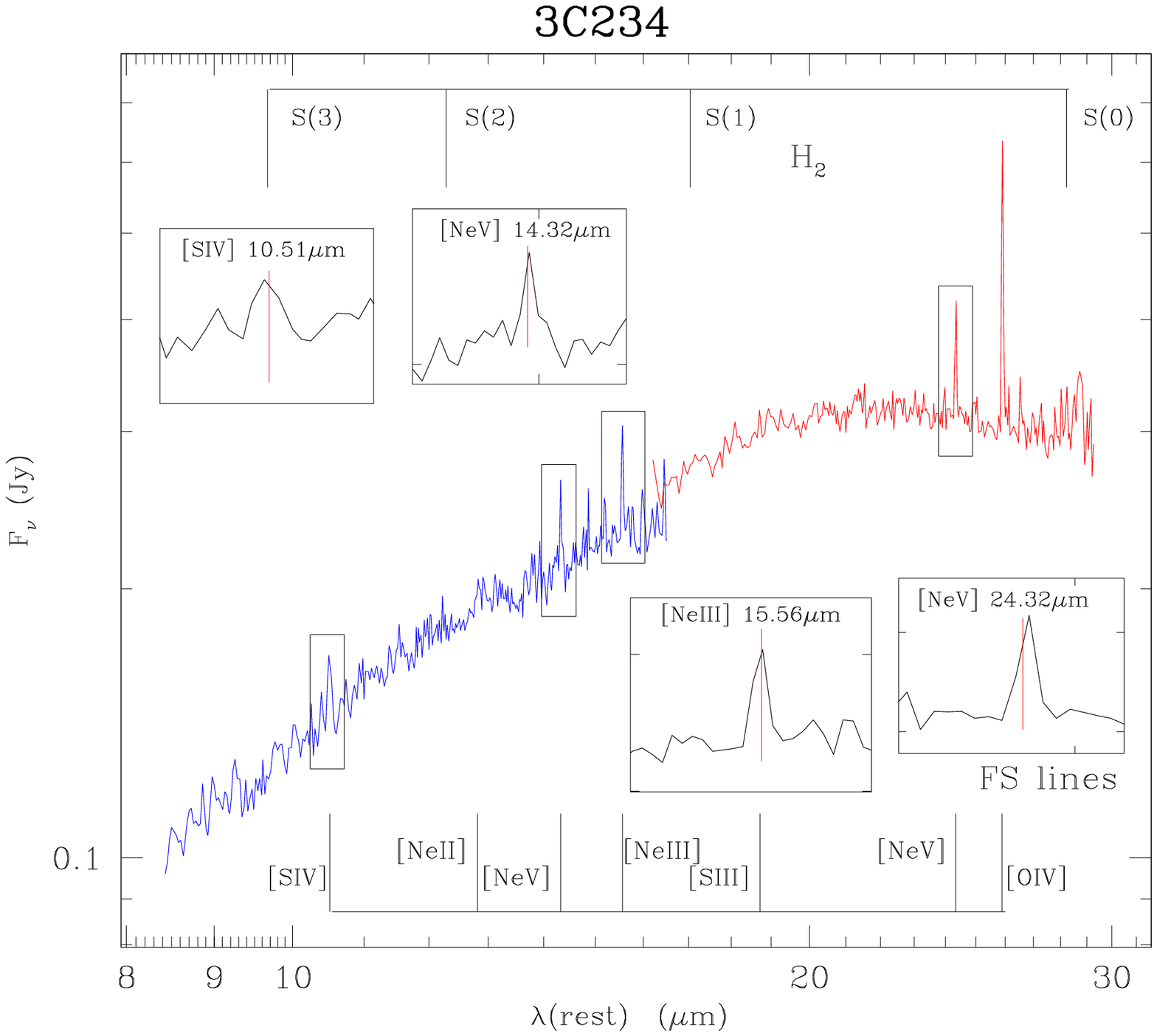}\includegraphics[width=8cm]{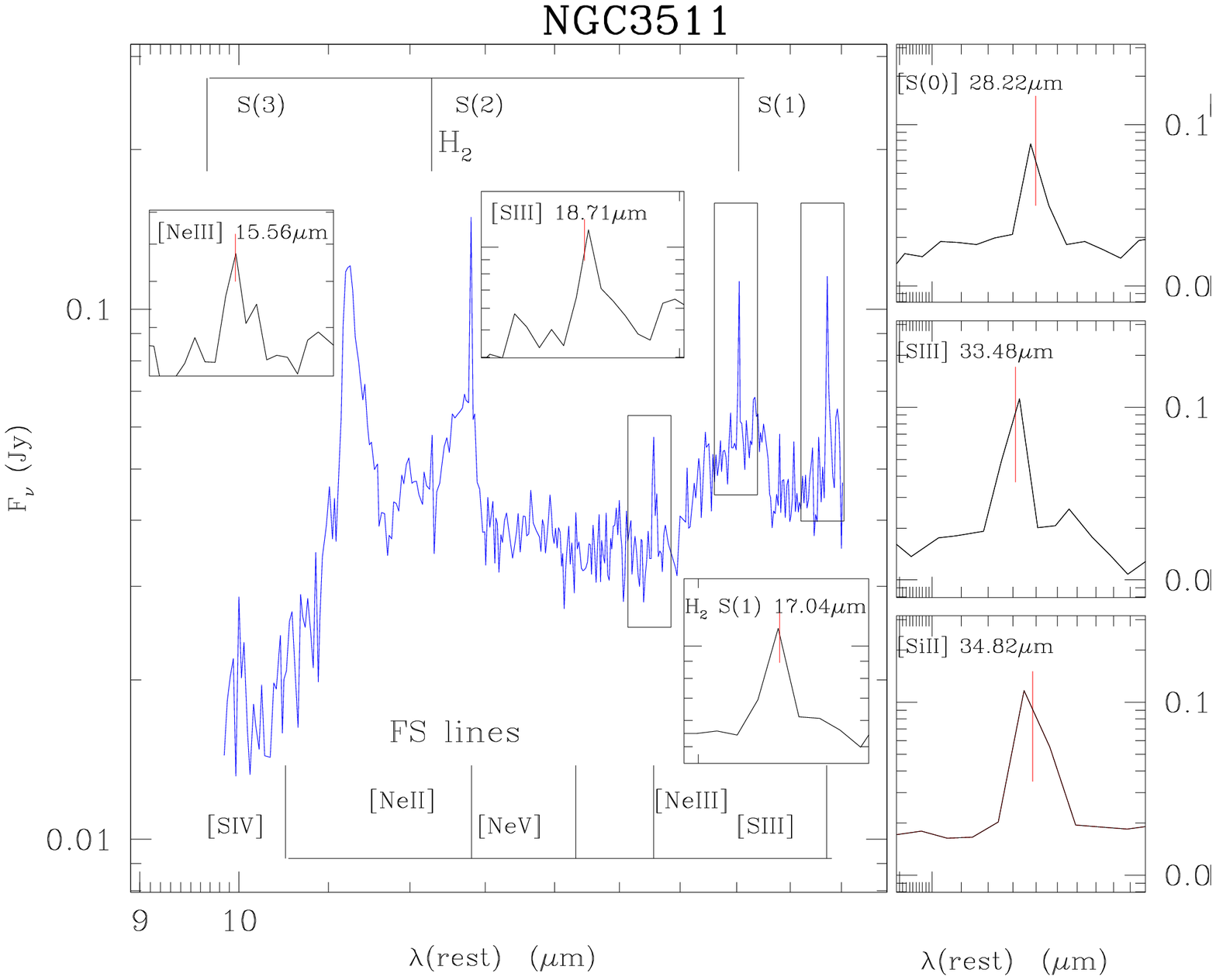}}

\centerline{\includegraphics[width=8cm]{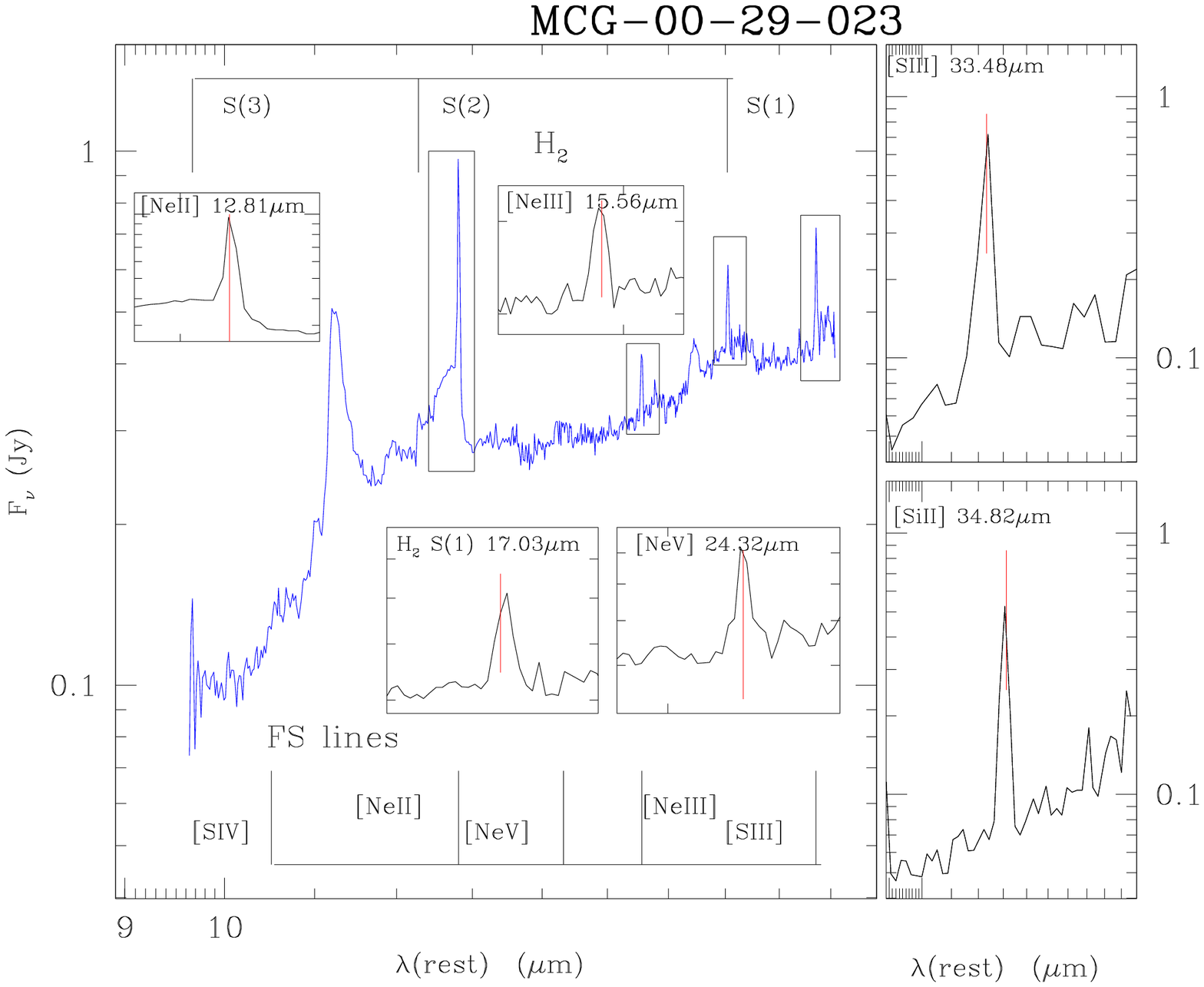}\includegraphics[width=8cm]{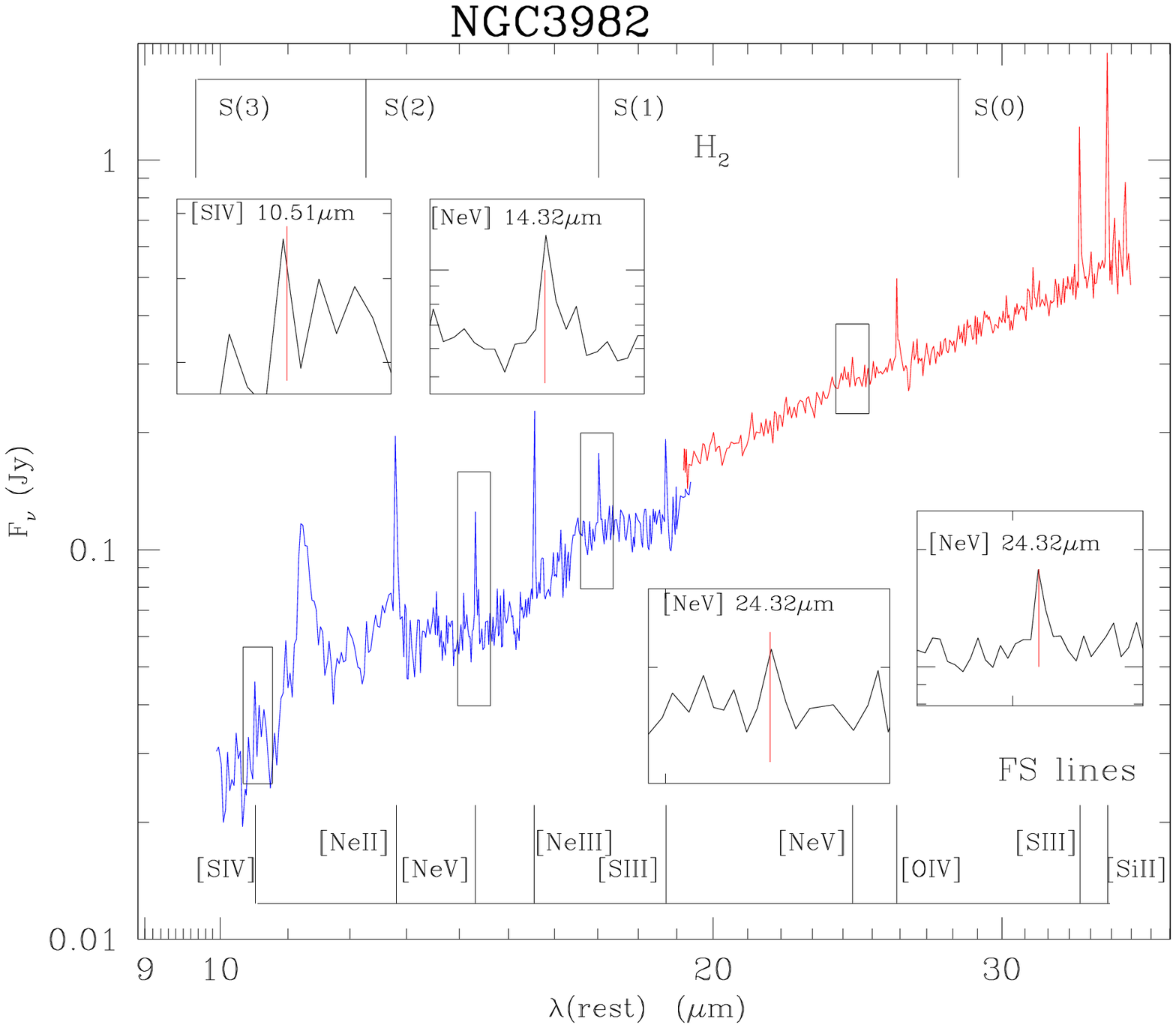}}

\end{figure}
\clearpage

\begin{figure}
\centerline{\includegraphics[width=8cm]{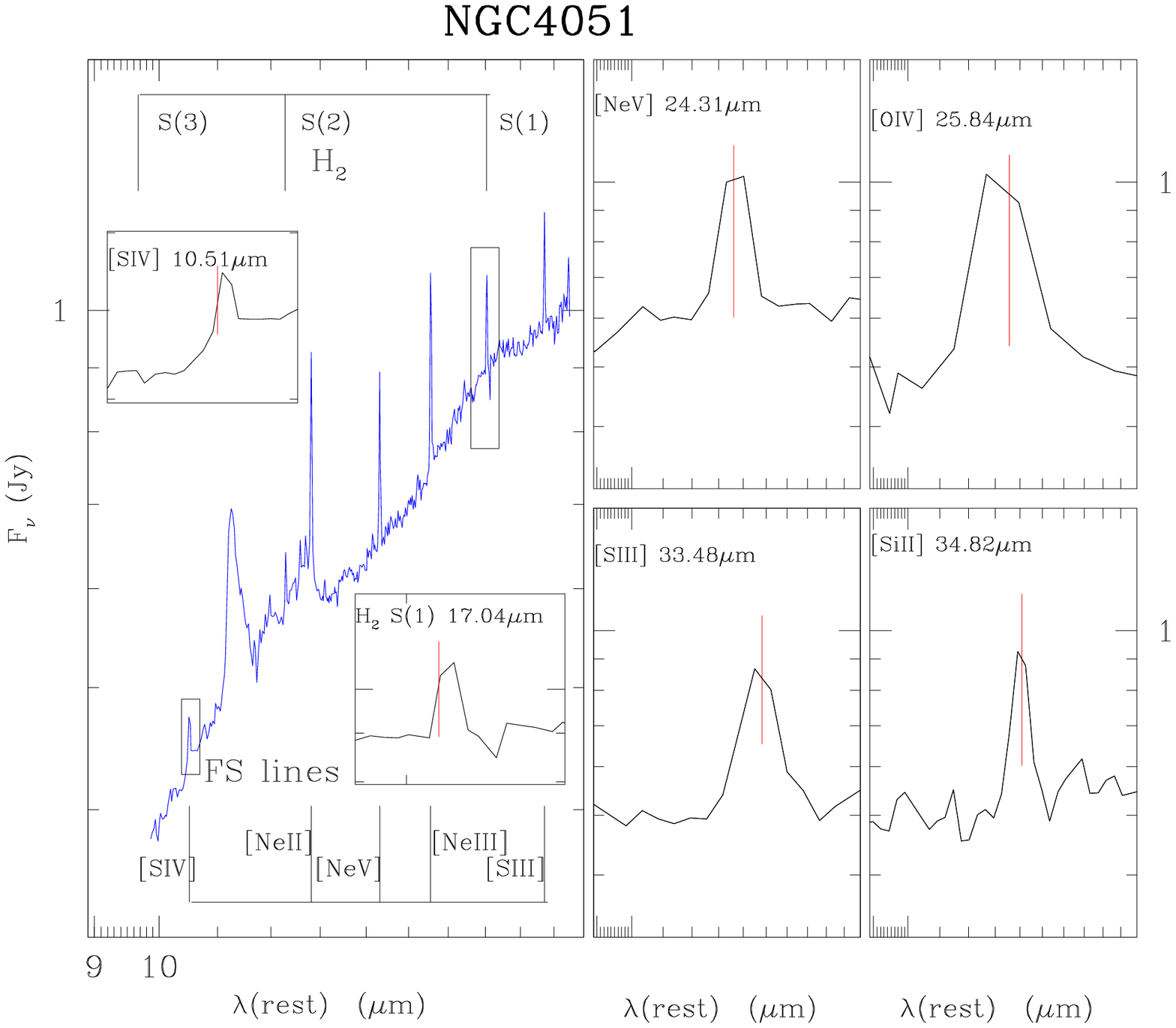}\includegraphics[width=8cm]{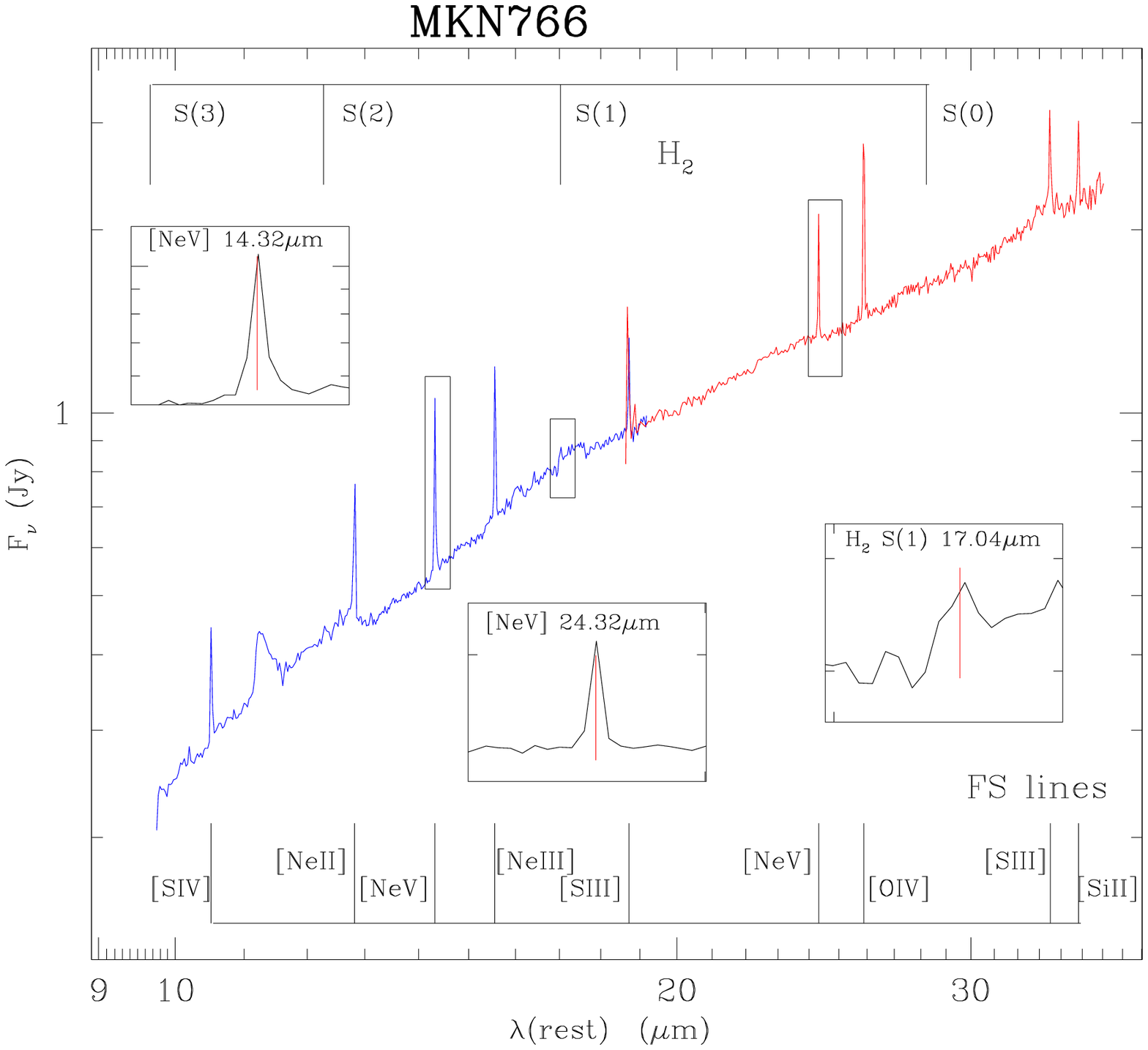}}

\centerline{\includegraphics[width=8cm]{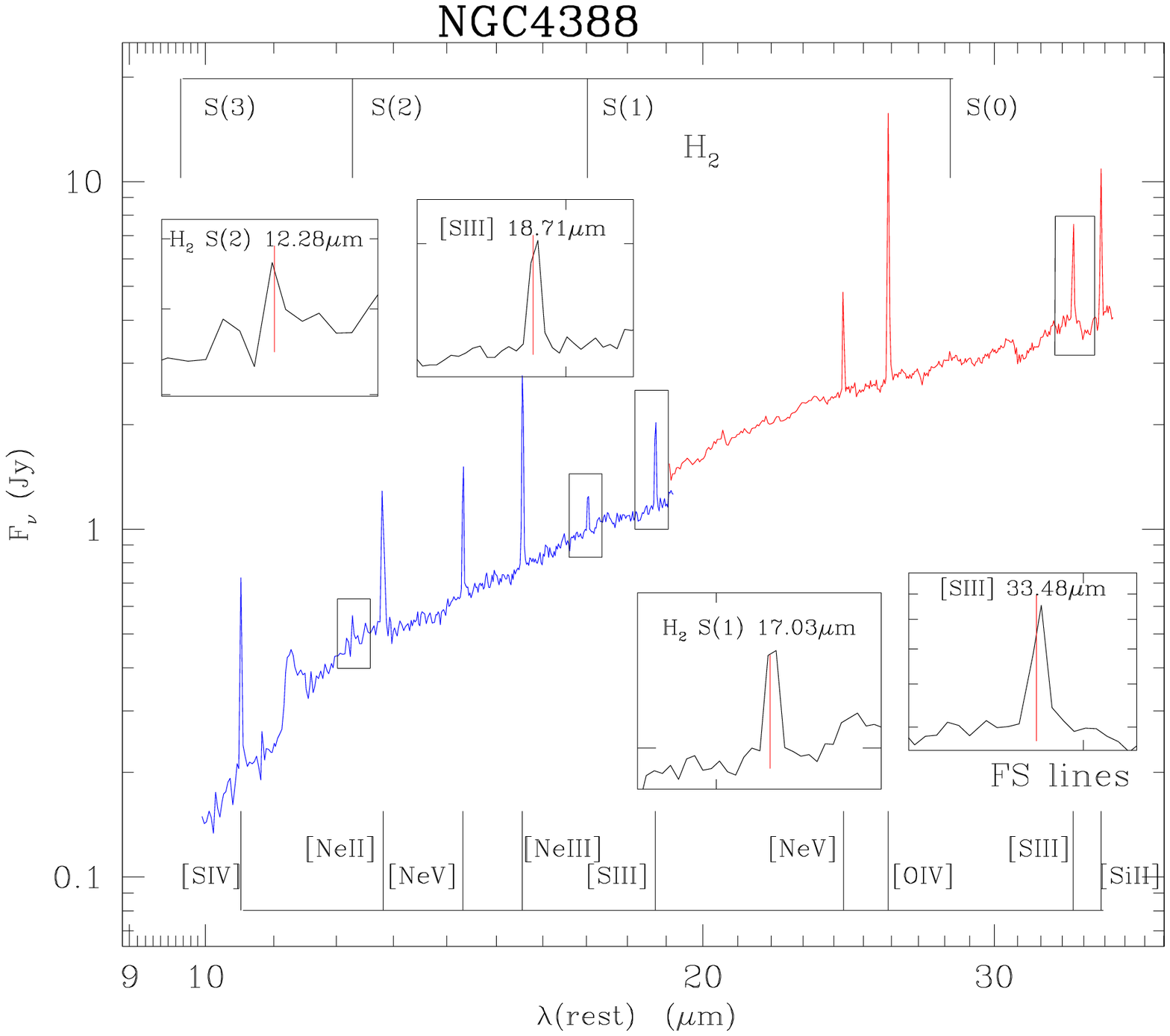}\includegraphics[width=8cm]{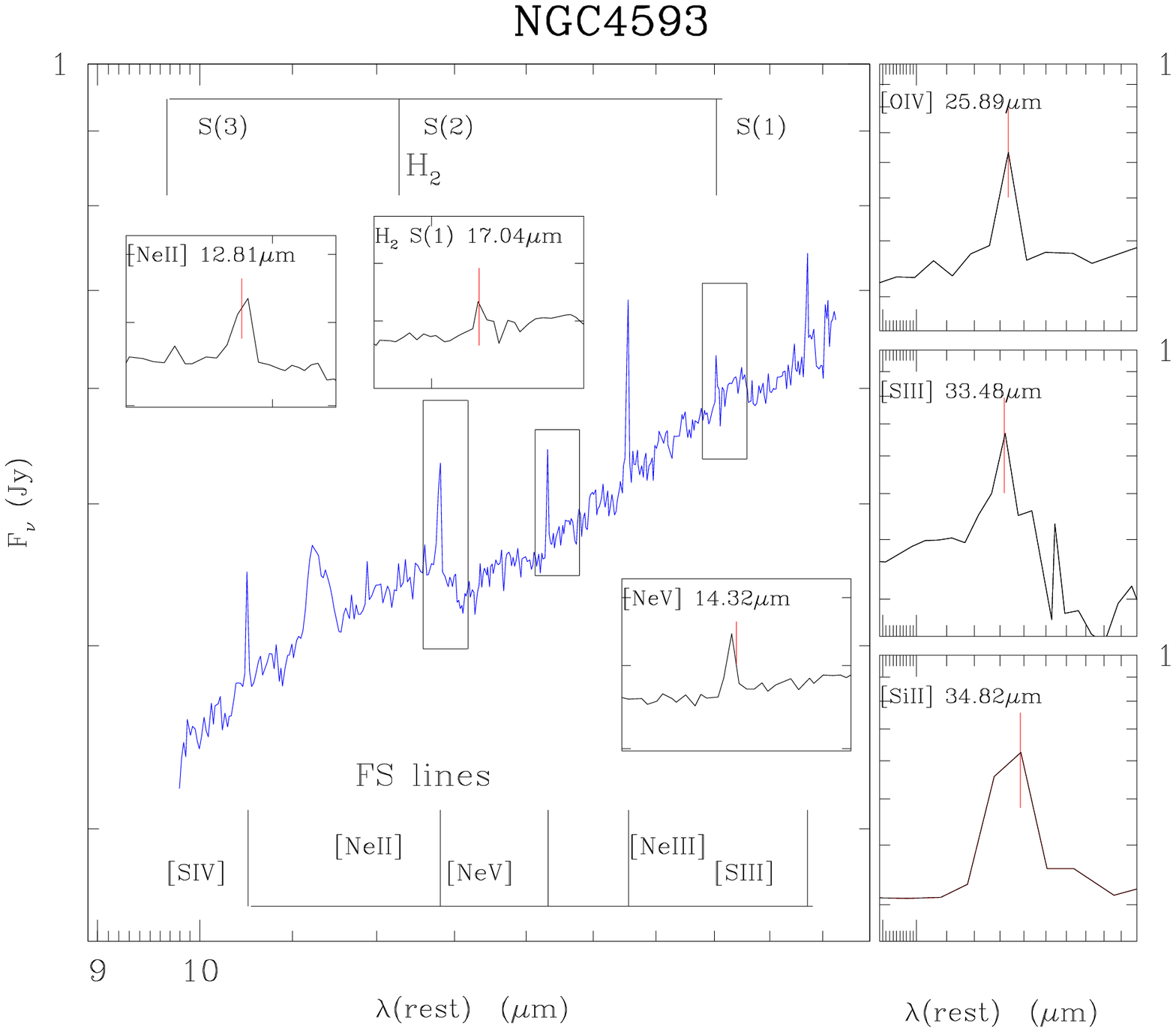}}

\centerline{\includegraphics[width=8cm]{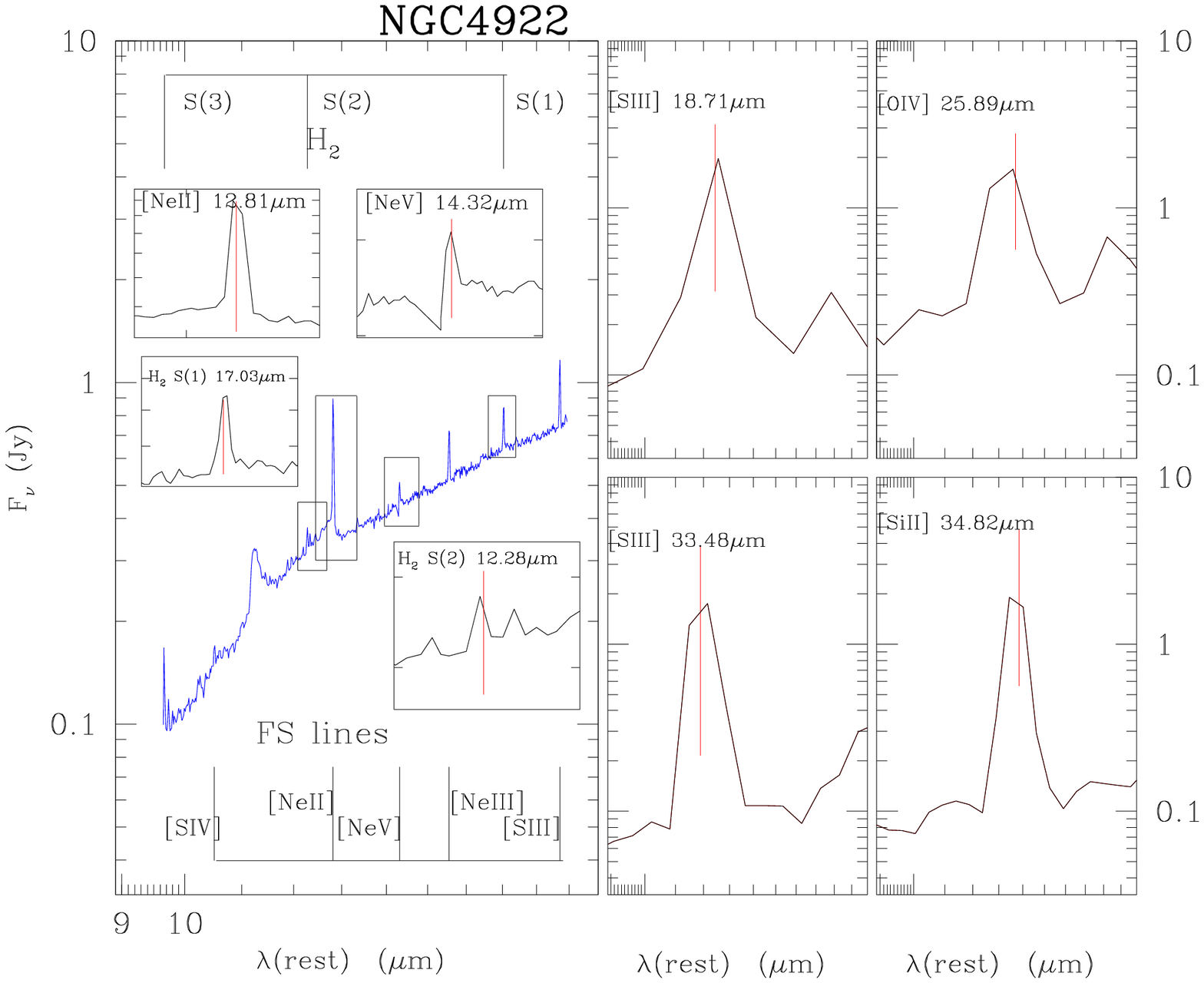}\includegraphics[width=8cm]{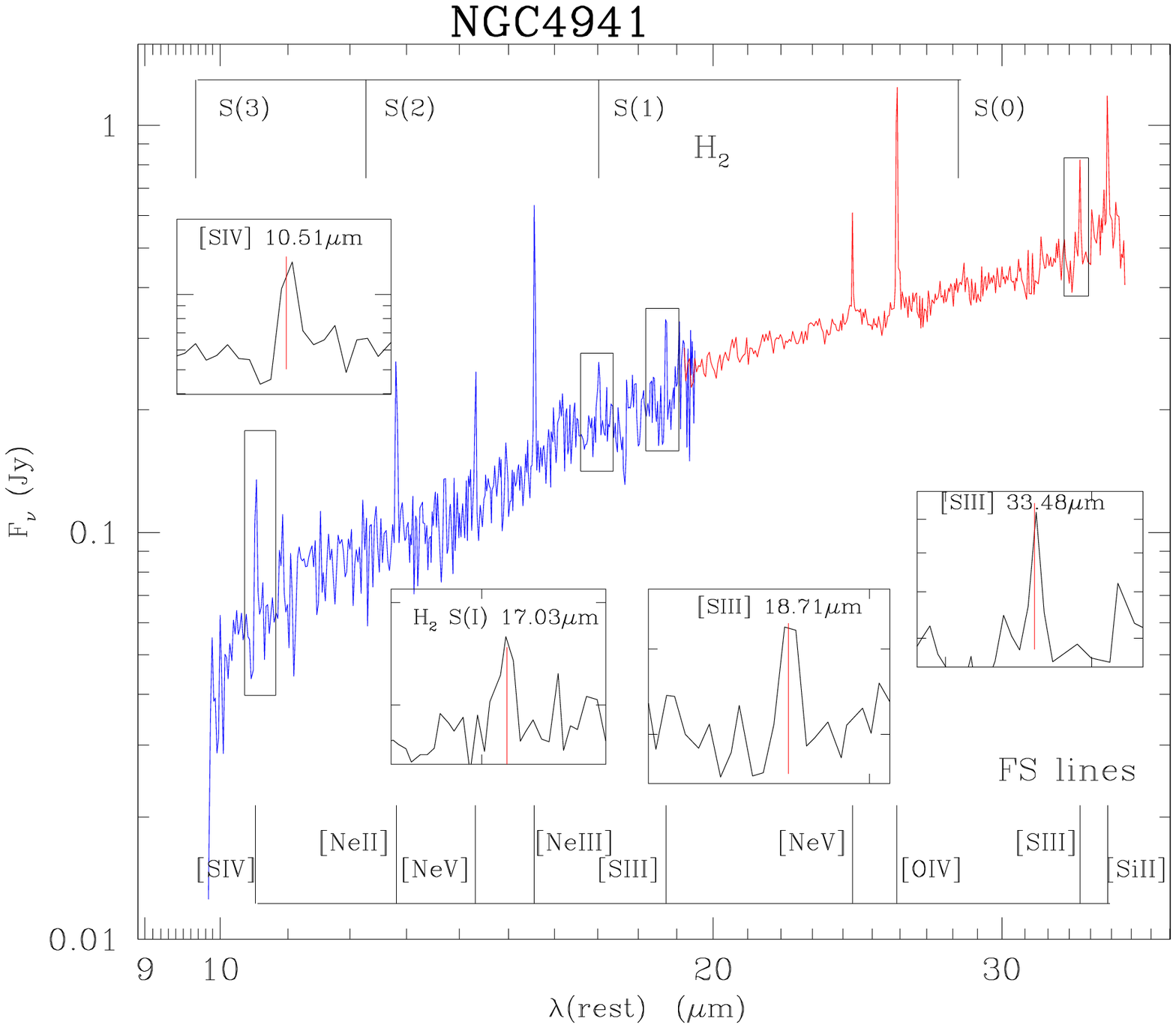}}

\end{figure}
\clearpage

\begin{figure}
\centerline{\includegraphics[width=8cm]{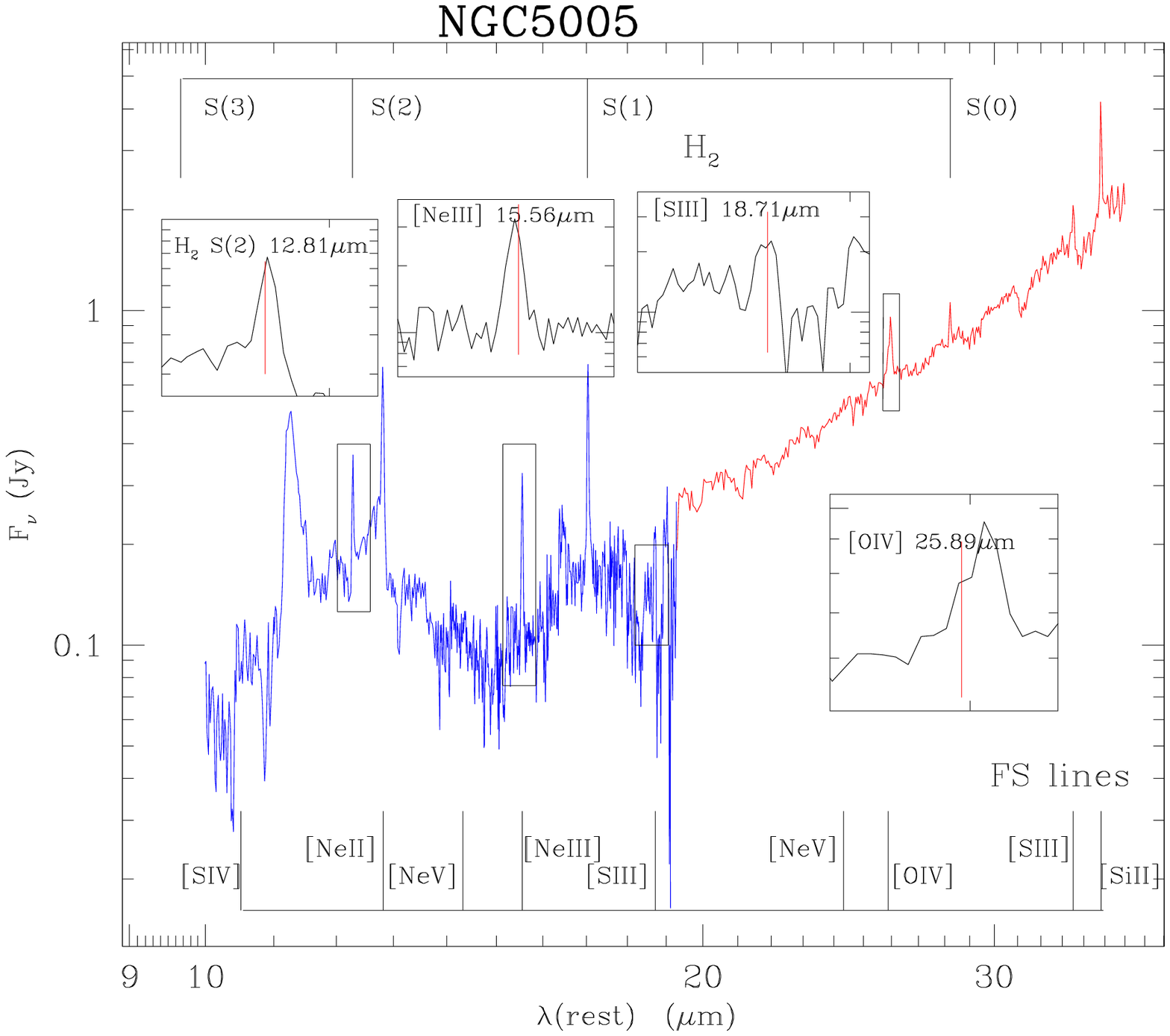}\includegraphics[width=8cm]{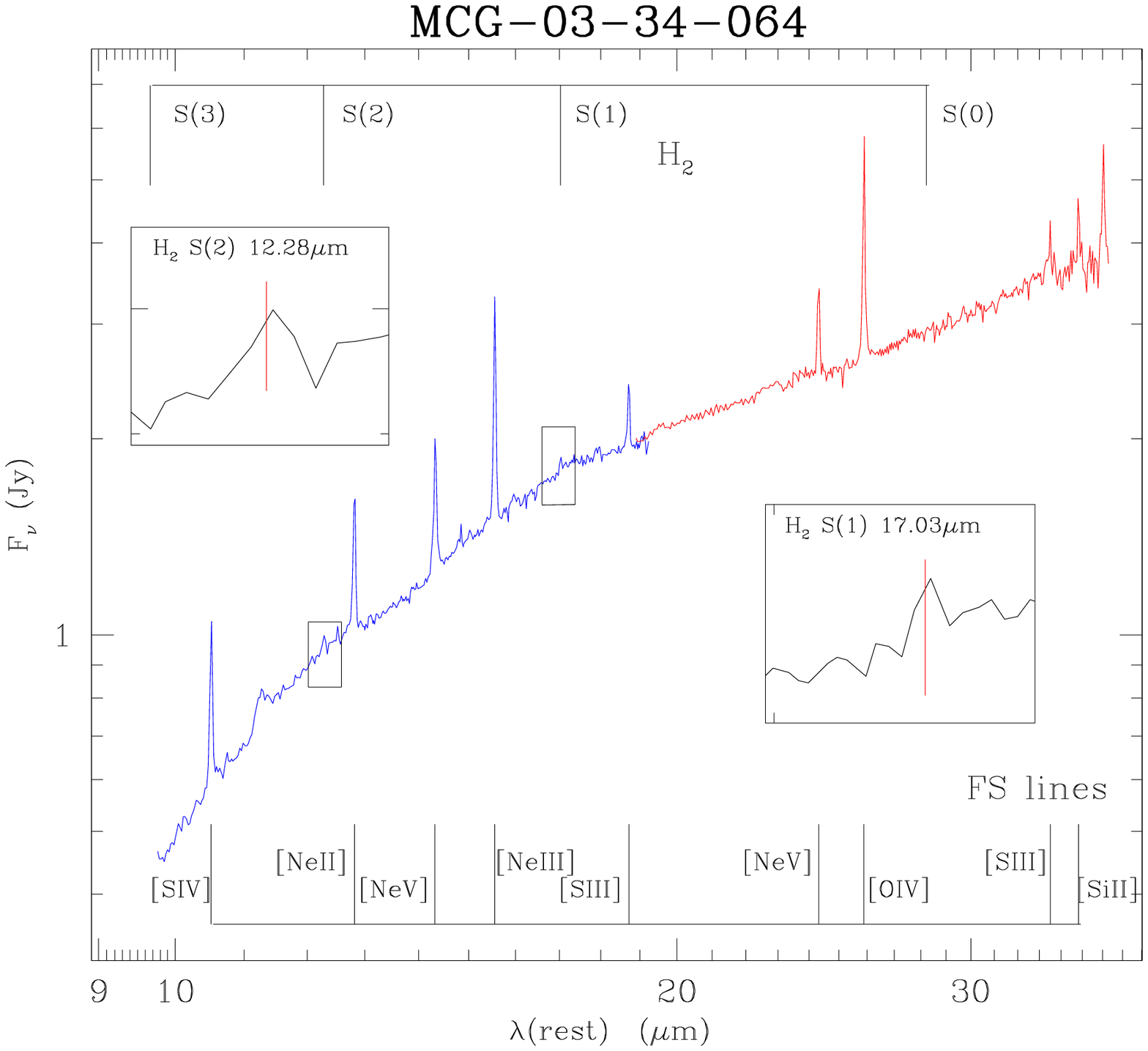}}

\centerline{\includegraphics[width=8cm]{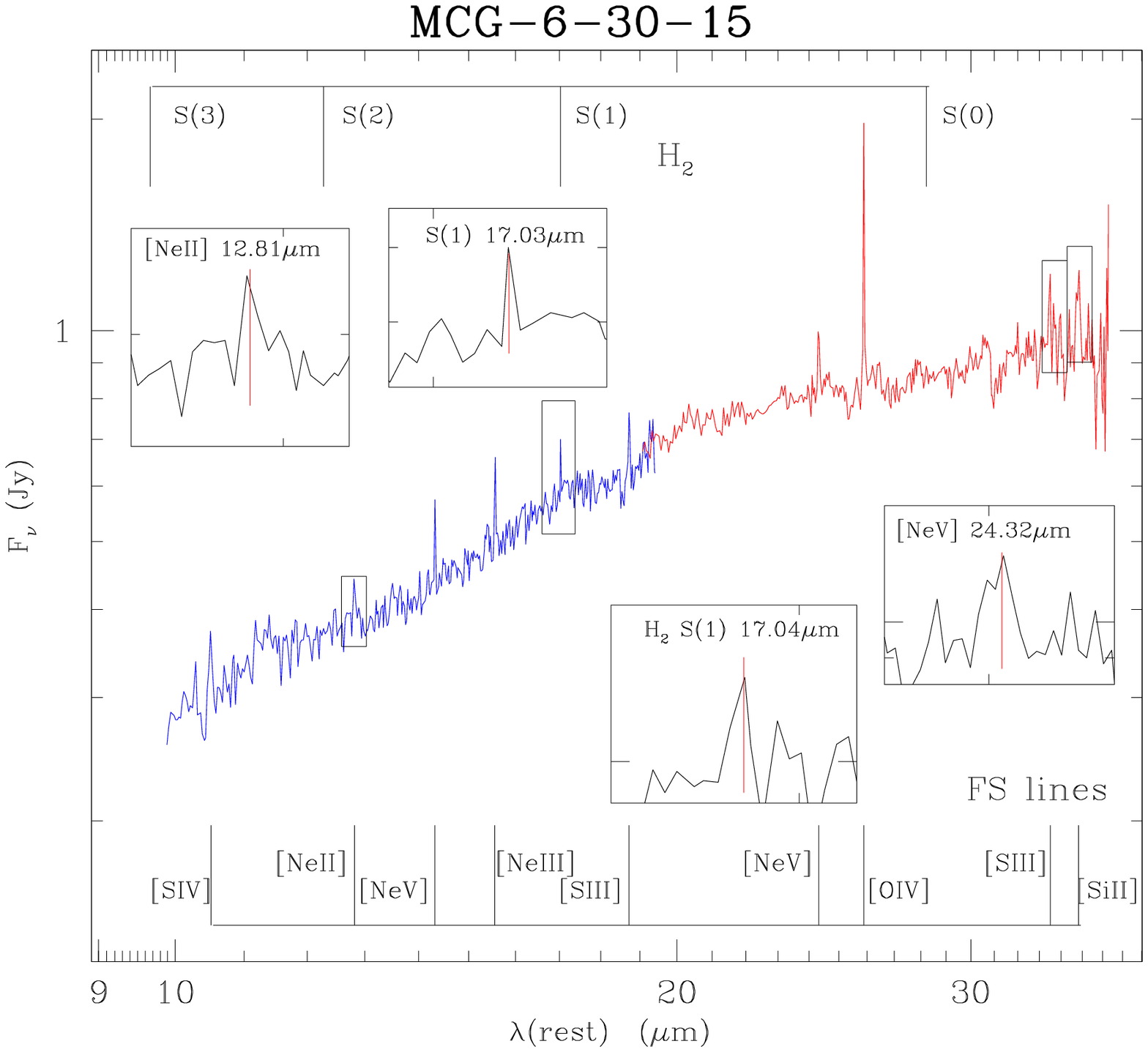}\includegraphics[width=8cm]{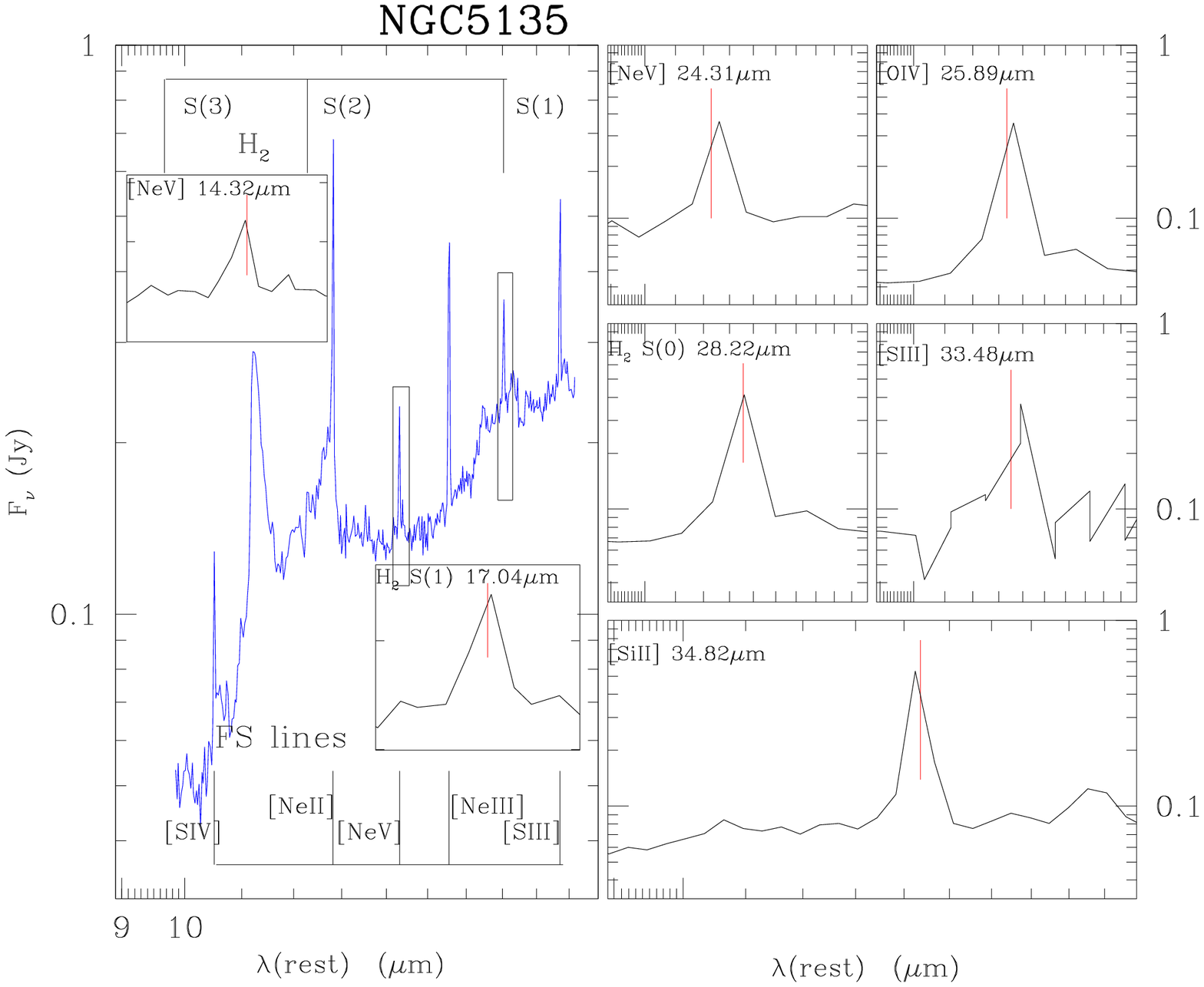}}

\centerline{\includegraphics[width=8cm]{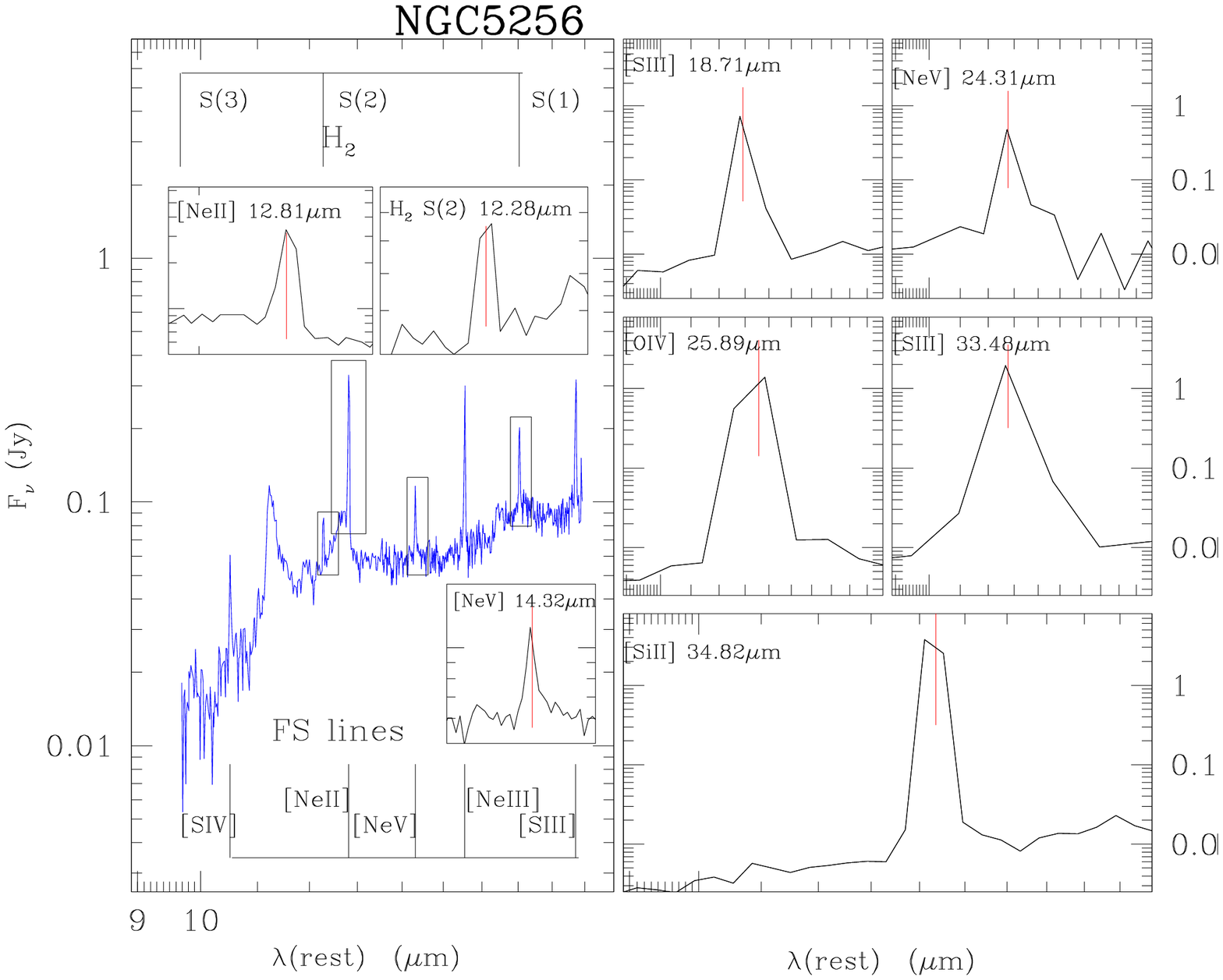}\includegraphics[width=8cm]{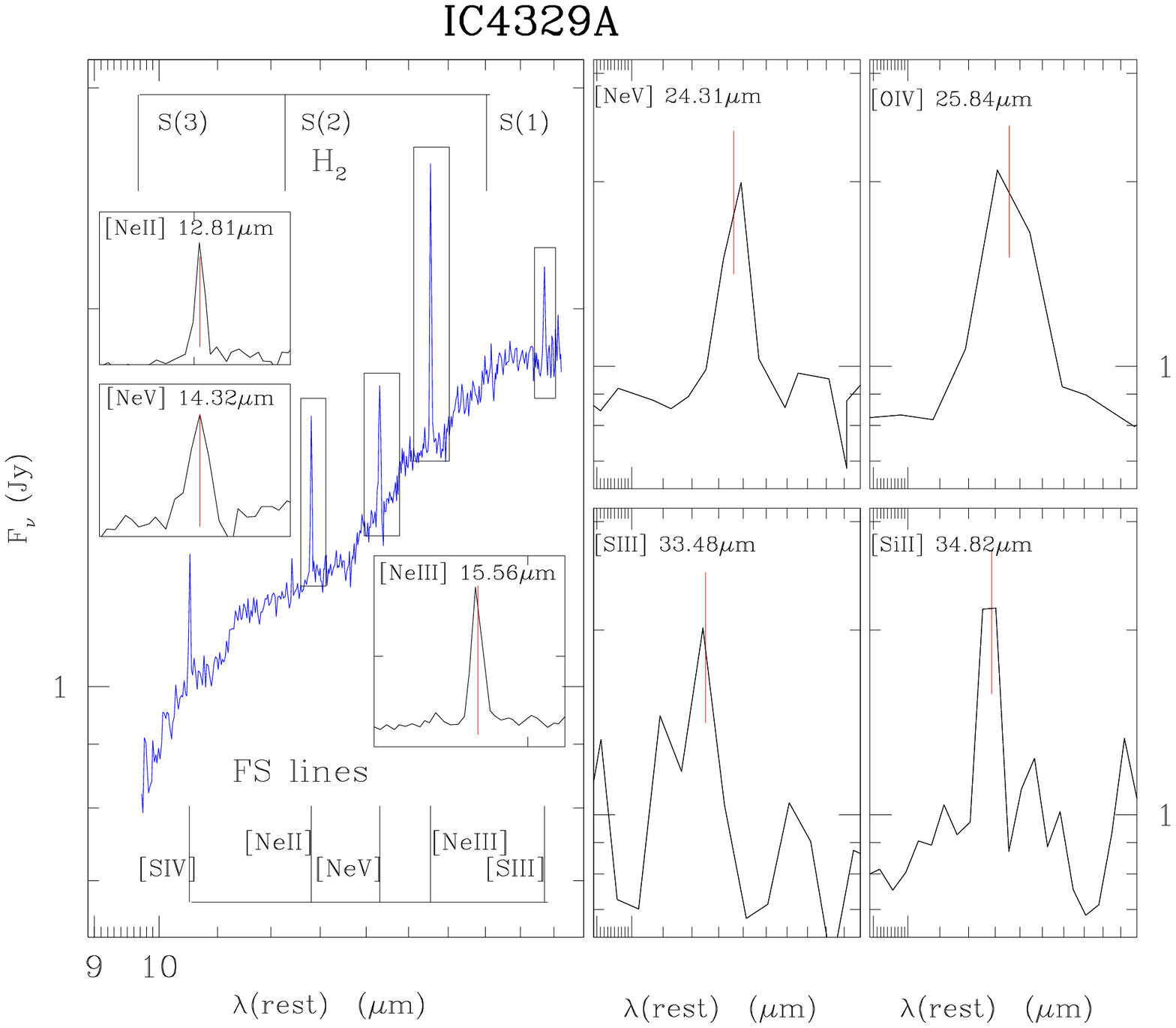}}
\end{figure}
\clearpage

\begin{figure}

\centerline{\includegraphics[width=8cm]{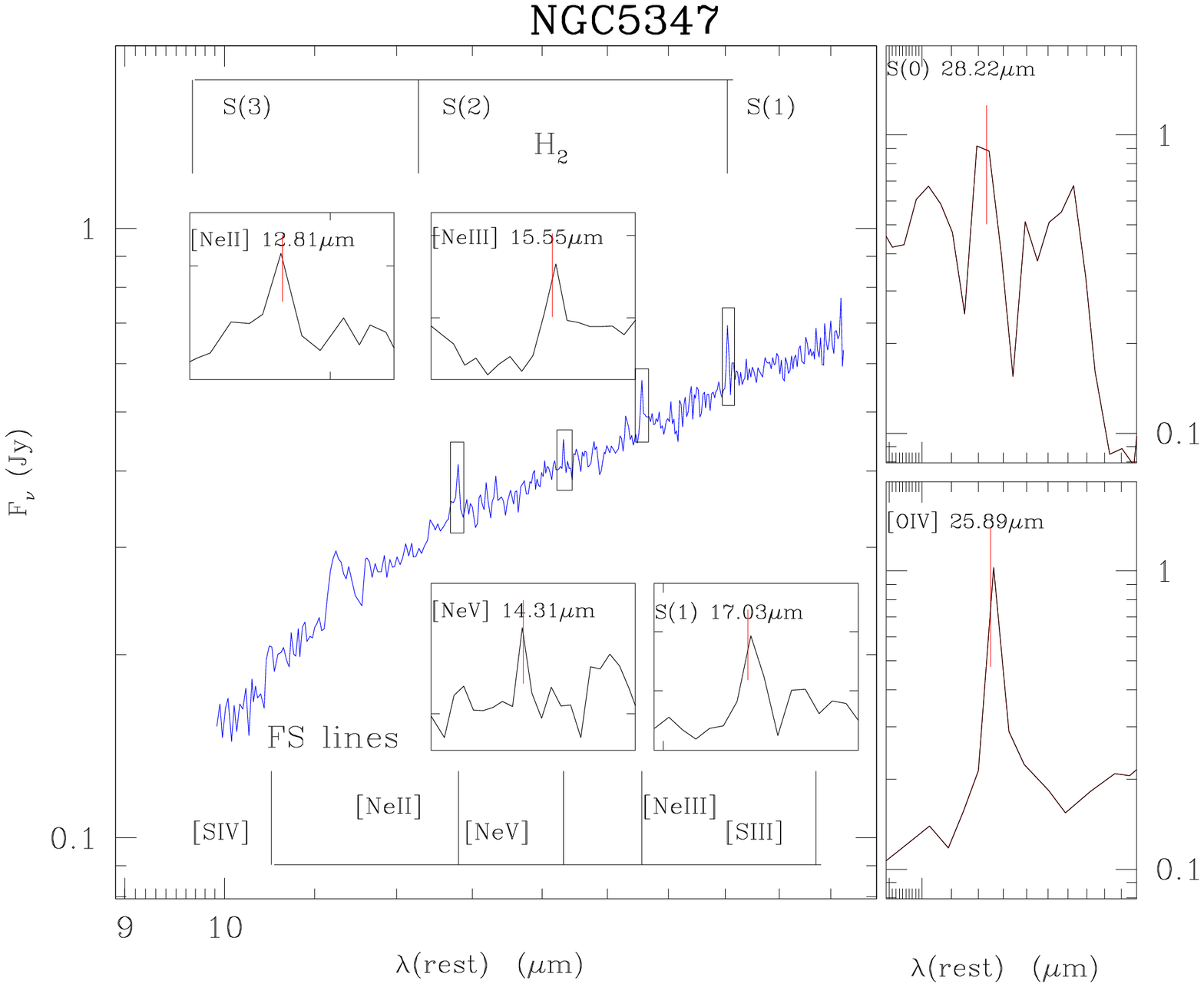}\includegraphics[width=8cm]{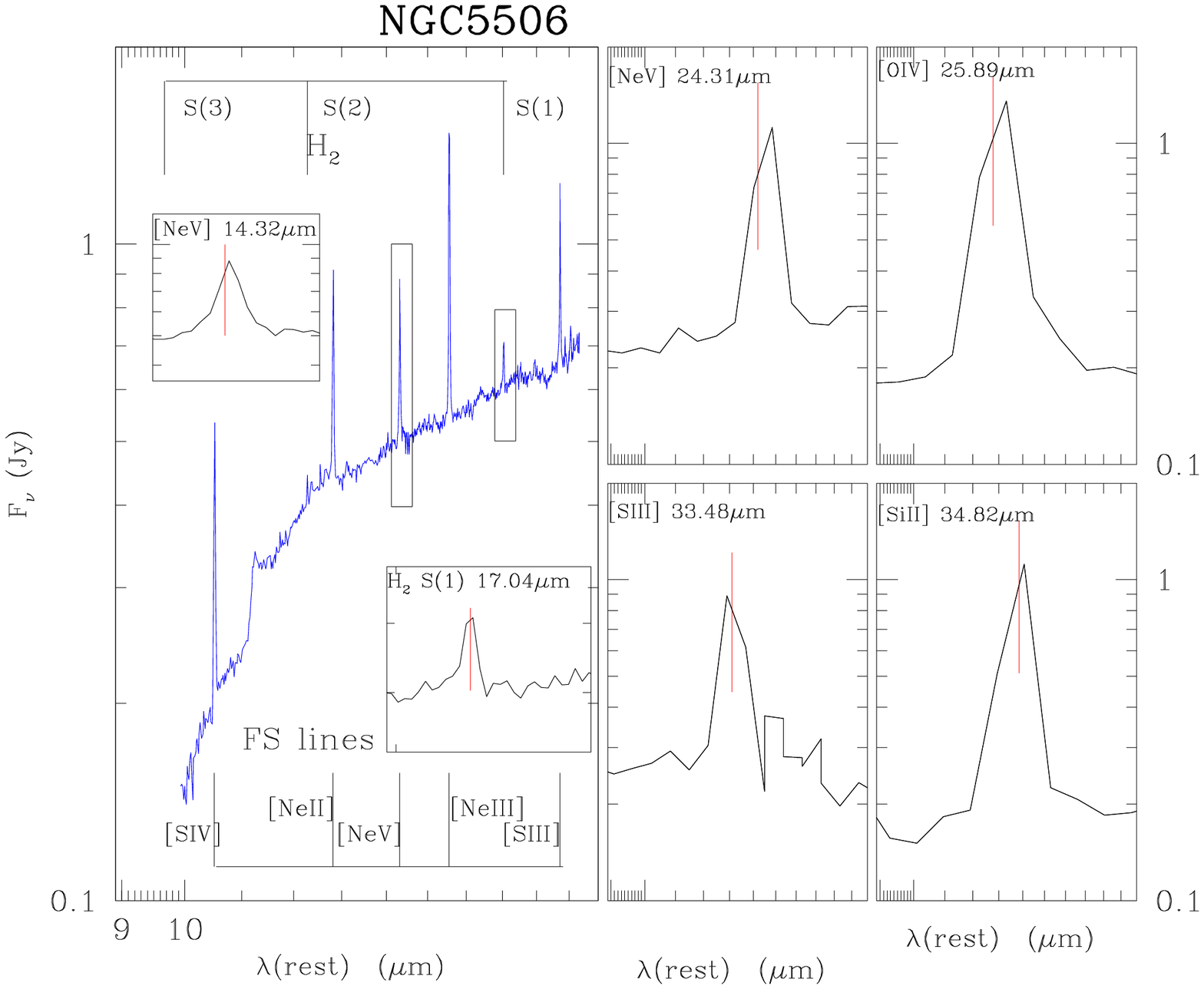}}

\centerline{\includegraphics[width=8cm]{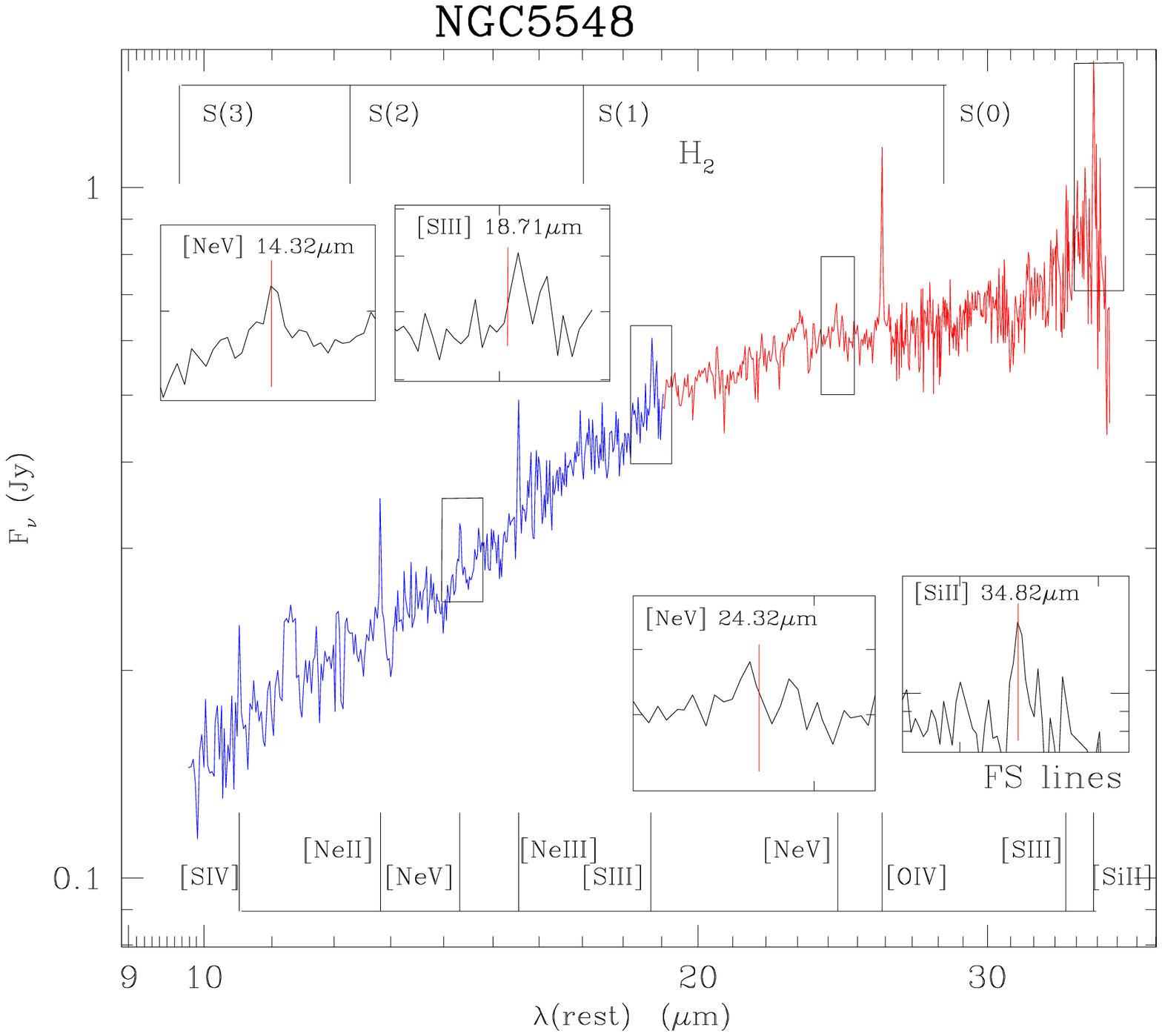}\includegraphics[width=8cm]{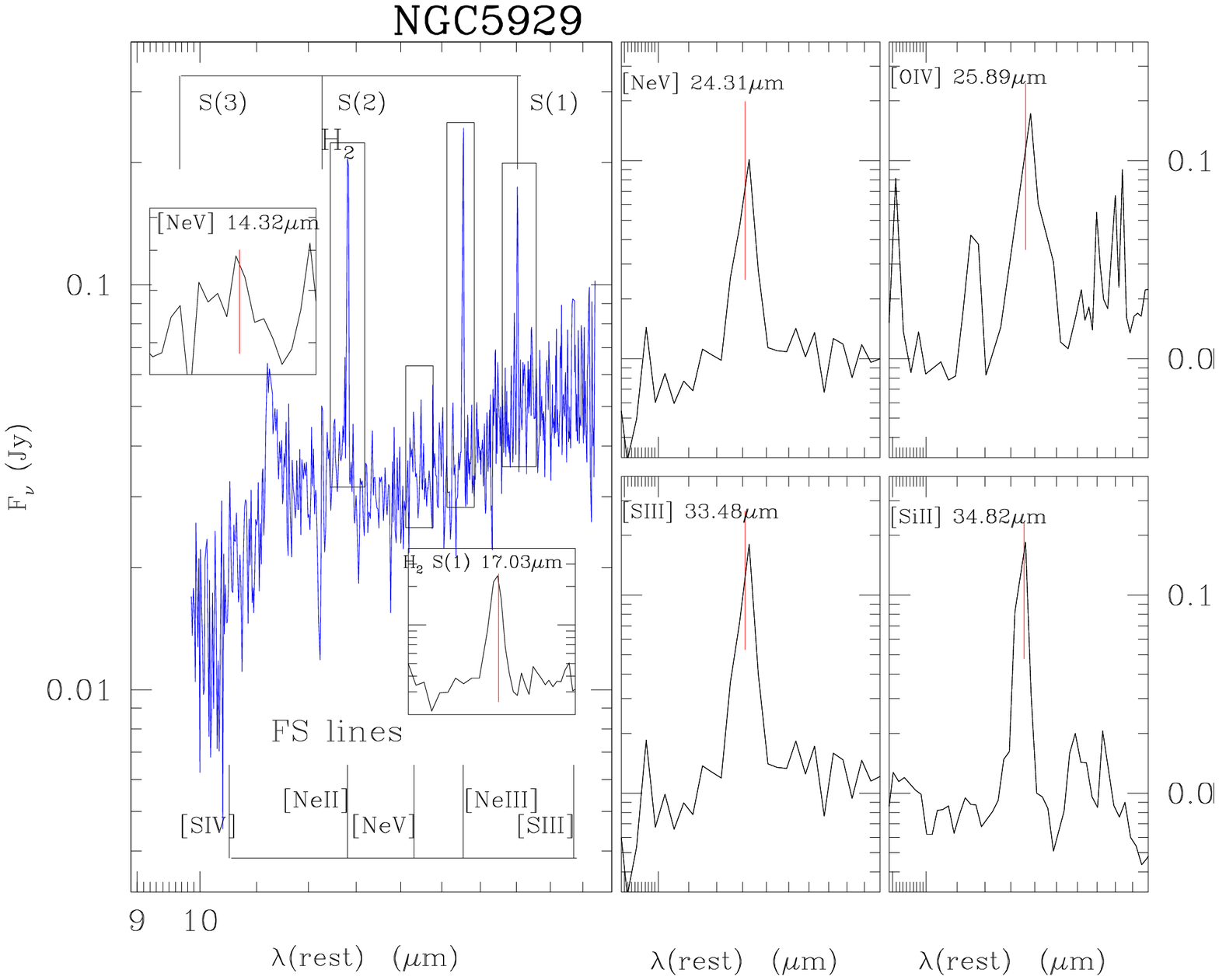}}

\centerline{\includegraphics[width=8cm]{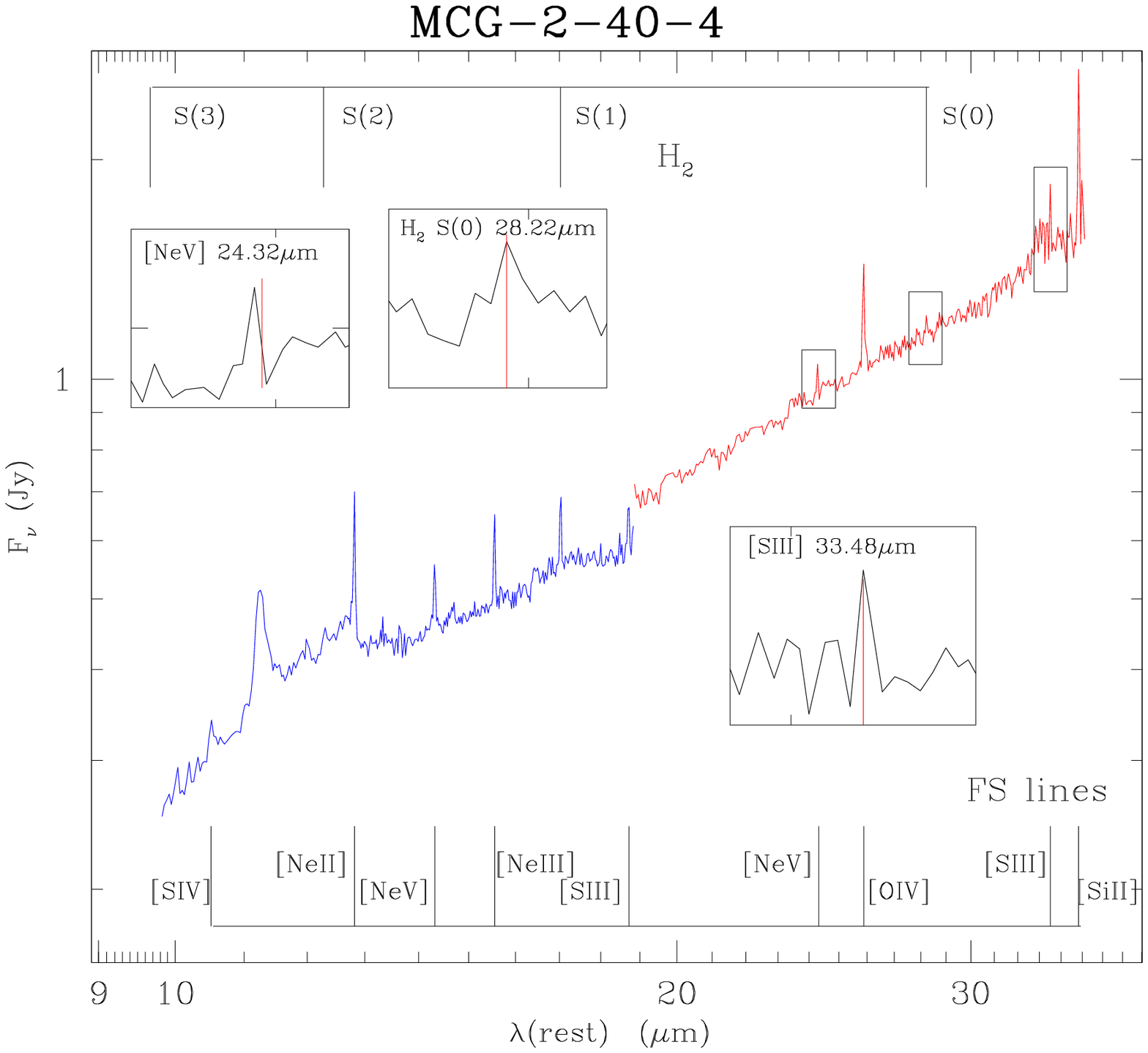}\includegraphics[width=8cm]{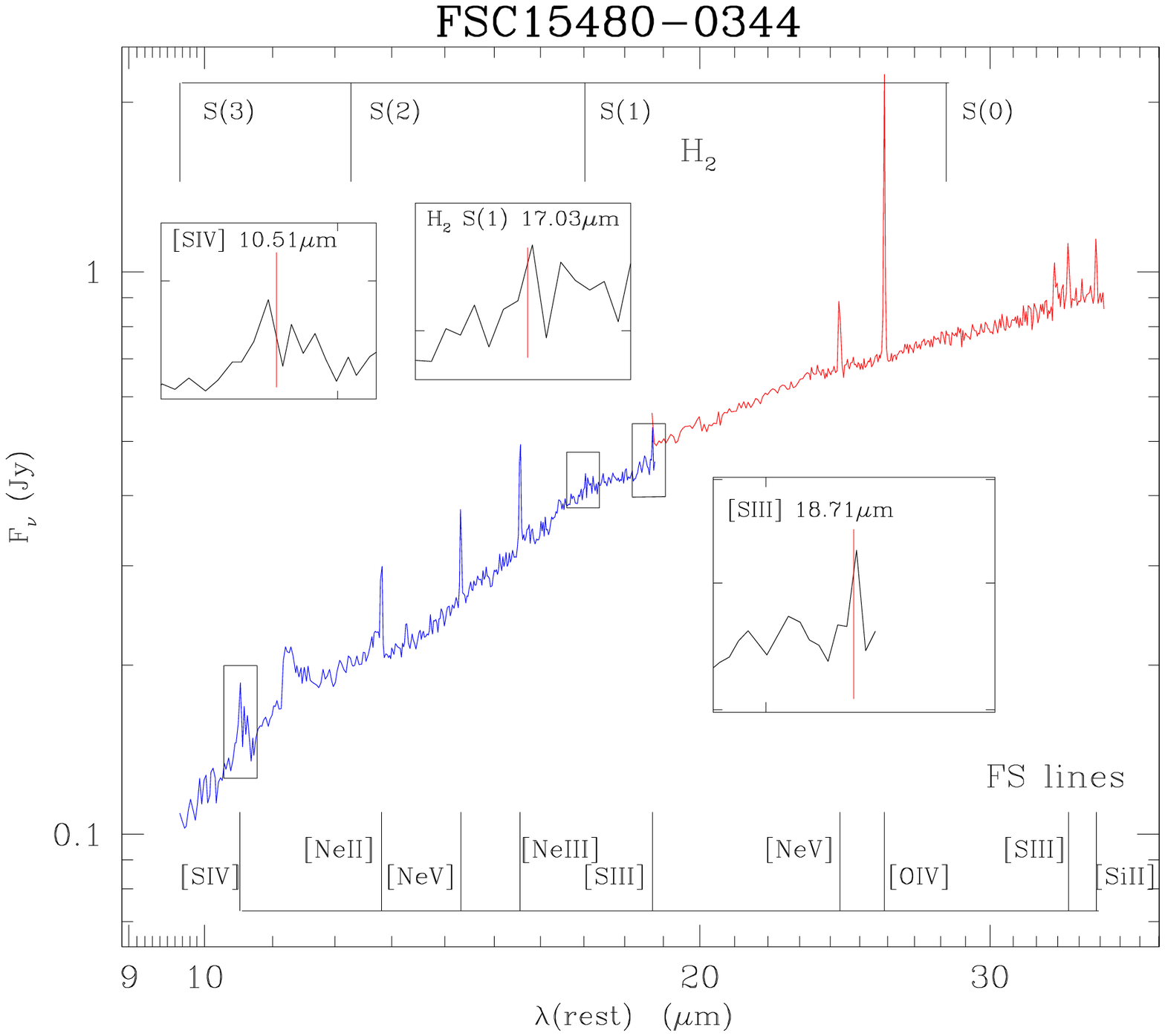}}
\end{figure}
\clearpage

\begin{figure}

\centerline{\includegraphics[width=8cm]{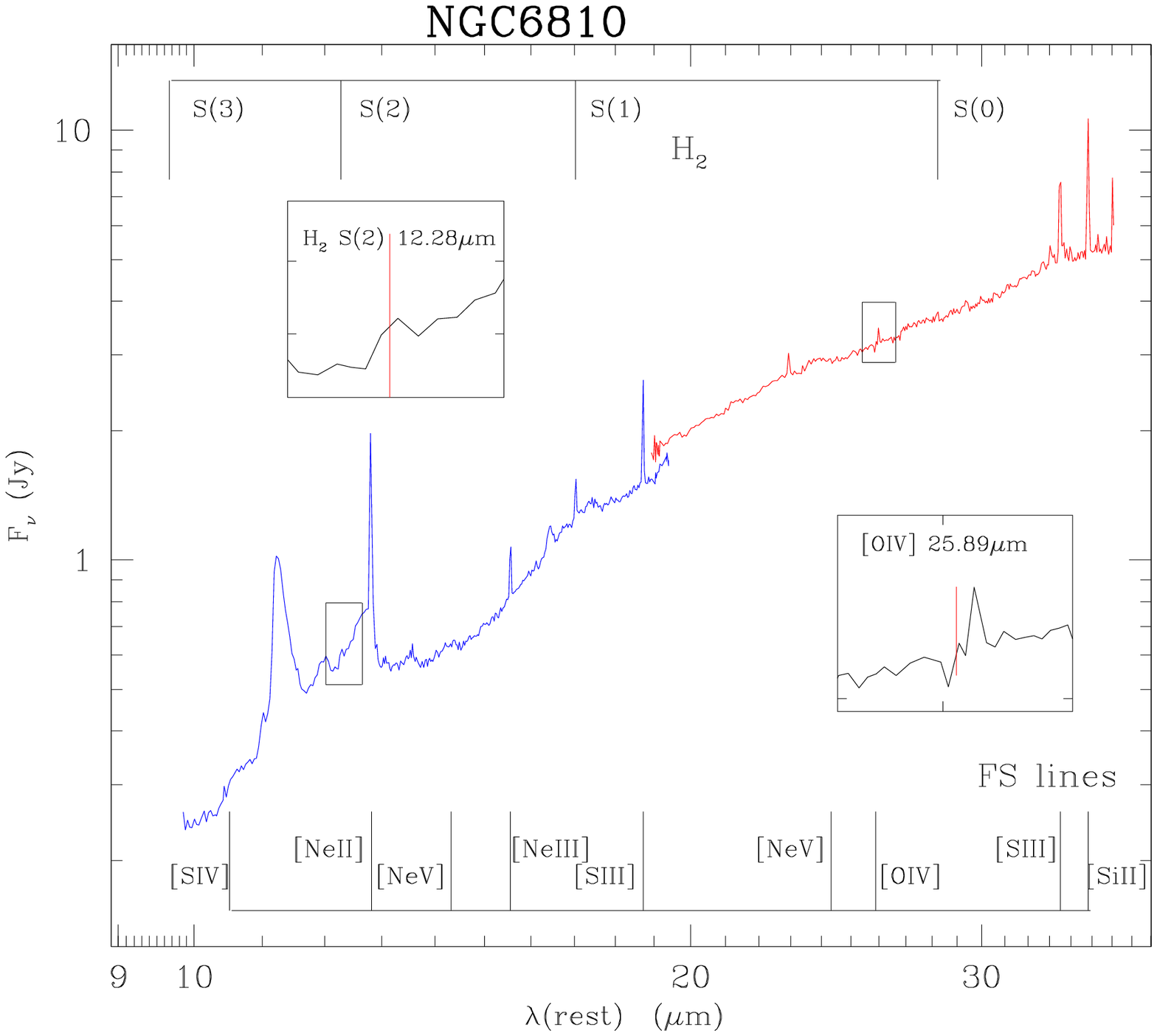}\includegraphics[width=8cm]{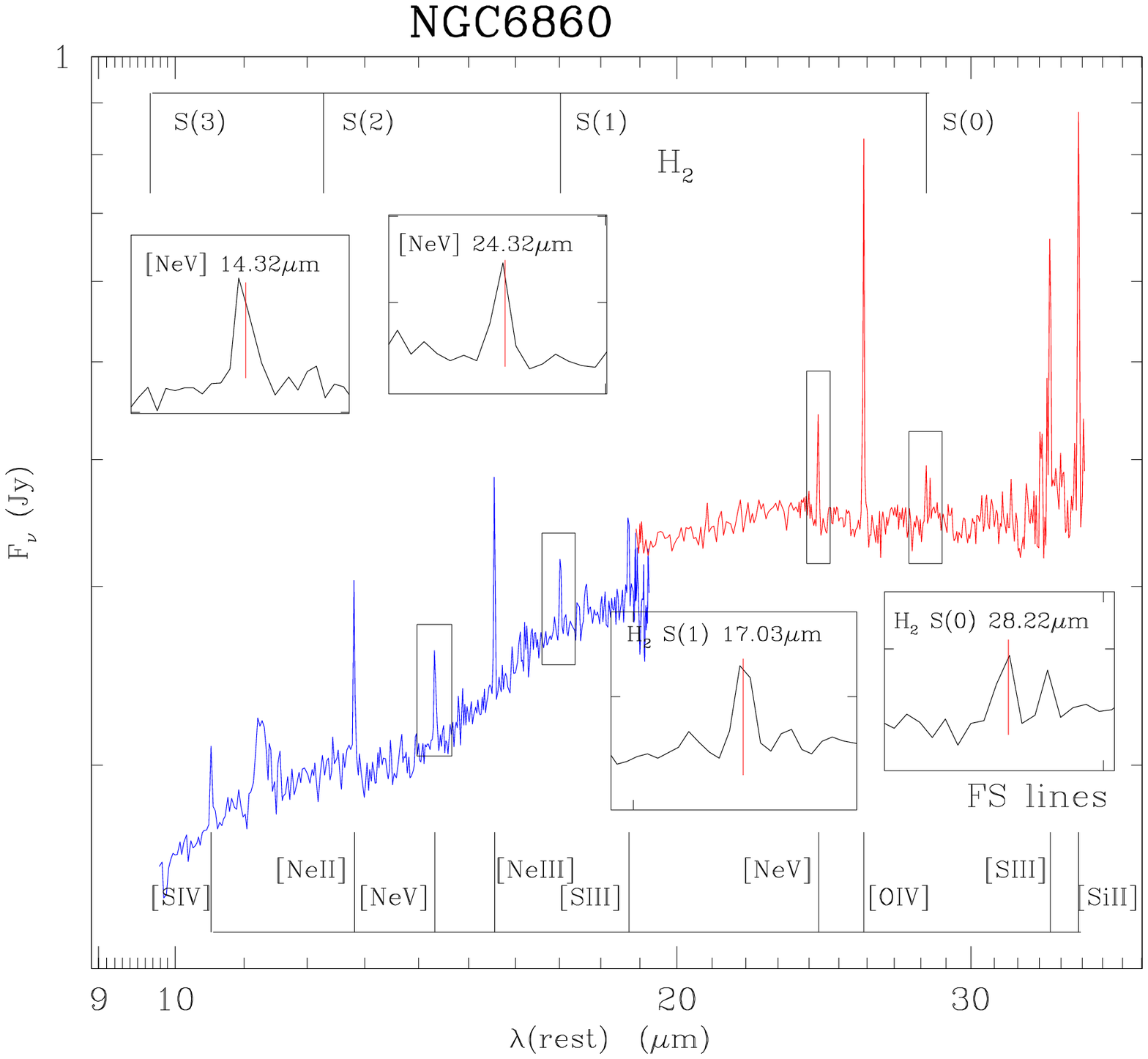}}

\centerline{\includegraphics[width=8cm]{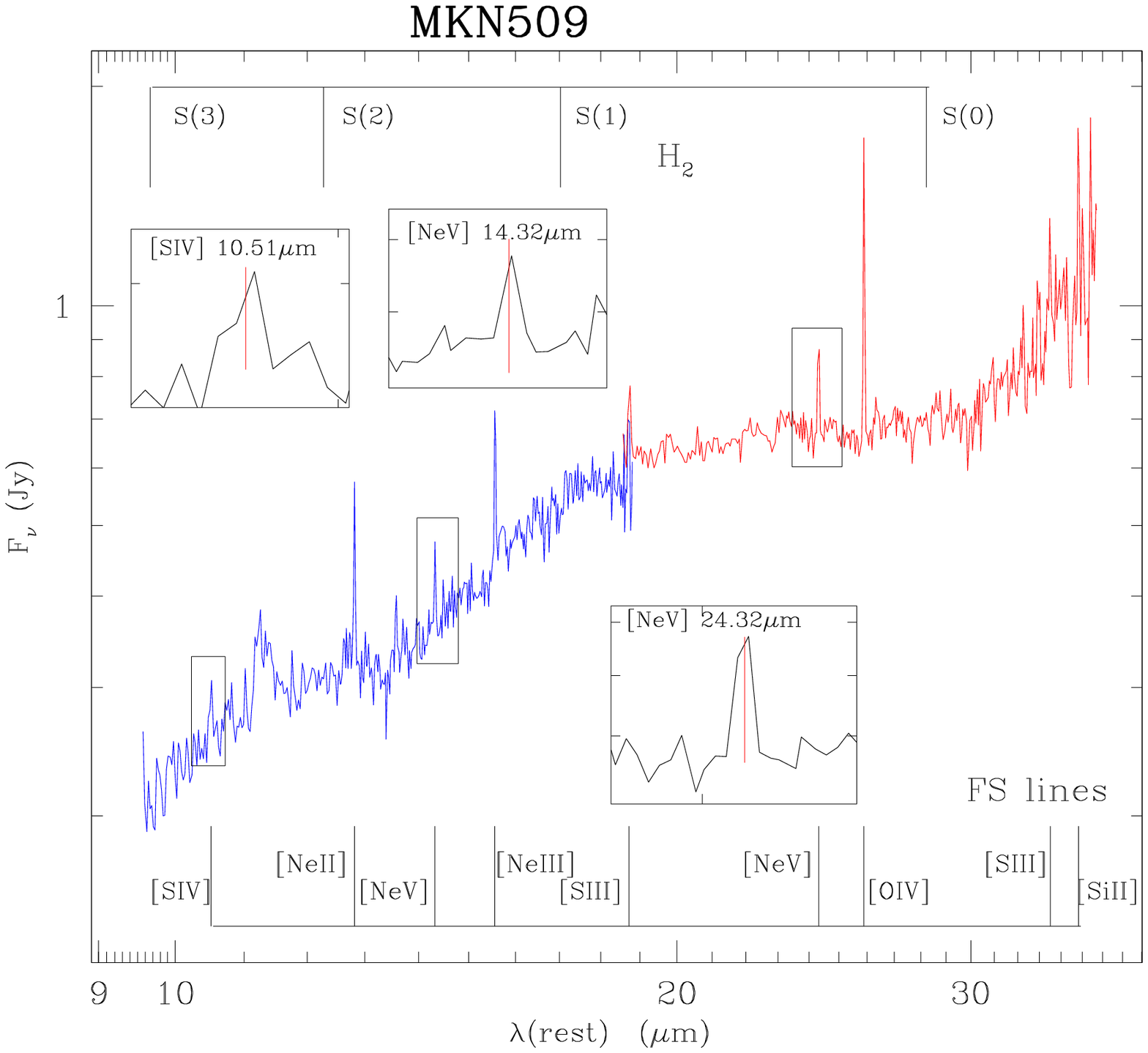}\includegraphics[width=8cm]{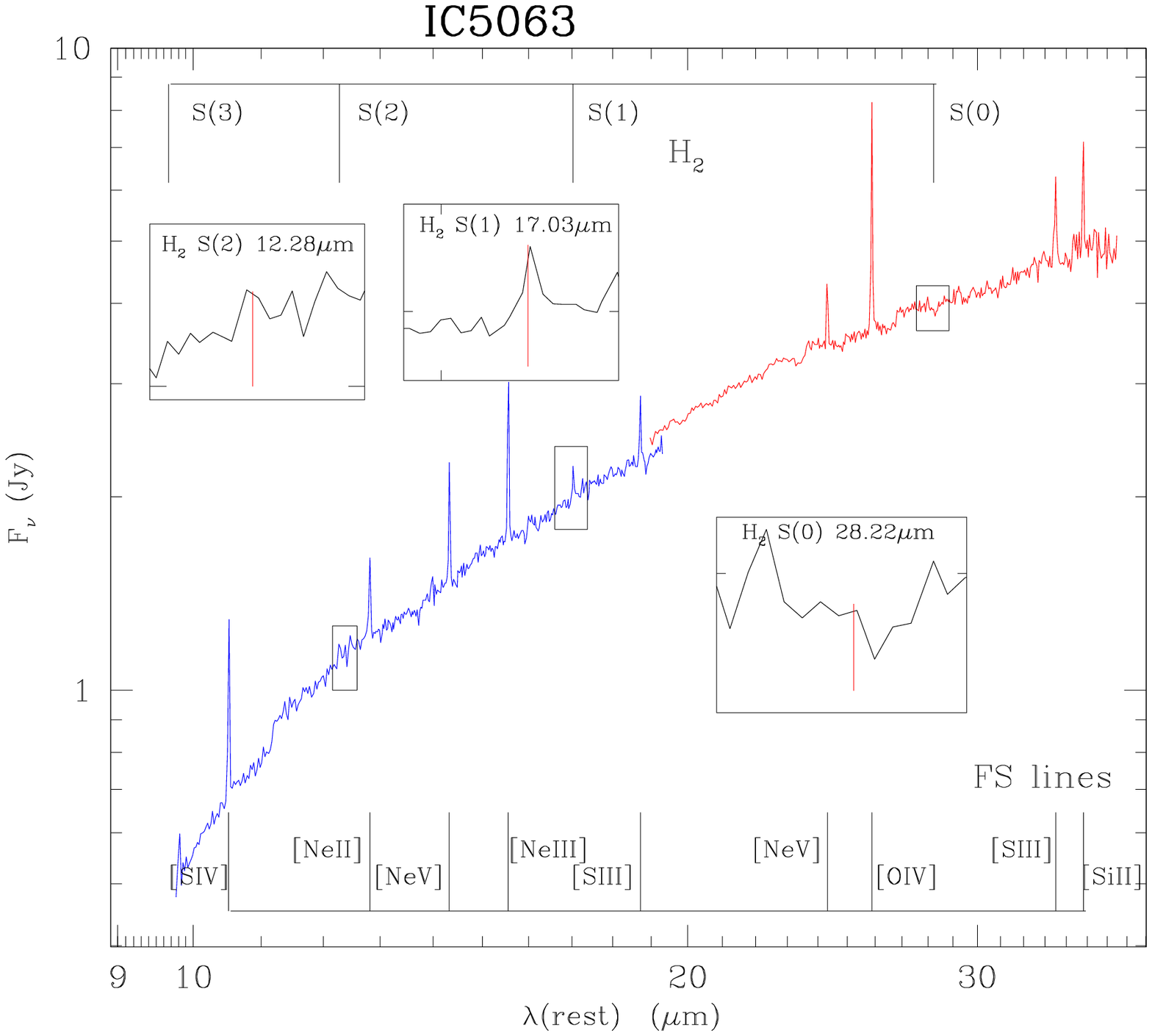}}

\centerline{\includegraphics[width=8cm]{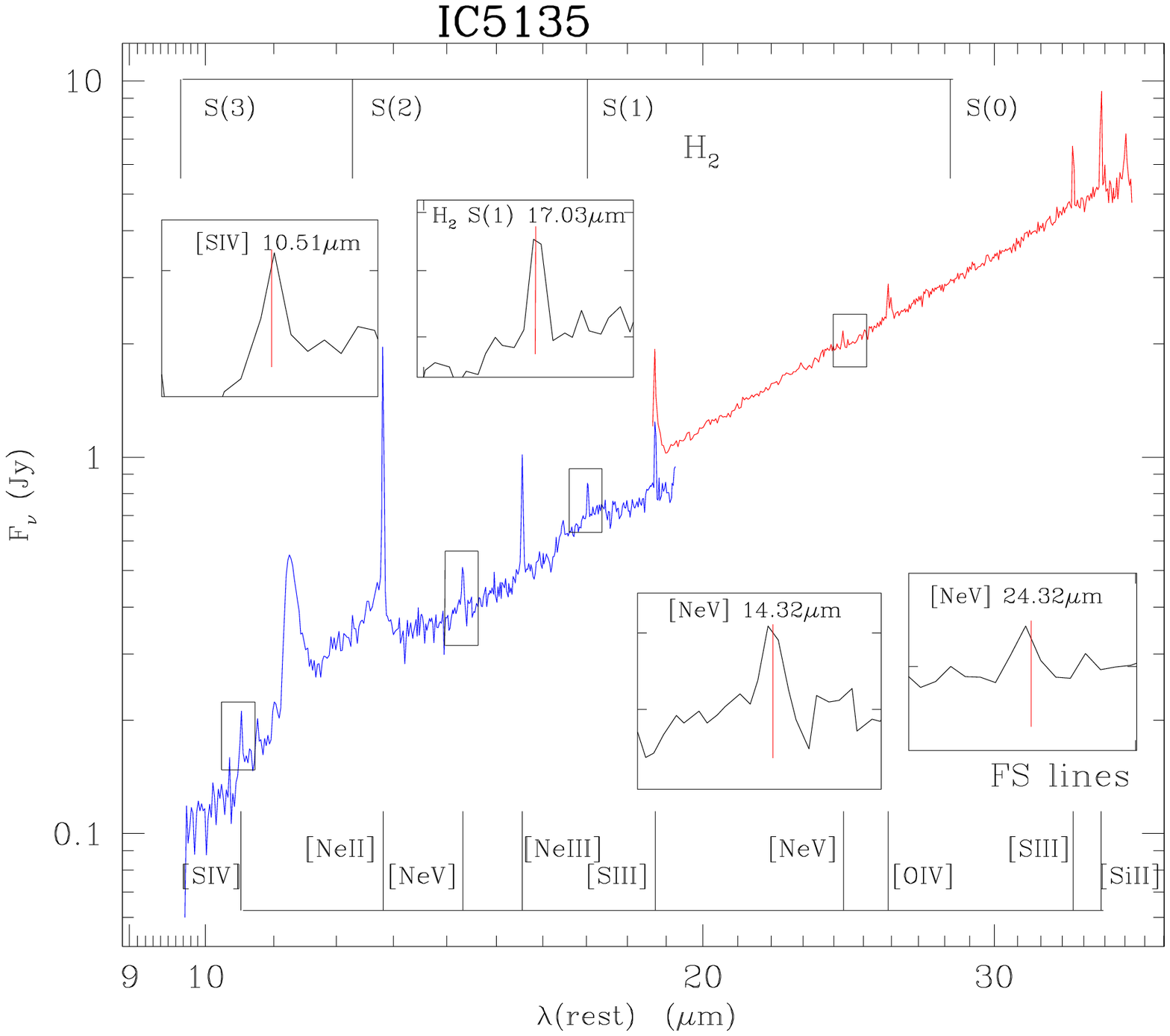}\includegraphics[width=8cm]{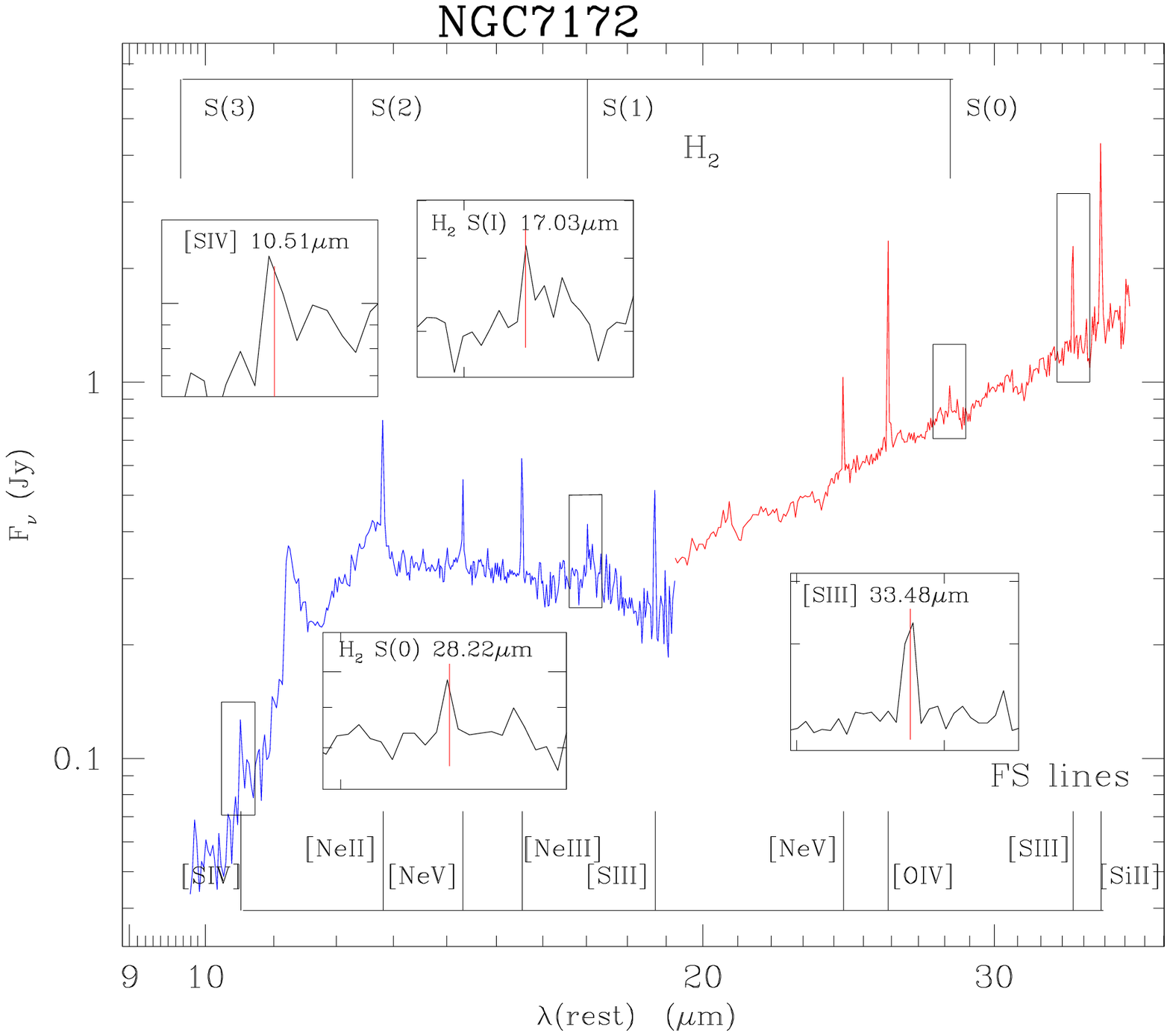}}
\end{figure}

\clearpage

\begin{figure}
\centerline{\includegraphics[width=8cm]{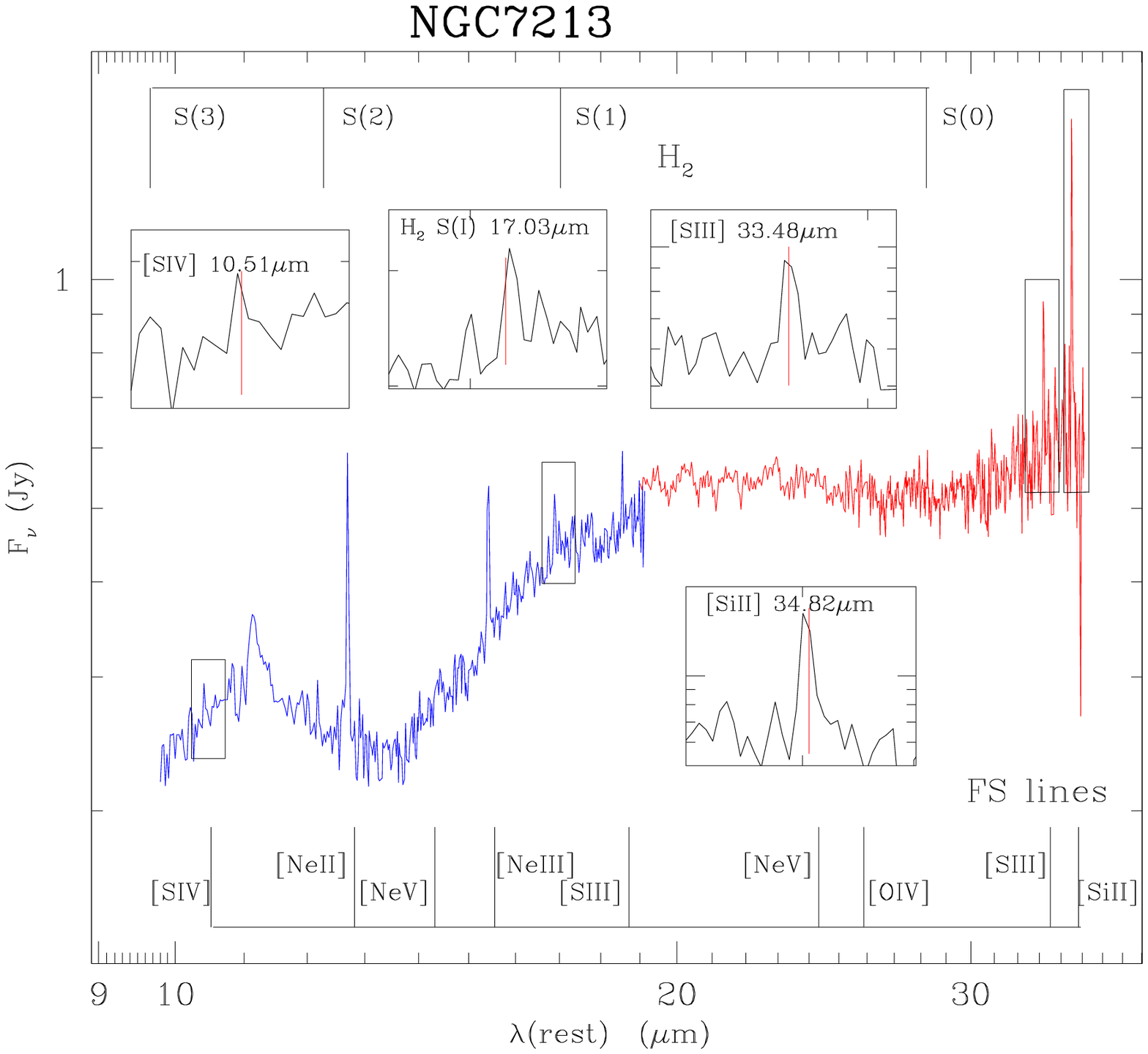}\includegraphics[width=8cm]{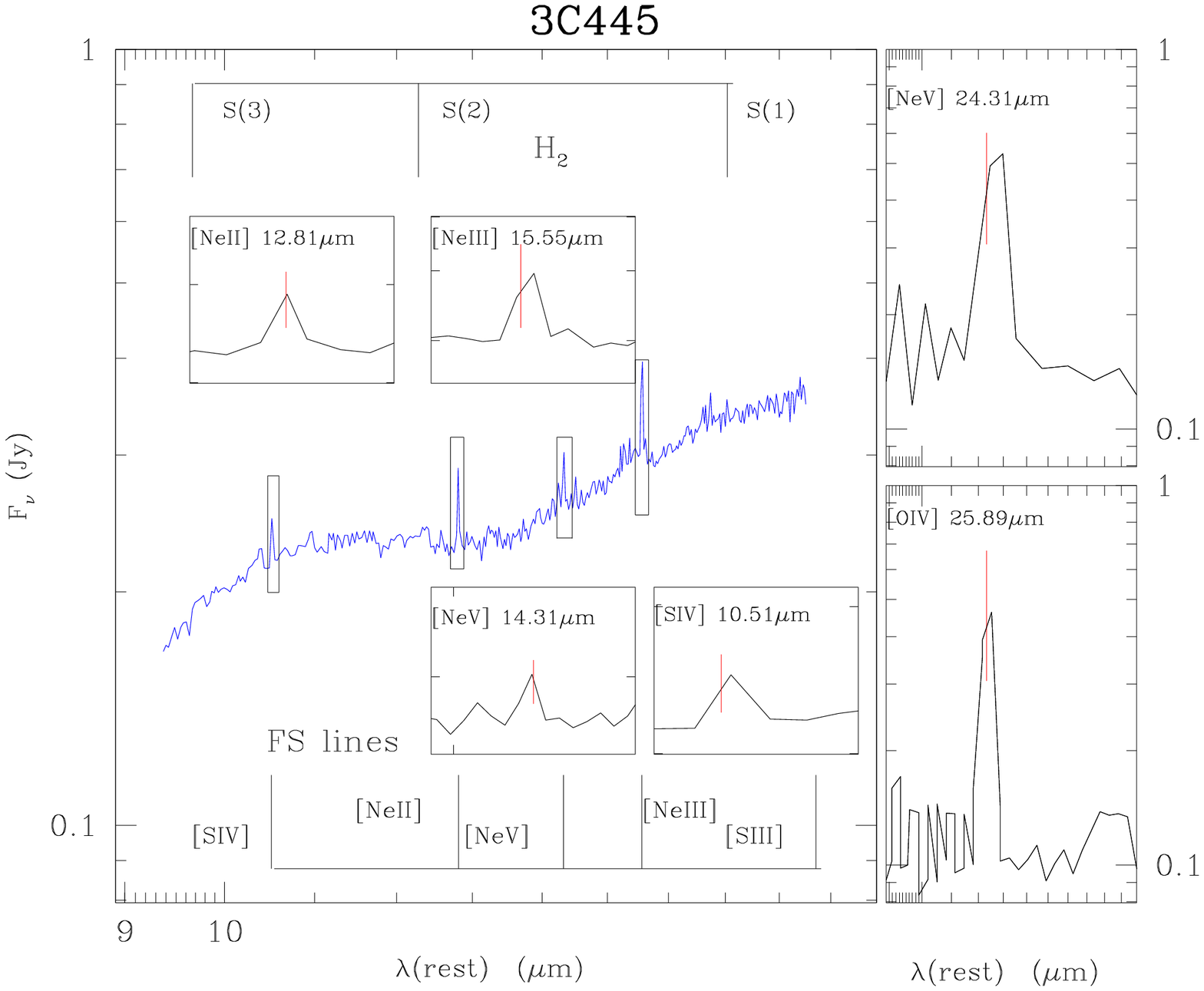}}

\centerline{\includegraphics[width=8cm]{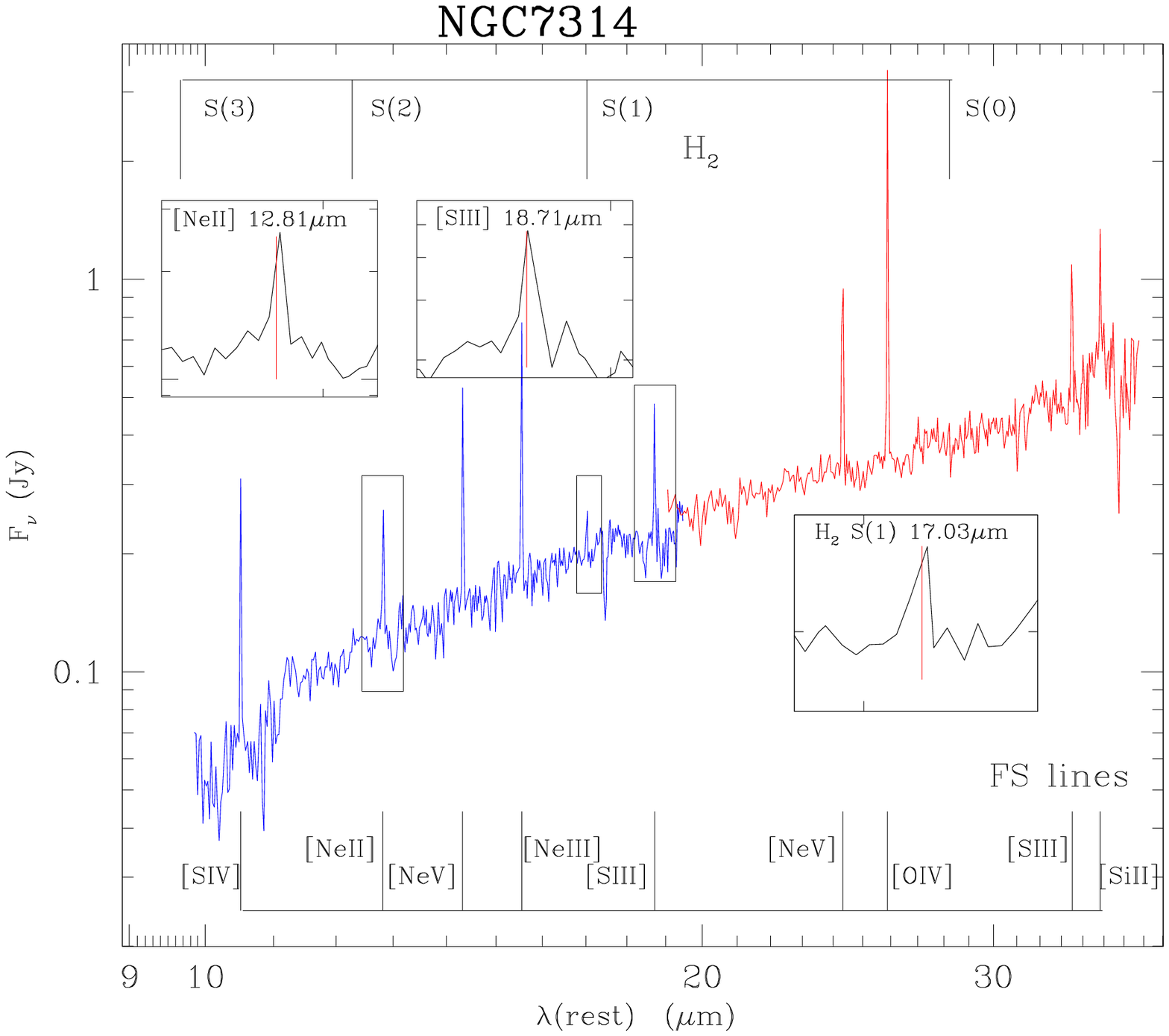}\includegraphics[width=8cm]{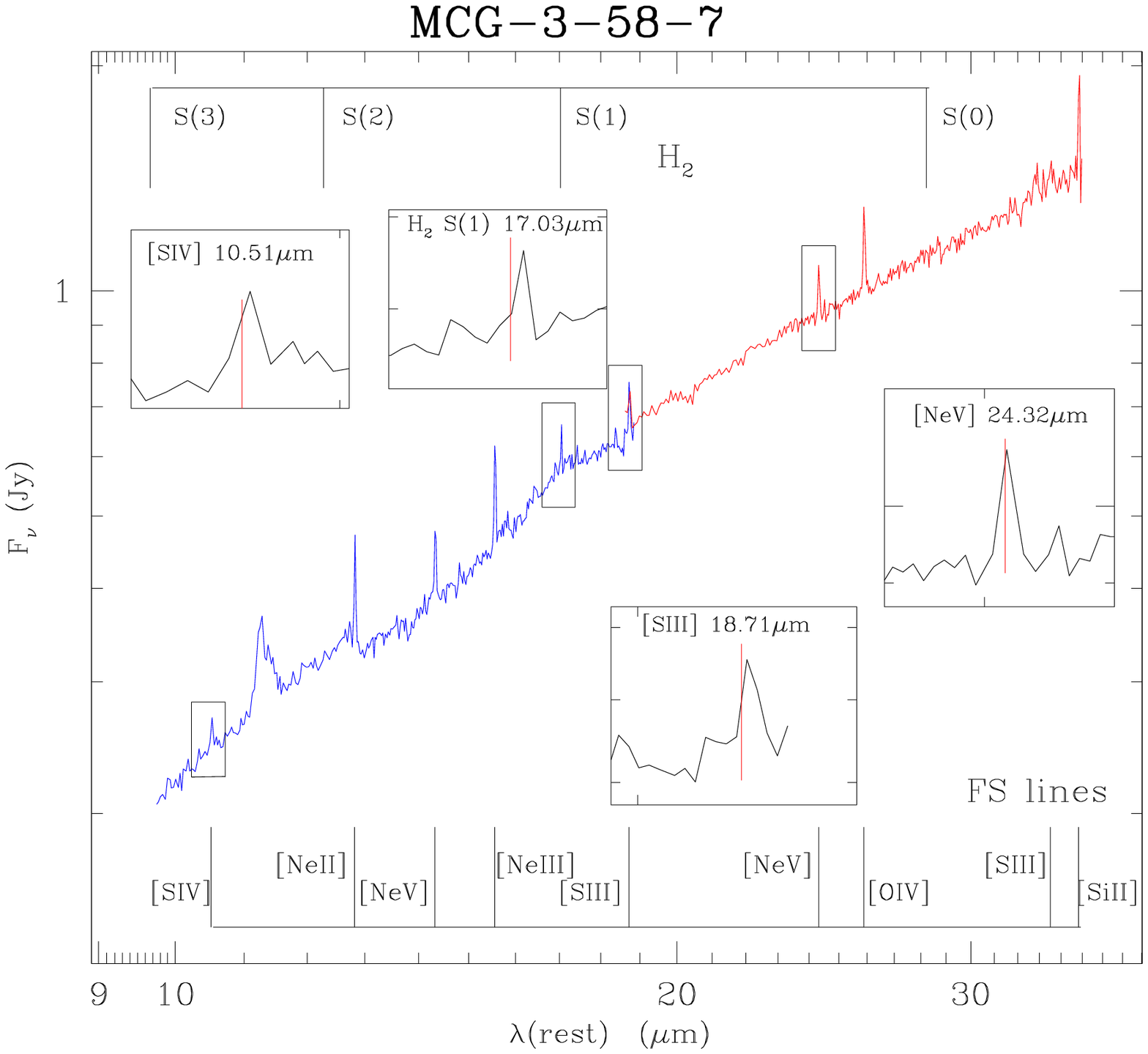}}

\centerline{\includegraphics[width=8cm]{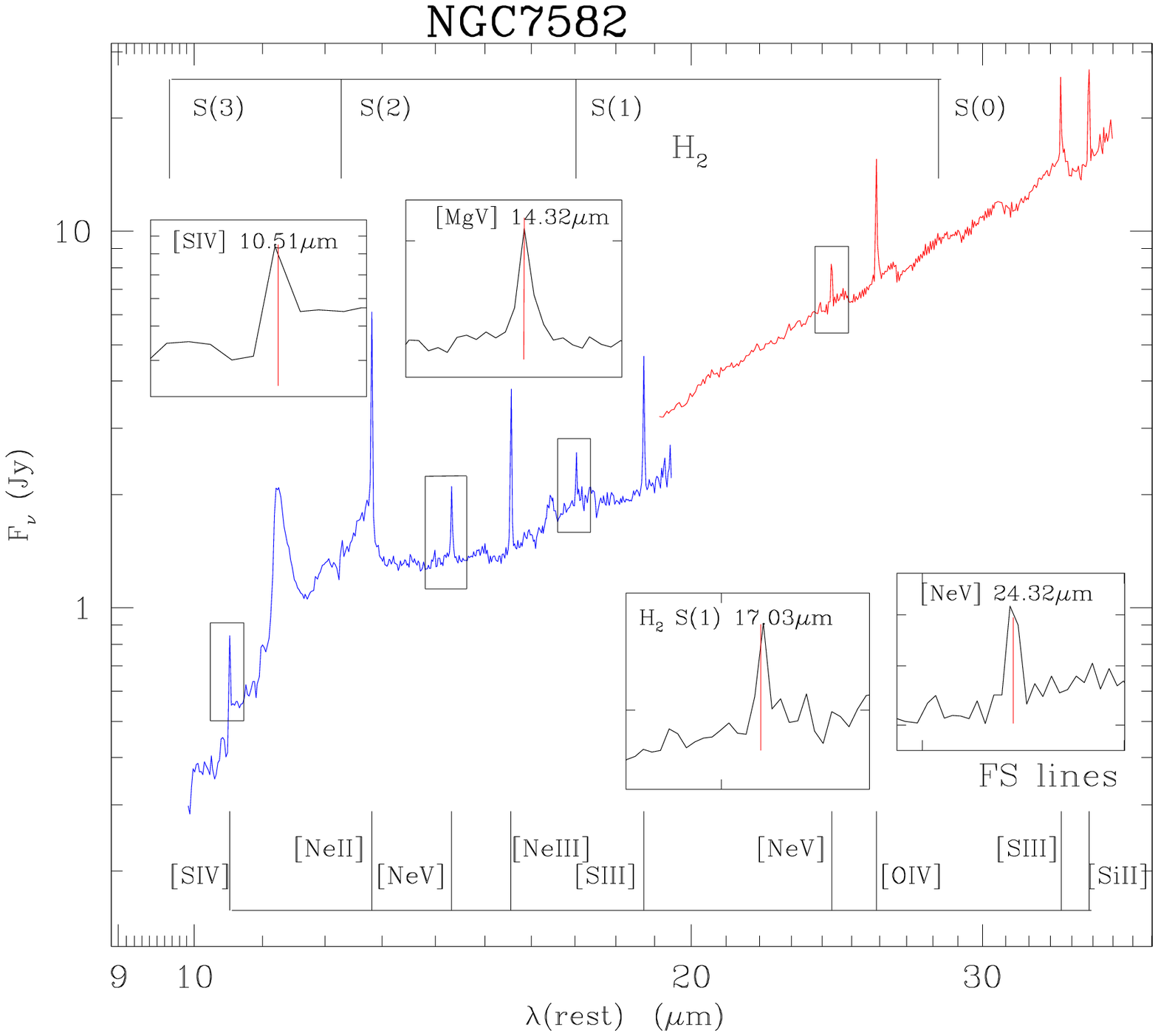}\includegraphics[width=8cm]{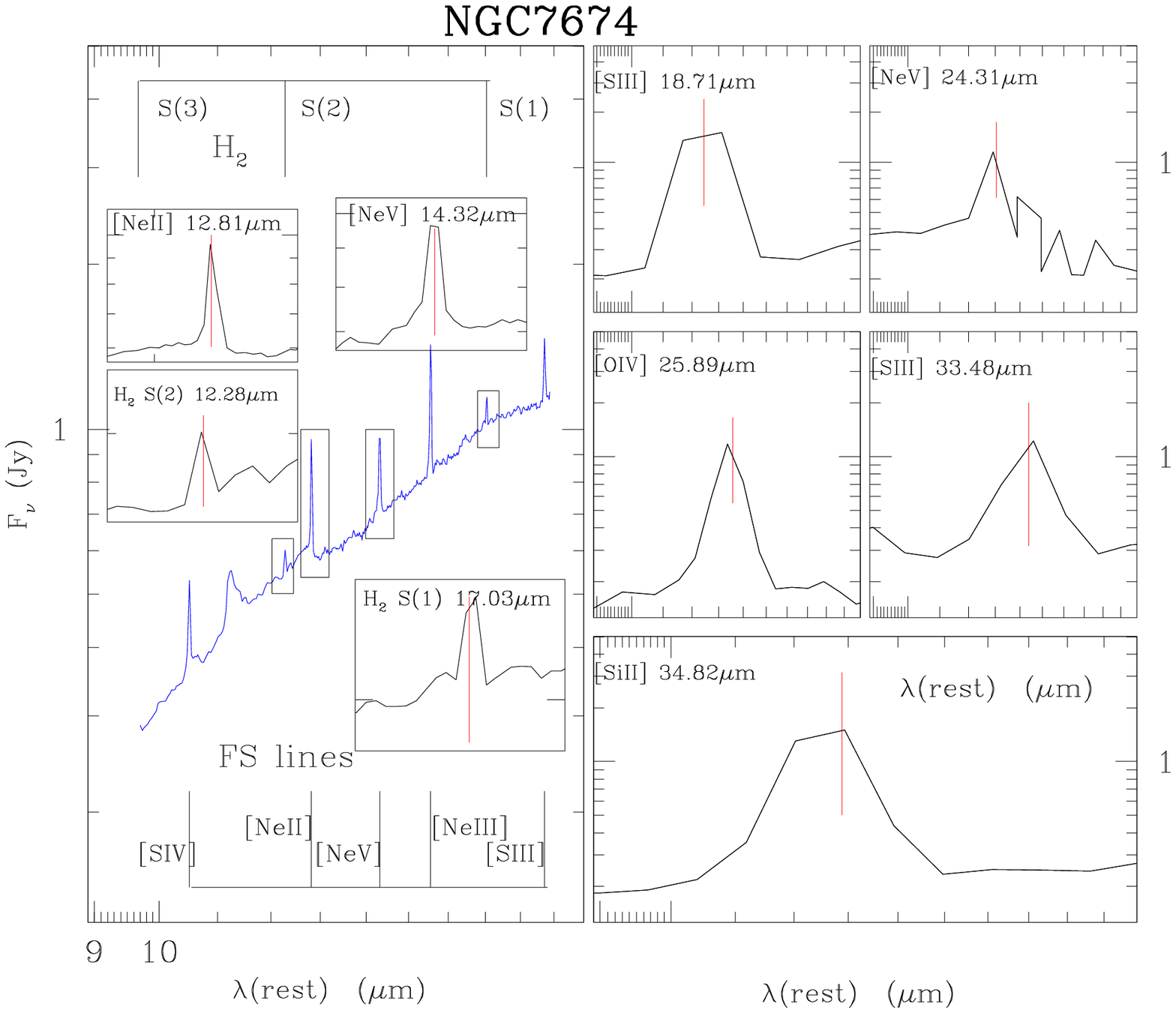}}
\end{figure}
\clearpage

\begin{figure}

\centerline{\includegraphics[width=8cm]{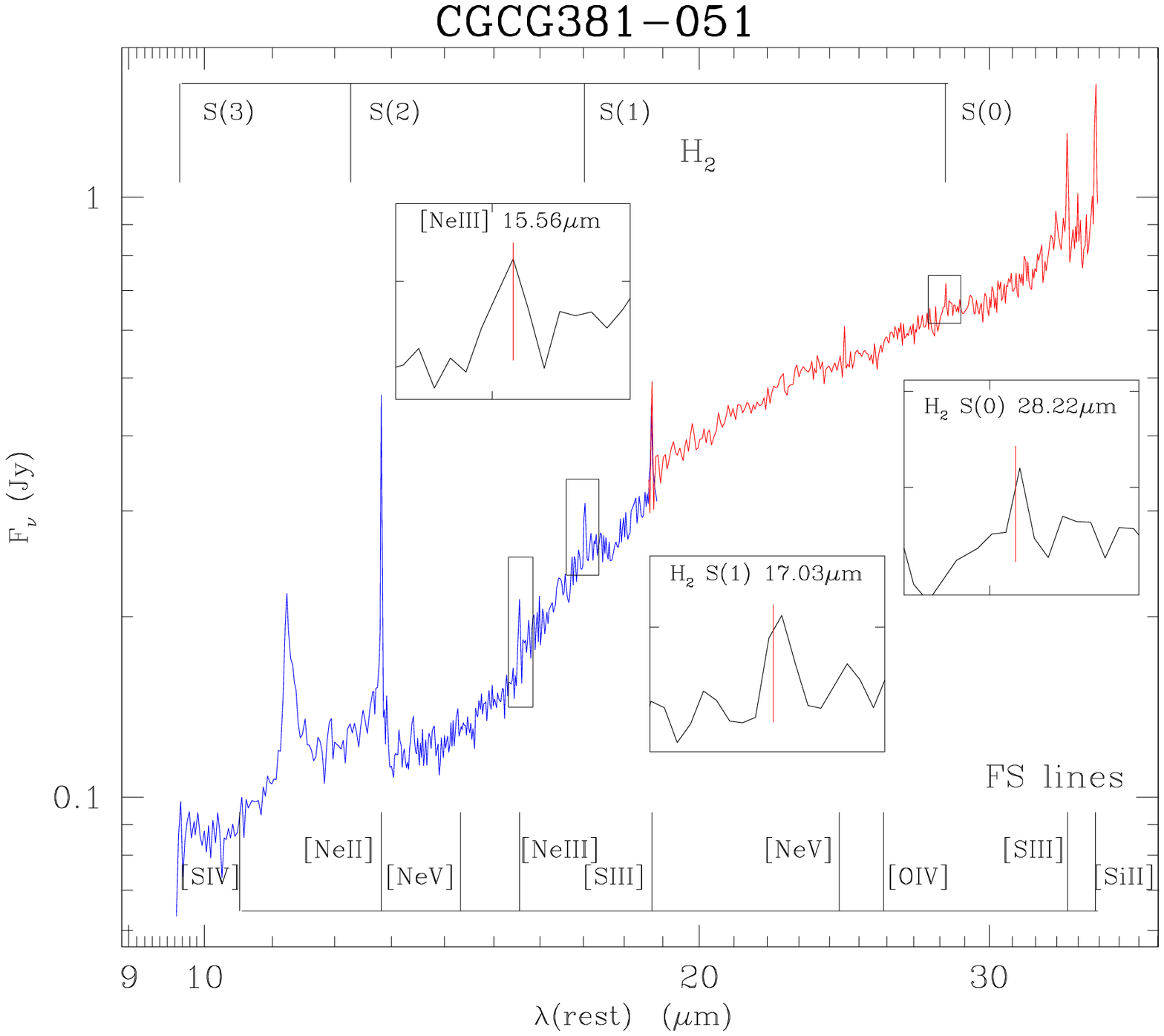}}
\end{figure}
\clearpage

\begin{figure}
\includegraphics[angle=0,scale=.80]{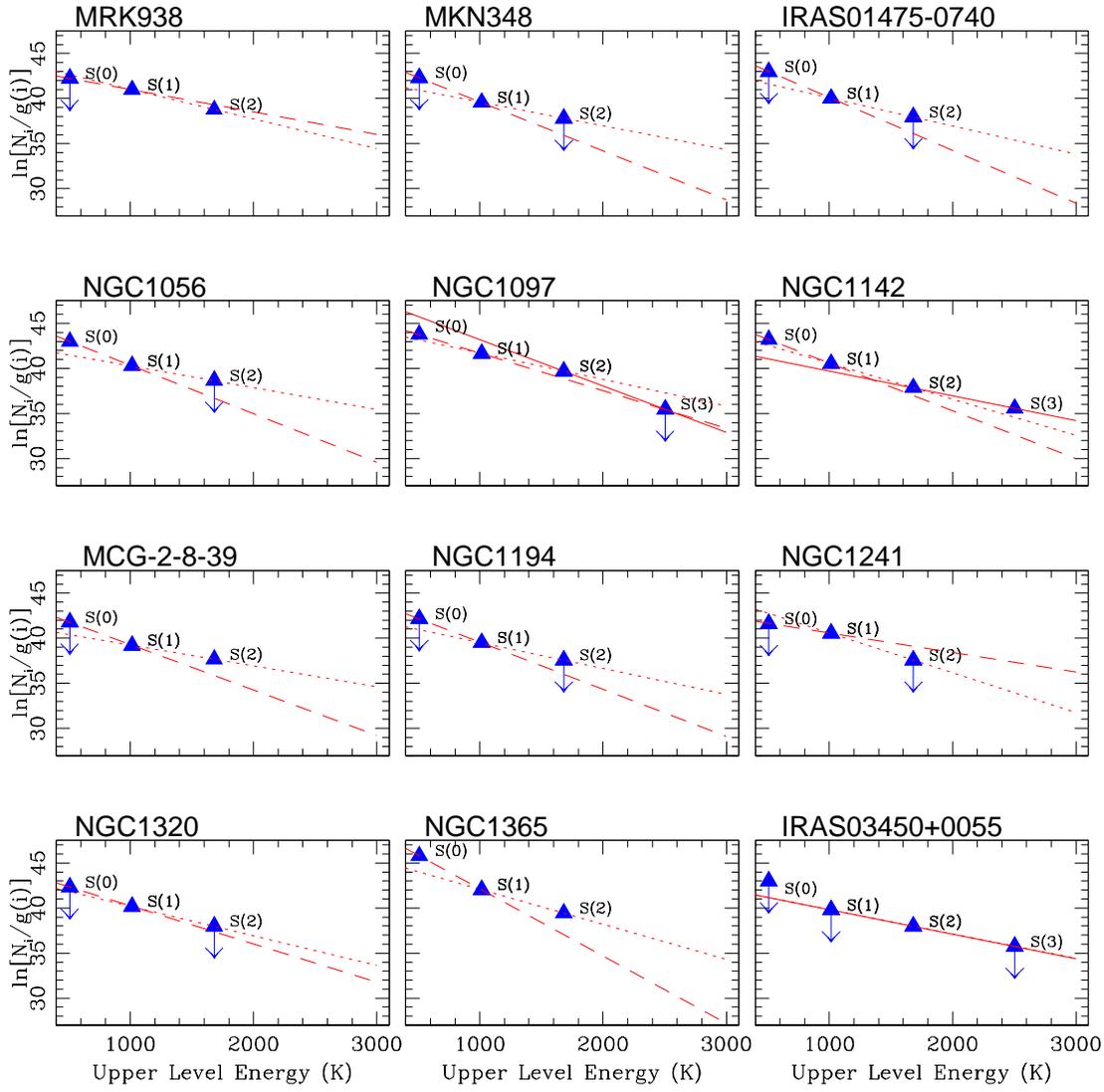}
\caption{H$_2$ excitation diagrams. For each measured line, the natural logarithm of the level population normalized to its statistical weight is plotted against the upper level  energy (in temperature units). For each pair of adjacent transitions the connecting line is shown, whose inverse value represents the gas temperature: the dashed line connects the S(0) and S(1) detections, the dotted line the S(1) and S(2), the solid line the S(2) and S(3). Upper limits have been used to obtain limiting slopes and hence limiting temperatures and masses (see text). Figures 2.12.43 are available in the online version of the Journal.}\label{fig2}
\end{figure}

\begin{figure}
\epsscale{.80}
\includegraphics[angle=0,scale=.80]{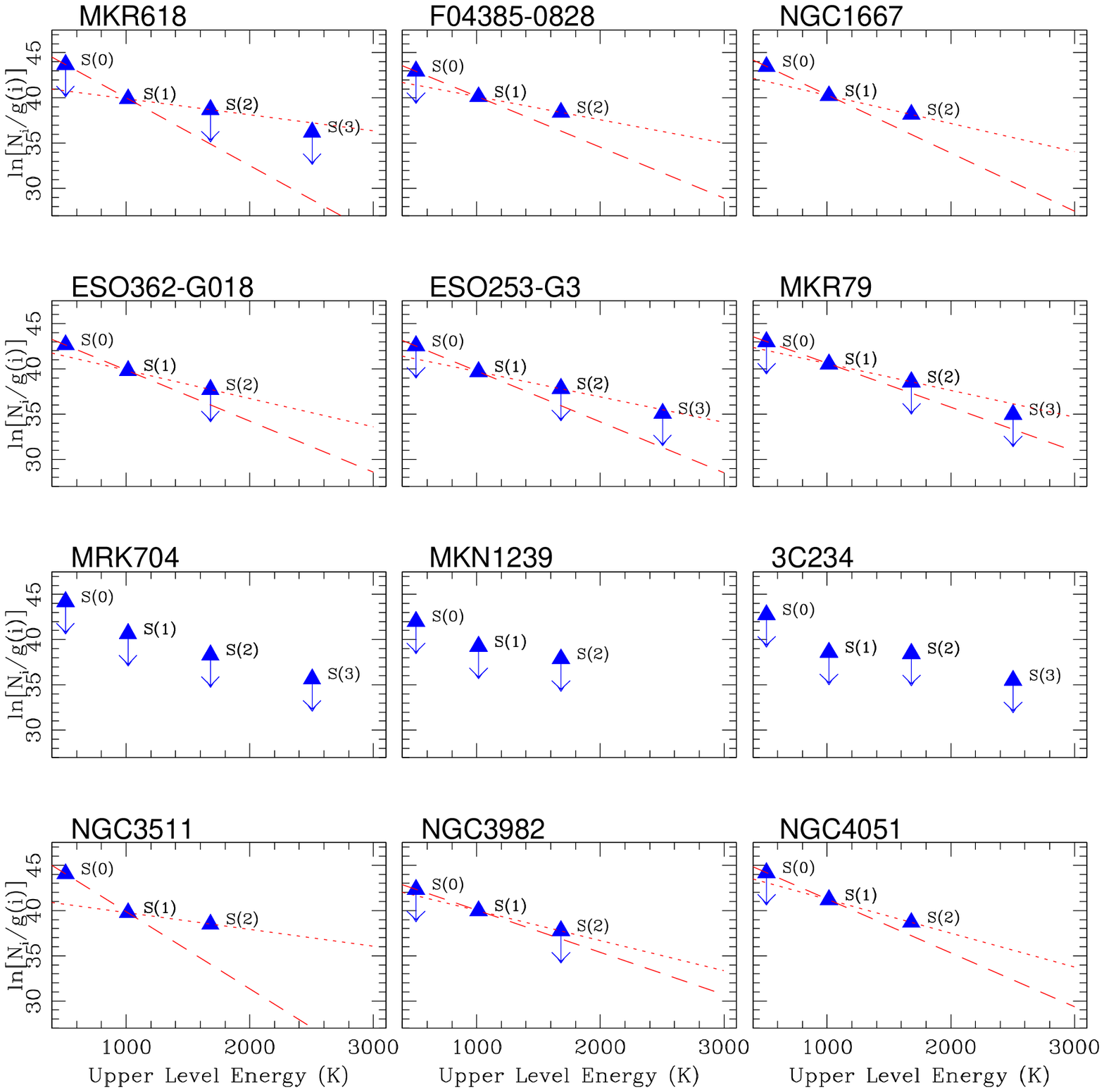}
\end{figure}

\begin{figure}
\includegraphics[angle=0,scale=.80]{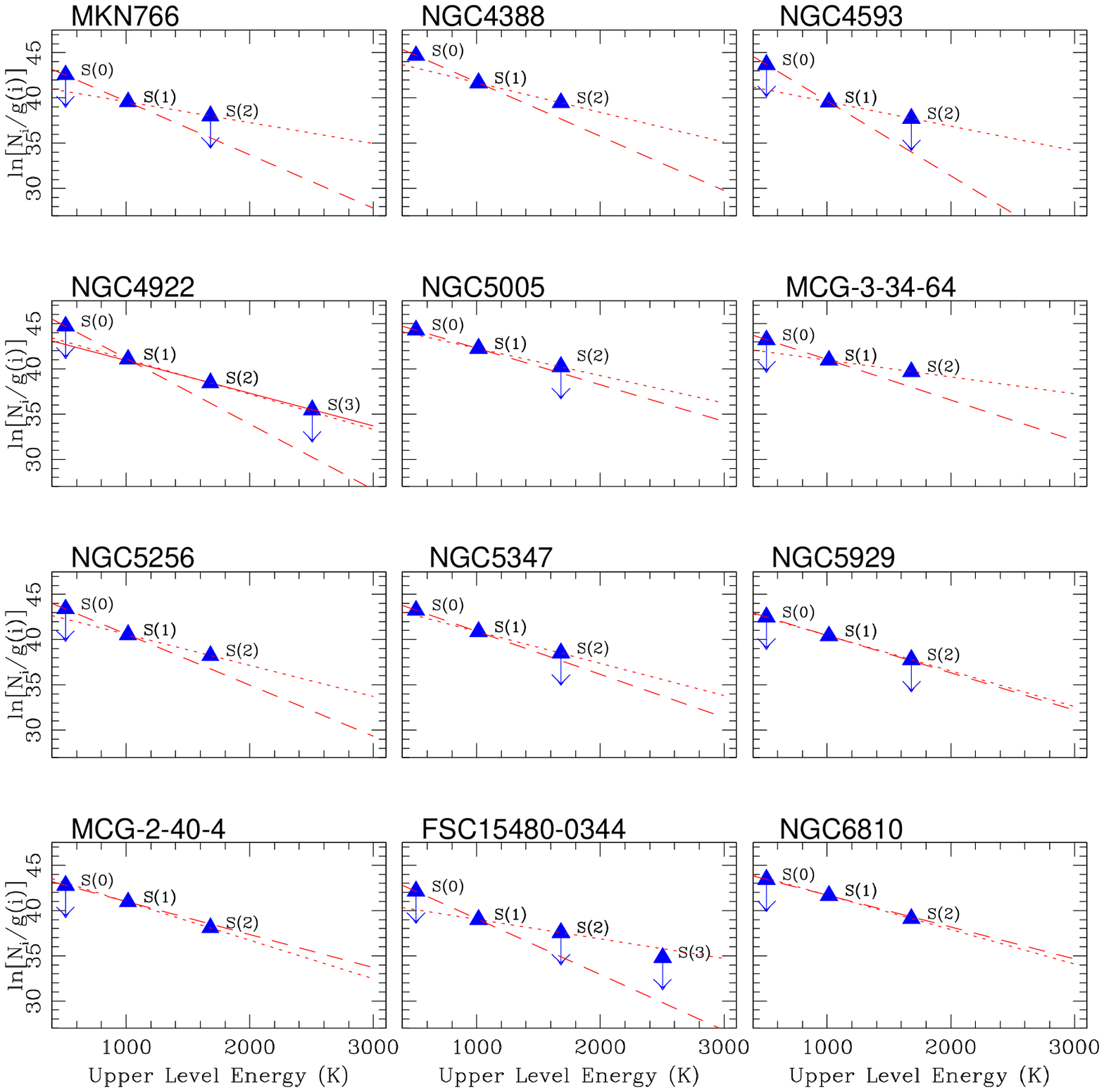}
\end{figure}

\begin{figure}
\includegraphics[angle=0,scale=.80]{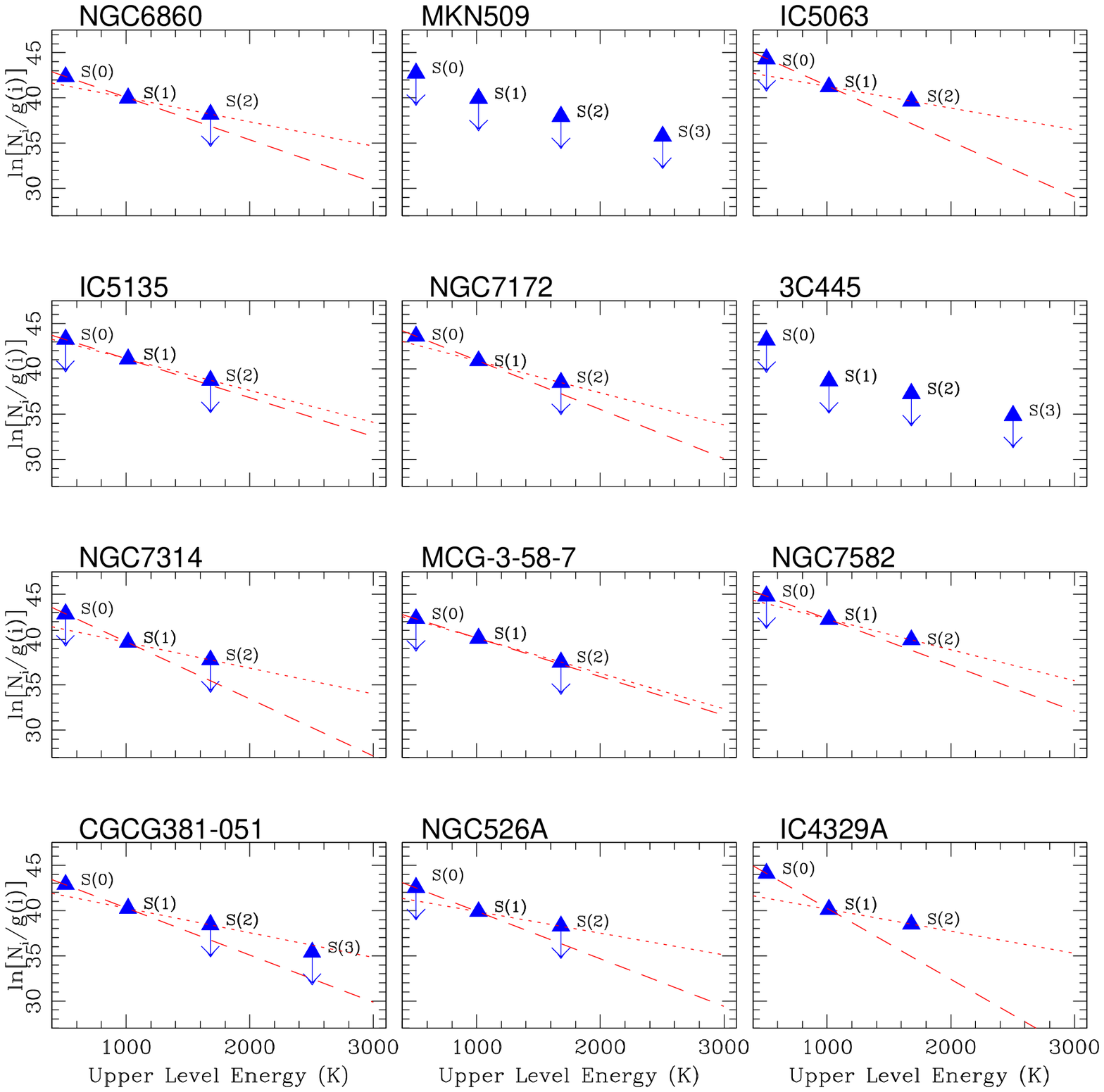}
\end{figure}

\begin{figure}
\includegraphics[angle=0,scale=.80]{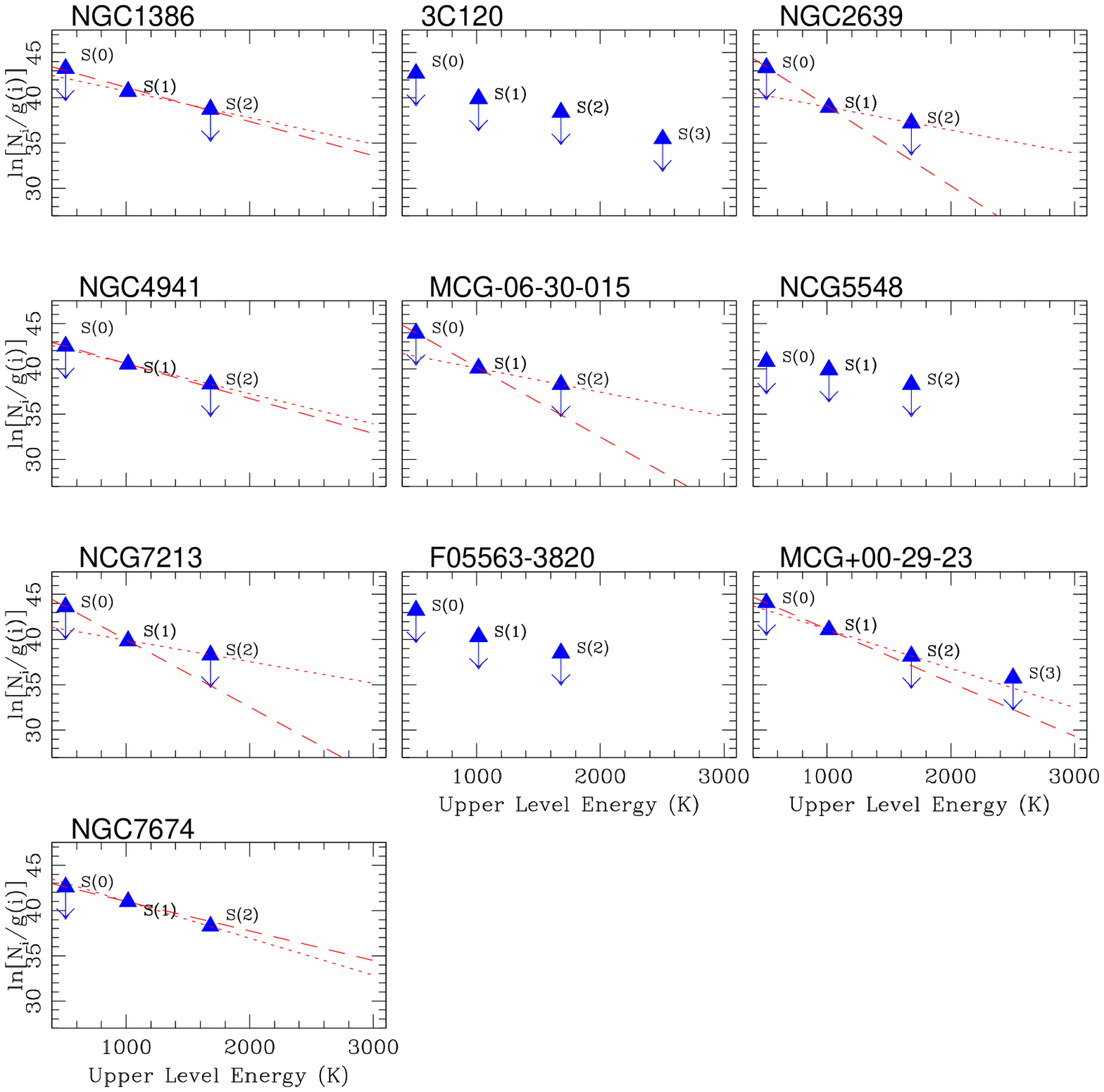}
\end{figure}

\clearpage

% FIG 3 A & B

\begin{figure}
\centerline{\includegraphics[width=9cm]{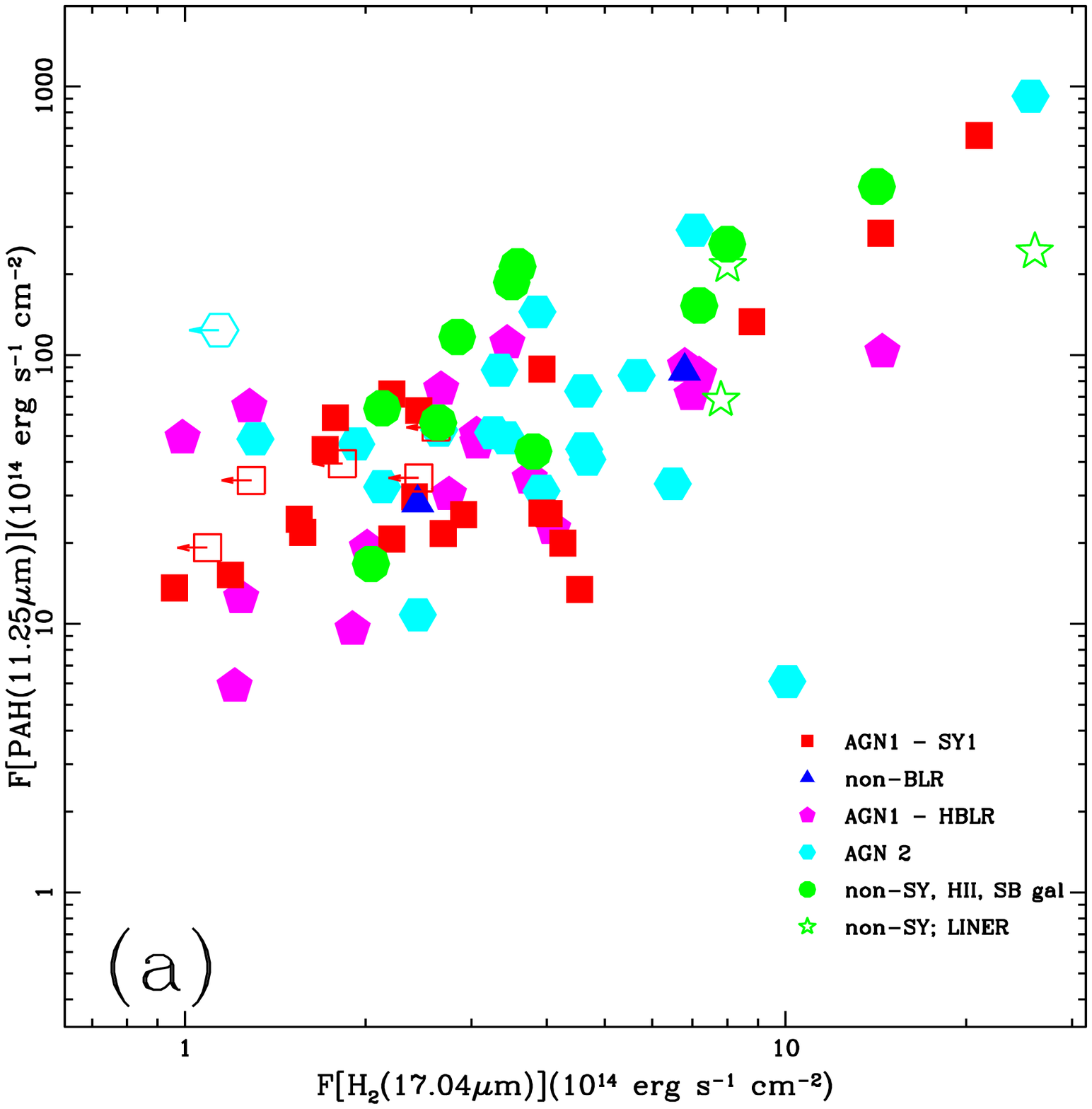}\includegraphics[width=9cm]{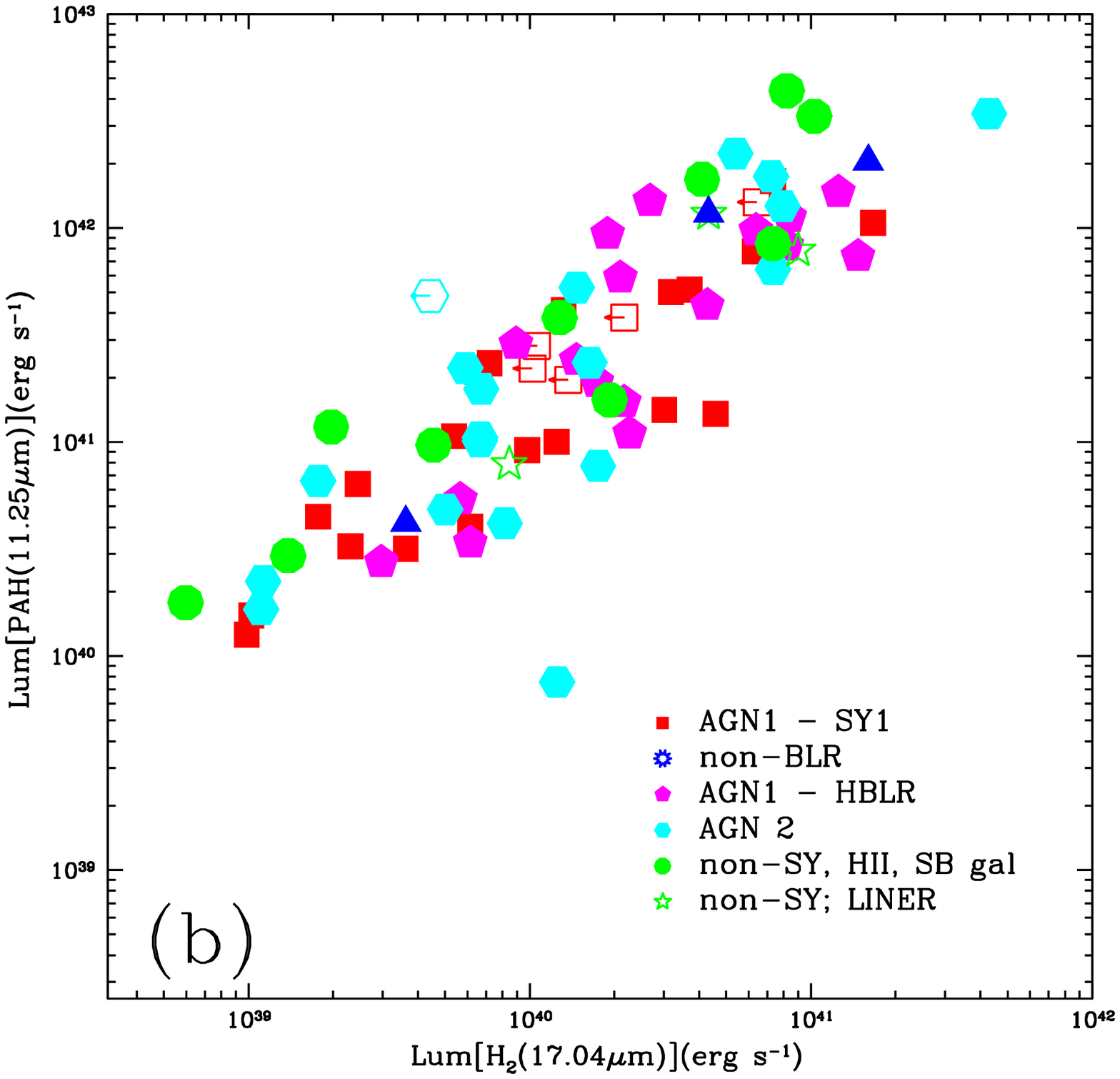}}
\caption{\textbf{a:} H$_2$ 17.04 $\mu$m line flux versus PAH 11.25$\mu$m integrated flux. 
\textbf{b:} H$_2$ 17.04 $\mu$m line luminosity 
versus PAH 11.25$\mu$m luminosity.\label{fig3}}
\end{figure}
% FIG 4 A & B
\begin{figure}
\centerline{\includegraphics[width=9cm]{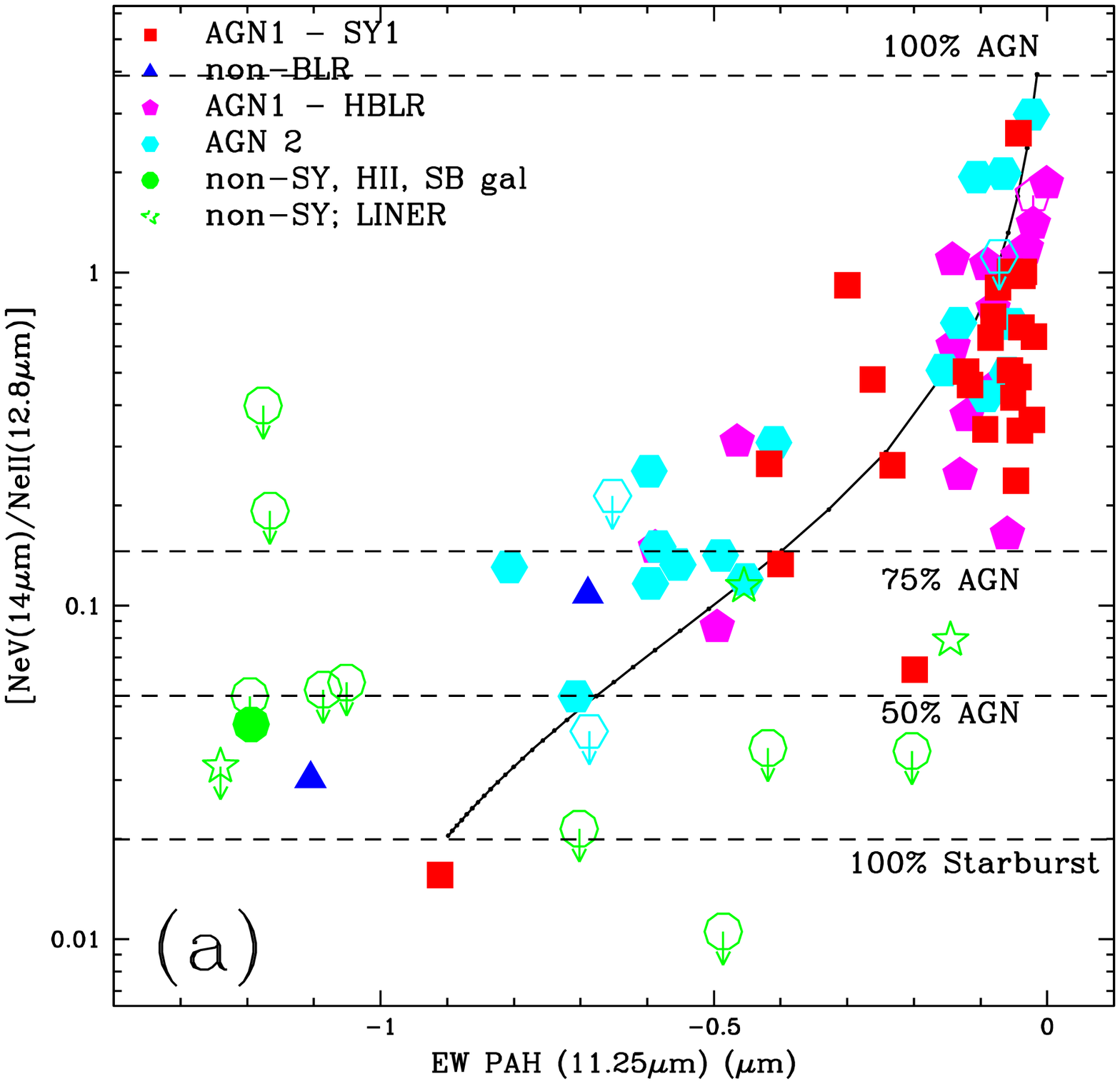}\includegraphics[width=9cm]{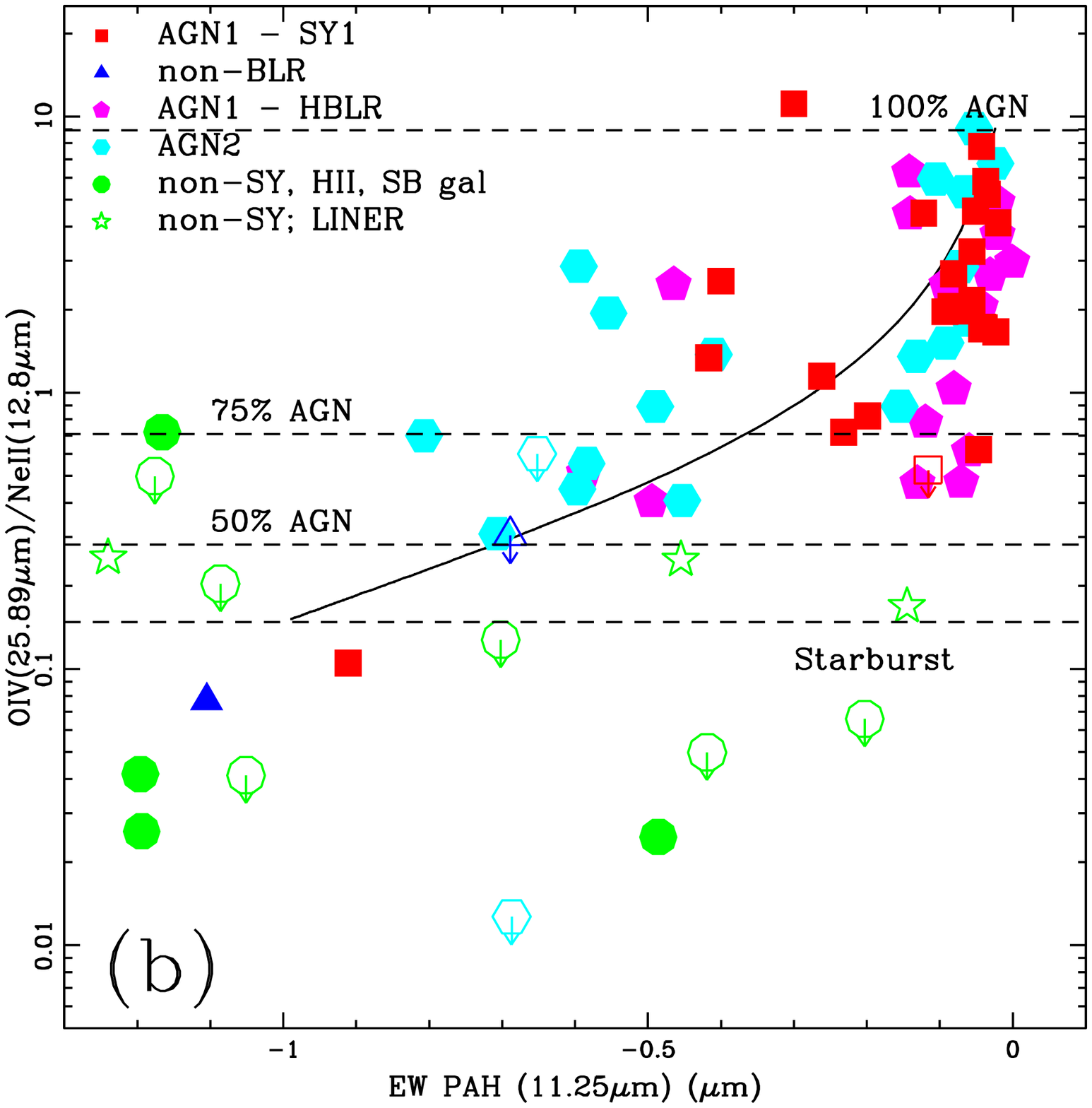}}
\caption{\textbf{a:} The PAH 11.25$\mu$m equivalent width versus [NeV]14.3$\mu$m/[NeII]12.8$\mu$m line ratio. 
%PAH EW 11.25$\mu$m of 84\% of the type 1's is $\vert$EW PAH$\vert<$0.2 and for the non-Seyfert and
%for the non bona fide Seyfert 1 is $\vert$EW PAH$\vert>$0.2, while the non-HBLR spread in -0.9$<$EW PAH$<$0
The black line shows the behaviour of the analytical model for this diagram. 
%The 92\% of the type 1's have the 75\% of the emission produced by AGN and the non-HBLR up to the 50\%. 
\textbf{b:}
The PAH 11.25$\mu$m equivalent width versus [OIV]25.9$\mu$m/[NeII]12.8$\mu$m line ratio. 
The black line 
as in Fig. 4a.
%shows 
%the behaviour of the analytical model for this diagram. %The 84\% of the type 1's 
%have the 75\% of the emission produced by AGN and the non-HBLR up to the 50\%. 
\label{fig4}}  
\end{figure}

% FIG 5 A & B
\begin{figure}
\centerline{\includegraphics[width=9cm]{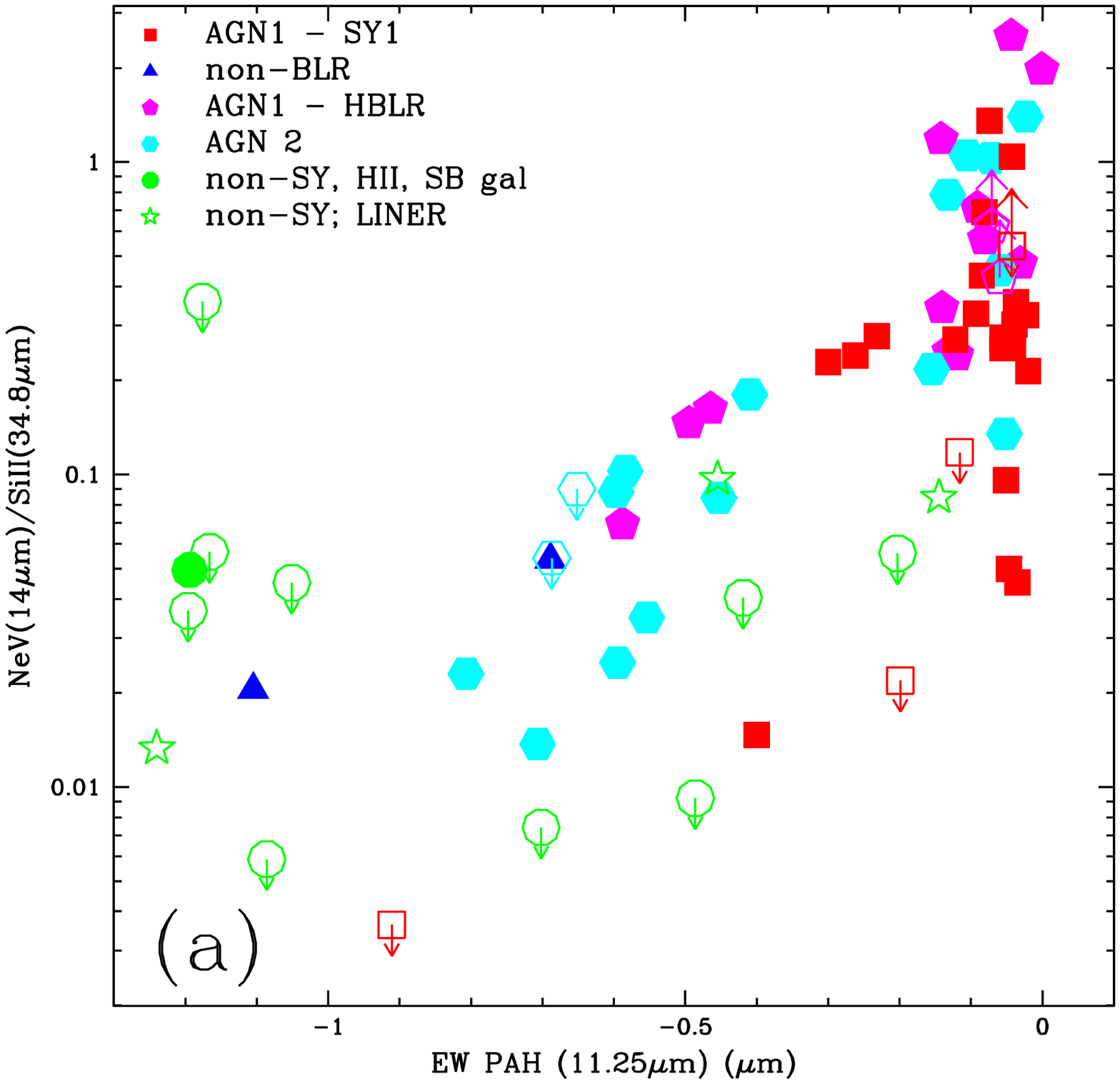}\includegraphics[width=9cm]{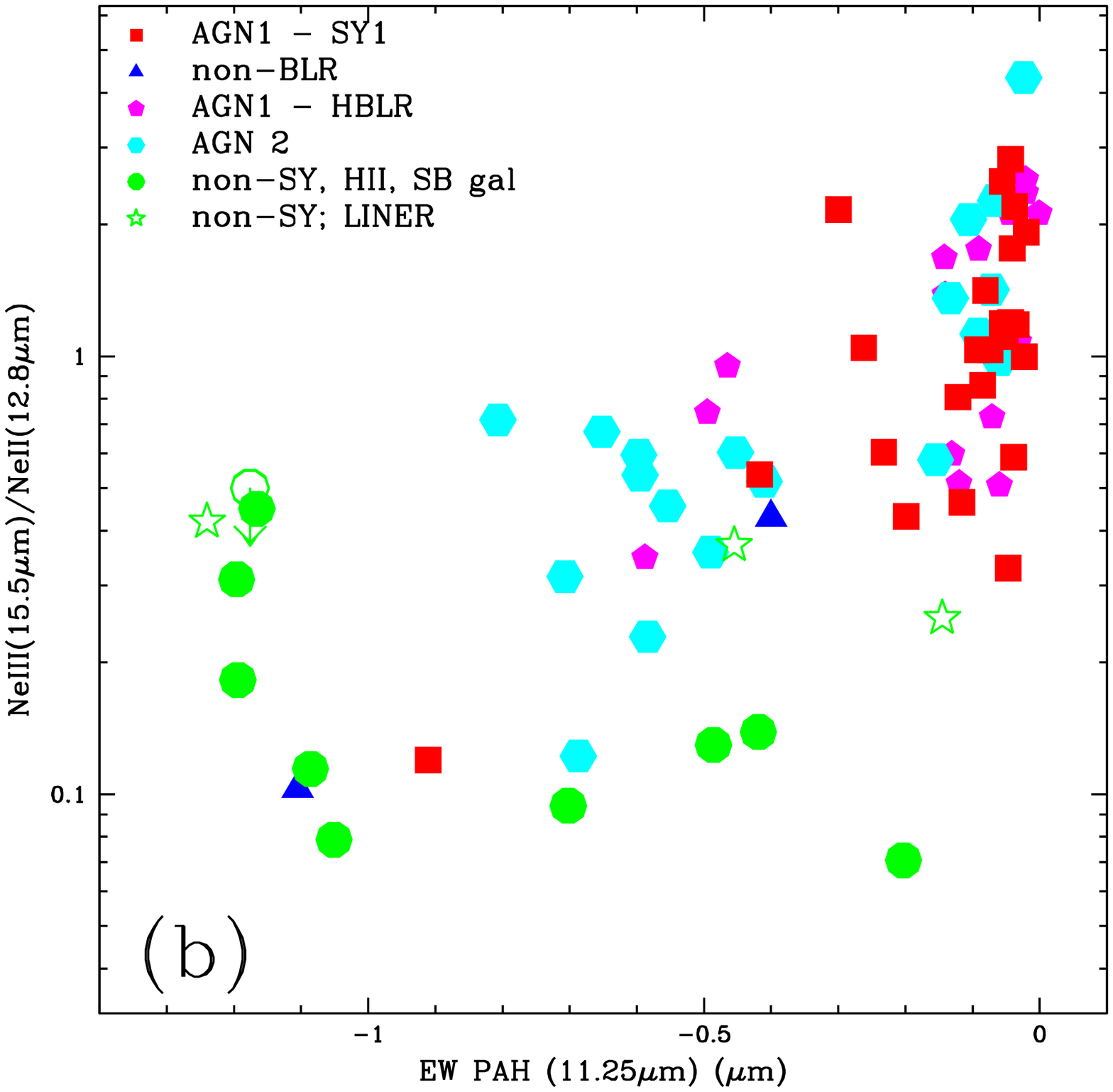}}
\caption{\textbf{a:} The PAH 11.25$\mu$m equivalent width versus [NeV]14.3$\mu$m/[SiII]34.8$\mu$m line ratio. 
 \textbf{b:} [NeIII]15.5$\mu$m/[NeII]12.8$\mu$m line ratio versus the PAH 11.25$\mu$m equivalent width.\label{fig5}}
\end{figure}

% FIG 6
\begin{figure}
\includegraphics[angle=0,scale=.80] {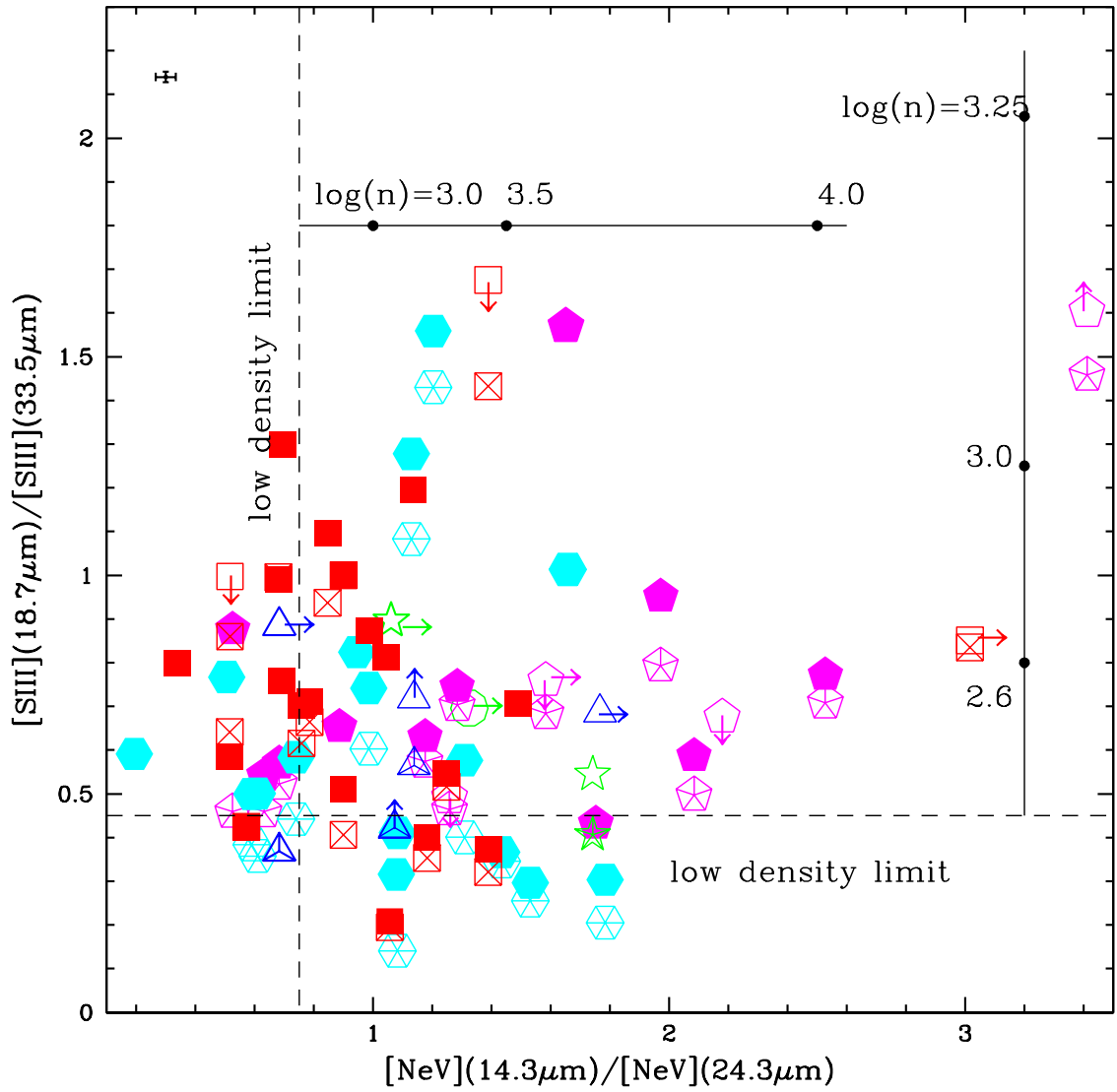}
\caption{[NeV]14.3$\mu$m/[NeV]24.3$\mu$m line ratio versus the 
[SIII]18.7$\mu$m/[SIII]33.5$\mu$m line ratio. Symbols as in
the previous figures, except for the open crossed symbols indicating 
the ratio of objects with small aperture [SIII]18.71$\mu$m measures
%objects 
%for which the [SIII]18.71$\mu$m line was measured only with the small (SH) aperture 
and the filled symbols directly above them showing the aperture corrected [SIII] line ratio. 
The dashed lines show the low density limits (see text). 
The solid lines at the top and at the right give the corresponding electron densities.\label{fig6}}
\end{figure}

% FIG 7 A & B
\begin{figure}
\centerline{\includegraphics[width=9cm]{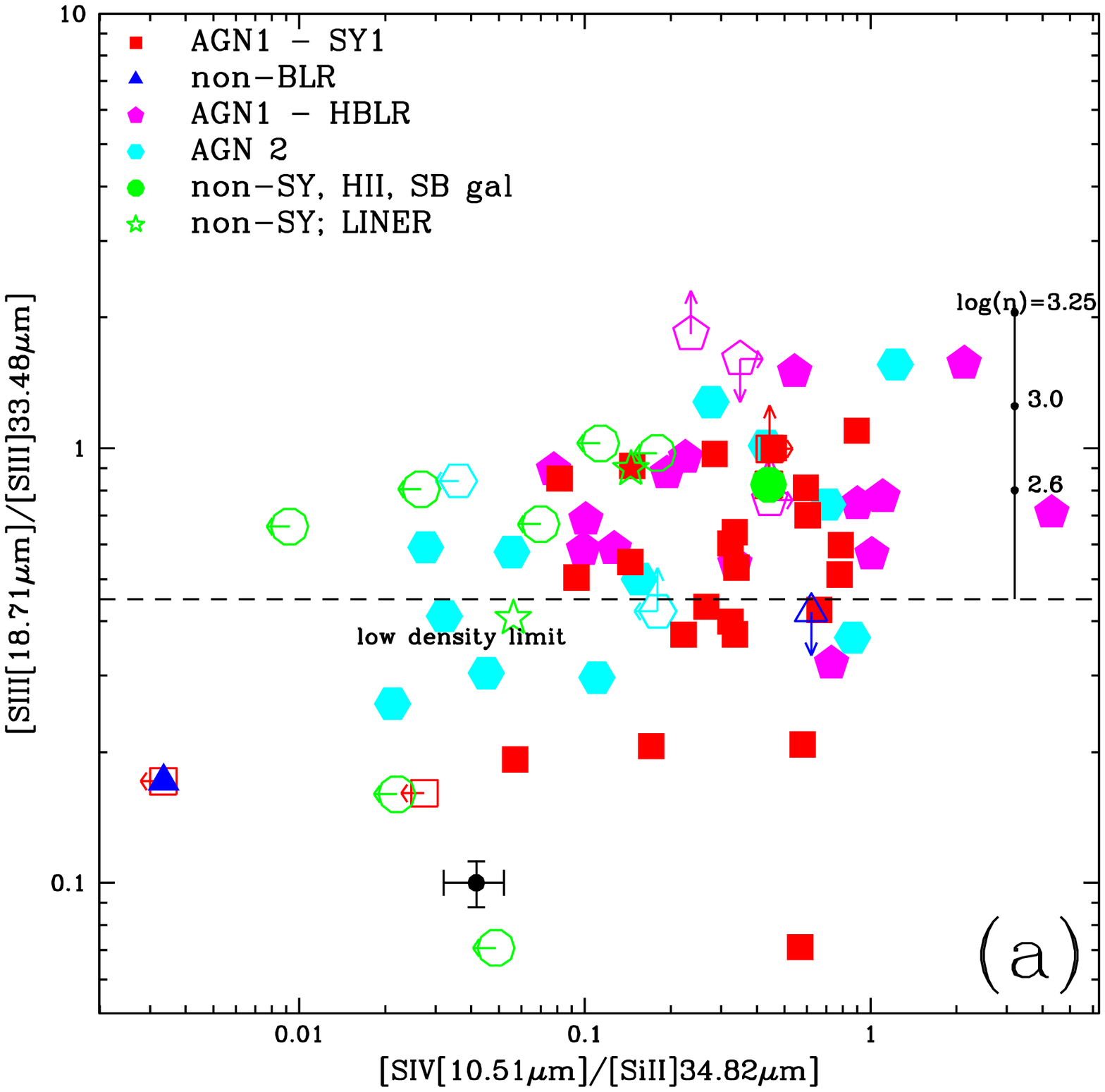}\includegraphics[width=9cm]{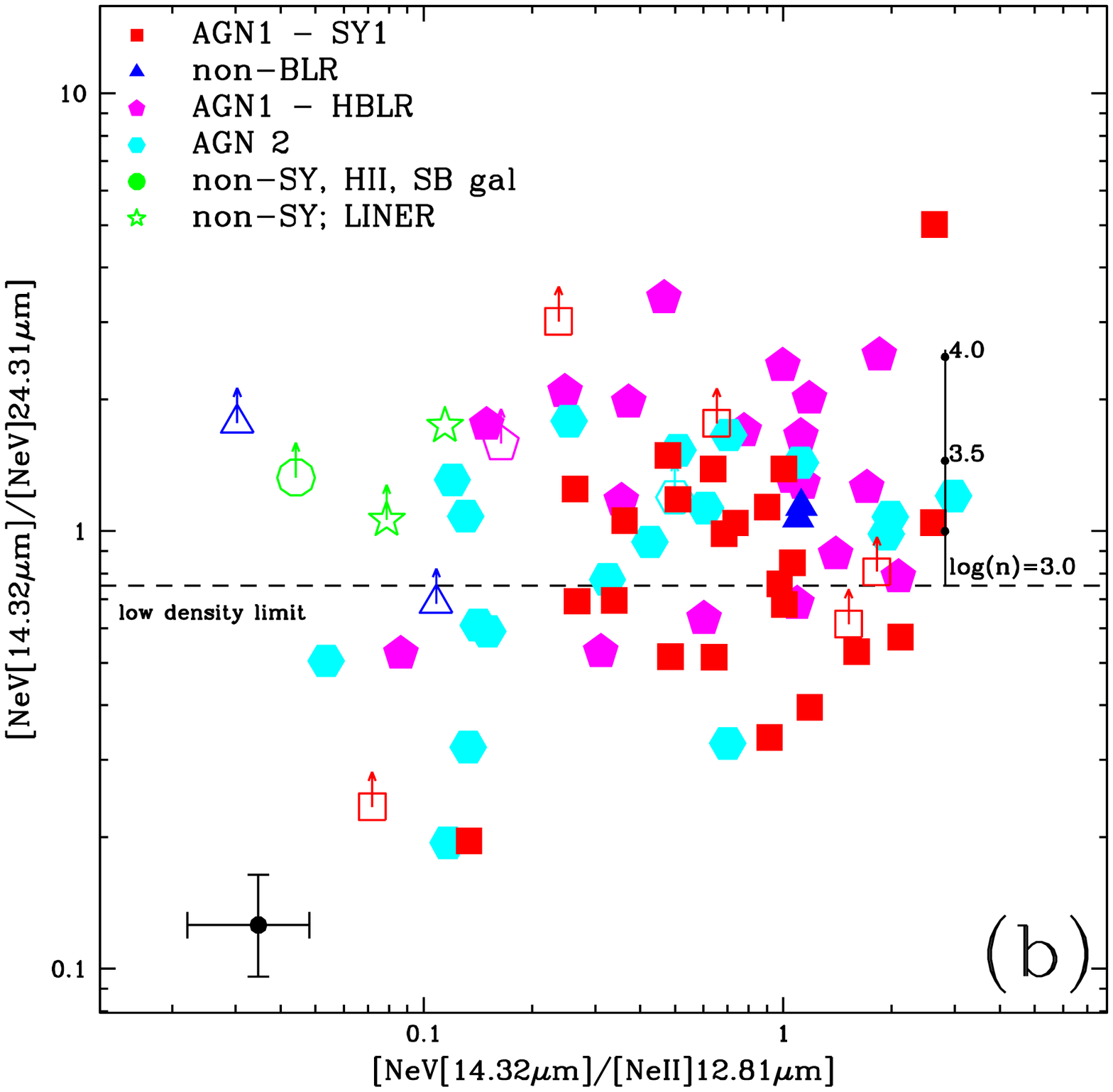}}
\caption{\textbf{a:} 
[SIV]10.51$\mu m$/[SiII]34.82$\mu m$ line ratio versus the 
[SIII]18.7$\mu$m/[SIII]33.5$\mu$m line ratio. The [SIII]18.7$\mu$m flux has been corrected 
by the extendedness \ref{ext}. The average error bars are calculated with 
the propagation of the uncertainty
from the mean line fluxes and their largest relative error %among the relative errors 
of the sample sources' line measurements. Dashed and solid lines are as in the previous figure.
 %\textbf{b:}[SIV]10.51$\mu m$/[SIII]18.7$\mu m$ line ratio versus the 
%[SIII]18.7$\mu$m/[SIII]33.5$\mu$m line ratio. Dashed and solid lines are as in the previous figure.  
\textbf{b:}[NeV]14.3$\mu$m/[NeII]12.8$\mu$m line ratio versus the [NeV]14.3$\mu$m/[NeV]24.3$\mu$m line ratio. 
Dashed and solid lines are as in the previous figure. \label{fig7}}
\end{figure}

%\begin{figure}
%\centerline{\includegraphics[width=9cm]{neon_err.eps}}
%\caption{[NeV]14.3$\mu$m/[NeII]12.8$\mu$m line ratio versus the [NeV]14.3$\mu$m/[NeV]24.3$\mu$m line ratio. Dashed and solid lines are as in the previous figure. \label{fig6a}}
%\end{figure}

% FIG 8 A & B
\begin{figure}
\centerline{\includegraphics[width=9cm]{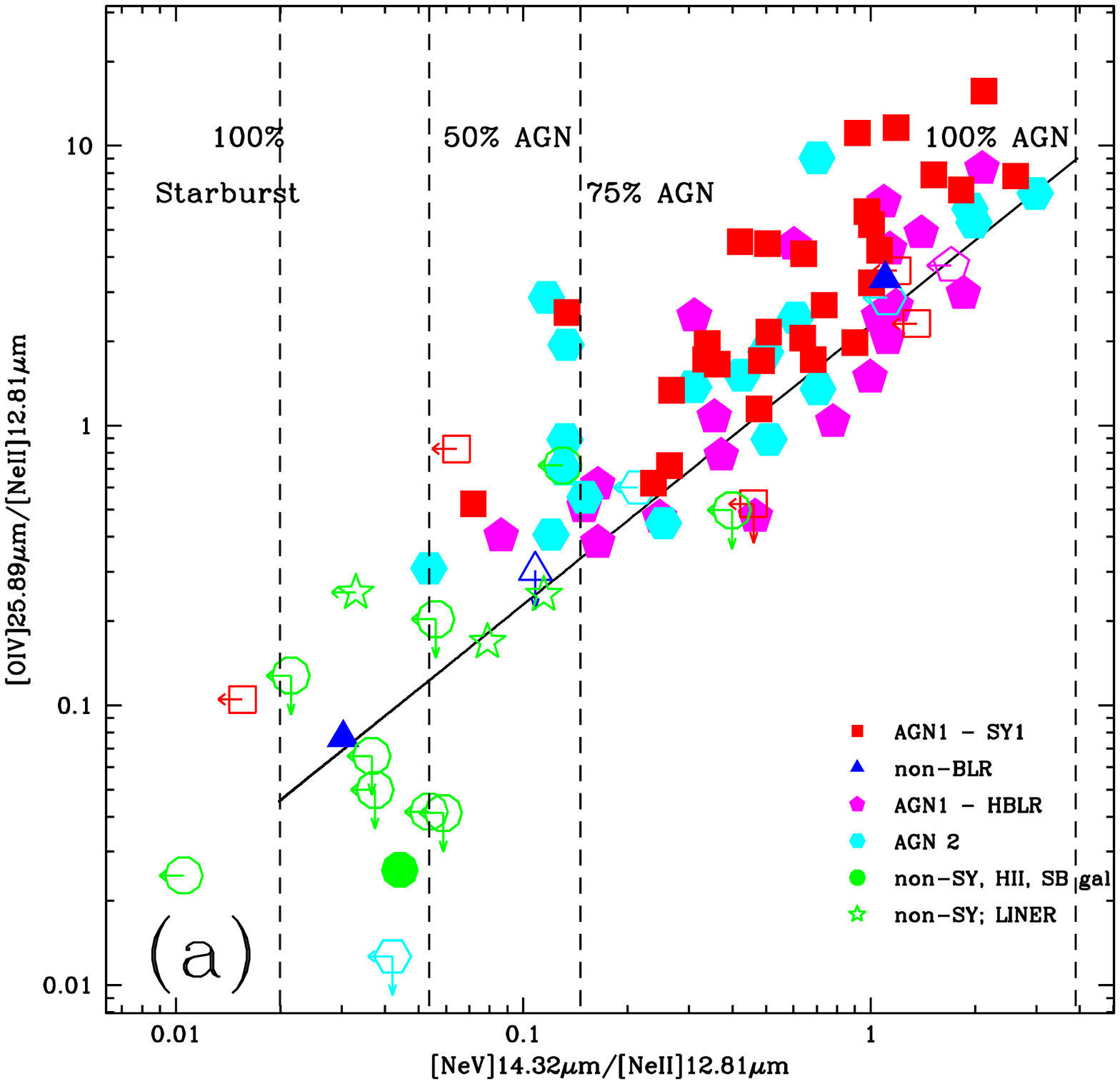}\includegraphics[width=9cm]{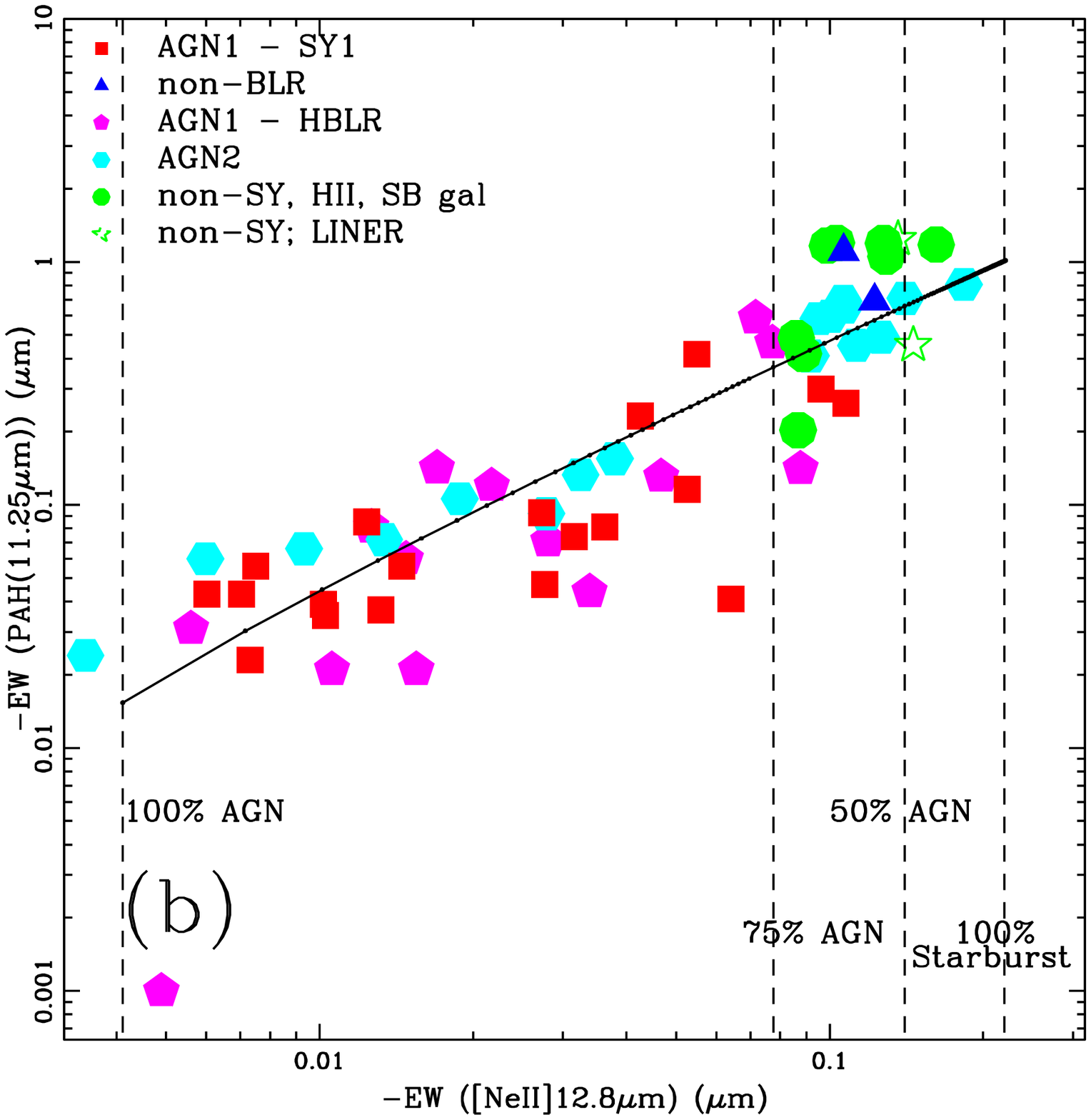}}
\caption{\textbf{a:} [NeV]14.3$\mu$m/[NeII]12.8$\mu$m line ratio versus the [OIV]25.9$\mu$m/[NeII]12.8$\mu$m line ratio. 
The black line shows the behaviour of the analytical model for this diagram. 
\textbf{b:} [NeII]12.8$\mu$m equivalent width versus the PAH 11.25$\mu$m equivalent width.  We note that, 
for graphical reasons, in this diagram and in the following ones in which the equivalent widths, 
covering a large range, the logarithm of the inverse of the actual 
equivalent width is plotted. 
The black line shows the behaviour of the analytical model for this diagram.
%The 92\% of the type 1's have the 75\% of the emission produced by AGN and the non-HBLR up to less then the 50\%.
\label{fig8}}
\end{figure}

% FIG 9 A & B
\begin{figure}
\centerline{\includegraphics[width=9cm]{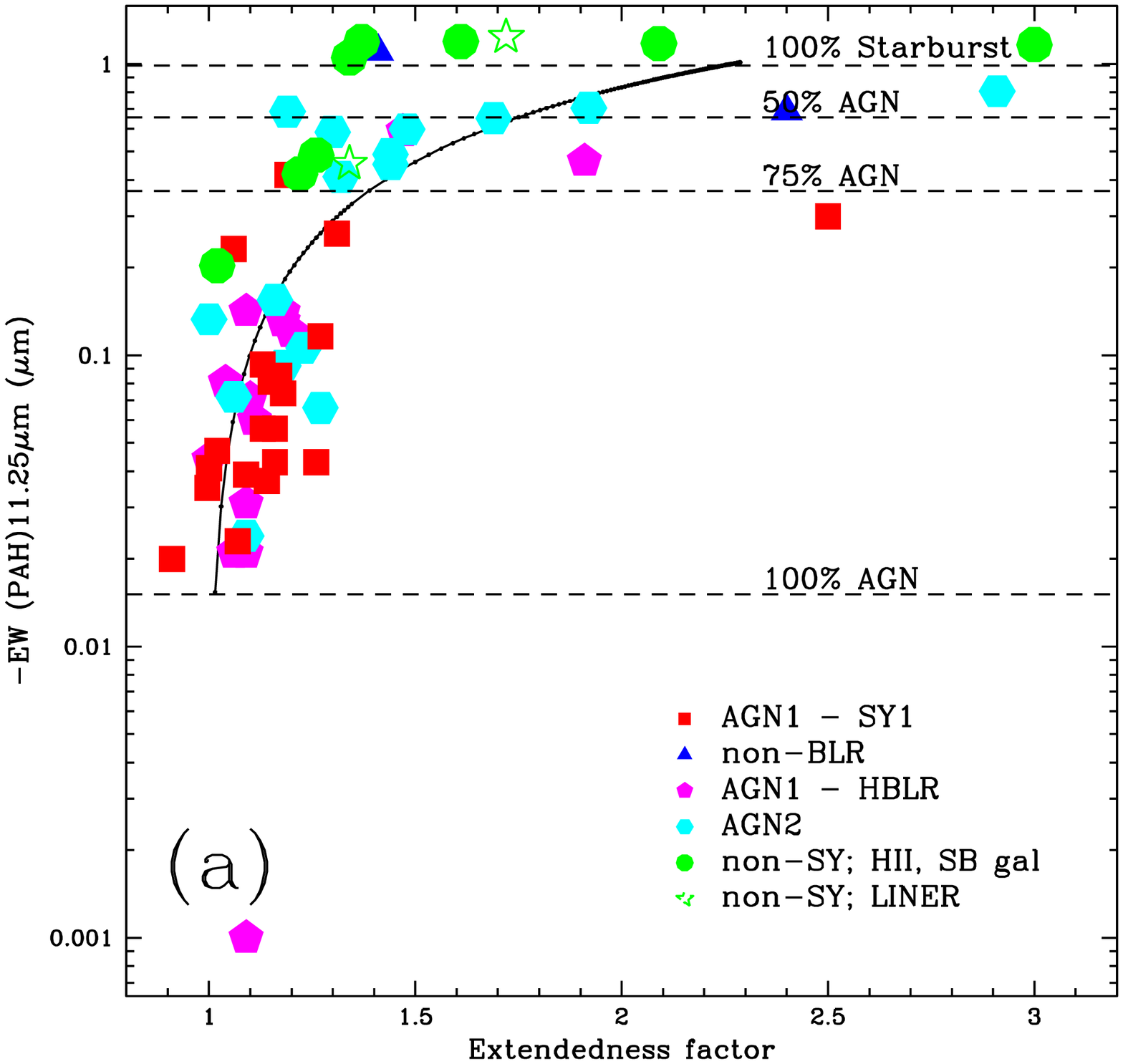}\includegraphics[width=9cm]{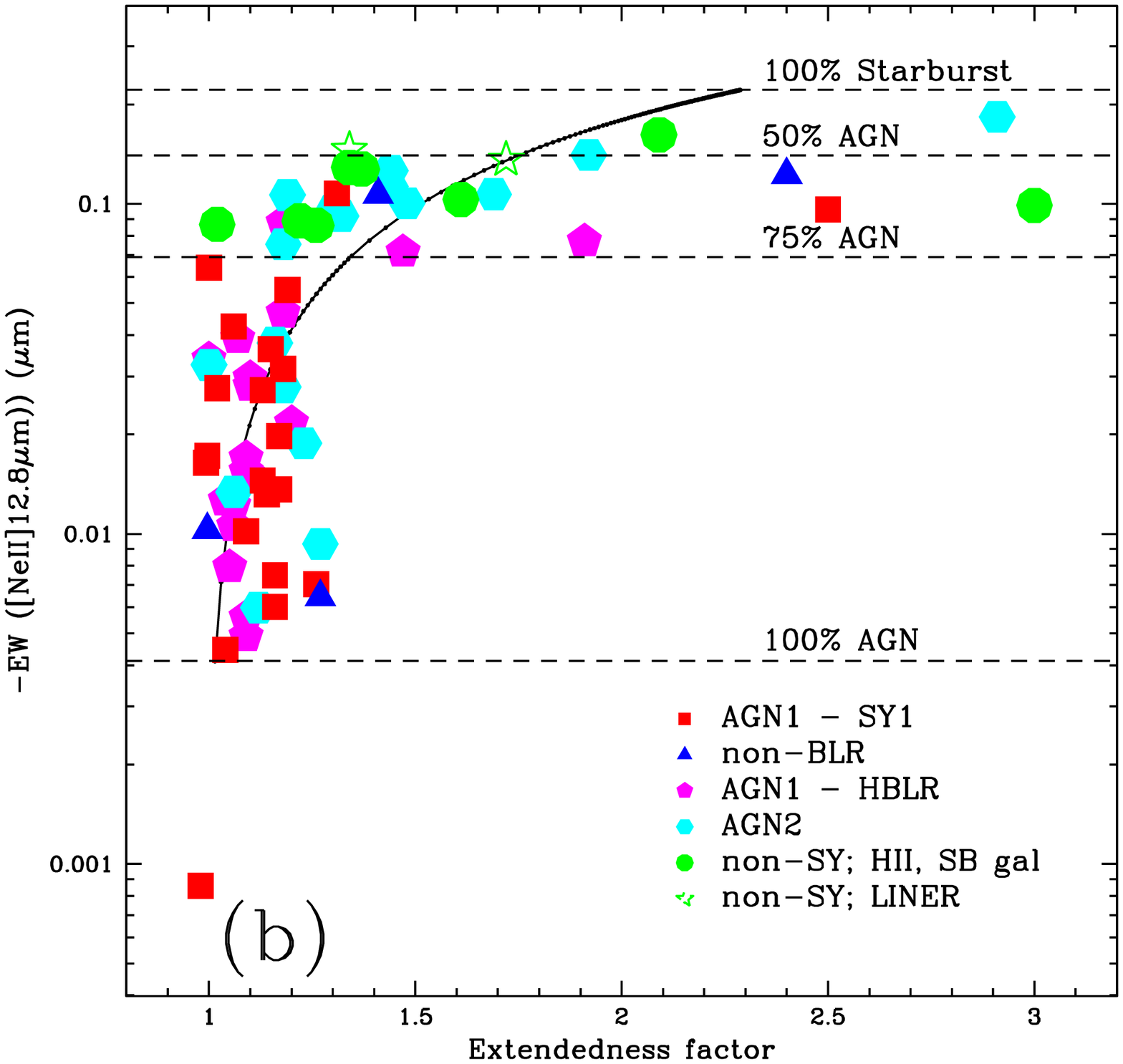}}
\caption{\textbf{a:} PAH 11.25$\mu$m equivalent widths versus extendness parameter.
The black line shows the behaviour of the analytical model for this diagram. 
\textbf{b:} [NeII]12.8$\mu$m equivalent widths versus extendness parameter.
The black line shows the behaviour of the analytical model for this diagram. 
%Extendness of 1.0 corresponds to completely unresolved (point-like) 19$\mu$m continuum,
%which is therefore completely AGN-dominated. In both figures the non-linear effects of
%extended emission from the host galaxy become important for extendness parameters above 1.3. 
%The black lines show the behaviour of the analytical models for these diagrams.
%The 92\% of the type 1's have the 75\% of the emission produced by AGN and the non-HBLR up to less than the 50\%.
\label{fig9}}
\end{figure}

% FIG 10
\begin{figure}
\centerline{\includegraphics[width=9cm]{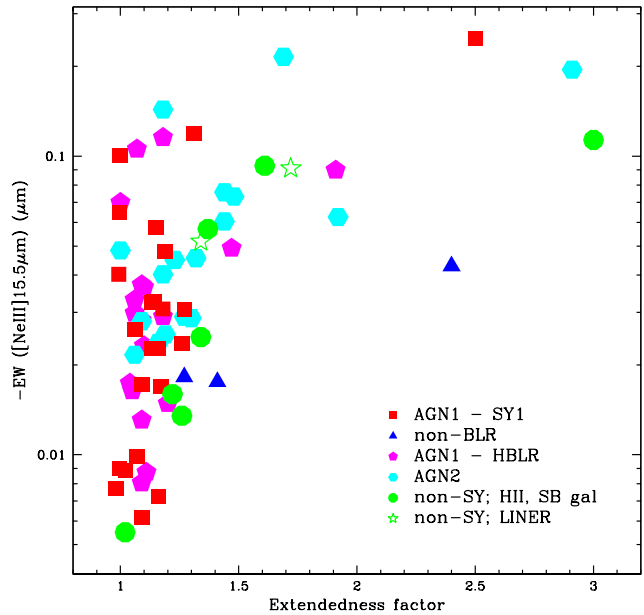}}
\caption{[NeIII]15.5$\mu$m line equivalent width versus extendness parameter.
\label{fig10}}
\end{figure}

%\begin{figure}
%\centerline{\includegraphics[width=9cm]{ext_ne5_star.eps}\includegraphics[width=9cm]{ext_o4_star.eps}}
%\caption{[NeV]14.3$\mu$, and [OIV]25.9$\mu$m line equivalent widths versus extendness parameter.
%As expected, there are no significant correlations, because the equivalent width
%is the ratio of high-ionization emission to the hot dust continuum, and both of
%these arise mostly from the AGN.\label{fig8}}
%\end{figure}

% FIG 11
\begin{figure}
\centerline{\includegraphics[width=9cm]{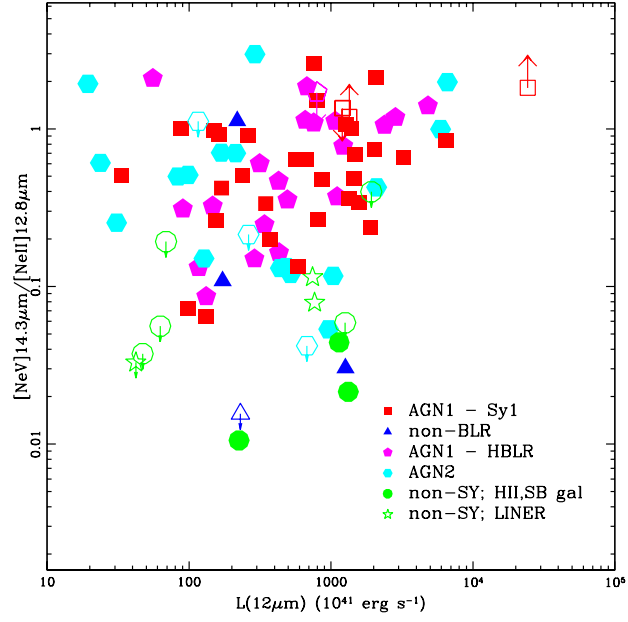}}
\caption{[NeV]14.3$\mu$m/[NeII]12.8$\mu$m line ratio versus 12$\mu$m luminosity. \label{fig11}}
\end{figure}

% FIG 12 A & B
\begin{figure}
\centerline{\includegraphics[width=9cm]{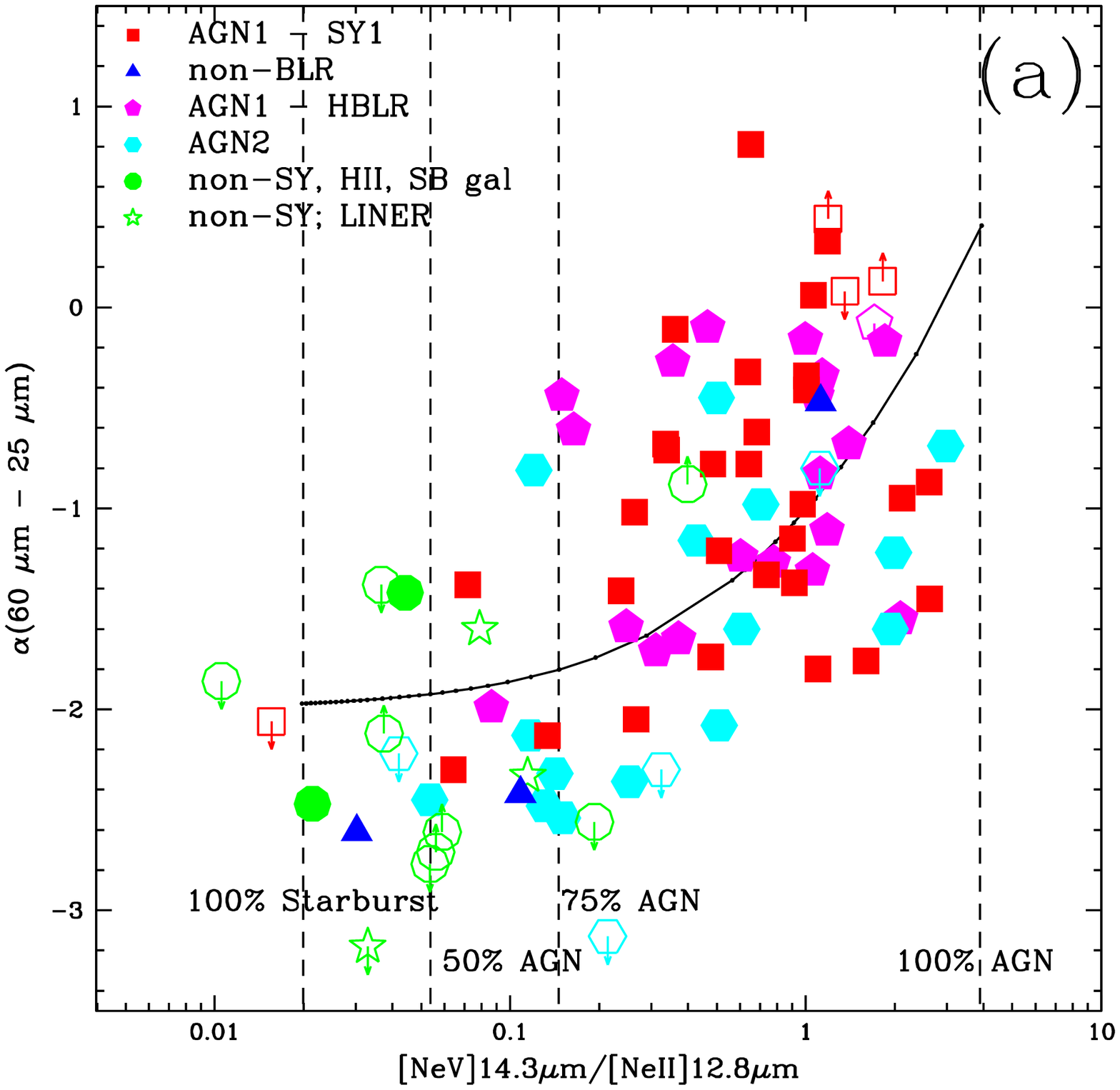}\includegraphics[width=9cm]{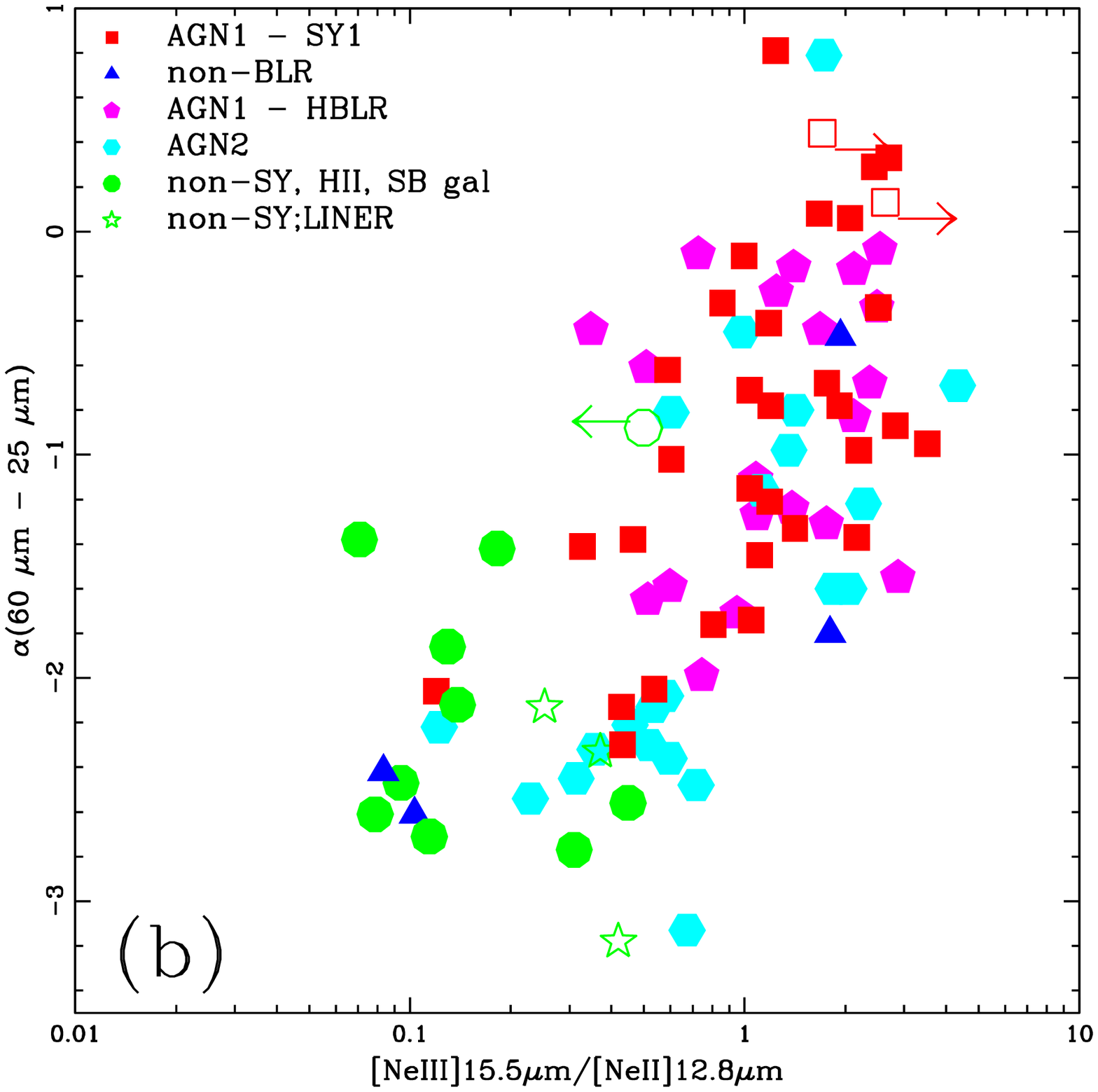}}
\caption{\textbf{a:} [NeV]14.3$\mu$m/[NeII]12.8$\mu$m  line ratio versus the 60-25$\mu$m spectral index. 
The black line shows the behaviour of the analytical model for this diagram. 
%The 92\% of the type 1's have the 75\% of the emission produced by AGN and the non-HBLR up to the 50\%.
\textbf{b:} [NeIII]15.5$\mu$m/[NeII]12.8$\mu$m line ratio versus the 60-25$\mu$m spectral index.\label{fig12}}
\end{figure}

%\begin{figure}
%\centerline{\includegraphics[width=9cm]{s_si_ne52_star.eps}\includegraphics[width=9cm]{s_si_ne32_star.eps}}
%\caption{\textbf{a:}[NeV]14.3$\mu$m/[NeII]12.8$\mu$m line ratio versus the 
%[SIII]33.5$\mu$m/[SiII]34.8$\mu$m line ratio. While the neon ratio distinguishes type 1 from the others,
%the [SIII] to [SiII] is less able to disentangle them. 
%\textbf{b:} [NeIII]15.5$\mu$m/[NeII]12.8$\mu$m line ratio versus the 
%[SIII]33.5$\mu$m/[SiII]34.8$\mu$m line ratio. 
%}\label{fig11}
%\end{figure}

% FIG 13
\begin{figure}
\centerline{\includegraphics[width=9cm]{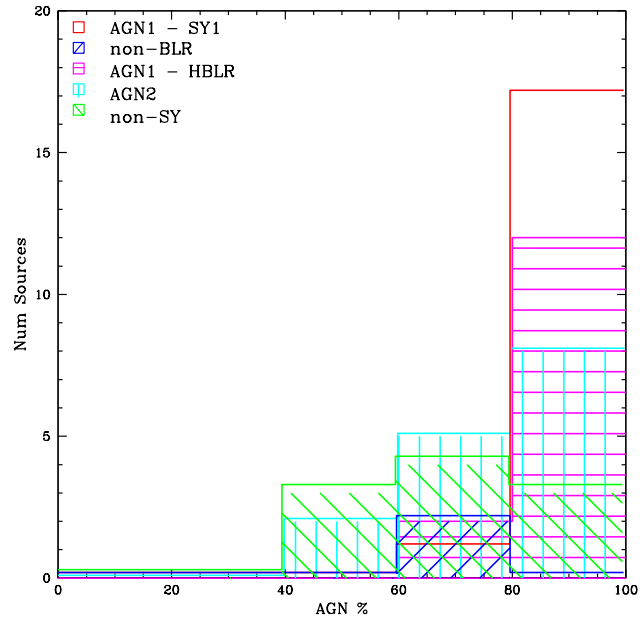}}
\caption{The histogram shows the percentage of AGN contribution.% on the total 19$\mu$m emission et on the number of the galaxies, distinguished type by type as told in the legend.
\label{fig13}}
\end{figure}

% FIG 14
\begin{figure}
\centerline{\includegraphics[width=18cm]{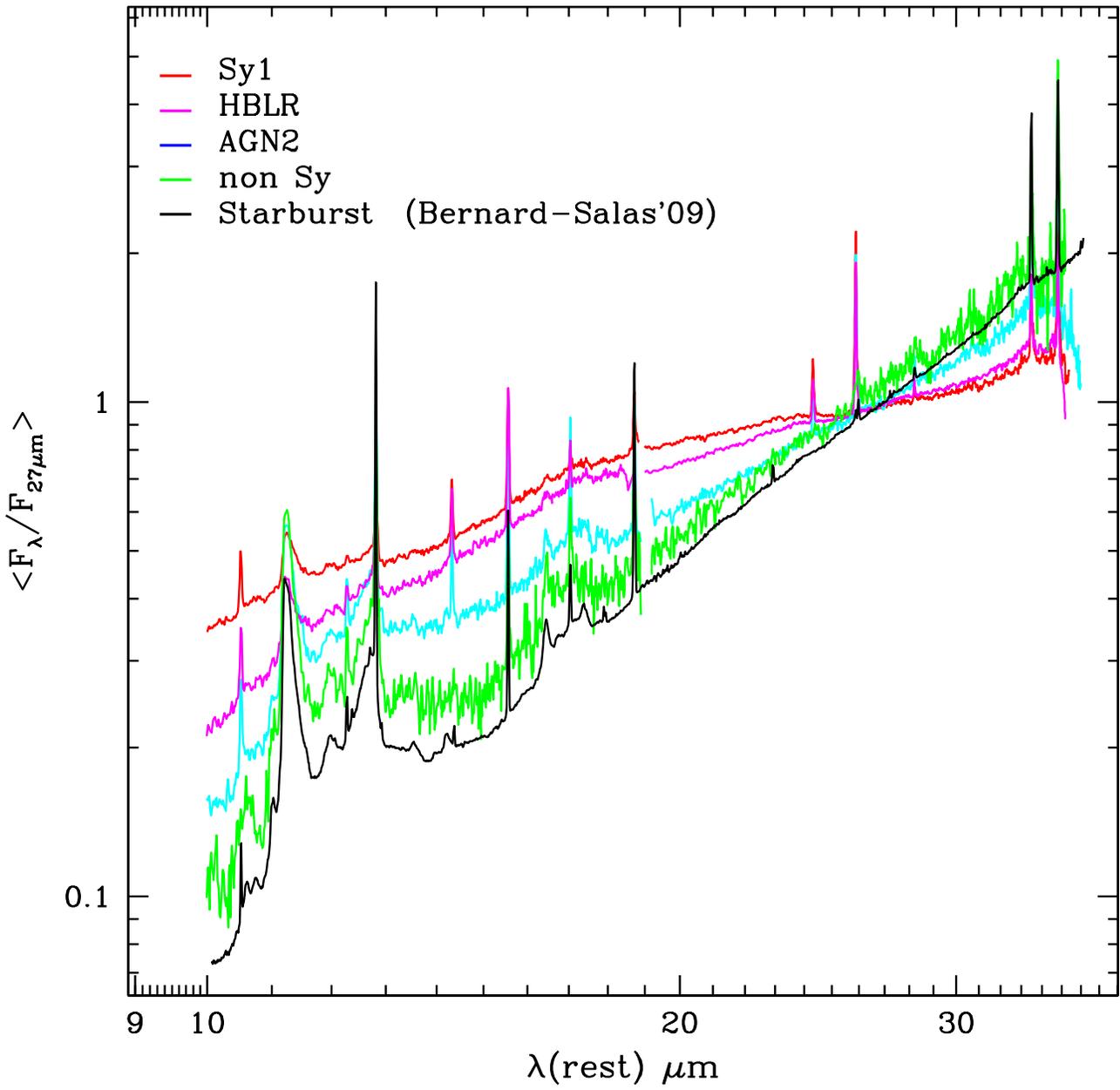}}
\caption{Average high resolution spectra for our classes of galaxies, compared with the mean high resolution spectrum of
starburst galaxies.\label{fig14}}
\end{figure}

% FIG 15 A & B
\clearpage
\begin{figure}
\centerline{\includegraphics[width=9cm]{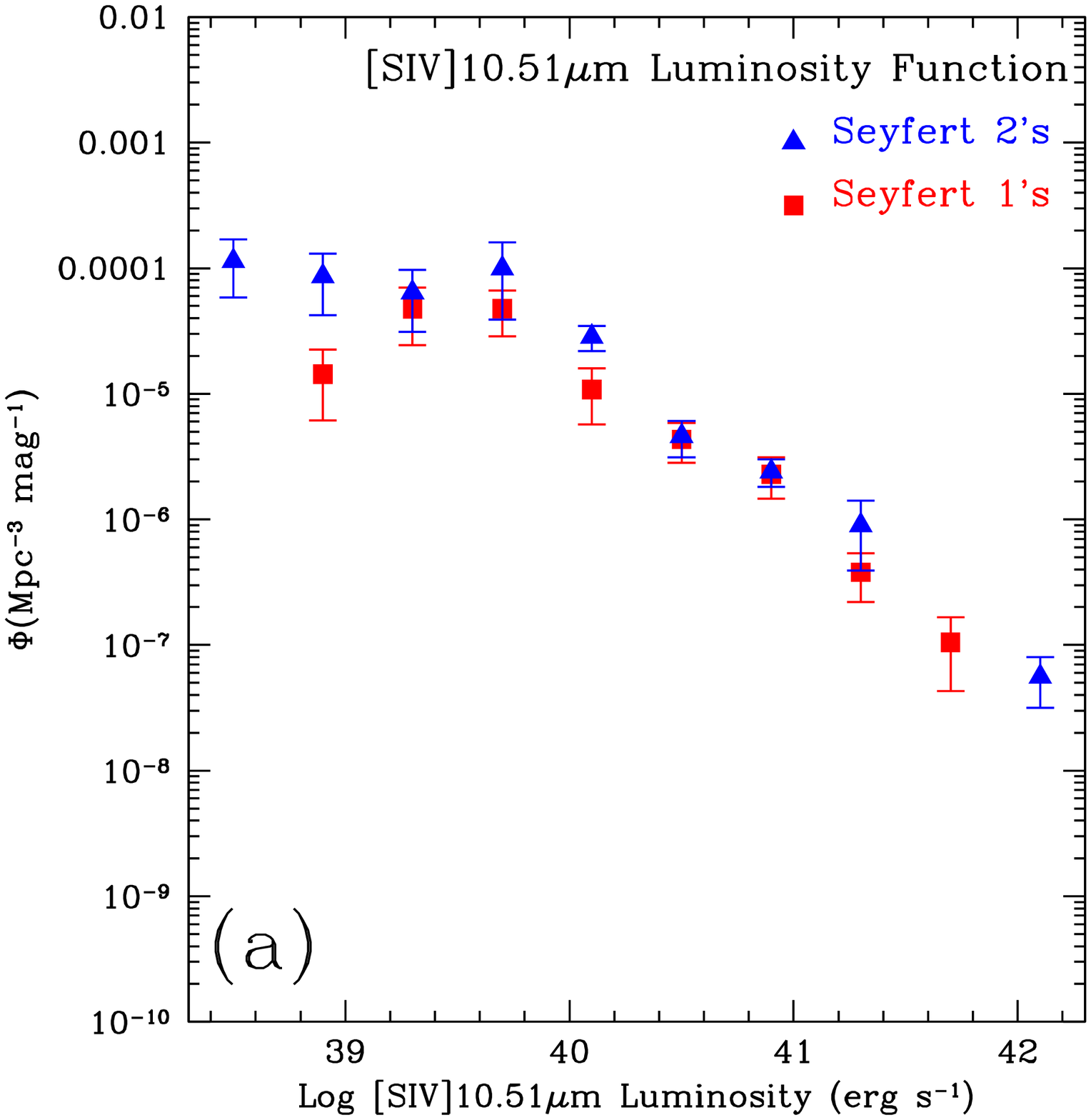}\includegraphics[width=9cm]{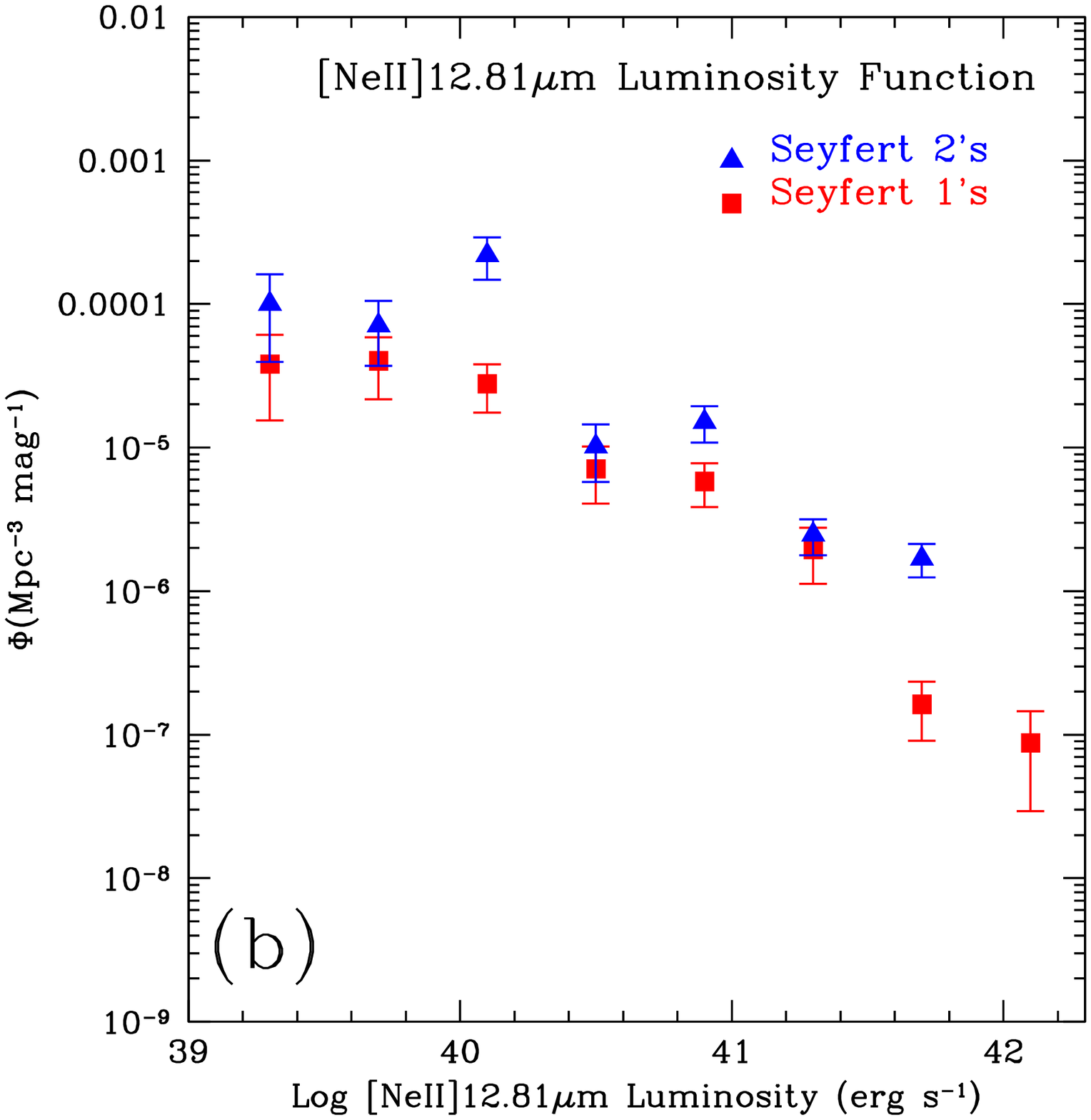}}
\caption{\textbf{a:} Line luminosity function of the [SIV]10.51$\mu m$, the red squares represent the Seyfert 1's and the blue triangles the Seyfert 2's.
\textbf{b:} Line luminosity function of the [NeII]12.81$\mu m$, symbols are as in the previous figure.\label{fig15}}
\end{figure}

% FIG 16 A & B
\begin{figure}
\centerline{\includegraphics[width=9cm]{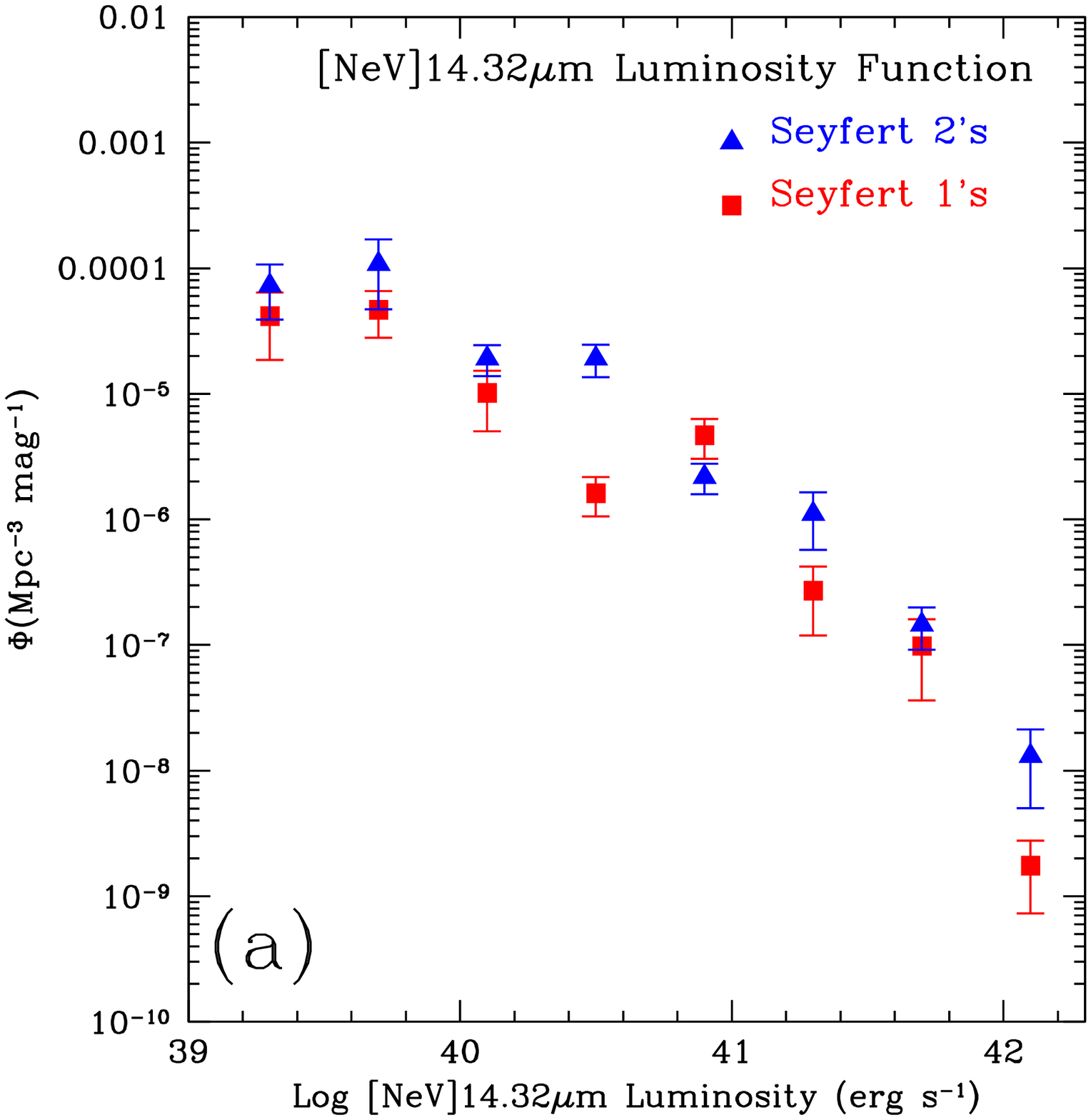}\includegraphics[width=9cm]{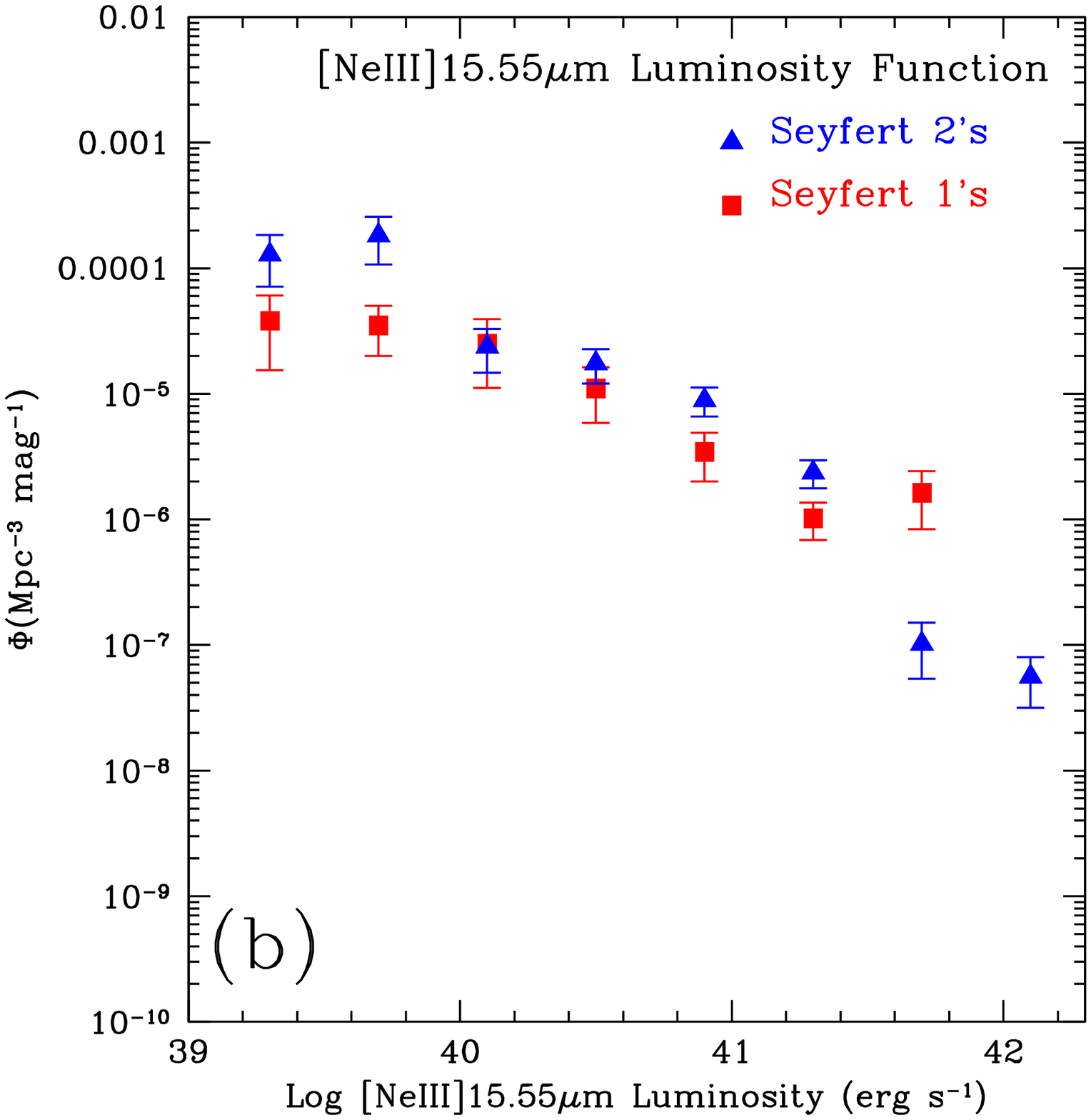}}
\caption{\textbf{a:} Line luminosity function of the [NeV]14.32$\mu m$, symbols are as in the previous figure.
\textbf{b:} Line luminosity function of the [NeIII]15.55$\mu m$, symbols are as in the previous figure. \label{fig16}} 
\end{figure}
\clearpage

% FIG 17 A & B
\begin{figure}
\centerline{\includegraphics[width=9cm]{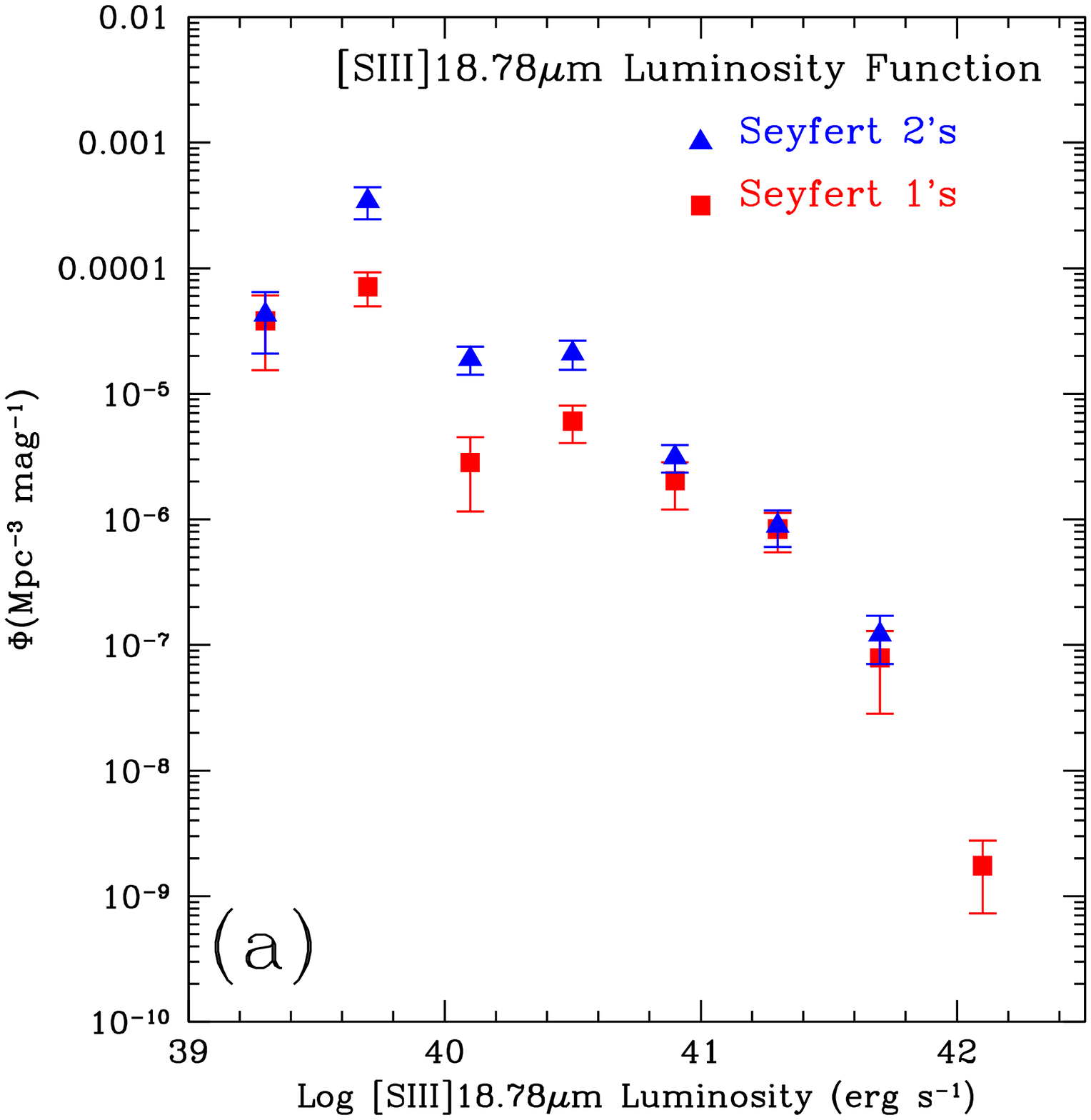}\includegraphics[width=9cm]{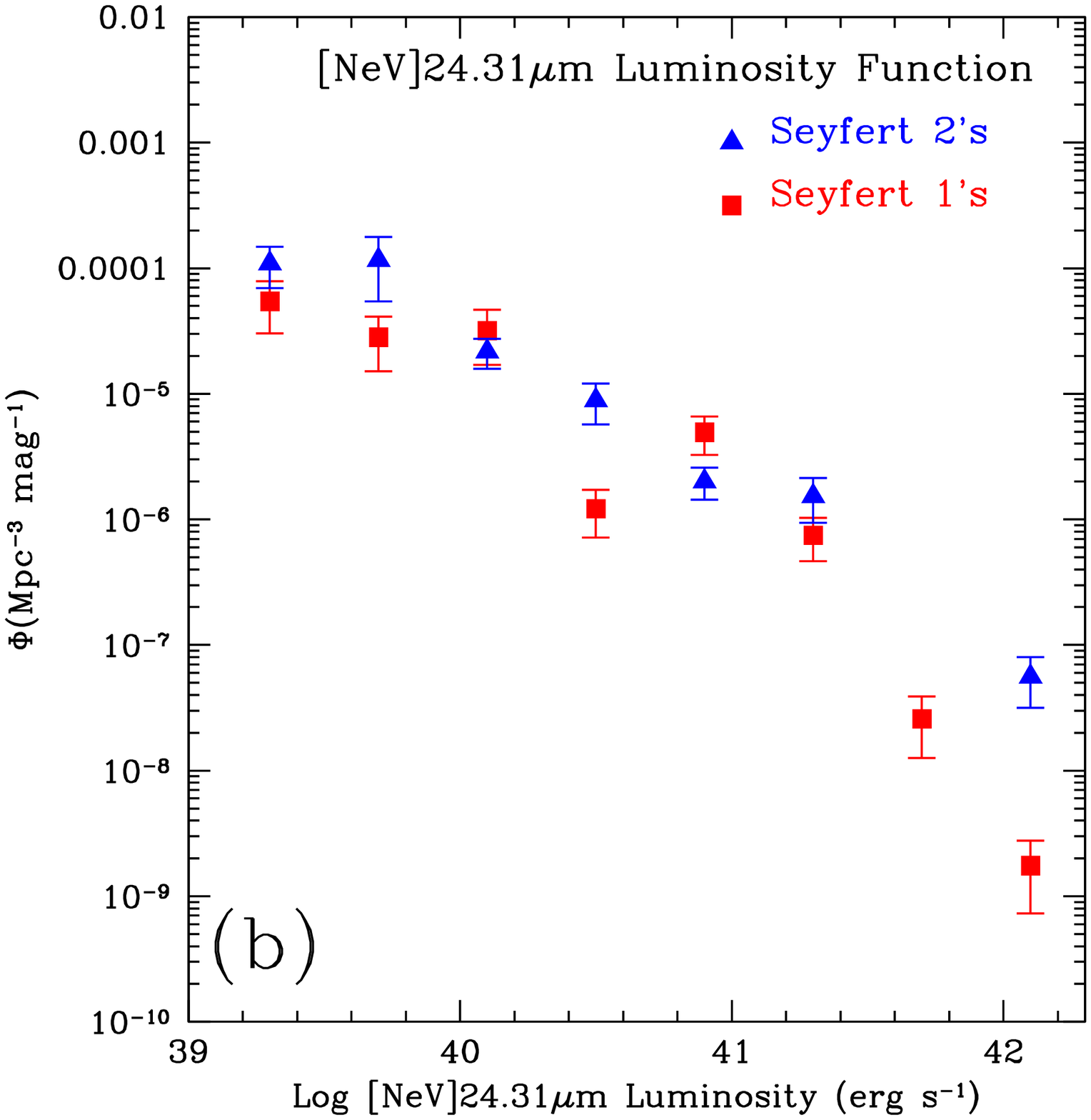}}
\caption{\textbf{a:} Line luminosity function of the [SIII]18.71$\mu m$, symbols are as in the previous figure.
\textbf{b:} Line luminosity function of the [NeV]24.31$\mu m$, symbols are as in the previous figure.\label{fig17}}
\end{figure}

% FIG 18 A & B
\begin{figure}
\centerline{\includegraphics[width=9cm]{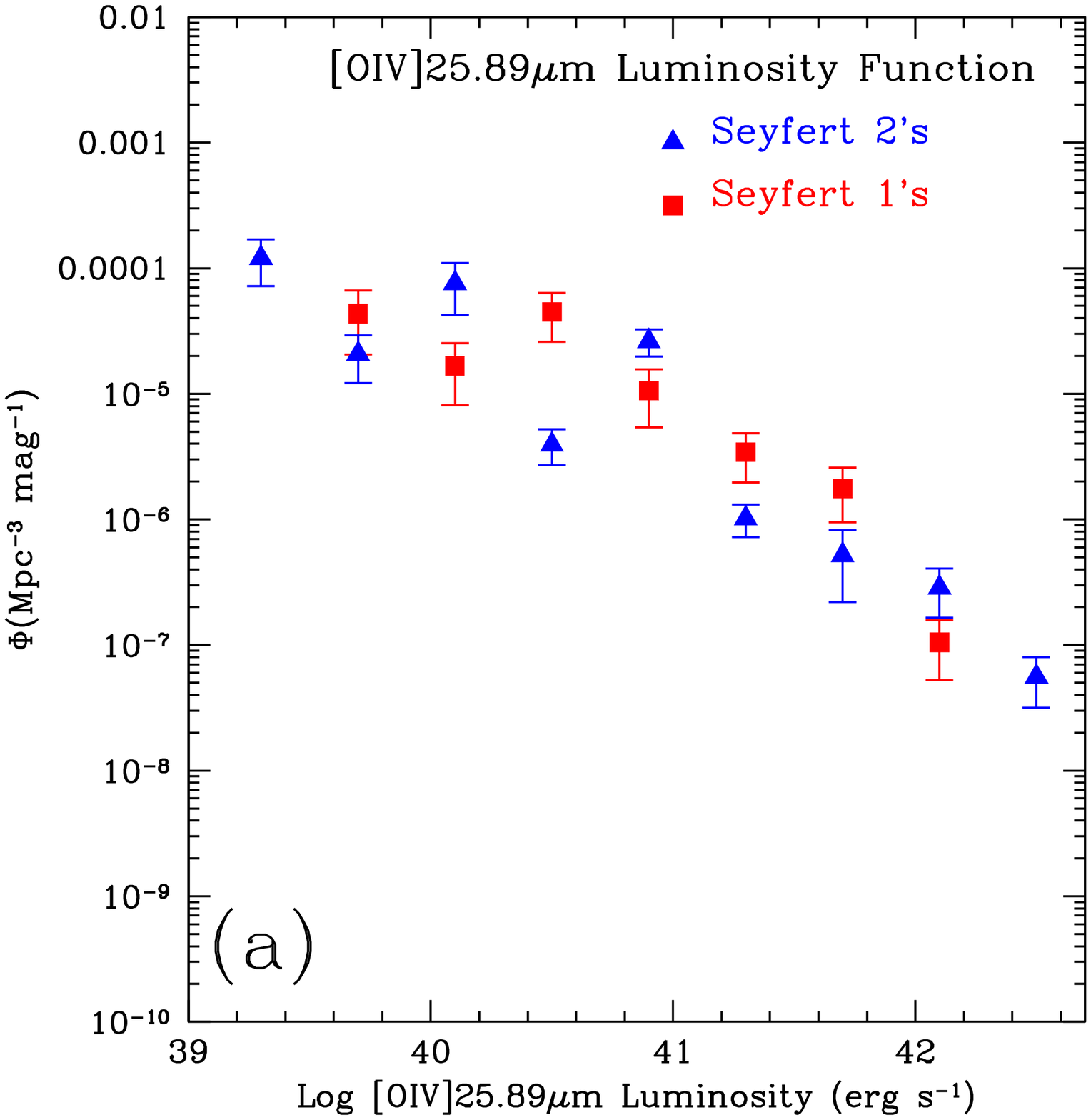}\includegraphics[width=9cm]{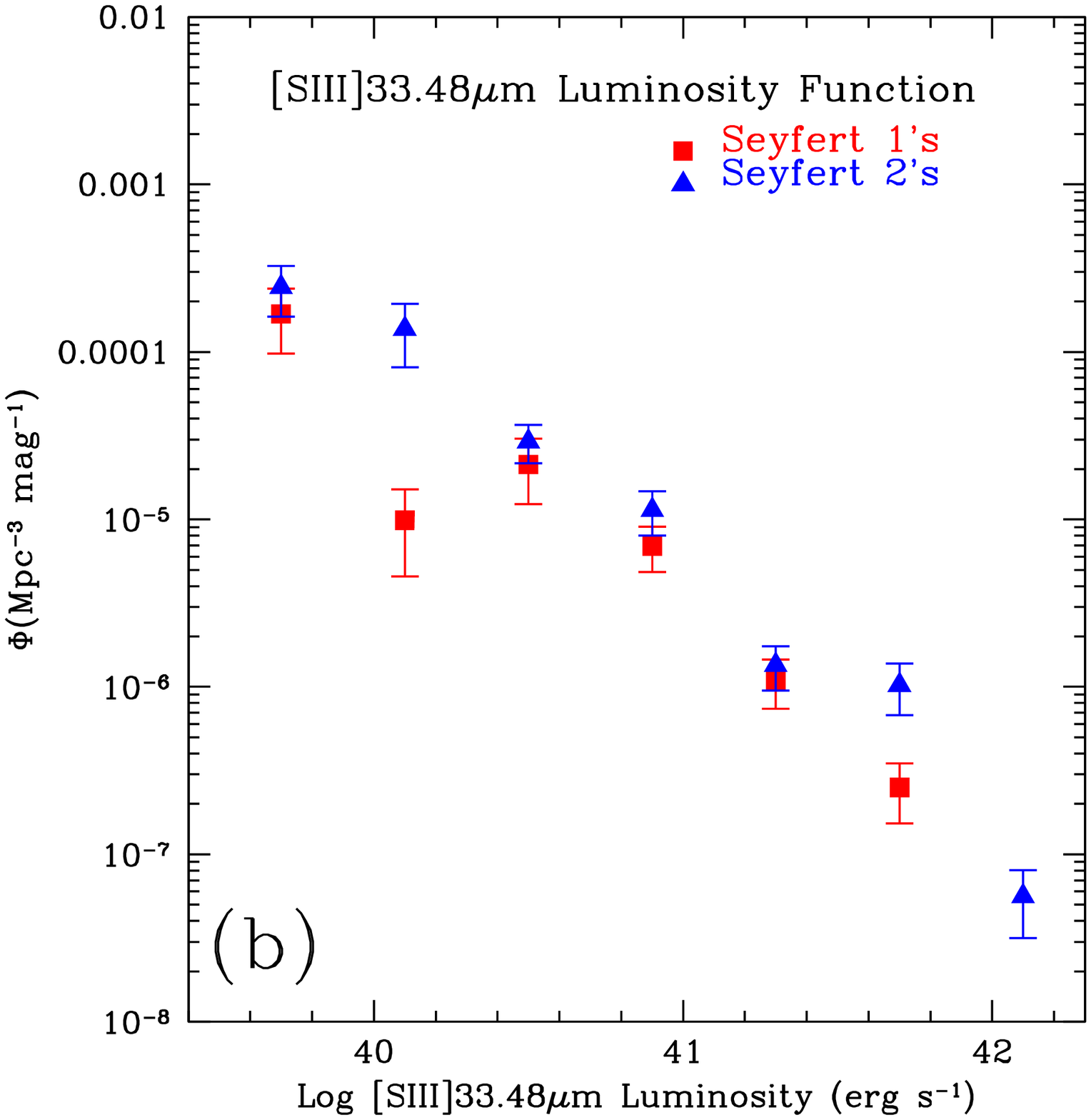}}
\caption{\textbf{a:} Line luminosity function of the [OIV]25.89$\mu m$, symbols are as in the previous figure.
\textbf{b:} Line luminosity function of the [SIII]33.48$\mu m$, symbols are as in the previous figure.\label{fig18}}
\end{figure}
\clearpage

% FIG 19
\begin{figure}
\centerline{\includegraphics[width=9cm]{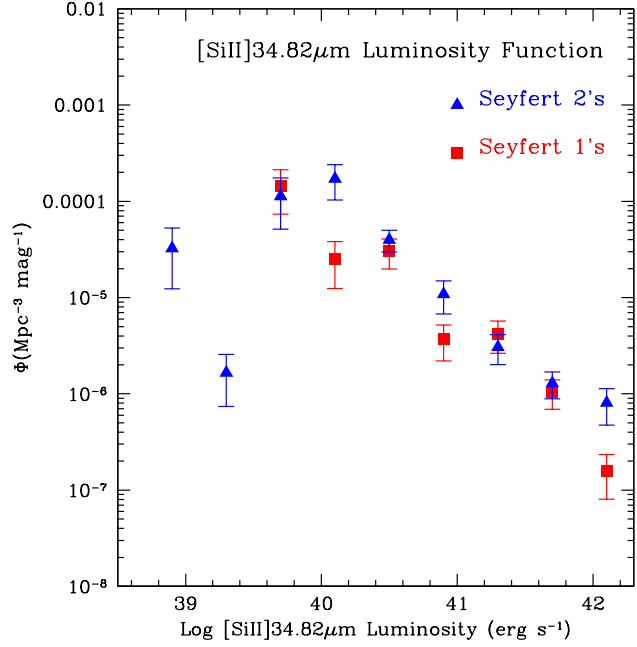}}
\caption{Line luminosity function of the [SiII]34.82$mu m$, symbols are as in the previous figure.\label{fig19}}
\end{figure}

% FIG 20 A & B
\begin{figure}
\centerline{\includegraphics[width=9cm]{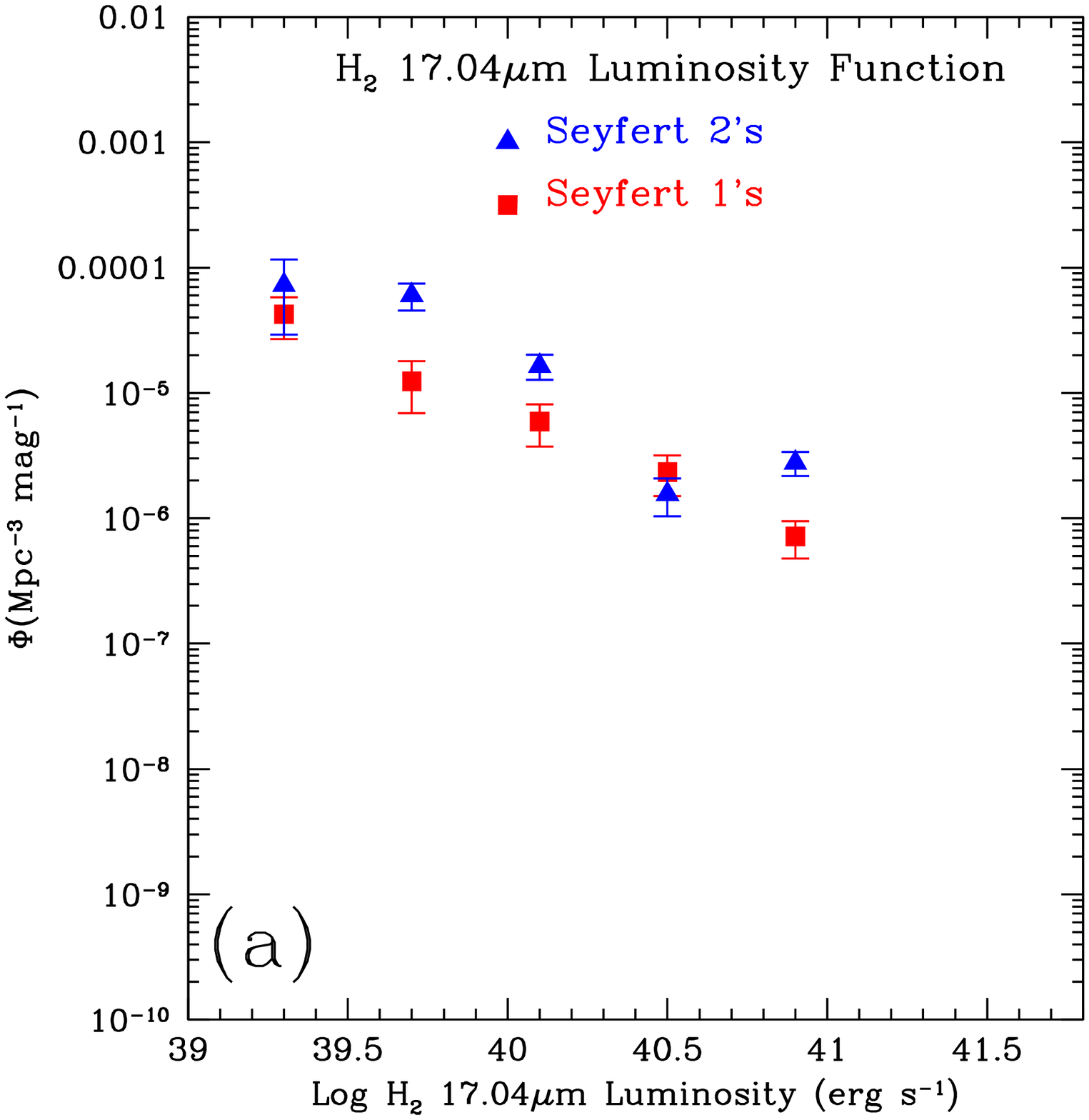}\includegraphics[width=9cm]{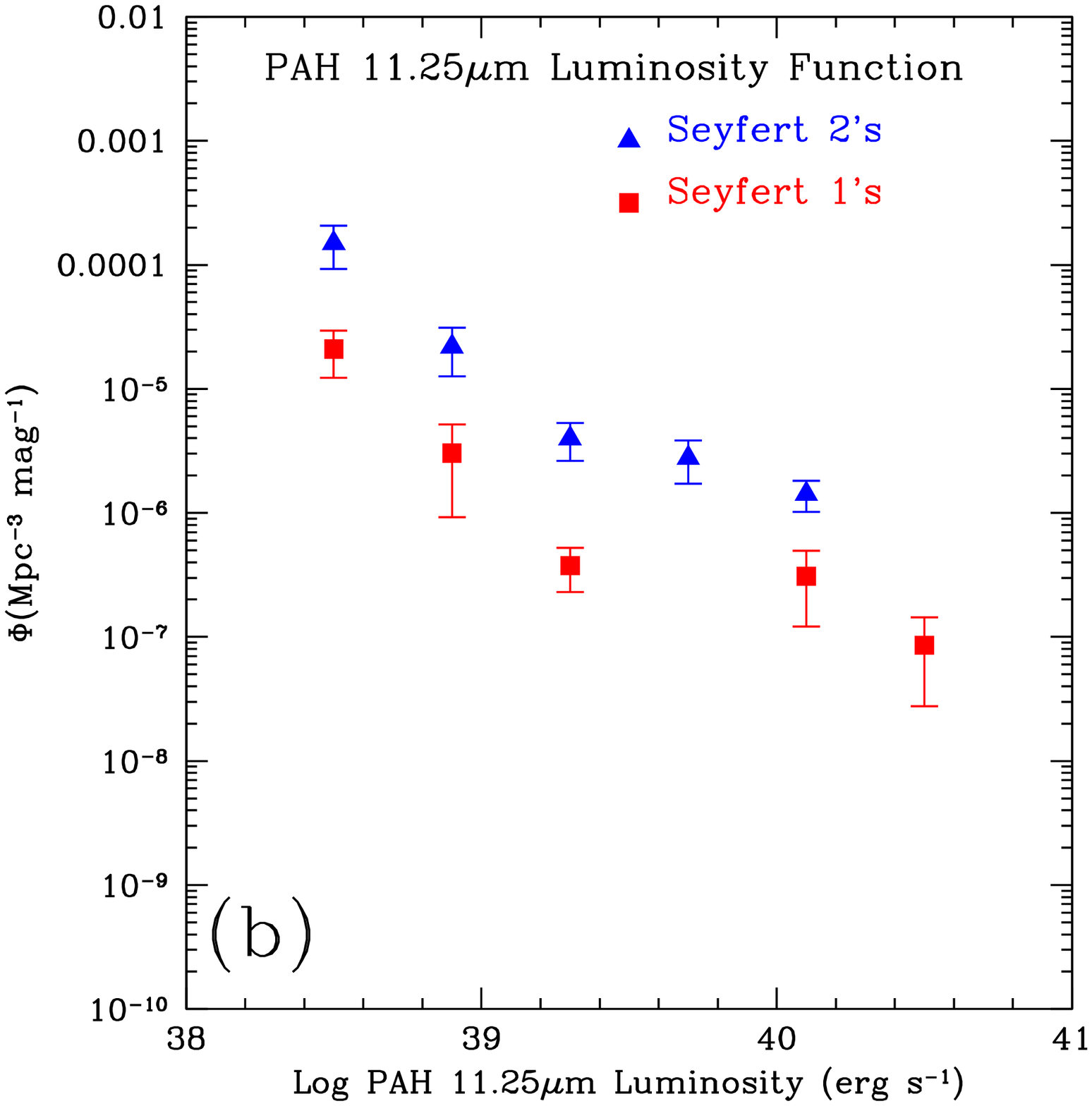}}
\caption{\textbf{a:} Line luminosity function of the H$_2$ 17.04$\mu m$, symbols are as in the previous figure.
\textbf{b:} Line luminosity function of the PAH 11.25$\mu m$, symbols are as in the previous figure.\label{fig20}}
\end{figure}
\clearpage

% FIG 21 A & B
\begin{figure}
\centerline{\includegraphics[width=9cm]{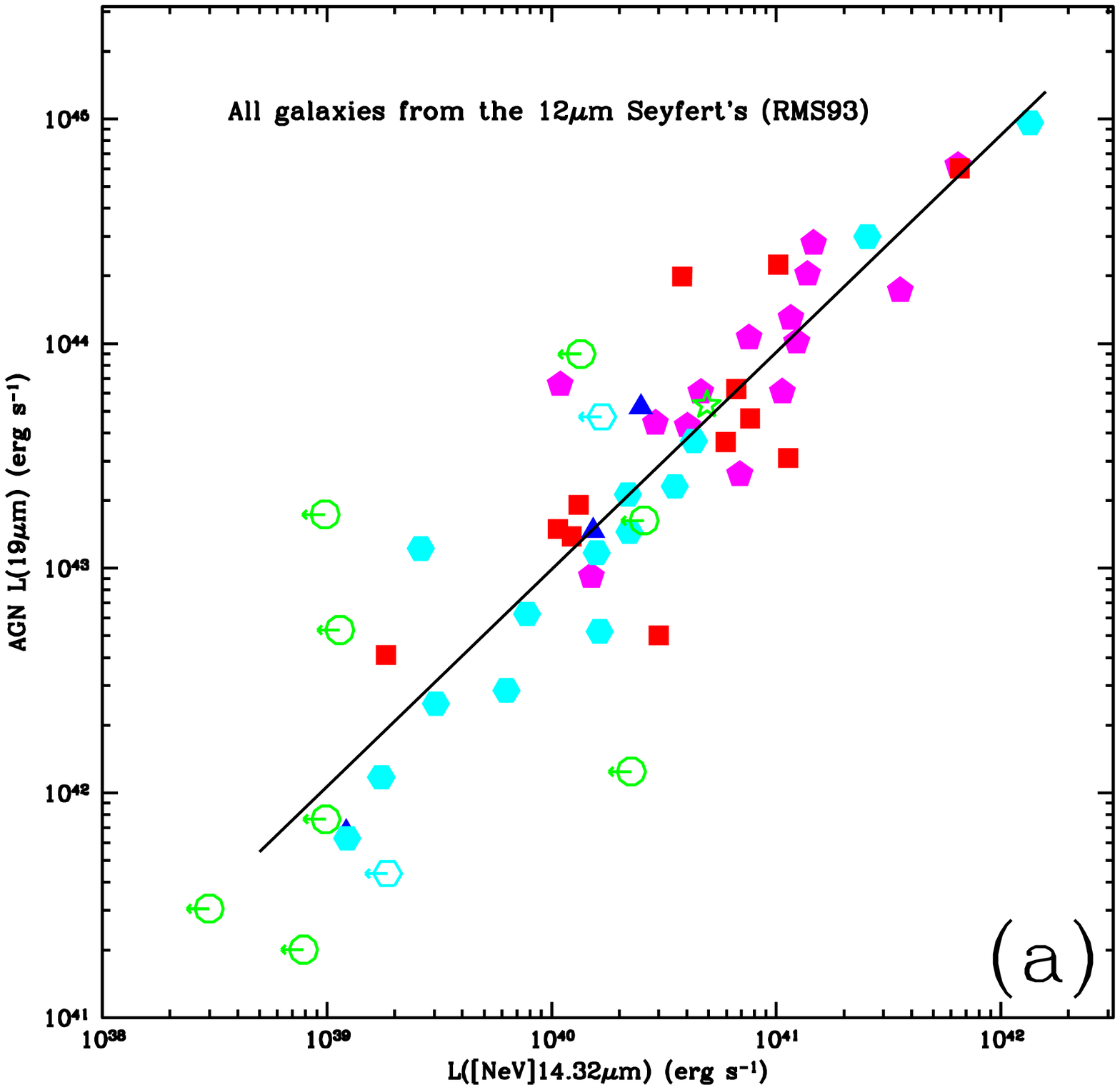}\includegraphics[width=9cm]{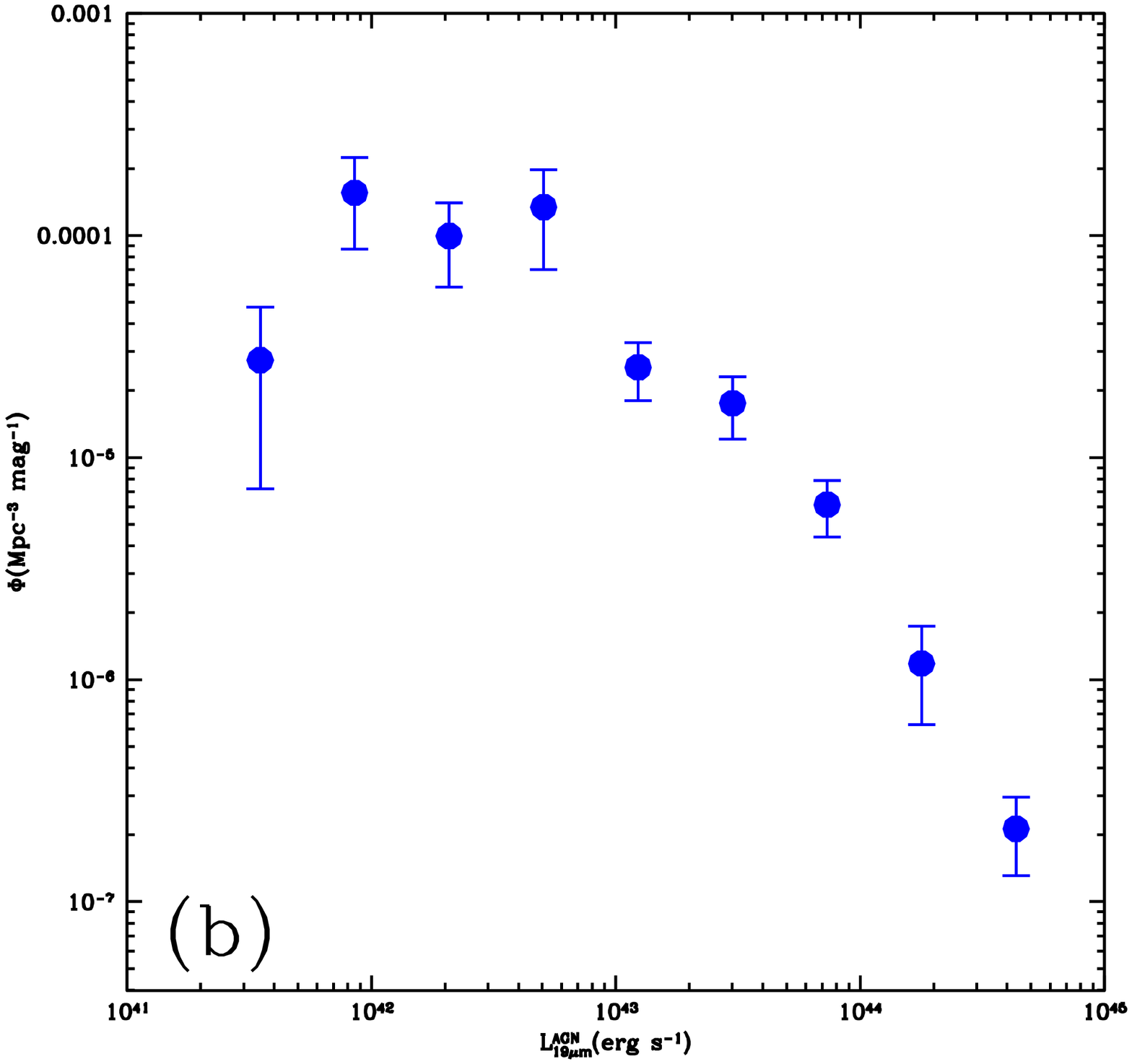}}
\caption{\textbf{a:} [NeV]14.3$\mu$m vs AGN 19$\mu$m luminosity, symbols are as in the previous figure.
\textbf{b:} AGN 19$\mu$m luminosity function.\label{fig21}}
\end{figure}

% FIG 22
\begin{figure}
\centerline{\includegraphics[width=9cm]{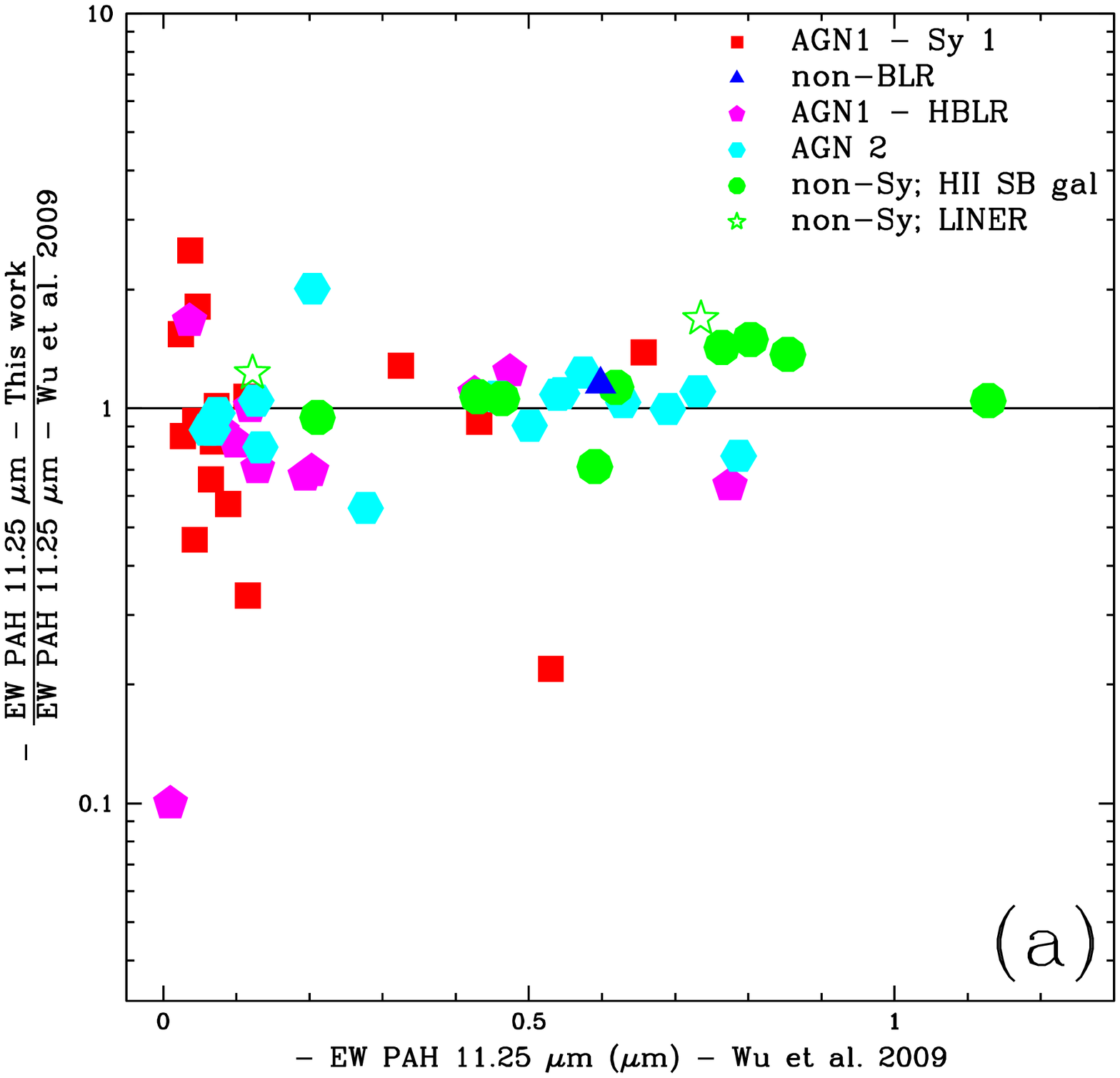}\includegraphics[width=9cm]{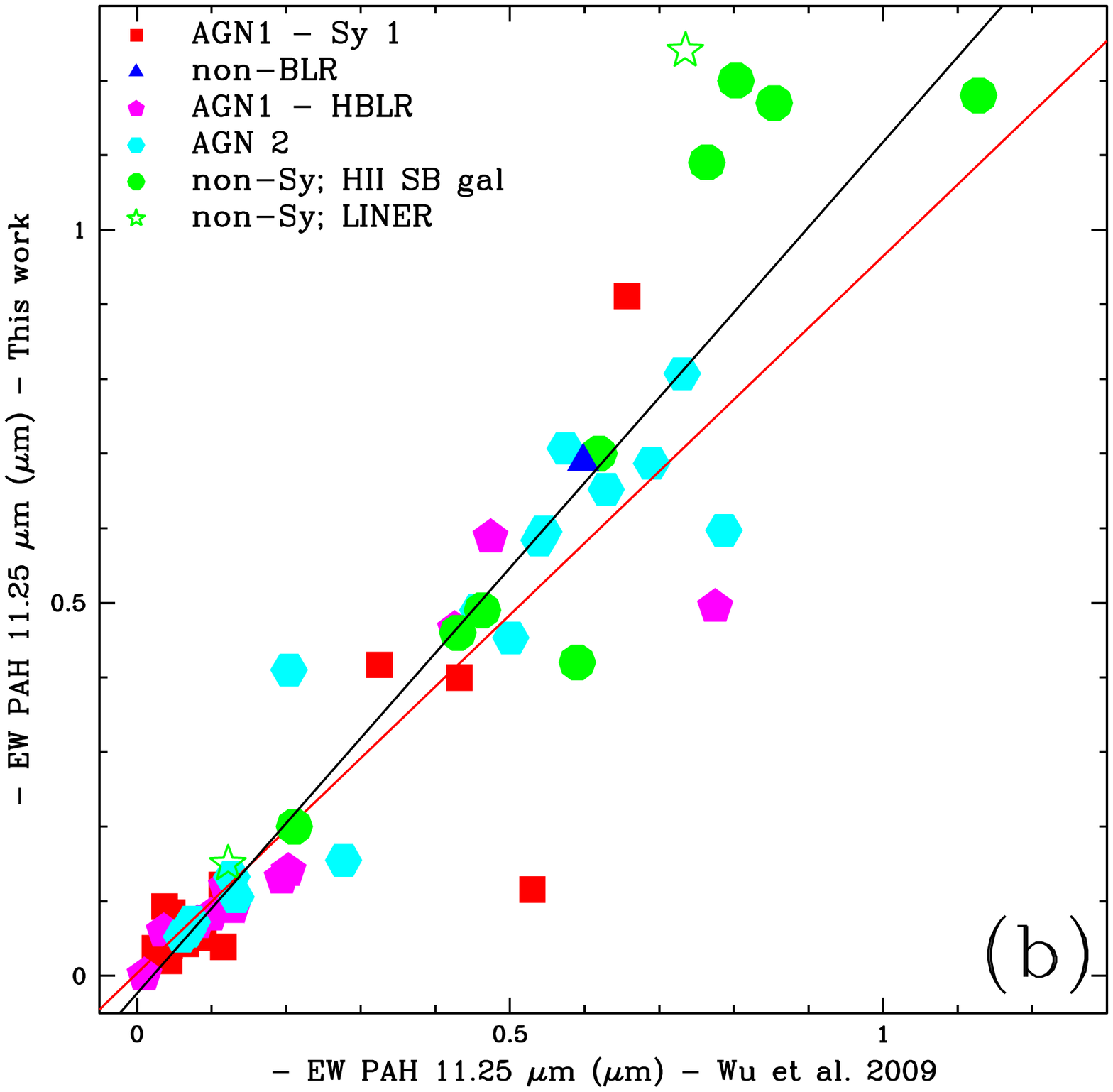}}
\caption{\textbf{a:} The ratio of the PAH 11.25$\mu m$ EW calculated by \citet{wu09} to ours as a function of
the values by \citet{wu09}.
\textbf{b:} PAH 11.25$\mu m$ EW calculated by \citet{wu09} versus ours. 
We computed a mean square fit among the AGN1 + AGN2 class (values are fitted by the red line) 
and all the galaxy types (black line).
For the slope values and confidence intervals, see the text.\label{fig22}}
\end{figure}
\end{document}